\documentclass{article}
\usepackage{amssymb}
\usepackage[linkcolor=black,colorlinks=true,urlcolor=]{hyperref}
\usepackage{mathtools}

\usepackage{color,xcolor,ucs}
\usepackage{mathtools}   \usepackage{tikz} 
\usepackage{ amssymb }
\usepackage{extarrows} 
\usepackage{natbib}
\usepackage{pgf,tikz}
\usepackage{float}
\usetikzlibrary{positioning}
\usetikzlibrary{shapes.geometric}
\usetikzlibrary{shapes.misc}
\usetikzlibrary{arrows}
\usepackage{caption}
\usepackage{mathrsfs}
\usetikzlibrary{arrows,shapes,automata,backgrounds,petri,positioning}
\usetikzlibrary{decorations.pathmorphing}
\usetikzlibrary{decorations.shapes}
\usetikzlibrary{decorations.text}
\usetikzlibrary{decorations.fractals}
\usetikzlibrary{decorations.footprints}
\usetikzlibrary{shadows}
\usetikzlibrary{calc}
\usetikzlibrary{spy}
\usepackage{amsmath}
\usepackage{array}
\usepackage{ amssymb }
\usepackage{braket}
\usepackage{qcircuit}
\usepackage{soul}
\usepackage{braket} 
\usepackage{relsize}

\usepackage{amsmath}
\usepackage{ amssymb }
\usepackage{braket}
\usepackage{qcircuit}
\usepackage{soul}
\usepackage{braket}
\usepackage{amsmath}

\title{{\LARGE Open boundary conditions of the $D^{(2)}_3$ spin chain and sectors of conformal field theories}}
\author{Pete Rigas \footnote{Newport Beach, CA, 92625, pbr43@cornell.edu}}
\date{}

\begin{document}

\noindent

\bigskip

\maketitle

\noindent \textbf{Abstract} We study open boundary conditions for the $D^{(2)}_3$ spin chain, which shares connections with the six-vertex model, under staggering, and also to the antiferromagnetic Potts model. By formulating a suitable transfer matrix that is related to $K$ matrices and to the Jimbo $R$-matrix, we obtain an analytical expression for the Hamiltonian, as a logarithmic derivative of the transfer matrix, under open boundary conditions. Such computations have connections with several objects studied in Integrable Probability, which underlie exactly solvable structures. \footnote{\textbf{MSC Class}: 34L25; 60K35}

\bigskip

\noindent \textbf{Keywords}. spin chain, open boundary conditions, CFT, conformal field theory sectors,  ground state, local Hamiltonian

\section{Introduction}

\subsection{Overview}

\noindent Spin chains have long been objects of study across the fields of quantum physics, high-energy physics, and statistical physics, for connections to computations of finite-size spectra (\citet{frahm2023}), staggered vertex models (\citet{Frahm2022}), integrable boundary conditions (\citet{frahm2022}), quantum R-matrices (\citet{jimbo1986}), the Bethe ansatz (\citet{nepomechie2021}), integrability, either through being able to completely solve spin chain models, or through boundary conditions, ({\citet{nepomechie2019}}, {\citet{Robertson2020}}), and conformal invariance (\cite{Robertson2019}). To further explore avenues of interest at the intersection of all of these fields, in the following we study the $D^{(2)}_2$, and $D^{(2)}_3$ spin chains. Despite having different rank, each of the two spin chains share similarities, not only from the fact that R-matrices can be constructed which satisfy the Yang Baxter equation, but also from the fact that open boundary conditions can be encoded from K-matrices satisfying variants of the Yang Baxter equation at the leftmost and rightmost endpoints of a finite volume. To determine which sections of the underlying conformal field theory (CFT) are selected depending upon the encoding of open boundary conditions, we introduce the lower, and higher, rank spin chains, from which we distinguish different sectors of the CFT depend on open boundary conditions. From an expansion of the Hamiltonian into a local Hamiltonian, we characterize the ground state with open boundary conditions about the point $\big( h_1 , h_2 \big) \equiv \big( 0 , 0 \big)$, and proceed to characterize other sectors of the CFT for $h_1 \equiv 0$, $h_2 \neq 0$, and for $h_1 \neq 0$, $h_2 \equiv 0$.

\bigskip

The root density approach has previous been applied for investigating several aspects of systems in the high-temperature limit, in which one often studies more general aspects of low temperature expansions of probability measures in the presence of different contributions to interactions term of the Hamiltonian. In the presence of different interactions, the root density approach for the Bethe equations implies that there is a change in the density of the roots of the Bethe equations, which not only has connections with conformal field theory, but also with interpretations surrounding excitations to the ground state.

\subsection{This paper's contributions}

\noindent We describe the contributions of this work below.

\subsubsection{Modifications to K-matrices for encoding open boundary conditions}

In comparison to K-matrices that have been obtained under quasi-periodic boundary conditions (\citet{frahm2023}), K-matrices for the higher rank generalization of the $D^{(2)}_2$ spin-chain, through the $D^{(2)}_3$ spin-chain, are of dimensions $36 \times 36$, instead of $6 \times 6$. Such classes of K-matrices take the form,

\[
K^{\mathrm{Open}}_{D^{(2)}_3,-} \equiv K^{\mathrm{Open}} \equiv   \begin{bmatrix}          \textit{36} \times \textit{36 finite dimensional representation of open boundary conditions} \\  \textit{on horizontal lines of } \textbf{Z}^2
\end{bmatrix}
\]

\noindent and,

\[
K^{\mathrm{Open}}_{D^{(2)}_3,+} \equiv K^{\mathrm{Open}} \equiv   \begin{bmatrix}          \textit{36} \times \textit{36 finite dimensional representation of open boundary conditions} \\  \textit{on vertical lines of } \textbf{Z}^2
\end{bmatrix}
\]

\subsubsection{Formulating the open-boundary transfer matrix in the higher-rank spin-chain}. Given the encoding of open-boundary conditions for the higher-rank spin chain introduced in the previous item, the transfer matrix asymptotically takes the form,

\begin{align*}
  \underset{N \longrightarrow + \infty}{\mathrm{lim}}  \textbf{T}_{N, D^{(2)}_3} \big( u \big) \equiv \underset{N \longrightarrow + \infty}{\mathrm{lim}} \bigg\{  \mathrm{Tr} \bigg[ K^{\mathrm{Open}}_{D^{(2)}_3} \bigg[ \underset{1 \leq j \leq N}{\prod} R_{D^{(2)}_3,j} \bigg]   \bigg]     \bigg\}  \equiv \underset{j,j^{\prime} \longrightarrow + \infty}{\mathrm{lim}} \bigg\{   \mathrm{Tr}_0 \bigg[ \mathcal{K}_{+,0} \big( u \big)  \\ \times   \bigg[ \prod_{1 \leq j \leq L} R_{0j} \big( u \big) \bigg]  \mathcal{K}_{-,0} \big( u \big) \bigg[   \prod_{1 \leq j^{\prime} \leq L} R_{j^{\prime}0} \big( u \big) \bigg]  \bigg] \bigg\}  , 
\end{align*}

\noindent for the $D^{(2)}_3$ R-matrix, $R_{D^{(2)}_3,j}$.

\subsection{Expressing conditions relating to symmetries of the higher-rank spin chain}

\noindent In the lower rank spin-chain, as recapitulated in the next section, $\mathrm{U} \big(1 \big)$ symmetry is captured through the family of relations,

\begin{align*}
        \big[ R \big( u \big) ,      \textbf{h}_j    \otimes \textbf{I} + \textbf{I} \otimes   \textbf{h}_j  \big]   \equiv 0            \text{, } 
\end{align*}

\noindent taken with respect to the Lie bracket. A family of symmetries captured through the relations,

\begin{align*}
        \big[ R_{D^{(2)}_3,j} \big( u \big) ,       K^{\mathrm{Open}}_{D^{(2)}_3}  \otimes K^{\mathrm{Open}}_{D^{(2)}_3}         \big]   \equiv 0          .
\end{align*}

\noindent As a higher-dimensional class of symmetries, the K-matrices from the $D^{(2)}_3$ spin-chain, in comparison to those for the quasi-periodic $D^{(2)}_3$ spin-chain, are dependent upon the following matrix basis,

\begin{align*}
\textit{K-matrix} \propto  \mathrm{span} \big\{ \textit{Diagonal matrices} , \textit{Antidiagonal matrices}, \mathcal{T}_1 , \mathcal{T}_2 \big\}  , 
\end{align*}

\noindent where, besides the diagonal and antidiagonal matrices in the spanning set, denote $\mathcal{T}$ with,

\begin{align*}
  \mathcal{T} \equiv \underset{1 \leq i \leq 2}{\bigcup} \mathcal{T}_i \equiv  \underset{1 \leq i \leq 2}{\bigcup}  \big\{ \textit{36 } \times \textit{36 dimensional matrices with a strictly positive  number,} \\ \textit{, of entries along the antidiagonal} \big\}  .
\end{align*}

{\tiny
\[
\begin{array}{cccccccccccccccccccccccccccccccccccc}
1 & 0 & 0 & 0 & 0 & 0 & 0 & 0 & 0 & 0 & 0 & 0 & 0 & 0 & 0 & 0 & 0 & 0 & 0 & 0 & 0 & 0 & 0 & 0 & 0 & \dots  \\
0 & 2 & 0 & 0 & 0 & 0 & 0 & 0 & 0 & 0 & 0 & 0 & 0 & 0 & 0 & 0 & 0 & 0 & 0 & 0 & 0 & 0 & 0 & 0 & 0 & \dots    \\
0 & 0 & 3 & 0 & 0 & 0 & 0 & 0 & 0 & 0 & 0 & 0 & 0 & 0 & 0 & 0 & 0 & 0 & 0 & 0 & 0 &  0 & 0 & 0 & 0 &  \dots  \\
0 & 0 & 0 & 4 & 0 & 0 & 0 & 0 & 0 & 0 & 0 & 0 & 0 & 0 & 0 & 0 & 0 & 0 & 0 & 0 & 0 & 0 & 0 & 0 & 0 &  \dots  \\
0 & 0 & 0 & 0 & 5 & 0 & 0 & 0 & 0 & 0 & 0 & 0 & 0 & 0 & 0 & 0 & 0 & 0 & 0 & 0 & 0 & 0 & 0 & 0 & 0 &  \dots   \\
0 & 0 & 0 & 0 & 0 & 6 & 0 & 0 & 0 & 0 & 0 & 0 & 0 & 0 & 0 & 0 & 0 & 0 & 0 & 0 & 0 & 0 & 0 & 0 & 0 &  \dots  \\ 0 & 0 & 0 & 0 & 0 & 0 & 7 & 0 & 0 & 0 & 0 & 0 & 0 & 0 & 0 & 0 & 0 & 0 & 0 & 0 & 0 & 0 & 0 & 0 & 0 &  \dots  \\
0 & 0 & 0 & 0 & 0 & 0 & 0 & 8 & 0 & 0 & 0 & 0 & 0 & 0 & 0 & 0 & 0 & 0 & 0 & 0 & 0 & 0 & 0 & 0 & 0 &   \dots   \\
\end{array}
\]}

{\tiny  \[
\begin{array}{cccccccccccccccccccccccccccccccccccc} 
\dots & 0 & 0 & 0 & 0 & 0 & 0 & 0 & 0 & 0 & 0 & 0 \\ \dots & 0 & 0 & 0 & 0 & 0 & 0 & 0 & 0 & 0 & 0 & 0  \\ \dots & 0 & 0 & 0 & 0 & 0 & 0 & 0 & 0 & 0 & 0 & 0  \\ \dots & 0 & 0 & 0 & 0 & 0 & 0 & 0 & 0 & 0 & 0 & 0  \\ \dots & 0 & 0 & 0 & 0 & 0 & 0 & 0 & 0 & 0 & 0 & 0  \\ \dots & 0 & 0 & 0 & 0 & 0 & 0 & 0 & 0 & 0 & 0 & 0  \\ \dots & 0 & 0 & 0 & 0 & 0 & 0 & 0 & 0 & 0 & 0 & 0  \\  \dots & 0 & 0 & 0 & 0 & 0 & 0 & 0 & 0 & 0 & 0 & 0  \\ \dots & 0 & 0 & 0 & 0 & 0 & 0 & 0 & 0 & 0 & 0 & 0  \\ \dots & 0 & 0 & 0 & 0 & 0 & 0 & 0 & 0 & 0 & 0 & 0 \\ \dots & 0 & 0 & 0 & 0 & 0 & 0 & 0 & 0 & 0 & 0 & 0  \\ \dots & 0 & 0 & 0 & 0 & 0 & 0 & 0 & 0 & 0 & 0 & 0  \\ \dots & 0 & 0 & 0 & 0 & 0 & 0 & 0 & 0 & 0 & 0 & 0  \\ \dots & 0 & 0 & 0 & 0 & 0 & 0 & 0 & 0 & 0 & 0 & 0  \\ \dots & 0 & 0 & 0 & 0 & 0 & 0 & 0 & 0 & 0 & 0 & 0  \\ \dots & 0 & 0 & 0 & 0 & 0 & 0 & 0 & 0 & 0 & 0 & 0  \\  \dots & 0 & 0 & 0 & 0 & 0 & 0 & 0 & 0 & 0 & 0 & 0  \\ \dots & 0 & 0 & 0 & 0 & 0 & 0 & 0 & 0 & 0 & 0 & 0   \\ \dots & 0 & 0 & 0 & 0 & 0 & 0 & 0 & 0 & 0 & 0 & 0 \\ \dots & 0 & 0 & 0 & 0 & 0 & 0 & 0 & 0 & 0 & 0 & 0  \\ \dots & 0 & 0 & 0 & 0 & 0 & 0 & 0 & 0 & 0 & 0 & 0  \\ \dots & 0 & 0 & 0 & 0 & 0 & 0 & 0 & 0 & 0 & 0 & 0  \\  \dots  & \vdots  & \vdots  & \vdots  & \vdots  & \vdots  & \vdots  & \vdots  & \vdots  &  \vdots  & \vdots  & \vdots \\      \dots & 0 & 0 & 0 & 0 & 0 & 0 & 0 & 0 & 0 & 0 &  36       
\end{array} , 
\]
}

{\tiny
\[
\begin{array}{cccccccccccccccccccccccccccccccccccc}
 0 & 0 & 0 & 0 & 0 & 0 & 0 & 0 & 0 & 0 & 0 & 0 & 0 & 0 & 0 & 0 & 0 & 0 & 0 & 0 & 0 & 0 &  \dots  \\
 0 & 0 & 0 & 0 & 0 & 0 & 0 & 0 & 0 & 0 & 0 & 0 & 0 & 0 & 0 & 0 & 0 & 0 & 0 & 0 & 0 & 0 &  \dots    \\
 0 & 0 & 0 & 0 & 0 & 0 & 0 & 0 & 0 & 0 & 0 & 0 & 0 & 0 & 0 & 0 & 0 & 0 & 0 & 0 &  0 & 0 &   \dots  \\
 0 & 0 & 0 & 0 & 0 & 0 & 0 & 0 & 0 & 0 & 0 & 0 & 0 & 0 & 0 & 0 & 0 & 0 & 0 & 0 & 0 & 0 &   \dots  \\
 0 & 0 & 0 & 0 & 0 & 0 & 0 & 0 & 0 & 0 & 0 & 0 & 0 & 0 & 0 & 0 & 0 & 0 & 0 & 0 & 0 & 0 &   \dots   \\
0 & 0 & 0 & 0 & 0 & 0 & 0 & 0 & 0 & 0 & 0 & 0 & 0 & 0 & 0 & 0 & 0 & 0 & 0 & 0 & 0 & 0  &   \dots  \\
0 & 0 & 0 & 0 & 0 & 0 & 0 & 0 & 0 & 0 & 0 & 0 & 0 & 0 & 0 & 0 & 0 & 0 & 0 & 0  &  0 & 0 &  \dots   \\
 0 & 0 & 0 & 0 & 0 & 0 & 0 & 0 & 0 & 0 & 0 & 0 & 0 & 0 & 0 & 0 & 0 & 0 & 0 & 0  & 0 & 0 &    \dots    \\
\end{array}
\]}

{\tiny  \[
\begin{array}{cccccccccccccccccccccccccccccccccccc} 
\dots & 0 & 0 & 0 & 0 & 0 & 0 & 0 & 0 & 0 & 0 & 1^{\prime} \\ \dots & 0 & 0 & 0 & 0 & 0 & 0 & 0 & 0 & 0 & 2^{\prime} & 0  \\ \dots & 0 & 0 & 0 & 0 & 0 & 0 & 0 & 0 & 3^{\prime} & 0 & 0  \\ \dots & 0 & 0 & 0 & 0 & 0 & 0 & 0 & 4^{\prime} & 0 & 0 & 0  \\  \dots & 0 & 0 & 0 & 0 & 0 & 0 & 5^{\prime} & 0 & 0 & 0 & 0  \\ \dots & 0 & 0 & 0 & 0 & 0 & 6^{\prime} & 0 & 0 & 0 & 0 & 0  \\  \dots & 0 & 0 & 0 & 0 & 7^{\prime} & 0 & 0 & 0 & 0 & 0 & 0  \\ \dots & 0 & 0 & 0 & 8^{\prime} & 0 & 0 & 0 & 0 & 0 & 0 & 0  \\
\end{array} , 
\]
}

\newpage

\noindent corresponding to the diagonal, and antidiagonal, matrices, respectively, with entries $\big\{ \textbf{1}, \cdots, \textbf{36} \big\}$, and with $\big\{ \textbf{1}^{\prime}, \cdots, \textbf{36}^{\prime} \big\}$. The remaining elements of the matrix basis in the spanning set for the K-matrices are expressed as,

{\tiny
\[
\begin{array}{cccccccccccccccccccccccccccccccccccc}
1 & 0 & 0 & 0 & 0 & 0 & 0 & 0 & 0 & 0 & 0 & 0 & 0 & 0 & 0 & 0 & 0 & 0 & 0 & 0 & 0 & 0 & 0 & 0 & 0 & \dots  \\
0 & 2 & * & * & * & * & * & * & * & * & * & * & * & * & * & * & * & * & * & * & * & * & * & * & * & \dots    \\
0 & 0 & 3 & * & * & * & * & * & * & * & * & * & * & * & * & * & * & * & * & * & * & * & * & * & *  &  \dots  \\
0 & 0 & 0 & 4 & * & * & * & * & * & * & * & * & * & * & * & * & * & * & * & * & * & * & * & * & * &  \dots  \\
0 & 0 & 0 & 0 & 5 & * & * & * & * & * & * & * & * & * & * & * & * & * & * & * & * & * & * & * & * &  \dots   \\
0 & 0 & 0 & 0 & 0 & 6 & * & * & * & * & * & * & * & * & * & * & * & * & * & * & * & * & * & * & * &   \dots  \\ 0 & 0 & 0 & 0 & 0 & 0 & 7 & * & * & * & * & * & * & * & * & * & * & * & * & * & * & * & * & * & * &   \dots  \\
0 & 0 & 0 & 0 & 0 & 0 & 0 & 8 & * & * & * & * & * & * & * & * & * & * & * & * & * & * & * & * & * &   \dots   \\
\end{array}
\]}

{\tiny  \[
\begin{array}{cccccccccccccccccccccccccccccccccccc} 
\dots & * & * & * & * & * & * & * & * & * & * & 0 \\ \dots & * & * & * & * & * & * & * & * & * & * & 0 \\ \dots & * & * & * & * & * & * & * & * & * & * & 0 \\ \dots & * & * & * & * & * & * & * & * & * & * & 0 \\ \dots & * & * & * & * & * & * & * & * & * & * & 0  \\ \dots & * & * & * & * & * & * & * & * & * & * & 0 \\ \dots & * & * & * & * & * & * & * & * & * & * & 0  \\ \dots & * & * & * & * & * & * & * & * & * & * & 0  \\ \dots & * & * & * & * & * & * & * & * & * & * & 0 \\ \dots & * & * & * & * & * & * & * & * & * & * & 0 \\ \dots & * & * & * & * & * & * & * & * & * & * & 0  \\ \dots & * & * & * & * & * & * & * & * & * & * & 0  \\ \dots & * & * & * & * & * & * & * & * & * & * & 0 \\ \dots & * & * & * & * & * & * & * & * & * & * & 0 \\ \dots & * & * & * & * & * & * & * & * & * & * & 0  \\ \dots & * & * & * & * & * & * & * & * & * & * & 0 \\ \dots & * & * & * & * & * & * & * & * & * & * & 0  \\ \dots & * & * & * & * & * & * & * & * & * & * & 0  \\ \dots & * & * & * & * & * & * & * & * & * & * & 0\\ \dots & * & * & * & * & * & * & * & * & * & * & 0  \\ \dots & * & * & * & * & * & * & * & * & * & * & 0  \\ \dots & * & * & * & * & * & * & * & * & * & * & 0  \\\dots & * & * & * & * & * & * & * & * & * & * & 0  \\ \dots & * & * & * & * & * & * & * & * & * & * & 0 \\ \dots & * & * & * & * & * & * & * & * & * & * & 0  \\  \dots & * & * & * & * & * & * & * & * & * & * & 0 \\ \dots & * & * & * & * & * & * & * & * & * & * & 0  \\  \dots & * & * & * & * & * & * & * & * & * & * & 0  \\ \dots & * & * & * & * & * & * & * & * & * & * & 0 \\  \dots & 26 & * & * & * & * & * & * & * & * & * & 0  \\  \vdots & \vdots & \vdots  & \vdots & \vdots & \vdots &  \vdots & \vdots & \vdots  & \vdots & \vdots & \vdots \\  \dots & 0 & 0 & 0 & 0 & 0 & 0 & 0 & 0 & 0 & 0 &  36       
\end{array} , 
\]
}

\noindent corresponding to $\mathcal{T}_1$ - the upper triangular matrices with off-diagonal entries denoted with $*$,

{\tiny
\[
\begin{array}{cccccccccccccccccccccccccccccccccccc}
1 & 0 & 0 & 0 & 0 & 0 & 0 & 0 & 0 & 0 & 0 & 0 & 0 & 0 & 0 & 0 & 0 & 0 & 0 & 0 & 0 & 0 & 0 & 0 & 0 & \dots  \\
0 & 2 & 0 & 0 & 0 & 0 & 0 & 0 & 0 & 0 & 0 & 0 & 0 & 0 & 0 & 0 & 0 & 0 & 0 & 0 & 0 & 0 & 0 & 0 & 0 & \dots    \\
0 & *^{\prime} & 3 & 0 & 0 & 0 & 0 & 0 & 0 & 0 & 0 & 0 & 0 & 0 & 0 & 0 & 0 & 0 & 0 & 0 & 0 &  0 & 0 & 0 & 0 &  \dots  \\
0 & *^{\prime} & *^{\prime} & 4 & 0 & 0 & 0 & 0 & 0 & 0 & 0 & 0 & 0 & 0 & 0 & 0 & 0 & 0 & 0 & 0 & 0 & 0 & 0 & 0 & 0 &  \dots  \\
0 & *^{\prime} & *^{\prime} & *^{\prime} & 5 & 0 & 0 & 0 & 0 & 0 & 0 & 0 & 0 & 0 & 0 & 0 & 0 & 0 & 0 & 0 & 0 & 0 & 0 & 0 & 0 &  \dots   \\
0 & *^{\prime} & *^{\prime} & *^{\prime} & *^{\prime} & 6 & 0 & 0 & 0 & 0 & 0 & 0 & 0 & 0 & 0 & 0 & 0 & 0 & 0 & 0 & 0 & 0 & 0 & 0 & 0 &  \dots  \\
0 & *^{\prime} & *^{\prime} & *^{\prime} & *^{\prime} & *^{\prime} & 7 & 0 & 0 & 0 & 0 & 0 & 0 & 0 & 0 & 0 & 0 & 0 & 0 & 0 & 0 & 0 & 0 & 0 & 0 &  \dots  \\
0 & *^{\prime} & *^{\prime} & *^{\prime} & *^{\prime} & *^{\prime} & *^{\prime} & 8 & 0 & 0 & 0 & 0 & 0 & 0 & 0 & 0 & 0 & 0 & 0 & 0 & 0 & 0 & 0 & 0 & 0 &   \dots   \\
\end{array}
\]}

{\tiny  \[
\begin{array}{cccccccccccccccccccccccccccccccccccc} 
\dots & 0 & 0 & 0 & 0 & 0 & 0 & 0 & 0 & 0 & 0 & 0 \\ \dots & 0 & 0 & 0 & 0 & 0 & 0 & 0 & 0 & 0 & 0 & 0  \\ \dots & 0 & 0 & 0 & 0 & 0 & 0 & 0 & 0 & 0 & 0 & 0  \\ \dots & 0 & 0 & 0 & 0 & 0 & 0 & 0 & 0 & 0 & 0 & 0  \\ \dots & 0 & 0 & 0 & 0 & 0 & 0 & 0 & 0 & 0 & 0 & 0  \\ \dots & 0 & 0 & 0 & 0 & 0 & 0 & 0 & 0 & 0 & 0 & 0  \\ \dots & 0 & 0 & 0 & 0 & 0 & 0 & 0 & 0 & 0 & 0 & 0  \\  \dots & 0 & 0 & 0 & 0 & 0 & 0 & 0 & 0 & 0 & 0 & 0  \\ \dots & 0 & 0 & 0 & 0 & 0 & 0 & 0 & 0 & 0 & 0 & 0  \\ \dots & 0 & 0 & 0 & 0 & 0 & 0 & 0 & 0 & 0 & 0 & 0 \\ \dots & 0 & 0 & 0 & 0 & 0 & 0 & 0 & 0 & 0 & 0 & 0  \\ \dots & 0 & 0 & 0 & 0 & 0 & 0 & 0 & 0 & 0 & 0 & 0  \\ \dots & 0 & 0 & 0 & 0 & 0 & 0 & 0 & 0 & 0 & 0 & 0  \\ \dots & 0 & 0 & 0 & 0 & 0 & 0 & 0 & 0 & 0 & 0 & 0  \\ \dots & 0 & 0 & 0 & 0 & 0 & 0 & 0 & 0 & 0 & 0 & 0  \\ \dots & 0 & 0 & 0 & 0 & 0 & 0 & 0 & 0 & 0 & 0 & 0  \\  \dots & 0 & 0 & 0 & 0 & 0 & 0 & 0 & 0 & 0 & 0 & 0  \\ \dots & 0 & 0 & 0 & 0 & 0 & 0 & 0 & 0 & 0 & 0 & 0   \\ \dots & 0 & 0 & 0 & 0 & 0 & 0 & 0 & 0 & 0 & 0 & 0 \\ \dots & 0 & 0 & 0 & 0 & 0 & 0 & 0 & 0 & 0 & 0 & 0  \\ \dots & 0 & 0 & 0 & 0 & 0 & 0 & 0 & 0 & 0 & 0 & 0  \\ \dots & 0 & 0 & 0 & 0 & 0 & 0 & 0 & 0 & 0 & 0 & 0  \\ \dots & 0 & 0 & 0 & 0 & 0 & 0 & 0 & 0 & 0 & 0 & 0  \\ \dots & 0 & 0 & 0 & 0 & 0 & 0 & 0 & 0 & 0 & 0 & 0  \\ \dots & 0 & 0 & 0 & 0 & 0 & 0 & 0 & 0 & 0 & 0 & 0  \\  \dots & 0 & 0 & 0 & 0 & 0 & 0 & 0 & 0 & 0 & 0 & 0  \\ \dots & 0 & 0 & 0 & 0 & 0 & 0 & 0 & 0 & 0 & 0 & 0   \\   \dots & 0 & 0 & 0 & 0 & 0 & 0 & 0 & 0 & 0 & 0 & 0  \\ \dots & 0 & 0 & 0 & 0 & 0 & 0 & 0 & 0 & 0 & 0 & 0  \\  \dots & 26 & 0 & 0 & 0 & 0 & 0 & 0 & 0 & 0 & 0 & 0  \\  \vdots & \vdots & \vdots  & \vdots & \vdots & \vdots &  \vdots & \vdots & \vdots  & \vdots & \vdots & \vdots \\  \dots & *^{\prime} & *^{\prime} & *^{\prime} & *^{\prime} & *^{\prime} & *^{\prime} & *^{\prime} & *^{\prime} & *^{\prime} & *^{\prime} &  36       
\end{array} , 
\]
}

\noindent corresponding to $\mathcal{T}_2$ - the lower triangular matrices with off-diagonal entries denoted with $*^{\prime}$. Altogether, taking a product of the four above matrix basis elements yields,

{\tiny
\[
\begin{array}{cccccccccccccccccccccccccccccccccccc}
1 & 0 & 0 & 0 & 0 & 0 & 0 & 0 & 0 & 0 & 0 & 0 & 0 & 0 & 0 & 0 & 0 & 0 & 0 & 0 & 0 & 0 & 0 & 0 & 0 & \dots  \\
0 & 2  & * & * & * & * & * & * & * & * & * & * & * & * & * & * & * & * & * & * & * & * & * & * & * &   \dots    \\
0 & *^{\prime} & 3  & * & * & * & * & * & * & * & * & * & * & * & * & * & * & * & * & * & * & * & * & * & *  &  \dots  \\
0 & *^{\prime} & *^{\prime} & 4 & *  & * & * & * & * & * & * & * & * & * & * & * & * & * & * & * & * & * & * & * & *  &  \dots  \\
0 & *^{\prime} & *^{\prime} & *^{\prime} & 5 & * & * & * & * & * & * & * & * & * & * & * & * & * & * & * & * & * & * & * & *  &  \dots   \\
0 & *^{\prime} & *^{\prime} & *^{\prime} & *^{\prime} & 6 & * & * & * & * & * & * & * & * & * & * & * & * & * & * & * & * & * & * & * &  \dots  \\
0 & *^{\prime} & *^{\prime} & *^{\prime} & *^{\prime} & *^{\prime} & 7 & * & * & * & * & * & * & * & * & * & * & * & * & * & * & * & * & * & * &  \dots  \\
0 & *^{\prime} & *^{\prime} & *^{\prime} & *^{\prime} & *^{\prime} & *^{\prime} & 8 & * & * & * & * & * & * & * & * & * & * & * & * & * & * & * & * & * &   \dots   \\
\end{array}
\]}

{\tiny  \[
\begin{array}{cccccccccccccccccccccccccccccccccccc} 
\dots & * & * & * & * & * & * & * & * & * & * & 0 \\ \dots & * & * & * & * & * & * & * & * & * & * & 0  \\ \dots & * & * & * & * & * & * & * & * & * & * & 0 \\ \dots & * & * & * & * & * & * & * & * & * & * & 0  \\ \dots & * & * & * & * & * & * & * & * & * & * & 0  \\ \dots & * & * & * & * & * & * & * & * & * & * & 0  \\ \dots & * & * & * & * & * & * & * & * & * & * & 0  \\  \dots & * & * & * & * & * & * & * & * & * & * & 0  \\ \dots & * & * & * & * & * & * & * & * & * & * & 0 \\ \dots & * & * & * & * & * & * & * & * & * & * & 0 \\ \dots & * & * & * & * & * & * & * & * & * & * & 0  \\ \dots & * & * & * & * & * & * & * & * & * & * & 0  \\ \dots & * & * & * & * & * & * & * & * & * & * & 0 \\ \dots & * & * & * & * & * & * & * & * & * & * & 0  \\ \dots & * & * & * & * & * & * & * & * & * & * & 0  \\ \dots & * & * & * & * & * & * & * & * & * & * & 0  \\  \dots & * & * & * & * & * & * & * & * & * & * & 0  \\ \dots & * & * & * & * & * & * & * & * & * & * & 0   \\ \dots & * & * & * & * & * & * & * & * & * & * & 0 \\ \dots & * & * & * & * & * & * & * & * & * & * & 0 \\ \dots & * & * & * & * & * & * & * & * & * & * & 0 \\ \dots & * & * & * & * & * & * & * & * & * & * & 0  \\ \dots & * & * & * & * & * & * & * & * & * & * & 0  \\ \dots & * & * & * & * & * & * & * & * & * & * & 0  \\ \dots & * & * & * & * & * & * & * & * & * & * & 0  \\  \dots & * & * & * & * & * & * & * & * & * & * & 0  \\ \dots & * & * & * & * & * & * & * & * & * & * & 0   \\  \vdots & \vdots & \vdots  & \vdots & \vdots & \vdots &  \vdots & \vdots & \vdots  & \vdots & \vdots & \vdots \\  \dots & *^{\prime} & *^{\prime} & *^{\prime} & *^{\prime} & *^{\prime} & *^{\prime} & *^{\prime} & *^{\prime} & *^{\prime} & *^{\prime} &  36       
\end{array} , 
\]
}

\noindent corresponding to the K-matrix, $\mathcal{K}_1 \equiv \mathcal{K}^{+}$, with entries above the diagonal given by $*$, and entries below the diagonal given by $*^{\prime}$. Besides the first K-matrix above which encodes boundary conditions on the left side of the boundary of a finite volume, one can obtain the remaining K-matrix, $\mathcal{K}_2 \equiv \mathcal{K}^{-}$, for encoding open boundary conditions on the right side of the boundary of a finite volume, where,

\begin{align*}
\mathcal{K}^{+ }_{\mathrm{Open},36} \equiv  \mathcal{K}^{+ } \equiv     %
   
\end{align*}

 \begin{align*} \subsetneq  \big\{ \textit{36 } \times \textit{36 finite-dimensional representations} \big\}  , 
\end{align*}

\noindent with,

\begin{align*}
k^n_{0,\pm} \big( u \big) \equiv \big( \mathrm{exp} \big( \pm 2 u \big) + \mathrm{exp} \big( 2 n \eta \big) \bigg[\xi^2_{-} \mathrm{exp} \big( u + 2 n \eta \big)   - \mathrm{exp} \big( - u \big) \bigg]  \text{, }   \\   k^n_{1,\pm} \big( u \big) \equiv  \frac{1}{2} \big( \mathrm{exp} \big(\pm 2 u \big) +1 \big) \bigg[     2 \xi_{-} \mathrm{exp} \big( 2 n \eta \big) \big( \mathrm{exp} \big( \pm 2 u \big) - 1 \big) - \mathrm{exp} \big( u \big) \big( 1 - \xi^2_{-} \mathrm{exp} \big( 2 n \eta \big) \big) \\ \times \big( 1 + \mathrm{exp} \big( 2 n \eta \big)  \big) \bigg]     \text{, }  \\   k^n_{2,\pm} \big( u \big)  \equiv  k^n_3 \big( u \big) \equiv \frac{1}{2} \mathrm{exp} \big( u \big) \big( \mathrm{exp} \big( \pm 2 u \big) - 1 \big) \big( 1 + \xi^{2}_{-} \mathrm{exp} \big( 2 n \eta \big) \big)  \\ \times \big( 1 - \mathrm{exp} \big( 2 n \eta \big) \big)       \text{, } \end{align*}

\begin{align*} 
k^n_{4,\pm} \big( u \big)  \equiv    \frac{1}{2} \big( \mathrm{exp} \big( \pm 2 u \big) +1 \big) \bigg[       - 2 \xi_{-} \mathrm{exp} \big( 2 n \eta \big) \big( \mathrm{exp} \big( \pm  2 u \big) - 1 \big) - \mathrm{exp} \big( u \big) \big( 1 - \xi^2_{-} \mathrm{exp} \big( 2 n \eta \big) \\ \times  \big( 1 + \mathrm{exp} \big( 2 n \eta \big)         \bigg]\text{, }   \\  k^n_{5,\pm} \big( u \big) \equiv       \big( \mathrm{exp} \big( \pm 2 u \big) + \mathrm{exp} \big( 2 n \eta \big) \big) \big( \xi^2_{-} \mathrm{exp} \big( u + 2 n \eta \big) - \mathrm{exp} \big( 3 u \big) \big)    \text{, }  
\end{align*}

\noindent which for $n \equiv 34$ equals,

\[   \left\{\!\begin{array}{ll@{}>{{}}l} 
k^{34}_{0,\pm} \big( u \big) , \\ k^{34}_{1,\pm} \big( u \big)   , \\ k^{34}_{2,\pm} \big( u \big) , \\ k^{34}_{3,\pm} \big( u \big) ,  \\ k^{34}_{4,\pm} \big( u \big) , \\ k^{34}_{5,\pm} \big( u \big)  , \\ k^{34}_{6,\pm} \big( u \big)   ,      \end{array}\right. 
\]

\noindent given boundary parameter $\xi^n_{-} \equiv \xi_{-}$. An overview of results involving computations with the Jimbo R-matrix is provided in the below table, \textit{Table 1}.

\bigskip

\begin{tabular}{|l|l|}
\hline\parbox[t]{0.25\textwidth}{
\begin{itemize}
\item \textbf{Claim 1} 
\item \textbf{Claim 2} 
\item \textbf{Claim 3} 
\item \textbf{Claim 4} 
\item \textbf{Claim 5} 
\item \textbf{Claim 6} 
\item \textbf{Claim 7} 
\item \textbf{Claim 8} 
\item \textbf{Claim 9} 
\item \textbf{Claim 10} 
\end{itemize}}& 
\parbox[t]{0.67\textwidth}{
\begin{itemize}
\item Term $\mathscr{R}_1$ of the Jimbo $R$-matrix
\item Terms $\mathscr{R}_2$, and $\mathscr{R}_3$, of the Jimbo $R$-matrix
\item Term $\mathscr{R}_4$ of the Jimbo $R$-matrix
\item  Term $\mathscr{R}_5$ of the Jimbo $R$-matrix
\item Finite-dimensional representation of the Jimbo $R$-matrix for general $n$
\item Approximation of transfer matrix from $\mathcal{K}$-matrices and the Jimbo $R$-matrix
\item Computation of derivative of transfer matrix
\item  Trace computation 
\item Derivative of $\mathcal{K}_{-}$ matrix
\item Logarithmic derivative of transfer matrix
\end{itemize}}\\
\hline
\end{tabular}
\noindent \textit{Table 1}. An overview of computations with the Jimbo R-matrix established later in the paper.

\subsection{Boundary Yang-Baxter, and reflection, equations for open boundary conditions from K-matrices}

The boundary Yang-Baxter equations (often abbreviated BYBE), take the form,

\begin{align*}
       R_{12} \big( u -v \big) \mathcal{K}^{\pm } \big( u \big)  R_{21} \big( u + v  \big) \mathcal{K}^{\pm} \big( v \big)   = \mathcal{K}^{\pm} \big( v \big)   R_{12} \big( u + v \big) \mathcal{K}^{\pm} \big( u \big)   R_{21} \big( u -v \big)       \text{, }  
\end{align*}

\noindent for the Jimbo R-matrix, up to an exponential prefactor, 

\begin{align*}
  R \big( u \big)  \equiv  \mathrm{exp} \big( - 2 u - 6 \eta \big) R_J \big( x \big)  \text{, }  
\end{align*}

\noindent up to an exponential prefactor, for strictly positive parameters $u$ and $\eta$. An identical R-matrix is described in the next section for quasi-periodic boundary conditions of the same spin-chain in \textit{1.7}. Moreover, the explicit form of the Jimbo R-matrix, provided in (\citet{nepomechie2021}) will be explicitly manipulated in \textit{3.0.1} for expressing the higher rank transfer matrix. In particular, by isolating components of the Jimbo R-matrix depending upon like terms in tensor products of elementary matrices, denoted $\mathscr{R}_1, \cdots, \mathscr{R}_5$, the transfer matrix, as a particular trace, is approximated from the following components:

\[   \left\{\!\begin{array}{ll@{}>{{}}l} 
  \mathrm{Tr}_0 \bigg[ \mathcal{K}_{+,0} \big( u \big)  \prod_{1 \leq j \leq L} \big( \mathscr{R}_1\big)_{0j} \big( u \big)  \mathcal{K}_{-,0} \big( u \big)   \prod_{1 \leq j^{\prime} \leq L} \big( \mathscr{R}_1\big)_{j^{\prime}0} \big( u \big)       \bigg]   , \\ \\  \mathrm{Tr}_0 \bigg[\mathcal{K}_{+,0} \big( u \big)  \prod_{1 \leq j \leq L} \big( \mathscr{R}_2\big)_{0j} \big( u \big)  \mathcal{K}_{-,0} \big( u \big)   \prod_{1 \leq j^{\prime} \leq L} \big( \mathscr{R}_2\big)_{j^{\prime}0} \big( u \big)   \bigg]    ,  \\ \\   \mathrm{Tr}_0 \bigg[ \mathcal{K}_{+,0} \big( u \big)  \prod_{1 \leq j \leq L} \big( \mathscr{R}_3\big)_{0j} \big( u \big)  \mathcal{K}_{-,0} \big( u \big)   \prod_{1 \leq j^{\prime} \leq L} \big( \mathscr{R}_3\big)_{j^{\prime}0} \big( u \big)  \bigg]   , \\ \\    \mathrm{Tr}_0 \bigg[ \mathcal{K}_{+,0} \big( u \big)  \prod_{1 \leq j \leq L} \big( \mathscr{R}_4\big)_{0j} \big( u \big)  \mathcal{K}_{-,0} \big( u \big)   \prod_{1 \leq j^{\prime} \leq L} \big( \mathscr{R}_4\big)_{j^{\prime}0} \big( u \big)  \bigg]  , \\ \\   \mathrm{Tr}_0 \bigg[ \mathcal{K}_{+,0} \big( u \big)  \prod_{1 \leq j \leq L} \big( \mathscr{R}_5\big)_{0j} \big( u \big)  \mathcal{K}_{-,0} \big( u \big)   \prod_{1 \leq j^{\prime} \leq L} \big( \mathscr{R}_5\big)_{j^{\prime}0} \big( u \big)  \bigg]  , \\ \\ \mathrm{Tr}_0 \bigg[      \bigg[ \mathcal{K}_{+,0} \big( u \big) \bigg[  \prod_{1 \leq j \leq L} \big( R_{0J} \big( u \big) - \underset{1 \leq k \leq 5}{\sum} \big( \mathscr{R}_k\big)_{0j} \big( u \big)  \bigg] \mathcal{K}_{-,0} \big( u \big)  \\ \times  \prod_{1 \leq j^{\prime} \leq L}  \bigg[ \big( R_{J0} \big( u \big)  - \underset{1 \leq  k \leq 5}{\sum} \big( \mathscr{R}_k\big)_{j^{\prime}0} \big( u \big)  \big)      \bigg]     \bigg] .
\end{array}\right. 
\]

\bigskip

\noindent Additionally, in the presence of open boundary conditions for the higher rank spin-chain, the fact that reflection equations, or equivalently, BYBE, take the form,

\begin{align*}
R_{1,2} \big( u - v \big)   \mathcal{K}^{-} \big( u \big)   R_{2,1} \big( u + v \big)      \mathcal{K}^{-} \big(  v \big) = \mathcal{K}^{-} \big(  v \big) R_{1,2} \big( u - v \big)   \mathcal{K}^{-} \big( u \big)   R_{2,1} \big( u + v \big) ,  
\end{align*}

\noindent corresponding to the reflection equation for the $-$ K-matrices, $\mathcal{K}_{-}$, and,

\begin{align*}
R_{1,2} \big( u - v \big)   \mathcal{K}^{+} \big( u \big)   R_{2,1} \big( u + v \big)      \mathcal{K}^{+} \big(  v \big) = \mathcal{K}^{+} \big(  v \big) R_{1,2} \big( u - v \big)   \mathcal{K}^{+} \big( u \big)   R_{2,1} \big( u + v \big) ,  
\end{align*}

\noindent \noindent corresponding to the reflection equation for the $+$ K-matrices, $\mathcal{K}_{+}$.

\subsection{Open boundary conditions through the nested Bethe equations}

\noindent Under the influence of open boundary conditions, the Bethe equations take the form,

\begin{align*}
\bigg[  \frac{\mathrm{sinh} \big( v_j  \big) }{\mathrm{sinh} \big( v_j - 3 \eta \big) }  \bigg]^L  = {\underset{1 \leq l \leq m, l \neq k, j \neq l \neq k}{\prod}}  \bigg[ \frac{\mathrm{sinh}\big( \frac{1}{3}  \big( v_j - v_k - v_l\big) + \eta \big) }{\mathrm{sinh} \big( \frac{1}{3}  \big( v_j - v_k - v_l\big) - \eta  \big) }        \bigg]    \text{. } 
\end{align*}

\noindent As was the case in previous expressions for Bethe equations that have been obtained for quasi-periodic boundary conditions for the higher rank spin-chain, (\citet{frahm2023}), one not only includes additional factors $v_l$ besides $v_j$ and $v_k$, but also the condition that $l \neq k$ and that $j \neq l \neq k$.

\subsection{Linearization of the open boundary condition Bethe equations}

The linearization of the equation provided in the previous item above takes the form,

\begin{align*}
  \bigg[  \frac{ v_j   }{ v_j - 3 \eta }  \bigg]^L  = {\underset{1 \leq l \leq m, l \neq k, j \neq l \neq k}{\prod}}  \bigg[ \frac{ \frac{1}{3}  \big( v_j - v_k - v_l\big) + \eta  }{\frac{1}{3}  \big( v_j - v_k - v_l\big) - \eta }    \bigg]         . 
\end{align*}

\subsubsection{Description  of the open-boundary condition root system}

\noindent For open boundary conditions of the higher rank spin-chain, one can determine the Bethe roots from eigenvalues of the transfer matrix. That is, by straightforward computations, one wishes to determine the open-boundary spectrum,$\mathcal{S}^{\mathrm{Open}}$, from the equation,

\begin{align*}
    \mathrm{det} \big\{ \textbf{T}^{\mathrm{Open}} - \lambda \textbf{I} \big\} \equiv  \mathrm{det} \bigg\{  \bigg[ \underset{j,j^{\prime} \longrightarrow + \infty}{\mathrm{lim}}  \mathrm{Tr}_0 \bigg[ \mathcal{K}_{+,0} \big( u \big)  \prod_{1 \leq j \leq L} R_{0j} \big( u \big)  \times   \mathcal{K}_{-,0} \big( u \big)   \prod_{1 \leq j^{\prime} \leq L} R_{j^{\prime}0} \big( u \big)  \bigg] \bigg] \\ - \lambda \textbf{I}  \bigg\}   = 0,  \end{align*}

\noindent for some $\lambda >0$. We provide such a computation for the characteristic polynomial of the open-boundary condition transfer matrix in \textit{3.0.2}.

\subsection{Application of the root-density apporoach under open-boundary conditions}

\noindent Under open-boundary conditions, it is expected that the Bethe equations of the higher rank spin-chain has roots belonging to the following system:

\begin{itemize}
    \item[$\bullet$] \textit{Roots of first type}. Fix $n_1 \in \textbf{Z}$. Finitely many roots solve the open boundary, higher-rank Bethe equations lie on $n_1 i \pi$.

      \item[$\bullet$] \textit{Roots of second type}. Fix $n_2 \in \textbf{Z}$. Finitely many roots solve the open boundary, higher-rank Bethe equations lie on $n_2 i \frac{\pi}{2}$.

        \item[$\bullet$] \textit{Roots of third type}. Finitely many roots solve the open boundary, higher-rank Bethe equations lie on the real line.
\end{itemize}

\subsubsection{Contributions Overview}

\noindent With the objects introduced above, after introducing quasi-periodic objects for the $D^{(2)}_3$ spin-chain in the next section, one is able to compute the:

\begin{itemize}
\item[$\bullet$] \textit{Local Hamiltonian encoding}, 
\item[$\bullet$] \textit{Logarithmic derivative of the transfer matrix},
\item[$\bullet$]  \textit{Linearization of the open-boundary reflection equation}
\end{itemize}

\subsection{Quasi-periodic Spin-chain objects}

\noindent We begin by providing an overview of the higher rank spin chain, and then proceed to describe its relations to the lower rank spin chain. To introduce such a model, define the $36 \times 36$ R-matrix, with,

\begin{align*}
  R \big( u \big)  \equiv  \mathrm{exp} \big( - 2 u - 6 \eta \big) R_J \big( x \big)  \text{, }  
\end{align*}

\noindent as a function of the single parameter $u$, where $R_J \big( x \big)$ denotes the Jimbo matrix {[4]}, and $x \equiv \mathrm{exp} \big( u \big)$ and $k \equiv \mathrm{exp} \big( 2 \eta \big)$. The R-matrix satisfies the Yang Baxter equation,

\begin{align*}
       R_{12} \big( u -v \big) R_{13} \big( u \big) R_{23} \big( v \big) = R_{23} \big( v \big) R_{13} \big( u \big) R_{12} \big( u - v \big)         \text{, }  
\end{align*}

\noindent for the anisotropy parameter $\eta \equiv i \gamma$, and another parameter $v$. Besides the R-matrix satisfying the Yang Baxter equation, it also possesses a $U \big( 1 \big)$ symmetry, which is captured by the condition,

\begin{align*}
        \big[ R \big( u \big) ,      \textbf{h}_j    \otimes \textbf{I} + \textbf{I} \otimes   \textbf{h}_j  \big]   \equiv 0            \text{, } 
\end{align*}

\noindent for $j=1$ and $j=2$, with,

\begin{align*}
   \textbf{h}_1 \equiv          \mathcal{M} \big( 1 , 1 \big) - \mathcal{M} \big( 6 , 6 \big)        \text{, }  \\ \textbf{h}_2 \equiv       \mathcal{M} \big( 2 , 2 \big) - \mathcal{M}  \big( 5 , 5 \big)    \text{, }  
\end{align*}

\noindent for the matrices $\mathcal{M} \big( 1 , 1 \big)$, $\mathcal{M} \big( 6 , 6 \big)$, $\mathcal{M} \big( 2,  2 \big)$ and $\mathcal{M} \big( 5,5\big)$, which are respectively given by the $6 \times 6$ matrices with nonzero entries at $\big( 1,1\big)$, $\big( 6 , 6 \big)$, $\big( 2,2\big)$, $\big(5,5\big)$, and the identity matrix $\textbf{I}$. The R-matrix also satisfies Parity-Time (PT) symmetry, in which,

\begin{align*}
   R_{21} \big( u \big) \equiv \mathcal{P}_{12} \mathcal{R}_{12} \big( u \big) \mathcal{P}_{12} \equiv     R^{t_1 , t_2}_{12} \big( u \big)       \text{, }  
\end{align*}

\noindent for the permutation matrix $\mathcal{P}$, for the transposition $t$. Additional properties, including braiding unitarity, regularity, crossing symmetry, quasi-periodicity, and $\textbf{Z}_2$ symmetries are also satisfied (\citet{frahm2023}). From the quantities introduced since the beginning of the section, the transfer matrix of the model takes the form,

\begin{align*}
   \textbf{T} \big( u \big) \equiv \mathrm{tr}_0 \big[    \textbf{K}_0   \big[  \textbf{T}_0 \big( u \big) \big]   \big] \equiv  \mathrm{tr} \bigg[ \textbf{K}_0    \bigg[  \prod_{1 \leq j \leq L}  \textbf{R}_{0j} \big( u \big)  \bigg] \bigg]  \text{, }  
\end{align*}

\noindent for the twist diagonal matrix,

\begin{align*}
 \textbf{K} \equiv \mathrm{diag} \big( \mathrm{exp} \big( i \phi_1 \big) , \mathrm{exp} \big( i \phi_2 \big) , 1 , 1 , \mathrm{exp} \big( - i \phi_2 \big) , \mathrm{exp} \big( - i \phi_1 \big)   \big)    \text{, }  
\end{align*}

\noindent given two angles $\phi_1$ and $\phi_2$, and product of R matrices for $1 \leq j \leq L$. The angles $\phi_1$ and $\phi_2$ determine the boundary conditions of the higher rank spin chain, as opposed to the open boundary conditions of the lower rank spin chain that is introduced in the remaining parts of this section.

\bigskip

\noindent To work towards introducing the higher rank spin chain and open boundary conditions for it, we start with defining the following R-matrix, and similar components, for the lower rank spin chain with the following. To construct the R-matrix, consider the $6 \times 6$ matrix, of the form,

\[
\widetilde{R}^{(\mathrm{XXZ})} \big( u \big) \equiv \begin{bmatrix}
  \mathrm{sinh} \big( - \frac{u}{2} + \eta \big)    & 0   & 0   & 0  \\   0& \mathrm{sinh} \big( \frac{u}{2} \big)  & \mathrm{exp} \big( - \frac{u}{2} \big) \mathrm{sinh} \big( \eta \big) & 0  \\  0 & \mathrm{exp} \big( \frac{u}{2} \big) \mathrm{sinh} \big( \eta \big)  & \mathrm{sinh} \big( \frac{u}{2} \big)  & 0  \\ 0 & 0 & 0 & \mathrm{sinh} \big( - \frac{u}{2} + \eta \big)   \\  
\end{bmatrix} \text{, }  
\]

\bigskip

\noindent from the R-matrix for the $A^{(1)}_1$ (XXZ) spin chain, which is related to the R-matrix of the lower rank spin chain from the fact that,

\begin{align*}
     \widetilde{R} \big( u \big) \propto B_{12} B_{34} 
         \textbf{R}^{\prime}_{12,34}\big( u \big)    B_{12} B_{34} \equiv B_{12} B_{34} 
           \bigg[ R_{14} \big( u \big) R_{13} \big( u \big) R_{24} \big( u \big) R_{23} \big( u \big)        \bigg] B_{12} B_{34}      \text{, }  
\end{align*}

\noindent and matrices $B$, which are given by,

\[
\begin{bmatrix}
    1 & 0 & 0    & 0 \\ 0 & \frac{\mathrm{cosh} \big( \frac{\eta}{2}\big) }{\sqrt{\mathrm{cosh} \big( \eta \big) }} &   - \frac{\mathrm{sinh} \big( \frac{\eta}{2}\big)}{\sqrt{\mathrm{cosh} \big( \eta \big)}}    &  0 \\ 0 & - \frac{\mathrm{sinh} \big( \frac{\eta}{2} \big)}{\sqrt{\mathrm{cosh} \big( \eta \big)}} & - \frac{\mathrm{cosh} \big( \frac{\eta}{2} \big) }{\sqrt{\mathrm{cosh}\big( \eta \big)}}  & 0 \\ 0 & 0 & 0 & 1
\end{bmatrix} \text{, }  
\]

\noindent satisfying,

\begin{align*}
 B^2 = \textbf{I}   \text{, }  
\end{align*}

\noindent and R-matrices solving the Yang Baxter equation,

\begin{align*}
          R_{12} \big( u - v \big) R_{13} \big( u \big) R_{23} \big( v \big) = R_{23} \big( v \big) R_{13} \big( u \big) R_{12} \big( u - v \big)      \text{. } 
\end{align*}

\noindent In contrast to the higher rank case, the R-matrix above for the lower rank spin chain satisfies,

\begin{align*}
 \textbf{R}^{\prime}_{12,34} \big( u \big) = R_{43} \big( - \theta \big) R_{13} \big( u \big) R_{14} \big( u + \theta \big) R_{23} \big( u - \theta \big) R_{24} \big( u \big) R_{34} \big( \theta \big)    \text{, }  
\end{align*}

\noindent which in turn implies,

\begin{align*}
  \widetilde{R} \big( u \big) \propto  B_{12} B_{34} 
           \bigg[ R_{14} \big( u \big) R_{13} \big( u \big) R_{24} \big( u \big) R_{23} \big( u \big)        \bigg] B_{12} B_{34}    \equiv  B_{12} B_{34} \bigg[ R_{43} \big( - \theta \big) \\ \times  R_{13} \big( u \big) R_{14} \big( u + \theta \big)  R_{23} \big( u - \theta \big) R_{24} \big( u \big) R_{34} \big( \theta \big)                   \bigg]B_{12} B_{34}      \text{. } 
\end{align*}

\noindent To encode open boundary conditions of the lower rank spin chain, we must further describe properties of the $\textbf{K}$ matrix, which was introduced earlier in the section with the definition of the transfer matrix $\textbf{T} \big( u \big)$ for the higher rank spin chain. In particular, in addition to the R matrices which satisfy the Yang Baxter equation for the lower rank spin chain, open boundary conditions of the chain are enforced from the fact that two other matrices, given by $K_{-} \big( u \big)$ and $K_{+} \big( u \big)$ below, satisfy, ({\citet{Robertson2020}}),

\begin{align*}
       R_{12} \big( u - v \big) K_{1,-} \big( u \big) R_{21} \big( u + v \big) K_{2,-} \big( v \big) =  K_{2,-} \big( u \big) R_{12} \big( u + v \big) K_{1,-} \big( u \big) R_{21} \big( u - v \big)    \text{, }  
\end{align*}

\noindent corresponding to the first, and second, boundary conditions which are reflected through the addition of the terms $K_{1,-} \big( u \big)$ and $K_{2,-} \big( v \big)$, as well as, {[2]},

\begin{align*}
      R_{1,2} \big(  - u + v     \big)    K^{t_1}_{1,+} \big( u \big) R_{1,2} \big( - u - v - 2 i \gamma \big)  K^{t_2}_{2,+} \big( v \big) = K^{t_2}_{2,+} \big( v \big)  \\ \times  R_{1,2} \big( - u - v - 2 i \gamma \big) K^{t_1}_{1,+} \big( u \big)R_{1,2} \big( - u + v \big)  \text{, }  
\end{align*}

\noindent corresponding to the Yang Baxter equation for parameters $t_1$ and $t_2$ from the PT symmetric property of the R-matrix satisfied by the higher rank spin chain, for the anisotropy parameter $\gamma$, where each matrix is respectively given by, ({\citet{Robertson2020}}),

\[
K_{-} \big( \lambda \big) \equiv \begin{bmatrix}
 - \mathrm{exp} \big( - \lambda \big) \big( \mathrm{exp} \big( 2 \lambda \big) + k \big) & 0   &  0 & 0  \\ 0 & (**)_1  & (**)_2 & 0 \\ 0 & (*)_1  & (*)_2 & 0 \\ 0 & 0 & 0 & (***)_1 
\end{bmatrix} \text{, }  
\]

\noindent where,

\begin{align*} (**)_1 \equiv - \frac{1}{2} \big( 1 + \mathrm{exp} \big( 2 \lambda \big) \big) \mathrm{exp} \big( \lambda \big) \big( 1 + k \big) \text{, } \\    (**)_2 \equiv  \frac{1}{2} \big( \mathrm{exp} \big( 2 \lambda \big) - 1 \big) \big( 1 - k \big) \mathrm{exp} \big( \lambda \big) \text{, }  \\ (*)_1 \equiv \frac{1}{2} \big( \mathrm{exp} \big( 2 \lambda \big) - 1 \big) \big( 1 - k \big) \mathrm{exp} \big( \lambda \big)\text{, } \\    (*)_2 \equiv - \frac{1}{2} \big( 1 + \mathrm{exp} \big( 2 \lambda \big) \big) \mathrm{exp} \big( \lambda \big) \big( 1 + k \big) \text{, } \\ (***)_1  \equiv - \mathrm{exp} \big( 3 \lambda \big) \big( \mathrm{exp} \big( 2 \lambda \big) + k \big) \text{, }  \end{align*}

\noindent which is equivalent to the matrix with symbols,

\[
\begin{bmatrix}
  Y_1 \big( \lambda \big)   &  0 & 0 & 0  \\ 0 & Y_2 \big( \lambda \big) & Y_5 \big( \lambda \big) & 0 \\ 0 & Y_6 \big( \lambda \big) & Y_3 \big( \lambda \big) & 0 \\ 0 & 0 & 0 & Y_4 \big( \lambda \big) 
\end{bmatrix} \text{, }  
\]

\noindent from the R-matrix basis,

\begin{align*}
  \bigg\{ \ket{1} , \ket{2} , \ket{3} , \ket{4}  \bigg\} \otimes \bigg\{  \ket{1} , \ket{2} , \ket{3} , \ket{4}     \bigg\}   \text{. } 
\end{align*}

\bigskip

\noindent With such an encoding of the boundary conditions of the spin chain with $K_{-} \big( u \big)$ and $K_{+} \big( u \big)$, the transfer matrix takes on a similar form, in which,

\begin{align*}
  \textbf{T}_{D^{(2)}_2} \big( u \big) \equiv \mathrm{tr}_a \big[  K_{+,a} \big( u \big) R_{a1} \big( u \big) \cdots \times R_{aL}  \big( u \big) K_{-,a} \big( u \big) R_{1a} \big( u \big) \cdots \times R_{La} \big( u \big)    \big] \\ \equiv  \mathrm{tr}_a \bigg[ K_{+,a} \big( u \big)  \bigg[ \prod_{1 \leq j \leq L} R_{aj} \big( u \big) \bigg]   K_{-,a} \big( u \big)  \bigg[ \ \prod_{1 \leq j^{\prime} \leq L} R_{j^{\prime}a} \big( u \big)  \bigg] \bigg]  \text{. } 
\end{align*}

\begin{figure}
\begin{align*}
\includegraphics[width=1.22\columnwidth]{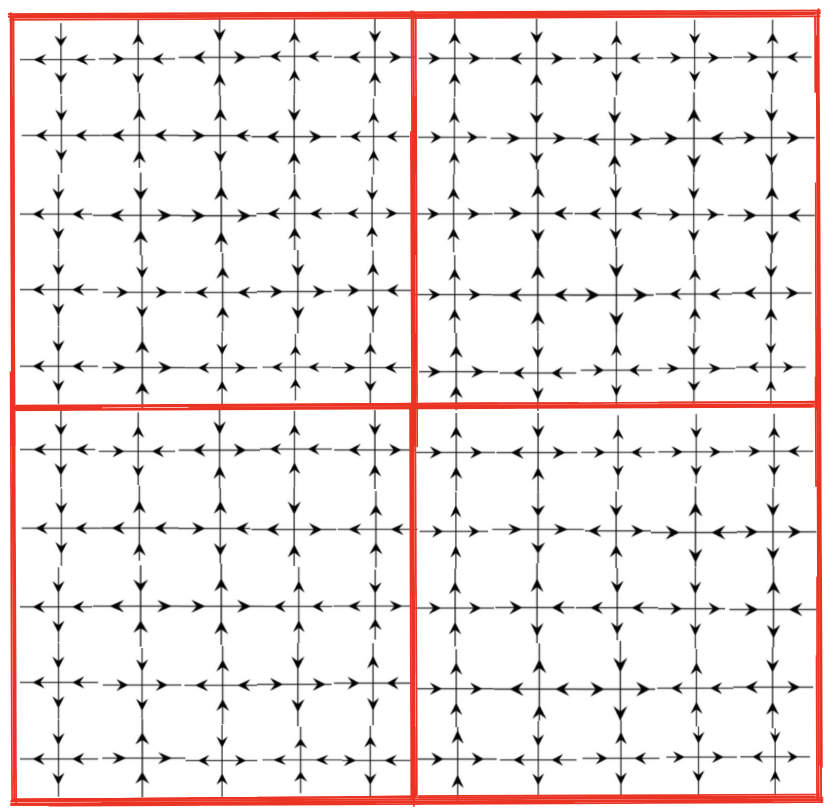}
\end{align*}
\caption{A depiction of a two-dimensional vertex configuration of the 6-vertex model sampled over $\textbf{Z}^2$. The box, whose boundary is outlined in red, is comprised of four equal boxes whose boundaries are also outlined in red within the interior.}
\end{figure}

\begin{figure}
\begin{align*}
\includegraphics[width=1.22\columnwidth]{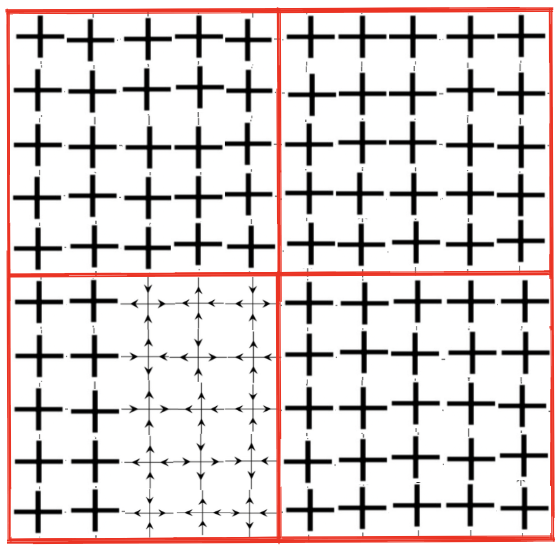}
\end{align*}
\caption{A depiction of a two-dimensional vertex configuration of the $D^{(2)}_3$ spin-chain sampled over $\textbf{Z}^2$. The box, whose boundary is outlined in red, is comprised of four equal boxes whose boundaries are also outlined in red within the interior.}
\end{figure}

\begin{figure}
\begin{align*}
\includegraphics[width=1.22\columnwidth]{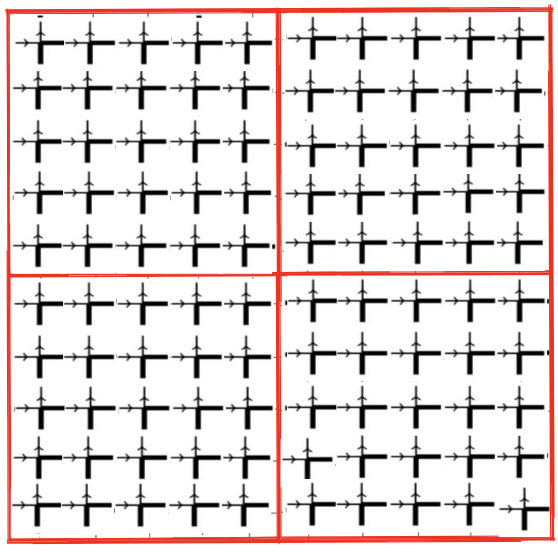}
\end{align*}
\caption{A second depiction of a two-dimensional vertex configuration of the $D^{(2)}_3$ spin-chain sampled over $\textbf{Z}^2$. The box, whose boundary is outlined in red, is comprised of four equal boxes whose boundaries are also outlined in red within the interior.}
\end{figure}

\begin{figure}
\begin{align*}
\includegraphics[width=1.3\columnwidth]{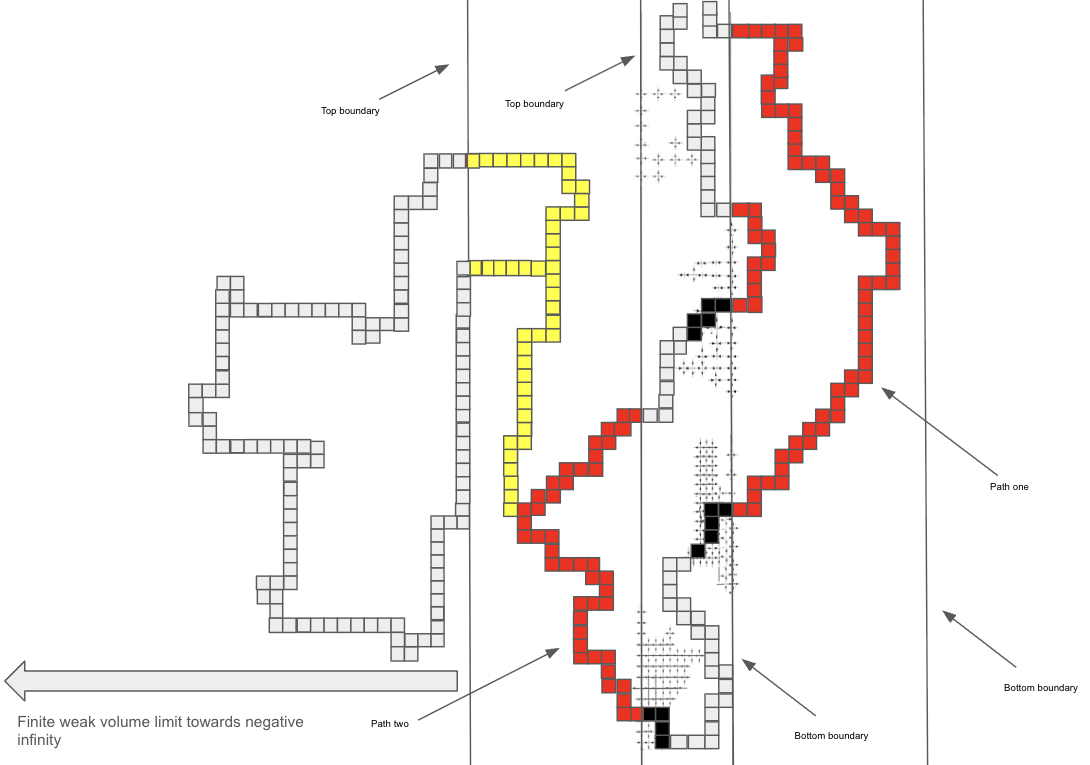}
\end{align*}
\caption{A depiction of taking the weak finite volume limit towards $- \infty$ along $\textbf{Z}^2$ for the height function of the 6-vertex model. The collection of faces highlighted in yellow above can be used to construct longer paths with faces highlighted in grey. The 6-vertex model, along with various formulations of the Bethe ansatz and BYBE, share in many similarities of the $D^{(2)}_3$ spin-chain considered in this work.}
\end{figure}

\begin{figure}
\begin{align*}
\includegraphics[width=1.3\columnwidth]{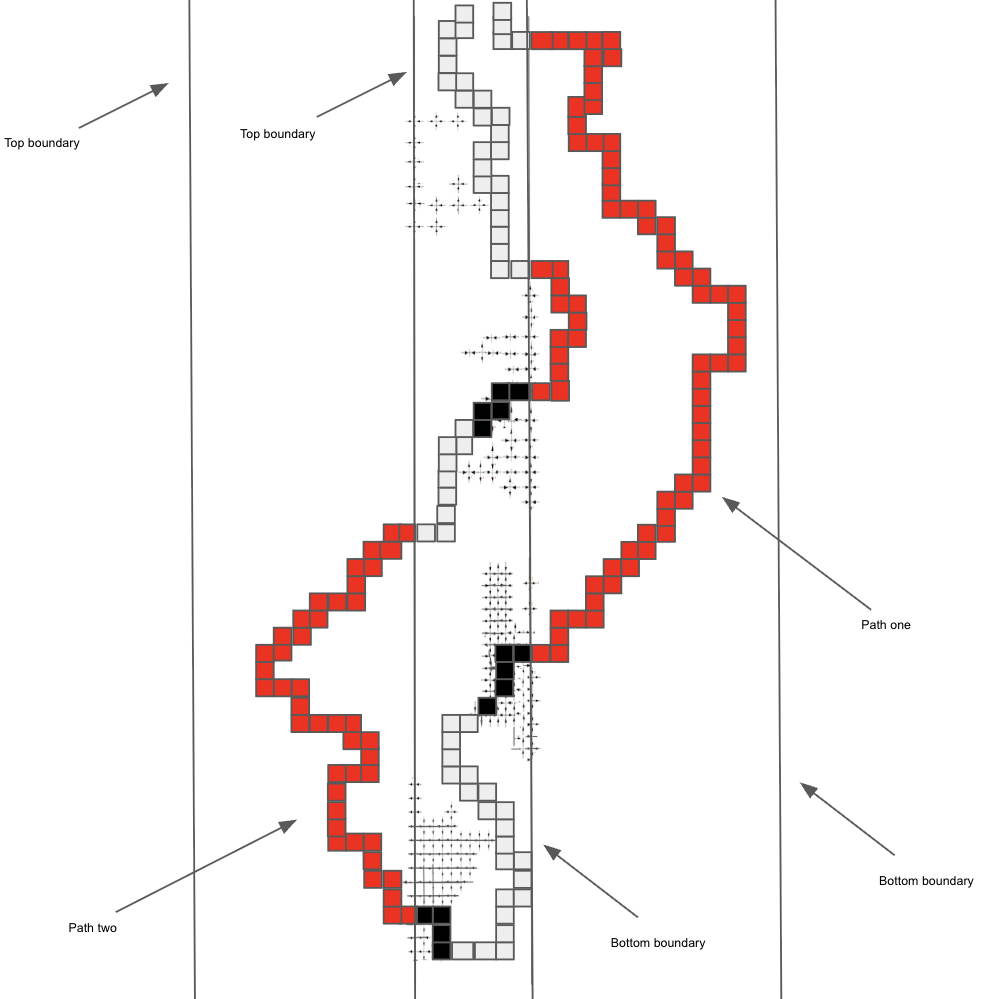}
\end{align*}
\caption{A depiction of a two-dimensional vertex configuration for the 6-vertex model with sloped boundary conditions, sampled over $\textbf{Z}^2$. For paths, including Path One, and Path Two, depicted above, arbitrarily long crossings across $\textbf{Z}^2$ can be obtained from paths which avoid collections of frozen faces.}
\end{figure}

\begin{figure}
\begin{align*}
\includegraphics[width=0.85\columnwidth]{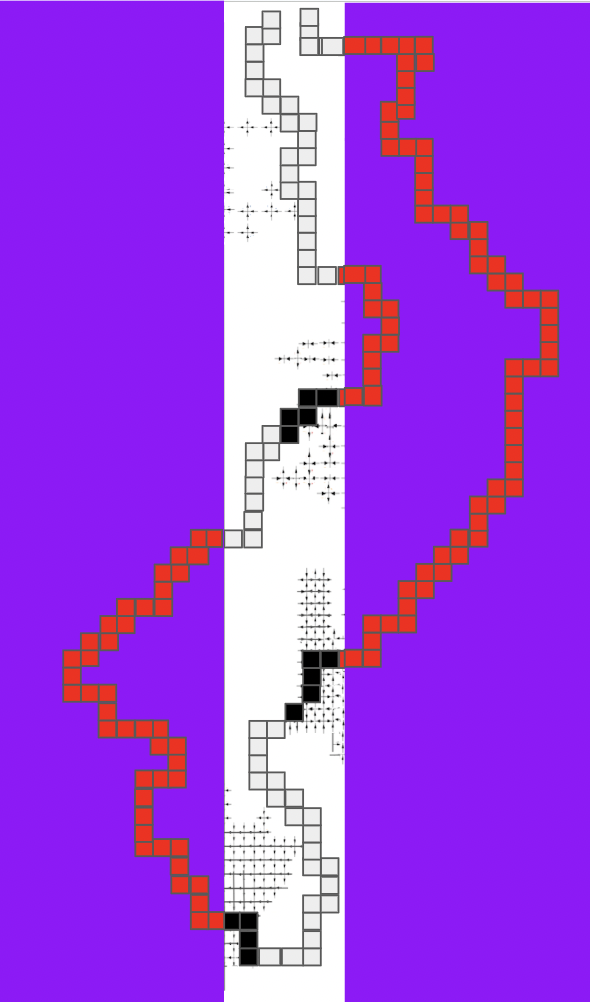}
\end{align*}
\caption{Over strips of $\textbf{Z}^2$, to take the finite weak volume limit over the entire lattice, blue regions to the left and right of the top, and bottom, boundaries, are included.}
\end{figure}

\noindent With $\textbf{T}_{D^{(2)}_2}  \big( u \big)$, which satisfies the condition $\big[ \textbf{T}_{D^{(2)}_2}  \big( u \big) , \textbf{T}_{D^{(2)}_2}  \big( v \big) \big] = 0$, we also stipulate, in order to properly construct open boundary conditions for the lower rank spin chain, that,

\begin{align*}
     K_{+,a} \big(     \lambda   \big)  =  K^{-t} \big( - \rho - \lambda \big) M \text{, }  
\end{align*}

\noindent where $t$ denotes the transposition of the matrix, and $\rho \equiv - \mathrm{log} \big( k \big)$, and $M \equiv \mathrm{diag} \big( k , 1 , 1 , \frac{1}{k} \big)$. Explicitly, the entries of $K_{+}$, from $K_{-}$ and the parameters $\rho$ and $M$, is given by,

\[
\begin{bmatrix} 
 Y_1 \big( \lambda \big) \big( - \rho - \lambda \big)  & 0 & 0 & 0    \\ 0  & Y_2 \big( \lambda \big)  \big( - \rho - \lambda \big)  & Y_6 \big( \lambda \big)  \big( - \rho - \lambda \big)  & 0 \\ 0 & Y_5 \big( \lambda \big)  \big( - \rho - \lambda \big)  & Y_3 \big( \lambda \big)   \big( - \rho - \lambda \big)  & 0 \\ 0 & 0 & 0  & Y_4 \big( \lambda \big)  \big( - \rho - \lambda \big) 
\end{bmatrix} \begin{bmatrix} k & 0 & 0 & 0 \\ 0 & 1 & 0 & 0 \\ 0 & 0 & 1 & 0 \\ 0 & 0 & 0 & \frac{1}{k}
\end{bmatrix}
\]

\subsection{Paper overview}

\noindent With the overview in \textit{1.1} and definitions of lower, and higher, rank spin chains in \textit{1.2}, in the remaining sections of the paper we apply the open boundary framework to the higher rank spin chain, in an effort to determine how the boundary conditions determine the CFT sector. From information on how open boundary conditions are encoded in the Yang-Baxter equation, and transfer matrix, for the lower rank spin chain, we incorporate open boundary conditions in the higher rank spin chain. To clearly demonstrate how encodings of open boundary conditions for the higher rank spin-chain relate to those of the lower rank spin-chain, namely the $D^{(2)}_2$ spin-chain, in the forthcoming subsection we reproduce the following objects for the $D^{(2)}_2$ spin-chain ({\citet{Robertson2020}}):

\begin{itemize}
    \item[$\bullet$] \textit{transfer matrix}. Introduce,

    \begin{align*}\textbf{T}^{\mathrm{Open}}_{D^{(2)}_3} \big( u \big) .
    \end{align*}

   \item[$\bullet$] \textit{K-matrices with open boundary conditions}. Introduce,

   \begin{align*}
       \textbf{K}^{\mathrm{Open}}_{-} \big( u \big) , 
   \end{align*}

\noindent and,

\begin{align*}
\textbf{K}^{\mathrm{Open}}_{+,a}  \big( u \big)  .
\end{align*}

      \item[$\bullet$] \textit{Logarithmic derivative of the open boundary condition transfer matrix}. Introduce,

\begin{align*}
   \frac{\mathrm{d}}{\mathrm{d} u} \bigg\{ \mathrm{log} \big( \textbf{T}^{\mathrm{Open}} \big( u \big) \big) \bigg\}  \bigg|_{u \equiv 0} .
\end{align*}

         \item[$\bullet$] \textit{Local Hamiltonian encoding}. Fix strictly positive parameters $k, \kappa$. Introduce,

             \begin{align*}
   \mathcal{H} \big( k , \kappa , \textbf{K}_{-} , \textbf{K}_{+} \big)  .    \end{align*}

\item[$\bullet$] \textit{Relating the computation of the local Hamiltonian encoding to the logarithmic derivative of the open boundary condition transfer matrix}. One has that,

\begin{align*}
\textbf{H}^{\mathrm{Open}}  \propto   \frac{\mathrm{d}}{\mathrm{d} u} \bigg\{ \mathrm{log} \big( \textbf{T}^{\mathrm{Open}} \big( u \big) \big) \bigg\}  \bigg|_{u \equiv 0} .
\end{align*}

\end{itemize}

\noindent Equipped with the lower rank spin-chain objects, we discuss how the higher-rank generalization can be used to classify sectors of the accompanying conformal field theory.

\section{Encoding quasi-periodic boundary conditions for the higher rank spin-chain}

\subsection{Obtaining the higher rank spin chain Hamiltonian from an expansion of the derivative of the transfer matrix about $u \equiv 0$}

\noindent In comparison to twisted boundary conditions encoded with the angles $\phi_1$ and $\phi_2$, open boundary conditions for the higher rank spin chain can be encoded by introducting a K matrix for the $36 \times 36$ R-matrix, from the basis,

\begin{align*}
  \bigg\{ \ket{1} , \ket{2} , \ket{3} , \ket{4}  , \ket{5} , \ket{6}  
 \bigg\} \otimes \bigg\{  \ket{1} , \ket{2} , \ket{3} , \ket{4} , \ket{5} , \ket{6}    \bigg\}   \text{, }  
\end{align*}

\noindent in which the trace would then take the form,

\begin{align*}
   \textbf{T}^{\mathrm{Quasi-periodic}} \big( u \big) \equiv \mathrm{tr}_0 \big[    \textbf{K}_{+,0} \big( u \big)   \big[ \textbf{T}_{+,0} \big( u \big) \big] \textbf{K}_{-,0} \big( u \big)       \big[ \textbf{T}_{-,0} \big( u \big)     \big]    \big] \\ \equiv  \mathrm{tr}_0 \bigg[\textbf{K}_{+,0}  \big( u \big)  \bigg[   \prod_{1 \leq j \leq L} R_{+,0j} \big( u \big)  \bigg]    \textbf{K}_{-,0} \big( u \big) \bigg[ \underset{1 \leq j^{\prime} \leq L}{\prod}      R_{-,j^{\prime}0}  \big( u \big)  \bigg]    \bigg] \\  \equiv \mathrm{tr}_0 \bigg[\textbf{K}^{\mathrm{Quasi-periodic}}_{+,0}  \big( u \big)    \bigg[ \prod_{1 \leq j \leq L}  R_{+,a0} \big( u \big)  \bigg]    \textbf{K}^{\mathrm{Quasi-periodic}}_{-,0} \big( u \big) \bigg[ \underset{1 \leq j^{\prime} \leq L}{\prod}      R_{-,j^{\prime}0}  \big( u \big)  \bigg]    \bigg] \text{, }  
\end{align*}

\bigskip

\noindent for the higher rank spin chain transfer matrix,

\begin{align*}
  \textbf{T}^{\mathrm{Open}}_{D^{(2)}_3} \big( u \big) \equiv \textbf{T}^{\mathrm{Open}} \big( u \big)   \text{, }  
\end{align*}

\noindent with open boundary conditions enforced through the K-matrix,

\[
\textbf{K}^{\mathrm{Open}}_{-} \big( u \big) \equiv \textbf{K}_{-} \big( u \big)  \equiv \begin{bmatrix}
k_0 \big( u \big)  & 0 & 0 & 0 & 0 & 0  \\ 0 & k_0 \big( u \big) & 0 & 0 & 0 & 0  \\ 0 & 0  & k_1 \big( u \big) & k_2 \big( u \big)   & 0 & 0 \\ 0 & 0 & k_3 \big( u \big)  & k_4 \big( u \big)  &  0 & 0  \\ 0 & 0 & 0 & 0 & k_5 \big( u \big) & 0  \\  0 & 0 & 0 & 0 & 0 & k_5 \big( u \big)  
\end{bmatrix} \text{, }  
\]

\bigskip

\noindent from the fact that the K-matrix is a special $n \equiv 2$ case of the matrix, (\citet{nepomechie2017}),

\[
\begin{bmatrix}
k_0 \big( u \big) \textbf{I}_{n \times n} & & &   \\ & k_1 \big( u \big) & k_2 \big( u \big) & \\ & k_3 \big( u \big) & k_4 \big( u \big) & \\ & & & k_5 \big( u \big) \textbf{I}_{n \times n}
\end{bmatrix} \text{, }  
\]

\noindent which amounts to the matrix,

\[
\begin{bmatrix}
k_0 \big( u \big) \textbf{I}_{2 \times 2} & & &   \\ & k_1 \big( u \big) & k_2 \big( u \big) & \\ & k_3 \big( u \big) & k_4 \big( u \big) & \\ & & & k_5 \big( u \big) \textbf{I}_{2 \times 2}
\end{bmatrix} \text{, }  
\]

\noindent for arbitrary boundary parameter $\xi_{-}$, and functions,

\begin{align*}
k_0 \big( u \big) \equiv \big( \mathrm{exp} \big( 2 u \big) + \mathrm{exp} \big( 2 n \eta \big) \bigg[\xi^2_{-} \mathrm{exp} \big( u + 2 n \eta \big)   - \mathrm{exp} \big( - u \big) \bigg]  \text{, }   \\   k_1 \big( u \big) \equiv  \frac{1}{2} \big( \mathrm{exp} \big( 2 u \big) +1 \big) \bigg[     2 \xi_{-} \mathrm{exp} \big( 2 n \eta \big) \big( \mathrm{exp} \big( 2 u \big) - 1 \big) - \mathrm{exp} \big( u \big) \big( 1 - \xi^2_{-} \mathrm{exp} \big( 2 n \eta \big) \big) \\ \times \big( 1 + \mathrm{exp} \big( 2 n \eta \big)  \big) \bigg]     \text{, }  \\    k_2 \big( u \big)  \equiv  k_3 \big( u \big) \equiv \frac{1}{2} \mathrm{exp} \big( u \big) \big( \mathrm{exp} \big( 2 u \big) - 1 \big) \big( 1 + \xi^{2}_{-} \mathrm{exp} \big( 2 n \eta \big) \big)  \\  \times \big( 1 - \mathrm{exp} \big( 2 n \eta \big) \big)       \text{, } \\  
k_4 \big( u \big)  \equiv    \frac{1}{2} \big( \mathrm{exp} \big( 2 u \big) +1 \big) \bigg[       - 2 \xi_{-} \mathrm{exp} \big( 2 n \eta \big) \big( \mathrm{exp} \big( 2 u \big) - 1 \big) - \mathrm{exp} \big( u \big) \big( 1 - \xi^2_{-} \mathrm{exp} \big( 2 n \eta \big) \\ \times  \big( 1 + \mathrm{exp} \big( 2 n \eta \big)         \bigg]\text{, }   \\  k_5 \big( u \big) \equiv       \big( \mathrm{exp} \big( 2 u \big) + \mathrm{exp} \big( 2 n \eta \big) \big) \big( \xi^2_{-} \mathrm{exp} \big( u + 2 n \eta \big) - \mathrm{exp} \big( 3 u \big) \big)    \text{, }  
\end{align*}

\bigskip

\noindent for a real parameter $\eta$. The trace of the product of matrices enforcing open boundary conditions, and the R matrices, is obtained by setting $a \equiv 0$ from,

\begin{align*}
  \mathrm{tr}_a \bigg[\textbf{K}^{\mathrm{Quasi-periodic}}_{+,a}  \big( u \big)    \bigg[ \prod_{1 \leq j \leq L}  R_{+,aj} \big( u \big)  \bigg]    \textbf{K}^{\mathrm{Quasi-periodic}}_{-,a} \big( u \big) \bigg[ \underset{1 \leq j^{\prime} \leq L}{\prod}      R_{-,j^{\prime}a}  \big( u \big)  \bigg]    \bigg] \text{. } 
\end{align*}

\bigskip

\noindent From the transfer matrix with open boundary conditions, one can introduce an open integrable Hamiltonian, which can be obtained from rearranging the expression above. To obtain the desired expression for the open, integrable Hamiltonian, we analyze the derivative of the transfer matrix above, upon set $u \equiv 0$,

\begin{align*}
 \bigg[\textbf{T}^{\mathrm{Quasi-periodic}} \big( 0 \big) \bigg]^{\prime} \equiv    \bigg[\mathrm{tr}_0 \bigg[\textbf{K}_{+,0}  \big( 0 \big)   \bigg[  \prod_{1 \leq j \leq L}  R_{+,j0} \big( 0 \big)    \bigg]  \textbf{K}_{-,0} \big( 0 \big) \bigg[ \underset{1 \leq j^{\prime} \leq L}{\prod}      R_{-,j^{\prime}0}  \big( 0 \big)   \bigg]   \bigg] \bigg]^{\prime}   \text{, }  
\end{align*}

\noindent from solutions to the Bethe equations, which can be formulated by observing that the transfer matrix, with open boundary conditions for the higher rank spin chain, satisfies, along the lines of arguments presented in (\citet{frahm2023}),

\begin{align*}
  \textbf{T} \big( u \big) \ket{\Lambda} = \Lambda \big( u \big)  \ket{\Lambda} \text{, }  \\    \textbf{h}_j \ket{\Lambda} = h_j \ket{\Lambda}  \text{, }  
\end{align*}

\noindent for $1 \leq j \leq 2$, where $\ket{\Lambda}$ denotes the normalized eigenstate of $\textbf{T} \big( u \big)$. From the two relations provided above, in the presence of twisted boundary conditions parameterized by the angles $\phi_1$ and $\phi_2$, the eigenvalues take the form, (\citet{frahm2023}),

\begin{align*}
  \Lambda \big( u \big) \equiv \big[ 4 \mathrm{sinh} \big( u - 2 i \gamma \big) \mathrm{sinh} \big( u - 4 i \gamma \big)]^L \mathrm{exp} \big( i \phi_1 \big) A \big( u \big)  +  \big[ \big( 4 \mathrm{sinh} \big( u  - 4 i \gamma \big) \\ \times \mathrm{sinh} \big( u \big) \big]^L \bigg[\mathrm{exp}  \big( i \phi_2 \big) B_1 \big( u \big) + B_2 \big( u \big) + B_3 \big( u \big) + \mathrm{exp} \big( - i \phi_2  \big) B_4 \big( u \big)  \bigg]\\ +   \big[  4 \mathrm{sinh} \big( u - 2 i \gamma \big) \mathrm{sinh} \big( u \big) \big]^L  \mathrm{exp} \big( - i \phi_1 \big) 
 C \big( u \big)    \text{, }  
\end{align*}

\noindent for quantities exhibiting the dependencies,

\begin{align*}
 A \big[ u , u^{[1]}_{j} , \gamma \big]   \equiv   A \big( u \big)    \text{, }  \\ B_1 \big[ u , u^{[1]}_j , \gamma \big]  \equiv B_1 \big( u \big) \text{, }   \\ B_2 \big[ u , u^{[2]}_j , \gamma \big] \equiv B_2 \big( u \big)  \text{, }  \\ 
 B_3 \bigg[ B_2 \big( u \big) , u , u^{[2]}_j , \gamma  \bigg]\equiv B_3 \big( u \big)  \text{, } \\  B_4 \big[ u , u^{[1]}_j , u^{[2]}_j , \gamma \big]  \equiv  B_4 \big( u  \big) \text{, }   \\  C  \bigg[A \big( u \big) , u , u^{[1]}_j , \gamma \bigg]\equiv   C \big( u \big) \text{, }  
\end{align*}

\noindent for the parameter $\gamma \in \big( 0 , \frac{\pi}{4} \big)$, and Bethe roots of the first, and second types,  $u^{[1]}_j$, and $u^{[2]}_j$, respectively. In the presence of twisted boundary conditions, the Bethe equations are,

\begin{align*}
 \bigg[ \frac{\mathrm{sinh} \big( u^{[1]}_j - i \gamma   \big) }{\mathrm{sinh}\big( u^{[1]}_j + i \gamma  \big) } \bigg]^L  =     \overset{m_1}{\underset{k \neq j}  {\prod}}     \overset{m_2}{\underset{k=1}{\prod} }         \bigg[  \bigg[  \frac{\mathrm{sinh} \big( u^{[1]}_j - u^{[1]}_k - 2 i \gamma \big) }{\mathrm{sinh}\big( u^{[1]}_j - u^{[1]}_k + 2 i \gamma  \big) }   \bigg]  \bigg[ \frac{\mathrm{sinh}\big( u^{[1]}_j - u^{[2]}_k + i \gamma \big) }{\mathrm{sinh}\big(  u^{[1]}_j - u^{[2]}_k - i \gamma \big) }    \bigg]   \bigg]    \text{. } 
\end{align*}

\bigskip

\noindent For the higher rank spin chain of the same length $L$ with open boundary conditions, the normalized eigenstates, $\ket{\Lambda^{\mathrm{Quasi-Periodic}}}$ would satisfy,

\begin{align*}
          \textbf{T}^{\mathrm{Quasi-periodic}} \big( u \big) \ket{\Lambda^{\mathrm{Quasi-periodic}}} = \Lambda^{\mathrm{Quasi-periodic}} \big( u \big)  \ket{\Lambda^{\mathrm{Quasi-periodic}}}  \text{, }  \\      \textbf{h}_j \ket{\Lambda^{\mathrm{Quasi-periodic}}} = h_j \ket{\Lambda^{\mathrm{Quasi-periodic}}}        \text{. } 
\end{align*}

\noindent Irrespective of an explicit form of the eigenstates from the first equality above, asymptotically the Hamiltonian takes the form,

\begin{align*}
   \frac{\mathrm{d}}{\mathrm{d} u} \bigg\{ \mathrm{log} \big( \textbf{T}^{\mathrm{Quasi-periodic}} \big( u \big) \big) \bigg\}  \bigg|_{u \equiv 0} \text{, } 
\end{align*}

\noindent from the logarithmic derivative of the higher rank spin chain transfer matrix with open boundary conditions, (\citet{nepomechie2017}),

\begin{align*}
   \mathcal{H} \big( k , \kappa , \textbf{K}_{-} , \textbf{K}_{+} \big)   \equiv \mathcal{H} \sim    \underset{1 \leq k \leq N-1}{\sum} h_{k,k+1} + \frac{1}{2\kappa} \bigg[\textbf{K}^{-}_1  \big( 0 \big) \bigg]^{\prime}   +   \frac{1}{\mathrm{tr} \big( \textbf{K}_{+} \big( 0 \big) \big)  }  \mathrm{tr}_0 \textbf{K}_{0,+} \big( 0 \big) h_{N0} \text{, }  
\end{align*}

\noindent for the two-site Hamiltonian appearing in the first term,

\begin{align*}
  h_{k,k+1} = \frac{1}{\xi \big( 0 \big)} \mathcal{P}_{k,k+1} \bigg[R_{k,k+1} \big( 0 \big)  \bigg]^{\prime}   \text{, }  
\end{align*}

\noindent and another Hamiltonian term appearing in the third term,

\begin{align*}
   h_{N0} \equiv  \frac{1}{\xi \big( 0 \big)} \mathcal{P}_{N,0} \bigg[R_{N,0} \big( 0 \big)  \bigg]^{\prime}\text{, }  
\end{align*}

\noindent for some parameter $\kappa$ and a permutation $\mathcal{P}$, given by,

\begin{align*}
   \mathcal{P} \equiv      \underset{1 \leq a, \beta \leq d}{\sum} e_{\alpha \beta}\otimes e_{\beta \alpha} \text{, }  
\end{align*}

\noindent over the basis for the tensor product of $d$-dimensional vector spaces, $\mathcal{V} \otimes \mathcal{V}$, and the function,

\begin{align*}
  \xi \big( u \big) \equiv       4 \mathrm{sinh} \big( u + 2 \eta \big) \mathrm{sinh} \big( u + 4 \eta \big)           \text{, } 
\end{align*}

\noindent and \textit{elementary matrices} $e$, where $e_{\alpha \beta} \equiv \delta_{\alpha,i} \delta_{\beta,j}$, namely the product of the two $\delta$ functions at row $i$ and column $j$, respectively. To perform several computations with the Jimbo R-matrix, whihc is ultimately reliant upon tensor products of elementary matrices of the above form defined with $\mathcal{P}$, we make use of the following series of correspondences:

  \[ \mathscr{R}_1 \propto   \left\{\!\begin{array}{ll@{}>{{}}l} 
    \underset{\alpha < n+1, \beta = n+1, n+2}{\sum} \big[ e_{\alpha \beta} \otimes e_{\beta \alpha} \big] \Longleftrightarrow \underset{\alpha < n+1, \beta = n+1, n+2}{\mathrm{span}} \big[ e_{\alpha \beta} \otimes e_{\beta \alpha} \big]  ,   \\ \\   \bigg[ \underset{\alpha < n+1, \beta = n+1, n+2}{\sum}  e_{\alpha \beta} \bigg]  \otimes \bigg[  \underset{\alpha < n+1}{\sum} e_{\beta^{\prime} \alpha} \bigg]     \Longleftrightarrow   \bigg[ \underset{\alpha < n+1, \beta = n+1, n+2}{\mathrm{span}}  e_{\alpha \beta} \bigg]  \\  \otimes \bigg[  \underset{\alpha < n+1}{\sum} e_{\beta^{\prime} \alpha} \bigg]   ,    \\ \\   - e_{\beta^{\prime} \alpha^{\prime}}   \otimes        \bigg[ \underset{\beta = n+1 , n+2}{\sum} e_{\alpha^{\prime}\beta}  \bigg] \Longleftrightarrow   - e_{\beta^{\prime} \alpha^{\prime}}   \otimes        \bigg[ \underset{\beta = n+1 , n+2}{\mathrm{span}} e_{\alpha^{\prime}\beta}  \bigg] ,  
\end{array}\right. 
\]

  \[ \mathscr{R}_2 \propto   \left\{\!\begin{array}{ll@{}>{{}}l} 
            \bigg[ \underset{\alpha \neq n+1, n+2}{\sum}     e_{\alpha \alpha} \bigg]  \bigotimes  \bigg[ \underset{\alpha \neq n+1, n+2}{\sum}   e_{\alpha \alpha} \bigg] \Longleftrightarrow       \underset{\alpha \neq n+1, n+2}{\mathrm{span}}      \big[ e_{\alpha \alpha} \otimes e_{\alpha \alpha} \big]   ,  \\ \\  \bigg[ \underset{\alpha = n+1 , n+2}{\sum} e_{\alpha \alpha}   \bigg]  \otimes e_{\alpha^{\prime} \alpha^{\prime}}  \Longleftrightarrow  \bigg[ \underset{\alpha = n+1 , n+2}{\mathrm{span}} e_{\alpha \alpha}   \bigg]  \otimes e_{\alpha^{\prime} \alpha^{\prime}}      ,  
\end{array}\right. 
\]

  \[  \mathscr{R}_3 \propto  \left\{\!\begin{array}{ll@{}>{{}}l}  
            \underset{\alpha \text{ } \text{or} \text{ } \beta, \beta^{\prime} \neq n+1, n+2}{\underset{\alpha \neq \beta, \beta^{\prime}}{\sum}}  e_{\alpha \alpha}  \otimes e_{\beta\beta} \Longleftrightarrow  \underset{\alpha \text{ } \text{or} \text{ } \beta, \beta^{\prime} \neq n+1, n+2}{\underset{\alpha \neq \beta, \beta^{\prime}}{\mathrm{span}}}  \big[ e_{\alpha \alpha}  \otimes e_{\beta\beta} \big] ,  \\ \\   \underset{\alpha = n+1, n+2}{\sum} e_{\alpha \alpha} \otimes e_{\alpha \alpha} \Longleftrightarrow      \underset{\alpha = n+1, n+2}{\mathrm{span}} \big[ e_{\alpha \alpha} \otimes e_{\alpha \alpha} \big] , 
\end{array}\right. 
\]

\[ \mathscr{R}_4 \propto   \left\{\!\begin{array}{ll@{}>{{}}l} 
      \bigg[ \bigg[ \underset{\alpha, \beta \neq n+1, n+2}{\sum}  e_{\alpha \beta} \bigg] \otimes \bigg[   \underset{\alpha, \beta \neq n+1, n+2}{\sum}     e_{\alpha^{\prime} \beta^{\prime}}   \bigg] \bigg]  \Longleftrightarrow   \bigg[ \bigg[ \underset{\alpha, \beta \neq n+1, n+2}{\mathrm{span}}  e_{\alpha \beta} \bigg] \\  \otimes \bigg[   \underset{\alpha, \beta \neq n+1, n+2}{\mathrm{span}}     e_{\alpha^{\prime} \beta^{\prime}}   \bigg] \bigg]  \\  \\  2 \bigg[ \bigg[    \underset{\alpha \neq n+1, n+2, \beta = n+1, n+2}{\sum} e_{\alpha \beta}     \bigg] \otimes  \bigg[   \underset{\alpha, \beta \neq n+1, n+2}{\sum}     e_{\alpha^{\prime} \beta^{\prime}}  \bigg]  \bigg]     \Longleftrightarrow  2 \bigg[ \bigg[    \underset{\alpha \neq n+1, n+2, \beta = n+1, n+2}{\mathrm{span}} e_{\alpha \beta}     \bigg] \\ \otimes  \bigg[   \underset{\alpha, \beta \neq n+1, n+2}{\mathrm{span}}     e_{\alpha^{\prime} \beta^{\prime}}  \bigg]  \bigg]  \\ \\   2  \bigg[ \bigg[ \underset{\alpha \neq n+1, n+2, \beta = n+1, n+2}{\sum}   e_{\alpha \beta} \bigg] \otimes  \bigg[  \underset{\alpha, \beta \neq n+1, n+2}{\sum}     e_{\alpha^{\prime} \beta^{\prime}}  \bigg]  \bigg]  \Longleftrightarrow   2  \bigg[ \bigg[ \underset{\alpha \neq n+1, n+2, \beta = n+1, n+2}{\mathrm{span}}   e_{\alpha \beta} \bigg] \\ \otimes  \bigg[  \underset{\alpha, \beta \neq n+1, n+2}{\mathrm{span}}     e_{\alpha^{\prime} \beta^{\prime}}  \bigg]  \bigg]      \\ \\  \bigg[ \bigg[   \underset{\alpha, \beta \neq n+1, n+2}{\sum}  e_{\alpha \beta}        \bigg] \otimes  \bigg[     \underset{\alpha \neq n+1, n+2, \beta = n+1,n+2}{\sum}         e_{\alpha^{\prime} \beta^{\prime}}        \bigg]  \bigg]   \Longleftrightarrow    \bigg[ \bigg[   \underset{\alpha, \beta \neq n+1, n+2}{\mathrm{span}}  e_{\alpha \beta}        \bigg] \\ \otimes  \bigg[     \underset{\alpha \neq n+1, n+2, \beta = n+1,n+2}{\mathrm{span}}         e_{\alpha^{\prime} \beta^{\prime}}        \bigg]  \bigg]   \\ \\ \bigg[ \bigg[       \underset{\alpha, \beta \neq n+1, n+2}{\sum}  e_{\alpha \beta}    \bigg] \otimes  \bigg[      \underset{\alpha \neq n+1, n+2, \beta = n+1, n+2}{\sum} e_{\alpha^{\prime} \beta}      \bigg]  \bigg]   \Longleftrightarrow  \bigg[ \bigg[       \underset{\alpha, \beta \neq n+1, n+2}{\mathrm{span}}  e_{\alpha \beta}    \bigg] \\ \otimes  \bigg[      \underset{\alpha \neq n+1, n+2, \beta = n+1, n+2}{\mathrm{span}} e_{\alpha^{\prime} \beta}      \bigg]  \bigg]   \\ \\  2 \bigg[ \bigg[  \underset{\alpha \neq n+1, n+2, \beta = n+1, n+2}{\sum}   e_{\alpha \beta}      \bigg] \otimes  \bigg[      \underset{\alpha \neq n+1, n+2, \beta = n+1,n+2}{\sum}         e_{\alpha^{\prime} \beta^{\prime}}     \bigg]  \bigg]       \\    \vdots    \end{array}\right. 
\]

  \[   \left\{\!\begin{array}{ll@{}>{{}}l}  \vdots \\ \Longleftrightarrow 2 \bigg[ \bigg[  \underset{\alpha \neq n+1, n+2, \beta = n+1, n+2}{\mathrm{span}}   e_{\alpha \beta}      \bigg]    \otimes  \bigg[      \underset{\alpha \neq n+1, n+2, \beta = n+1,n+2}{\mathrm{span}}         e_{\alpha^{\prime} \beta^{\prime}}     \bigg]  \bigg]    \\ \\   2 \bigg[ \bigg[  \underset{\alpha \neq n+1, n+2, \beta = n+1, n+2}{\sum}   e_{\alpha \beta}      \bigg] \otimes  \bigg[   \underset{\alpha \neq n+1, n+2, \beta = n+1, n+2}{\sum} e_{\alpha^{\prime} \beta}        \bigg]  \bigg]  \\ \Longleftrightarrow  2 \bigg[ \bigg[  \underset{\alpha \neq n+1, n+2, \beta = n+1, n+2}{\mathrm{span}}   e_{\alpha \beta}      \bigg] \otimes  \bigg[   \underset{\alpha \neq n+1, n+2, \beta = n+1, n+2}{\mathrm{span}} e_{\alpha^{\prime} \beta}        \bigg]  \bigg] ,  
\end{array}\right. 
\]

\[ \mathscr{R}_5 \propto   \left\{\!\begin{array}{ll@{}>{{}}l}     \bigg[ \bigg[  \underset{\alpha \neq n+1, n+2, \beta = n+1, n+2}{\sum} b^+_{\alpha} \big( u \big) e_{\beta^{\prime} \alpha^{\prime}}  \bigg]   \otimes       \big[ e_{\beta \alpha} \big] \bigg] \Longleftrightarrow   \underset{\alpha \neq n+1, n+2, \beta = n+1, n+2}{\mathrm{span}}  \bigg[   b^+_{\alpha} \big( u \big) e_{\beta^{\prime} \alpha^{\prime}}  \\    \otimes       \big[ e_{\beta \alpha} \big] \bigg]  ,  \\ \\  \bigg[ \bigg[  \underset{\alpha \neq n+1, n+2, \beta = n+1, n+2}{\sum} b^-_{\alpha} \big( u \big) e_{\beta \alpha^{\prime}} \bigg]  \otimes       \big[ e_{\beta \alpha} \big] \bigg]   \Longleftrightarrow   \underset{\alpha \neq n+1, n+2, \beta = n+1, n+2}{\mathrm{span}}   \bigg[ b^-_{\alpha} \big( u \big) e_{\beta \alpha^{\prime}}  \\ \otimes       \big[ e_{\beta \alpha} \big] \bigg] 
  ,  
\end{array}\right. 
\]

\noindent $\mathscr{R}_1, \cdots, \mathscr{R}_5$ are defined, and further manipulated, in \textit{3.0.1}. These terms denote components of the Jimbo R-matrix, $R_J$, which are used to reconstruct the transfer matrix of the $D^{(2)}_3$ spin-chain under open boundary conditions, and exhibit the following dependencies,

\begin{align*}
    \mathscr{R}_1 \equiv \mathscr{R}_1 \big( \alpha , \beta , \beta^{\prime} \big) , \\ \\  \mathscr{R}_2 \equiv \mathscr{R}_2 \big( \alpha , \alpha^{\prime} \big) ,\\ \\  \mathscr{R}_3 \equiv \mathscr{R}_3 \big(    \alpha  , \beta  \big) , \\ \\  \mathscr{R}_4 \equiv \mathscr{R}_4 \big( \alpha , \alpha^{\prime} , \beta , \beta^{\prime} \big) ,\\ \\  \mathscr{R}_5 \equiv \mathscr{R}_5 \big( \alpha^{\prime} , \beta , \beta^{\prime}  \big) ,
\end{align*}

\noindent Explicitly, in \textit{3.0.1}, each $\mathscr{R}$ corresponding to terms of the Jimbo R-matrix will be shown to take the following form,

 { \tiny   \begin{align*} \mathscr{R}_1 \equiv   \left\{\!
  .  
\end{array}\right. 
\end{align*} }

\subsection{Integrability of the Hamiltonian under higher rank, quasi-periodic boundary conditions}

\noindent Equipped with the transfer matrix under open boundary conditions, and the corresponding Hamiltonian, we identify eigenvectors of the Hamiltonian obtained in the previous section. To do this, observe,

\begin{align*}
 E \propto \big( \Lambda^{\mathrm{Quasi-periodic}} \big( 0 \big) \big)^{\prime\prime}   \text{, }  
\end{align*}

\noindent from which we write, {[1]},

\begin{align*}
   E  \equiv   - \underset{1 \leq k \leq m_1}{\sum}  \frac{2 \text{ } \mathrm{sinh}^2 \big( 2 i \gamma \big) }{\mathrm{cosh} \big( 2 u^{[1]}_k \big) - \mathrm{cosh} \big( 2 i \gamma \big) }  \text{, }  
\end{align*}

\noindent corresponding to the energy of the eigenvalues, termed the eigenenergies. To establish the connection between the transfer matrix, integrable Hamiltonian, and boundary CFT, the summation for $E$ above can be expressed as, (\citet{Robertson2019}),

\begin{align*}
  E = f_0 L         +   f_s - \frac{\pi v_F  \big( \frac{c}{24} - h \big) }{L} + \mathcal{O} \big( L^{-2} \big) \text{, }  
\end{align*}

\noindent for the length $L$ of the chain, which coincides with the system size, the central charge, conformal weight of the field, the bulk energy density $f_0$, surface energy $f_s$, and Fermi velocity,

\begin{align*}
 v_F \equiv \frac{2 \pi \text{ }  \mathrm{sin} \big( \pi - 2 \gamma \big) }{\pi - 2 \gamma}   \text{. } 
\end{align*}

\noindent Furthermore, observe from the equation,

\begin{align*}
   \textbf{h}_j \ket{\Lambda^{\mathrm{Quasi-Periodic}}} = h_j \ket{\Lambda^{\mathrm{Quasi-Periodic}}}     \text{, }  
\end{align*}

\noindent that the eigenvalues are given by,

\begin{align*}
  h^{\mathrm{Quasi-periodic}}_1 \equiv h_1 \equiv        L - m_1     \text{, }  \\ h^{\mathrm{Quasi-periodic}}_2 \equiv h_2  \equiv          m_1 - m_2       \text{, } 
\end{align*}

\noindent for parameters $m_1 \geq  m_2 \geq 0$. The expression for the summation over $k$ given for $E$ above is obtained from the leading term of an expansion of the transfer matrix,

\begin{align*}
 \textbf{T}^{\mathrm{Quasi-periodic}} \big( 0 \big)  \approx \big[  4 \mathrm{sinh} \big( 2 i \gamma \big) \mathrm{sinh} \big( 4 i \gamma \big)             \big]^L  \mathrm{exp} \big(  i        \textbf{P}         \big) \text{, }  
\end{align*}

\noindent where the term multiplying the $L$ th power is given by,

\begin{align*}
 \underset{1 \leq i \leq L}{\prod}          \delta^{b_{a+i}}_{a_i}                  \text{, }  
\end{align*}

\noindent the translation operator under open boundary conditions, under the equivalence, for some $j>0$,

\begin{align*}
      \big(   a_{L+i+j} \big) \mathrm{mod} \text{ } L   \equiv  a_{i+j}  \text{, }  \\ \big(   b_{L+i+j}         \big) \mathrm{mod} \text{ } L   \equiv b_{i+j}   \text{, }   \\          a_L         \equiv       b_1        \text{. } 
\end{align*}

\noindent In turn, substituting the leading order term for the natural logarithm of $\textbf{T}^{\mathrm{Open}} \big( 0 \big)$ into the expansion for the Hamiltonian,

\begin{align*}
     \textbf{H}^{\mathrm{Quasi-periodic}} \approx - \mathrm{sinh} \big( 2 i \gamma \big) \bigg[ \frac{\mathrm{d}}{\mathrm{d} u} \bigg\{  \mathrm{log} \big( \textbf{T}^{\mathrm{Quasi-periodic}} \big( u \big)\big)  \bigg\}  \bigg|_{u \equiv 0}\bigg] + L \text{ } \mathrm{sinh} \big( 2 i \gamma \big) \\ \times  \big[             \mathrm{coth} \big( 2 i \gamma \big) + \mathrm{coth} \big( 4 i \gamma \big)        \big] \textbf{I}^{\otimes L}          \text{, }  
\end{align*}

\noindent yields an expression for a local Hamiltonian, 

\begin{align*}
 - \mathrm{sinh} \big( 2 i \gamma \big) \bigg\{          \frac{\mathrm{d}}{\mathrm{d} u} \bigg[\mathrm{log} \bigg[    \mathrm{tr}_0 \bigg[\textbf{K}^{\mathrm{Quasi-periodic}}_{+,0}  \big( u \big)   \bigg[  \prod_{1 \leq j \leq L}  R_{+,a0} \big( u \big)   \bigg]   \textbf{K}^{\mathrm{Quasi-periodic}}_{-,0} \big( u \big) \\ \times  \bigg[ \underset{1 \leq j^{\prime} \leq L}{\prod}      R_{-,j^{\prime}0}  \big( u  \big) \bigg]  \bigg]         \bigg]  \bigg\}   L \text{ } \mathrm{sinh} \big( 2 i \gamma \big)  \big[             \mathrm{coth} \big( 2 i \gamma \big)  +  \mathrm{coth} \big( 4 i \gamma \big)        \big] \textbf{I}^{\otimes L}  \end{align*}
 
 \noindent which is equivalent to, after collecting like terms,
 
 \begin{align*}
 - \mathrm{sinh} \big( 2 i \gamma \big) \bigg[          \frac{\mathrm{d}}{\mathrm{d} u} \bigg[\mathrm{log} \bigg[    \mathrm{tr}_0 \bigg[\textbf{K}^{\mathrm{Quasi-periodic}}_{+,0}  \big( u \big)  \bigg[   \prod_{1 \leq j \leq L}  R_{+,a0} \big( u \big)  \bigg]    \textbf{K}^{\mathrm{Quasi-periodic}}_{-,0} \big( u \big) \\ \times  \bigg[ \underset{1 \leq j^{\prime} \leq L}{\prod}      R_{-,j^{\prime}0}  \big( u \big)   \bigg]        \bigg]    \bigg]   + L  \text{ }           \big[             \mathrm{coth} \big( 2 i \gamma \big)  +  \mathrm{coth} \big( 4 i \gamma \big)        \big] \textbf{I}^{\otimes L}                \bigg]            \text{, }  
\end{align*}

\noindent in terms of the site translation operator. Computing the derivative of the natural logarithm of the transfer matrix for the lower rank spin chain under open boundary conditions, and evaluating at $u \equiv 0$, yields approximately to first order,

\begin{align*}
     \bigg[        \mathrm{tr}_0 \bigg[\textbf{K}^{\mathrm{Quasi-periodic}}_{+,0}  \big( u \big)    \bigg[  \prod_{1 \leq j \leq L}  R_{+,a0} \big( u \big) \bigg]     \textbf{K}^{\mathrm{Quasi-periodic}}_{-,0} \big( u \big) \bigg[ \underset{1 \leq j^{\prime} \leq L}{\prod}      R_{-,j^{\prime}0}  \big( u \big) \bigg]     \bigg]   
 \bigg]^{-1}   \\ \times        \bigg[                   \big[  4 \mathrm{sinh} \big( 2 i \gamma \big) \mathrm{sinh} \big( 4 i \gamma \big)             \big]^L  \mathrm{exp} \big(  i        \textbf{P}         \big)           \bigg]      \text{. } 
\end{align*}

\noindent implies the approximation,

\begin{align*}
       - \mathrm{sinh} \big( 2 i \gamma \big) \bigg[            \bigg[        \mathrm{tr}_0 \bigg[\textbf{K}^{\mathrm{Quasi-periodic}}_{+,0}  \big( u \big)    \bigg[ \prod_{1 \leq j \leq L}  R_{+,a0} \big( u \big)    \bigg]  \textbf{K}^{\mathrm{Quasi-periodic}}_{-,0} \big( u \big) \bigg[ \underset{1 \leq j^{\prime} \leq L}{\prod}      R_{-,j^{\prime}0}  \big( u \big) \bigg]     \bigg]   
 \bigg]^{-1}       \\ \times   \bigg[                   \big[  4 \mathrm{sinh} \big( 2 i \gamma \big) \mathrm{sinh} \big( 4 i \gamma \big)             \big]^L  \mathrm{exp} \big(  i        \textbf{P}         \big)           \bigg] +  L \text{ }          \big[             \mathrm{coth} \big( 2 i \gamma \big) + \mathrm{coth} \big( 4 i \gamma \big)        \big] \textbf{I}^{\otimes L}                  \bigg]      \text{. } 
\end{align*}

\noindent for the quasi-periodic boundary condition Hamiltonian.

\subsection{Statement of the Bethe equations for anisotropy parameters approaching $0$, and the root density approach}

\noindent For anisotropy parameters that are very close to $0$, the Bethe equations,

\begin{align*}
 \bigg[ \frac{\mathrm{sinh} \big( u^{[1]}_j - i \gamma   \big) }{\mathrm{sinh}\big( u^{[1]}_j + i \gamma  \big) } \bigg]^L  =     \overset{m_1}{\underset{k \neq j}  {\prod}}  \text{ }     \overset{m_2}{\underset{k=1}{\prod} }         \bigg[    \text{ } \bigg[ \frac{\mathrm{sinh} \big( u^{[1]}_j - u^{[1]}_k - 2 i \gamma \big) }{\mathrm{sinh}\big( u^{[1]}_j - u^{[1]}_k + 2 i \gamma  \big) }   \bigg] \text{ }      \bigg[ \frac{\mathrm{sinh}\big( u^{[1]}_j - u^{[2]}_k + i \gamma \big) }{\mathrm{sinh}\big(  u^{[1]}_j - u^{[2]}_k - i \gamma \big) }    \bigg]   \bigg]   \text{, }  
\end{align*}

\noindent can be approximated with the relations,

\begin{align*}
       \bigg[           \frac{u^{[1]}_j - i}{u^{[1]}_j + i}        \bigg]^L       =        \overset{m_1}{\underset{k \neq j}  {\prod}}  \text{ }     \overset{m_2}{\underset{k=1}{\prod} }       \bigg[         \text{ } \bigg[ \frac{u^{[1]}_j - u^{[k]}_k - 2 i }{u^{[1]}_j - u^{[2]}_k + 2 i }              \bigg] \text{ } \bigg[          \frac{u^{[1]}_j - u^{[2]}_k + i }{u^{[1]}_j - u^{[2]}_k - i }                       \bigg]             \bigg]        \text{. } 
\end{align*}

\noindent From the fact that the spin-chain has rank two, there exists a mapping between pairs $\big\{ \lambda_j , - \lambda_j  \big\}$, and two possible solutions to the Bethe equations,

\begin{align*}
\lambda_j   \Longleftrightarrow       u^{[1]}_j       \text{, }  \\  - \lambda_j \Longleftrightarrow        -   u^{[1]}_j         \text{, }  \\ \lambda_k   \Longleftrightarrow       u^{[1]}_k   \text{, }  \\   - \lambda_k \Longleftrightarrow        -   u^{[1]}_k  \text{, } 
\end{align*}

\noindent take the form,

\begin{align*}
    \bigg[           \frac{\lambda_j - i}{\lambda_j + i}        \bigg]^L       \approx        \overset{m_1}{\underset{k \neq j}  {\prod}}  \text{ }     \overset{m_2}{\underset{k=1}{\prod} } \bigg[ \text{ }       \bigg[         \frac{\lambda_j - \lambda_k - 2 i }{\lambda_j - \lambda_k + 2 i }              \bigg] \text{ } \bigg[          \frac{\lambda_j - \lambda_k + i }{\lambda_j - \lambda_k - i }                       \bigg]    \bigg]                 \text{, }  
\end{align*}

\noindent under the assumption that,

\begin{align*}
 \mathrm{sin} \big( u^{[1]}_j - i \gamma \big)  \approx   u^{[1]}_j - i  \text{, }  \\   \mathrm{sin} \big( u^{[1]}_j + i \gamma \big)  \approx   u^{[1]}_j + i  \text{, }   
\end{align*}

\noindent for $\gamma \approx 0$. Under the identification, (\citet{frahm2023}),

\begin{align*}
      u^{[1]}_j \longrightarrow     x_j + \delta^{[1]}_j + i \frac{\pi}{2} - i \big( \gamma - \epsilon^{[1]}_j \big)                 \text{, }  \\   u^{[2]}_j \longrightarrow      x_j + \delta^{[2]}_j + i  \big( \frac{\pi}{2} +  \epsilon^{[2]}_j      \big)          \text{, }  
\end{align*}

\noindent of the first and second roots of the Bethe equation, for sufficiently small parameters,

\begin{align*}
 \delta^{[1]}_j , \delta^{[2]}_j , \epsilon^{[1]}_j , \epsilon^{[2]}_j  \in \textbf{R}  \text{, } 
\end{align*}

\noindent whose complex conjugates satisfy,

\begin{align*}
     \bar{u^{[1]}_j} \longrightarrow     x_j + \delta^{[1]}_j - i \frac{\pi}{2} + i \big( \gamma - \epsilon^{[1]}_j \big)                 \text{, }  \\   \bar{u^{[2]}_j }\longrightarrow      x_j + \delta^{[2]}_j - i  \big( \frac{\pi}{2} +  \epsilon^{[2]}_j      \big)          \text{, }  
\end{align*}

\noindent one can substitute these expressions for the first and second root types appearing in the Bethe equation, with the following rearrangements.

\bigskip

\noindent Given the two possible root types for solutions to the Bethe equation, for even $L$, 

\begin{align*}
  \mathrm{log} \bigg[ \text{ }  \bigg|       \frac{\mathrm{sinh} \big(  u^{[1]}_j - i     \big) }{\mathrm{sinh} \big(  u^{[1]}_j + i    \big) }   \bigg|^L  \text{ } \bigg]  \approx   \mathrm{log} \bigg[ \text{ }  \bigg|       \frac{ u^{[1]}_j - i    }{ u^{[1]}_j + i     }   \bigg|^L  \text{ } \bigg] =    \mathrm{log} \bigg[ \text{ }  \bigg|       \frac{    x_j + \delta^{[1]}_j + i \frac{\pi}{2} - i \big( \gamma - \epsilon^{[1]}_j \big)  }{   x_j + \delta^{[2]}_j + i  \big( \frac{\pi}{2} +  \epsilon^{[2]}_j      \big)     }   \bigg|^L  \text{ } \bigg] \\  = L \bigg[ \text{ } \mathrm{log} \big[ \big|      x_j + \delta^{[1]}_j + i \frac{\pi}{2} - i \big( \gamma - \epsilon^{[1]}_j \big)                  \big|  \big]  -  \mathrm{log} \big[ \big|   x_j + \delta^{[2]}_j + i  \big( \frac{\pi}{2} +  \epsilon^{[2]}_j      \big)             \big|  \big] \text{ }  \bigg] \\  =  L \bigg[ \text{ }  \mathrm{log} \big[          x_j + \delta^{[1]}_j + i \frac{\pi}{2} - i \big( \gamma - \epsilon^{[1]}_j \big)         \big]  -  \mathrm{log} \big[          x_j + \delta^{[2]}_j + i  \big( \frac{\pi}{2} +  \epsilon^{[2]}_j      \big)          \big] \text{ } \bigg] \text{, }  
\end{align*}

\noindent corresponding to terms on LHS of the Bethe equations, and,

\begin{align*}
\mathrm{log} \bigg[ \text{ }  \overset{m_1}{\underset{k \neq j}  {\prod}}  \text{ }     \overset{m_2}{\underset{k=1}{\prod} }       \bigg[         \text{ } \bigg[ \frac{\mathrm{sinh} \big( u^{[1]}_j - u^{[k]}_k - 2 i \big) }{ \mathrm{sinh} \big( u^{[1]}_j - u^{[2]}_k + 2 i \big)  }              \bigg] \text{ } \bigg[          \frac{ \mathrm{sinh} \big( u^{[1]}_j - u^{[2]}_k + i \big) }{ \mathrm{sinh} \big( u^{[1]}_j - u^{[2]}_k - i \big) }                       \bigg]         \bigg]            \bigg] \\  =   \mathrm{log} \bigg[   \text{ }  \overset{m_1}{\underset{k \neq j}  {\prod}}  \text{ }     \overset{m_2}{\underset{k=1}{\prod} }       \bigg[  \frac{\mathrm{sinh} \big( u^{[1]}_j - u^{[k]}_k - 2 i \big) }{ \mathrm{sinh} \big( u^{[1]}_j - u^{[2]}_k + 2 i \big)  }     \bigg] \text{ }                  \bigg]  + \mathrm{log}  \bigg[ \text{ }          \overset{m_1}{\underset{k \neq j}  {\prod}}  \text{ }     \overset{m_2}{\underset{k=1}{\prod} }       \bigg[  \frac{ \mathrm{sinh} \big( u^{[1]}_j - u^{[2]}_k + i \big) }{ \mathrm{sinh} \big( u^{[1]}_j - u^{[2]}_k - i \big) }   \bigg]    \bigg]  \\ \approx               \mathrm{log} \bigg[  \overset{m_1}{\underset{k \neq j}  {\prod}}  \text{ }     \overset{m_2}{\underset{k=1}{\prod} }       \bigg[                  \frac{u^{[1]}_j - u^{[k]}_k - 2 i }{u^{[1]}_j - u^{[2]}_k + 2 i }                          \bigg] \text{ }                    \bigg] +  \mathrm{log} \bigg[      \overset{m_1}{\underset{k \neq j}  {\prod}}  \text{ }     \overset{m_2}{\underset{k=1}{\prod} }       \bigg[              \frac{u^{[1]}_j - u^{[2]}_k + i }{u^{[1]}_j - u^{[2]}_k - i }      \bigg]   \bigg]       \text{, }  
\end{align*}

\noindent corresponding to terms on the RHS of the Bethe equations, which can be expressed as,

\begin{align*}
    \mathrm{log} \bigg[  \overset{m_1}{\underset{k \neq j}  {\prod}}  \text{ }     \overset{m_2}{\underset{k=1}{\prod} }       \bigg[                   \frac{ x_j - x_k + \delta^{[1]}_j - \delta^{[2]}_k + i \big( 1 + \frac{\pi}{2} \big)  - i \big( \gamma - \epsilon^{[1]}_j  - \frac{\pi}{2} - \epsilon^{[2]}_k \big) }{ x_j - x_k + \delta^{[1]}_j - \delta^{[2]}_k + i \big( -  1 + \frac{\pi}{2} \big)  - i \big( \gamma - \epsilon^{[1]}_j  - \frac{\pi}{2} - \epsilon^{[2]}_k \big)   }                         \bigg]                  \bigg] \\ +  \mathrm{log} \bigg[      \overset{m_1}{\underset{k \neq j}  {\prod}}  \text{ }     \overset{m_2}{\underset{k=1}{\prod} }       \bigg[              \frac{ x_j - x_k + \delta^{[1]}_j - \delta^{[2]}_k + i \big( - 2  + \frac{\pi}{2} \big)  - i \big( \gamma - \epsilon^{[1]}_j  - \frac{\pi}{2} - \epsilon^{[2]}_k \big) }{ x_j - x_k + \delta^{[1]}_j - \delta^{[2]}_k + i \big( 2  + \frac{\pi}{2} \big)  - i \big( \gamma - \epsilon^{[1]}_j  - \frac{\pi}{2} - \epsilon^{[2]}_k \big)   }       \bigg]    \bigg]           \text{. } 
\end{align*}

\noindent Hence,

\begin{align*}
    L \bigg[ \text{ }  \mathrm{log} \big[          x_j + \delta^{[1]}_j + i \frac{\pi}{2} - i \big( \gamma - \epsilon^{[1]}_j \big)         \big] - \mathrm{log} \big[          x_j + \delta^{[2]}_j + i  \big( \frac{\pi}{2} +  \epsilon^{[2]}_j      \big)          \big] \text{ } \bigg]    \end{align*}

    \begin{align*} \approx    \mathrm{log} \bigg[  \overset{m_1}{\underset{k \neq j}  {\prod}}  \text{ }     \overset{m_2}{\underset{k=1}{\prod} }       \bigg[                   \frac{ x_j - x_k + \delta^{[1]}_j - \delta^{[2]}_k + i \big( 1 + \frac{\pi}{2} \big)  - i \big( \gamma - \epsilon^{[1]}_j  - \frac{\pi}{2} - \epsilon^{[2]}_k \big) }{ x_j - x_k + \delta^{[1]}_j - \delta^{[2]}_k + i \big( -  1 + \frac{\pi}{2} \big)  - i \big( \gamma - \epsilon^{[1]}_j  - \frac{\pi}{2} - \epsilon^{[2]}_k \big)   }                         \bigg]                  \bigg] \\ +  \mathrm{log} \bigg[      \overset{m_1}{\underset{k \neq j}  {\prod}}  \text{ }     \overset{m_2}{\underset{k=1}{\prod} }       \bigg[              \frac{ x_j - x_k + \delta^{[1]}_j - \delta^{[2]}_k + i \big( - 2  + \frac{\pi}{2} \big)  - i \big( \gamma - \epsilon^{[1]}_j  - \frac{\pi}{2} - \epsilon^{[2]}_k \big) }{ x_j - x_k + \delta^{[1]}_j - \delta^{[2]}_k + i \big( 2  + \frac{\pi}{2} \big)  - i \big( \gamma - \epsilon^{[1]}_j  - \frac{\pi}{2} - \epsilon^{[2]}_k \big)   }       \bigg]    \bigg]          \text{. } 
\end{align*}

\noindent In terms of $\lambda_j$ and $\lambda_k$, the approximate relation for the Bethe equations for anisotropy parameters that are approximately $0$ reads,

\begin{align*}
  L \text{ } \mathrm{log}   \bigg[           \frac{\lambda_j - i}{\lambda_j + i}        \bigg]       \approx    \mathrm{log} \bigg[ \text{ }     \overset{m_1}{\underset{k \neq j}  {\prod}}  \text{ }     \overset{m_2}{\underset{k=1}{\prod} } \bigg[ \text{ }       \bigg[         \frac{\lambda_j - \lambda_k - 2 i }{\lambda_j - \lambda_k + 2 i }              \bigg] \text{ } \bigg[          \frac{\lambda_j - \lambda_k + i }{\lambda_j - \lambda_k - i }                       \bigg]     \bigg]   \bigg]           \text{, }  
\end{align*}

\noindent under the identification,

\begin{align*}
\lambda_j \longrightarrow            \hat{x_j} + \hat{\delta^{[1]}_j} + i \frac{\pi}{2} - i \big( \gamma - \hat{\epsilon^{[1]}_j} \big)                           \text{, }  \\   \lambda_k \longrightarrow               \hat{x_j} + \hat{\delta^{[2]}_j} + i  \big( \frac{\pi}{2} +  \hat{\epsilon^{[2]}_j}      \big)          \text{, }  
  \end{align*}

  \noindent for sufficiently small parameters,

\begin{align*}
        \hat{x_j} , \hat{\delta^{[1]}_j} ,  \hat{\epsilon^{[1]}_j} ,    \hat{x_j}   , \hat{\delta^{[2]}_j} , \hat{\epsilon^{[2]}_j}   \in \textbf{R}  \text{. } 
\end{align*}

  \noindent Under invariance of solutions to the Bethe equations, in which solutions come in pairs $\big\{ \lambda_j , - \lambda_j \big\}$ and $\big\{ \lambda_k , - \lambda_k \big\}$, the Bethe equations take the form,

\begin{align*}
    L \bigg[ \text{ }  \mathrm{log} \bigg[          - \bigg[x_j + \delta^{[1]}_j + i \frac{\pi}{2} - i \big( \gamma - \epsilon^{[1]}_j \big)  \bigg]       \bigg] - \mathrm{log} \bigg[   - \bigg[       x_j + \delta^{[2]}_j + i  \big( \frac{\pi}{2} +  \epsilon^{[2]}_j      \big)  \bigg]        \bigg]  \bigg]    \\  \approx    \mathrm{log} \bigg[  \overset{m_1}{\underset{k \neq j}  {\prod}}  \overset{m_2}{\underset{k=1}{\prod} }       \bigg[                   \frac{ - \bigg[x_j - x_k + \delta^{[1]}_j - \delta^{[2]}_k + i \big( 1 + \frac{\pi}{2} \big)  - i \big( \gamma - \epsilon^{[1]}_j  - \frac{\pi}{2} - \epsilon^{[2]}_k \big) \bigg] }{ - \bigg[x_j - x_k + \delta^{[1]}_j - \delta^{[2]}_k + i \big( -  1 + \frac{\pi}{2} \big)  - i \big( \gamma - \epsilon^{[1]}_j  - \frac{\pi}{2} - \epsilon^{[2]}_k \big)  \bigg] }                         \bigg]                 \bigg] \\   +  \mathrm{log} \bigg[      \overset{m_1}{\underset{k \neq j}  {\prod}}  \text{ }     \overset{m_2}{\underset{k=1}{\prod} }       \bigg[              \frac{  - \bigg[ x_j - x_k + \delta^{[1]}_j - \delta^{[2]}_k + i \big( - 2  + \frac{\pi}{2} \big)  - i \big( \gamma - \epsilon^{[1]}_j  - \frac{\pi}{2} - \epsilon^{[2]}_k \big) \bigg]}{ - \bigg[x_j - x_k + \delta^{[1]}_j - \delta^{[2]}_k + i \big( 2  + \frac{\pi}{2} \big)  - i \big( \gamma - \epsilon^{[1]}_j  - \frac{\pi}{2} - \epsilon^{[2]}_k \big) \bigg]  }       \bigg]     \bigg]          \text{, }  
\end{align*}

  \noindent from the fact that the identification from {[1]} also takes the form,

\begin{align*}
      -  u^{[1]}_j \longrightarrow  -   x_j - \delta^{[1]}_j - i \frac{\pi}{2} + i \big( \gamma - \epsilon^{[1]}_j \big)            \text{, }  \\   - u^{[2]}_j \longrightarrow  -    x_j - \delta^{[2]}_j - i  \big( \frac{\pi}{2} +  \epsilon^{[2]}_j      \big)      \text{, }  
\end{align*}

  \noindent for pairs of solutions, $\big\{ u^{[1]}_j ,  - u^{[1]}_j \big\}$ and $\big\{ u^{[1]}_k ,  - u^{[1]}_k \big\}$. From the set of relations above for the Bethe equations after taking natural logarithms, one obtains the density for roots of the Bethe equations which can be used to study the ground state, (\citet{frahm2023}),

\begin{align*}
         \rho^x \big( x \big) \equiv      \frac{1}{2 \big( \pi - 4 \gamma \big) } \bigg[  \mathrm{cosh}  \big( \frac{\pi x}{\pi - 4 \gamma } \big) \bigg]^{-1}              \text{, }  
\end{align*}

\noindent from the expression for the centers of the counting function, {[1]}, from the root density approach,

\begin{align*}
      z^x \big( x \big) \equiv  \frac{1}{2 \pi }  \bigg[\psi \big( x , 2 \gamma \big) + \frac{1}{L}    \underset{1 \leq k \leq \frac{L}{2}}{\sum}   \chi \big( x - x_k , 4 \gamma \big)      \bigg]               \text{, }  
\end{align*}

\noindent for roots of the Bethe equation. For the counting function above, the two functions are given by,

\begin{align*}
     \chi \big( x , y \big)    \equiv  2 \text{ } \mathrm{arctan} \bigg[\mathrm{tanh } \big( x \big) \mathrm{cot} \big( y \big)  \bigg] \text{, }    \\       \psi \big( x , y \big)     \equiv  2 \text{ } \mathrm{arctan} \bigg[\mathrm{tanh} \big( x \big) \mathrm{tan} \big( y \big)  \bigg]\text{. } 
\end{align*}

\section{Integrable, and exactly solvable, objects for the open boundary condition higher rank spin chain}

\subsection{Overview}

\noindent In the presence of quasi-periodic boundary conditions, in the previous section, recall that the following objects were introduced:

\begin{itemize}
    \item[$\bullet$] \textit{transfer matrix}:

    \begin{align*}
        \mathrm{tr}_0 \bigg[\textbf{K}^{\mathrm{Quasi-periodic}}_{+,0}  \big( u \big)    \bigg[ \prod_{1 \leq j \leq L}  \textbf{R}_{+,a0} \big( u \big)  \bigg]    \textbf{K}^{\mathrm{Quasi-Periodic}}_{-,0} \big( u \big) \bigg[ \underset{1 \leq j^{\prime} \leq L}{\prod}      \textbf{R}_{-,j^{\prime}0}  \big( u \big)   \bigg]   \bigg] \text{. }  
\end{align*}

     \item[$\bullet$] \textit{Derivative of the transfer matrix}:

\begin{align*}
    \bigg[\mathrm{tr}_0 \bigg[\textbf{K}^{\mathrm{Quasi-periodic}}_{+,0}  \big( 0 \big)    \bigg[ \prod_{1 \leq j \leq L}  \textbf{R}_{+,j0} \big( 0 \big) \bigg]     \textbf{K}^{\mathrm{Quasi-periodic}}_{-,0} \big( 0 \big) \bigg[ \underset{1 \leq j^{\prime} \leq L}{\prod}      \textbf{R}_{-,j^{\prime}0}  \big( 0 \big)     \bigg]\bigg]  \bigg]^{\prime}   \text{, }  
\end{align*}

      \item[$\bullet$] \textit{Logarithmic derivative of the transfer matrix}:

\begin{align*}
   \frac{\mathrm{d}}{\mathrm{d} u} \bigg\{ \mathrm{log} \big( \textbf{T}^{\mathrm{Quasi-periodic}} \big( u \big) \big) \bigg\}  \bigg|_{u \equiv 0} \text{. } 
\end{align*}

\item[$\bullet$] \textit{Boundary Yang-Baxter equation}:

\begin{align*}
       R_{12} \big( u -v \big) \mathcal{K}^{\pm }_{\mathrm{Open}, 6} \big( u \big)  R_{21} \big( u + v  \big) \mathcal{K}^{\pm}_{\mathrm{Open}, 6} \big( v \big)   = \mathcal{K}^{\pm}_{\mathrm{Open}, 6} \big( v \big)  R_{12} \big( u + v \big) \\ \times   \mathcal{K}^{\pm}_{\mathrm{Open}, 6} \big( u \big)   R_{21} \big( u -v \big)       \text{. }  
\end{align*}

       \item[$\bullet$] \textit{Bethe equation in the presence of anistropy}:

\begin{align*}
 \bigg[ \frac{\mathrm{sinh} \big( u^{[1]}_j - i \gamma   \big) }{\mathrm{sinh}\big( u^{[1]}_j + i \gamma  \big) } \bigg]^L  =     \overset{m_1}{\underset{k \neq j}  {\prod}}  \text{ }     \overset{m_2}{\underset{k=1}{\prod} }         \bigg[    \text{ } \bigg[ \frac{\mathrm{sinh} \big( u^{[1]}_j - u^{[1]}_k - 2 i \gamma \big) }{\mathrm{sinh}\big( u^{[1]}_j - u^{[1]}_k + 2 i \gamma  \big) }   \bigg] \text{ }      \bigg[ \frac{\mathrm{sinh}\big( u^{[1]}_j - u^{[2]}_k + i \gamma \big) }{\mathrm{sinh}\big(  u^{[1]}_j - u^{[2]}_k - i \gamma \big) }    \bigg]   \bigg]   \text{. }  
\end{align*}

        \item[$\bullet$] \textit{Linearization of the Bethe equation in the presence of anisotropy}:

\begin{align*}
    \bigg[           \frac{\lambda_j - i}{\lambda_j + i}        \bigg]^L       \approx        \overset{m_1}{\underset{k \neq j}  {\prod}}  \text{ }     \overset{m_2}{\underset{k=1}{\prod} } \bigg[ \text{ }       \bigg[         \frac{\lambda_j - \lambda_k - 2 i }{\lambda_j - \lambda_k + 2 i }              \bigg] \text{ } \bigg[          \frac{\lambda_j - \lambda_k + i }{\lambda_j - \lambda_k - i }                       \bigg]    \bigg]                 \text{. }  
\end{align*}
        
\end{itemize}

\noindent In the forthcoming subsections of the final section, we compute each one of the quantities provided in the list above for the $D^{(2)}_3$ spin-chain, from the objects previously obtained for the $D^{(2)}_2$ spin-chain.

\subsubsection{Higher rank transfer matrix}

\noindent From the K-matrices $K_{\pm}$ introduced in \textit{1.2}, one can straightforwardly obtain the desired higher rank transfer matrix by recalling that, (\citet{nepomechie2021}), the Jimbo R-matrix, $R_J \big( u \big)$, is composed of the following contributions:

\begin{itemize}
    \item[$\bullet$] \textit{Difference of exponentials}:

    \begin{align*}
      e^{2u} - e^{4 \eta}  , \\ \\ e^{2\eta} \big( e^{2u} - 1 \big) \big( e^{2u} - e^{4n \eta} \big) , \\ \\ - \big( e^{4 \eta} - 1 \big) \big( e^{2u} - e^{4 n \eta} \big) , \\ \\ - \frac{1}{2} \big( e^{4 \eta } - 1 \big) \big( e^{2u} - e^{4n \eta  } \big) \big( e^u + 1 \big) , 
    \end{align*}

     \item[$\bullet$] \textit{Tensor product of $\alpha$ elementary matrices}:

    \begin{align*}
     \underset{\alpha \neq n+1, n+2}{\sum}   e_{\alpha\alpha} \otimes e_{\alpha\alpha} , 
    \end{align*}

      \item[$\bullet$] \textit{Tensor product of $\alpha$ and $\beta$ elementary matrices}:

      \begin{align*}
      \underset{ \alpha , \beta \neq n+1, n+2}{\underset{\alpha\neq \beta, \beta^{\prime}}{\sum}}   e_{\alpha\alpha} \otimes e_{\beta\beta} ,
      \end{align*}

       \item[$\bullet$] \textit{Tensor product of elementary matrices with interchanged indices, $\alpha \beta$ and $\beta \alpha$}:

       \begin{align*}
     \bigg[  \underset{\alpha, \beta \neq n+1, n+2}{\underset{\alpha < \beta, \alpha \neq \beta^{\prime}}{\sum}}  +   e^{2u} \underset{\alpha, \beta \neq n+1, n+2}{\underset{\alpha > \beta, \alpha \neq \beta^{\prime}}{\sum}}    \bigg] e_{\alpha \beta} \otimes e_{\beta \alpha} ,
       \end{align*}

        \item[$\bullet$] \textit{Tensor products of linear combinations of elementary matrices with interchanged indices, $\alpha \beta$ and $\beta \alpha$, with $\beta^{\prime}\alpha^{\prime}$ and $\alpha^{\prime} \beta^{\prime}$}:

\begin{align*}
 \bigg[  \underset{a < n+ 1, \beta = n+1 , n+2}{\sum} + e^u \underset{\alpha > n+2, \beta = n+1, n+2}{\sum}  \bigg]  \big( e_{\alpha \beta} \otimes e_{\beta \alpha} + e_{\beta^{\prime} \alpha^{\prime}} \otimes e_{\alpha^{\prime} \beta^{\prime}}    \big)   ,
\end{align*}
           \item[$\bullet$] \textit{Tensor product of linear combinations of elementary matrices with interchanged indices, $\alpha \beta$ and $\beta^{\prime} \alpha$, and $\beta^{\prime} \alpha^{\prime}$ and $\alpha^{\prime} \beta$}:

\begin{align*}
 \bigg[ -  \underset{a < n+ 1, \beta = n+1 , n+2}{\sum} + e^u \underset{\alpha > n+2, \beta = n+1, n+2}{\sum}  \bigg]  \big( e_{\alpha \beta} \otimes e_{\beta^{\prime} \alpha } + e_{\beta^{\prime} \alpha^{\prime}} \otimes e_{\alpha^{\prime} \beta} \big)   ,
\end{align*}

              \item[$\bullet$] \textit{Tensor product of elementary matrices}:

              \begin{align*}
                 \underset{\alpha , \beta \neq n+1,n+2}{\sum} a_{\alpha \beta} \big( u \big) e_{\alpha \beta} \otimes e_{\alpha^{\prime} \beta^{\prime}} ,
              \end{align*}

              \item[$\bullet$] \textit{Tensor product of elementary matrices, and $a$, interchanged orders of basis elements}:

              \begin{align*}
             \frac{1}{2}    \underset{\alpha \neq n+1, n+2, \beta = n+1, n+2}{\sum} \bigg[ b^+_{\alpha} \big( u \big)  \big( e_{\alpha \beta} \otimes e_{\alpha^{\prime} \beta^{\prime}} + e_{\beta^{\prime} \alpha^{\prime}} \otimes e_{\beta \alpha} \big) + b^-_{\alpha} \big( u \big)  \big( e_{\alpha \beta} \otimes e_{\alpha^{\prime} \beta} \\ + e_{\beta \alpha^{\prime}} \otimes e_{\beta \alpha} \big)\bigg]     , 
              \end{align*}

\item[$\bullet$] \textit{Tensor product of elementary matrices dependent upon $\alpha$ and $\alpha^{\prime}$ only}:

\begin{align*}
  \underset{\alpha = n+1, n+2}{\sum} \bigg[  c^+ \big( u \big) e_{\alpha \alpha} \otimes e_{\alpha^{\prime} \alpha^{\prime}} + c^{-} \big( u \big) e_{\alpha  \alpha} \otimes e_{\alpha \alpha} \bigg]  , 
\end{align*}

\item[$\bullet$] \textit{Remaining tensor product elements}:

\begin{align*}
   R_J \big( x \big)  - \underset{1 \leq i \leq 5}{\sum} \mathscr{R}_i \big( u \big) \equiv   R_J   - \underset{1 \leq i \leq 5}{\sum} \mathscr{R}_i  ,
\end{align*}

\noindent equals,

\begin{align*}
            \underset{\alpha , \beta \neq n+1, n+2}{\sum} a_{\alpha \beta} \big( u \big) e_{\alpha \beta} \otimes e_{\alpha^{\prime} \beta^{\prime}}           \\ + \underset{\alpha = n+1 , n+2}{\sum} \bigg[ d^+ \big( u \big) e_{\alpha \alpha^{\prime}} \otimes e_{\alpha^{\prime} \alpha^{\prime}} + c^- \big( u \big) e_{\alpha \alpha} \otimes e_{\alpha \alpha^{\prime}} \bigg] , 
\end{align*}

\noindent where the remaining coefficients, $a_{\alpha\beta} \big(u\big)$, along with $d^{\pm} \big( u \big)$, are defined below.

\end{itemize}

\noindent for,

\begin{align*}
    \mathscr{R}_1 \equiv \mathscr{R}_1 \big( \alpha , \beta , \beta^{\prime} \big) , \\ \\  \mathscr{R}_2 \equiv \mathscr{R}_2 \big( \alpha , \alpha^{\prime} \big) ,\\ \\  \mathscr{R}_3 \equiv \mathscr{R}_3 \big(    \alpha  , \beta  \big) , \\ \\  \mathscr{R}_4 \equiv \mathscr{R}_4 \big( \alpha , \alpha^{\prime} , \beta , \beta^{\prime} \big) ,\\ \\  \mathscr{R}_5 \equiv \mathscr{R}_5 \big( \alpha^{\prime} , \beta , \beta^{\prime}  \big) ,
\end{align*}

\noindent and, (\citet{jimbo1986}), 

\begin{align*}
   b^{\pm}_{\alpha} \big( u \big) \equiv \pm \mathrm{exp} \bigg[ 2 \eta \big( \alpha - \frac{1}{2} \big) \bigg]   \bigg[   \big( e^{4 \eta} - 1 \big) \big( e^{2u} - 1 \big) \big( e^u \pm e^{2n \eta} \big) \bigg] \Longleftrightarrow \alpha < n+1    , \\ \\    b^{\pm}_{\alpha} \big( u \big) \equiv           \mathrm{exp} \bigg[ 2 \eta \big( \alpha - n - \frac{5}{2} \big)    \bigg]   \bigg[    \big( e^{4 \eta} - 1 \big) \big( e^{2u} - 1\big) e^u \big( e^u \pm e^{2n \eta} \big)     \bigg]                \Longleftrightarrow \alpha >  n+2    , \\  \\ c^{\pm} \big( u \big) \equiv \pm \frac{1}{2} \bigg[ \big( e^{2 \eta} - 1 \big) \big( e^{2n \eta} + 1 \big) e^u \big( e^u \mp 1 \big) \big( e^u \pm e^{2n\eta} \big) + e^{2\eta} \big( e^{2u} - 1 \big) \big( e^{2u } - e^{4n \eta} \big) \bigg]  , \end{align*}

   \begin{align*} d^{\pm} \big( u \big) \equiv \pm \frac{1}{2} \bigg[ \big( e^{4 \eta} - 1 \big) \big( e^{2n \eta} - 1 \big) e^u \big( e^u \pm 1 \big) \big( e^u \pm e^{2n \eta} \big) \bigg] ,  
\end{align*}

\begin{align*}
   a_{\alpha \beta} \big( u \big)  \equiv  \big( e^{4 \eta} e^{2u} - e^{4 n \eta} \big) \big( e^{2u} - 1 \big)         \Longleftrightarrow  \alpha = \beta , \\ \\  a_{\alpha \beta} \big( u \big)  \equiv              \big( e^{4 \eta} - 1 \big)   \big( e^{4n \eta} e^{2 \eta ( \bar{\alpha} - \bar{\beta} ) } \big( e^{2u} - 1 \big) - \delta_{\alpha, \beta^{\prime}} \big( e^{2u} - e^{4 n \eta} \big) \big)    \Longleftrightarrow \alpha < \beta , \\ \\ a_{\alpha \beta} \big( u \big) \equiv                                  \big( e^{4 \eta} - 1 \big)               e^{2u} \big( e^{2 \eta ( \bar{\alpha} - \bar{\beta} ) } \big( e^{2u} - 1 \big) - \delta_{\alpha, \beta^{\prime}} \big( e^{2u} - e^{4 n \eta} \big) \big)      \Longleftrightarrow \alpha > \beta  , 
\end{align*}

\begin{align*}
  \bar{\alpha} =   \alpha + 1     \Longleftrightarrow  1 \leq \alpha \leq  n+1 , \\ \\ \bar{\alpha} =   n + \frac{3}{2}    \Longleftrightarrow  \alpha = n+1  ,\\ \\ \bar{\alpha} =   n     + \frac{3}{2}    \Longleftrightarrow  \alpha = n+2 ,  \\ \\ \bar{\alpha} =   \alpha - 1     \Longleftrightarrow    n + 2 < \alpha \leq 2n +2 , \\ \\ \alpha^{\prime} = 2 n + 3 - \alpha      . 
\end{align*}

\noindent corresponding to the R-matrix, 

\begin{align*}
  R \big( u \big)  \equiv  \mathrm{exp} \big( - 2 u - 6 \eta \big) R_J \big( x \big)  \text{, }  
\end{align*}

\noindent defined from the Jimbo R-matrix, up to an exponential prefactor. Hence, the fact that the desired \textit{double row} transfer matrix takes the form,

\begin{align*}
\textbf{T}^{\mathrm{Open}}\big( u \big)  \equiv \textbf{T}^{\mathrm{Open}}_{D^{(2)}_3} \big( u \big) \equiv  \mathrm{Tr}_0 \bigg[ \mathcal{K}_{+,0} \big( u \big) \bigg[  \prod_{1 \leq j \leq L} R_{0j} \big( u \big)   \bigg] \mathcal{K}_{-,0} \big( u \big)   \bigg[ \prod_{1 \leq j^{\prime} \leq L} R_{j^{\prime}0} \big( u \big)  \bigg]   \bigg]  , 
\end{align*}

\noindent readily implies that the Jimbo R-matrix, when multiplied by $K$-matrices of the form indicated in \textit{1.3}, can be grouped into the following terms. By the bilinearity property of the tensor product, given two real parameters $a \neq b$,

\begin{align*}
\big(  a u \otimes b v \big)+  \big(   a w \otimes b v \big)  = a \big( u +  w \big) \otimes b v   , \\ \\ \big( a u \otimes b v \big)  + \big(  a u \otimes b w \big)  = a u \otimes b \big( v + w \big)  .
\end{align*}

\noindent We make use of the two above straightforward properties of the tensor product, particularly in the below series of claims, in order to aggregate tensor products of elementary matrices, provided below, with $\mathscr{R}_1, \mathscr{R}_2, \mathscr{R}_3, \mathscr{R}_4, \mathscr{R}_5$. In particular, for computing the partial trace,

\begin{align*}
  \underset{N \longrightarrow + \infty}{\mathrm{lim}}  \textbf{T}_{N, D^{(2)}_3} \big( u \big) \equiv \underset{N \longrightarrow + \infty}{\mathrm{lim}} \bigg\{  \mathrm{Tr} \bigg[ K^{\mathrm{Open}}_{D^{(2)}_3} \bigg[ \underset{1 \leq j \leq N}{\prod} R_{D^{(2)}_3,j} \bigg]    \bigg]     \bigg\}  \equiv \underset{j,j^{\prime} \longrightarrow + \infty}{\mathrm{lim}} \bigg\{   \mathrm{Tr}_0 \bigg[ \mathcal{K}_{+,0} \big( u \big)  \\ \times   \bigg[ \prod_{1 \leq j \leq L} R_{0j} \big( u \big) \bigg]   \mathcal{K}_{-,0} \big( u \big)  \bigg[  \prod_{1 \leq j^{\prime} \leq L} R_{j^{\prime}0} \big( u \big) \bigg]  \bigg] \bigg\}  , 
\end{align*}

\noindent to obtain the desired expression for the higher transfer matrix under open boundary conditions, we make use of the following series of expressions for tensor products of elementary matrices $e$:

\begin{itemize}
\item[$\bullet$] \textit{Tensor product of elementary matrices for $\mathscr{R}_1$}. Disregarding prefactors, one has that,

\begin{align*}
   \underset{\alpha < n+1, \beta = n+1, n+2}{\sum}  e_{\alpha \beta} \otimes \big( e_{\beta \alpha}  + e_{\beta^{\prime} \alpha} \big) -   \underset{ \alpha > n+2 , \beta = n+1 , n+2}{\sum} e_{\beta^{\prime} \alpha^{\prime}} \otimes \big( e_{\alpha^{\prime} \beta^{\prime}}  \\  + e_{\alpha^{\prime}\beta}   \big)  \\ \\ =   \underset{\alpha < n+1, \beta = n+1, n+2}{\sum}  e_{\alpha \beta} \otimes \bigg[ \underset{\alpha < n+1, \beta = n+1, n+2}{\sum} e_{\beta \alpha}  + \underset{\alpha < n+1, \beta = n+1, n+2}{\sum} e_{\beta^{\prime} \alpha} \bigg] \\  -   \underset{ \alpha > n+2 , \beta = n+1 , n+2}{\sum} e_{\beta^{\prime} \alpha^{\prime}} \otimes \bigg[    \underset{ \alpha > n+2 , \beta = n+1 , n+2}{\sum} e_{\alpha^{\prime} \beta^{\prime}}  \end{align*}

   \begin{align*} +    \underset{ \alpha > n+2 , \beta = n+1 , n+2}{\sum} e_{\alpha^{\prime}\beta}   \bigg]   \\ \\ =   \underset{\alpha < n+1, \beta = n+1, n+2}{\sum}  e_{\alpha \beta} \otimes \bigg[ \underset{\alpha < n+1, \beta = n+1, n+2}{\sum} e_{\beta \alpha}  + \underset{\alpha < n+1}{\sum} e_{\beta^{\prime} \alpha} \bigg]  -   e_{\beta^{\prime} \alpha^{\prime}} \\ \otimes \bigg[    e_{\alpha^{\prime} \beta^{\prime}}     +    \underset{\beta = n+1 , n+2}{\sum} e_{\alpha^{\prime}\beta}   \bigg]  \\ \\ =   \bigg[  \underset{\alpha < n+1, \beta = n+1, n+2}{\sum}  e_{\alpha \beta} \bigg]  \otimes  \bigg[ \underset{\alpha < n+1, \beta = n+1, n+2}{\sum} e_{\beta \alpha} \bigg]  + \bigg[ \underset{\alpha < n+1, \beta = n+1, n+2}{\sum}  e_{\alpha \beta} \bigg] \\  \otimes \bigg[  \underset{\alpha < n+1}{\sum} e_{\beta^{\prime} \alpha} \bigg]  -    e_{\beta^{\prime} \alpha^{\prime}}  \otimes     e_{\alpha^{\prime} \beta^{\prime}}    - e_{\beta^{\prime} \alpha^{\prime}}   \otimes  \bigg[       \underset{\beta = n+1 , n+2}{\sum} e_{\alpha^{\prime}\beta}   \bigg]    \\ \\ =   \underset{\alpha < n+1, \beta = n+1, n+2}{\sum} \big[ e_{\alpha \beta} \otimes e_{\beta \alpha} \big] +  \bigg[ \underset{\alpha < n+1, \beta = n+1, n+2}{\sum}  e_{\alpha \beta} \bigg]  \otimes \bigg[  \underset{\alpha < n+1}{\sum} e_{\beta^{\prime} \alpha} \bigg]           \end{align*}

   \begin{align*}    -    e_{\beta^{\prime} \alpha^{\prime}}  \otimes     e_{\alpha^{\prime} \beta^{\prime}}    - e_{\beta^{\prime} \alpha^{\prime}}   \otimes        \bigg[ \underset{\beta = n+1 , n+2}{\sum} e_{\alpha^{\prime}\beta}  \bigg]  \\ \\ =  \underset{\alpha < n+1, \beta = n+1, n+2}{\mathrm{span}} \big[ e_{\alpha \beta} \otimes e_{\beta \alpha} \big] +  \bigg[ \underset{\alpha < n+1, \beta = n+1, n+2}{\mathrm{span}}  e_{\alpha \beta} \bigg]  \otimes \bigg[  \underset{\alpha < n+1}{\mathrm{span}} e_{\beta^{\prime} \alpha} \bigg]          \\   -    e_{\beta^{\prime} \alpha^{\prime}}  \otimes     e_{\alpha^{\prime} \beta^{\prime}}    - e_{\beta^{\prime} \alpha^{\prime}}   \otimes        \bigg[ \underset{\beta = n+1 , n+2}{\mathrm{span}} e_{\alpha^{\prime}\beta}  \bigg]      
\end{align*}

 { \tiny  \begin{align*} =  \begin{array}{cccccccccccccccccccccccccccccccccccc}        e_{11} \big[ \underset{\alpha < n+1, \beta = n+1, n+2}{\mathrm{span}} e_{\beta \alpha}    \big]  & \dots &  e_{n1} \big[ \underset{\alpha < n+1, \beta = n+1, n+2}{\mathrm{span}} e_{\beta \alpha}    \big] & 0 & 0 &  e_{(n+3)1} \big[ \underset{\alpha < n+1, \beta = n+1, n+2}{\mathrm{span}} e_{\beta \alpha}    \big]   & \dots    \\     \vdots   & \dots &  \vdots  & 0 & 0 &  \vdots   & \dots  \\    \vdots   & \dots &  \vdots  &  \vdots  &  \vdots  &  \vdots   & \dots           \\     e_{1(2n+2)} \big[ \underset{\alpha < n+1, \beta = n+1, n+2}{\mathrm{span}} e_{\beta \alpha}    \big]  & \dots &  e_{n(2n+2)} \big[ \underset{\alpha < n+1, \beta = n+1, n+2}{\mathrm{span}} e_{\beta \alpha}    \big] & 0 & 0 &  e_{(n+3)(2n+2)} \big[ \underset{\alpha < n+1, \beta = n+1, n+2}{\mathrm{span}} e_{\beta \alpha}    \big]   & \dots      \end{array}      
\end{align*} }

 { \tiny  \begin{align*}   \begin{array}{cccccccccccccccccccccccccccccccccccc}                                   \dots &     e_{(n+4)1} \big[ \underset{\alpha < n+1, \beta = n+1, n+2}{\mathrm{span}} e_{\beta \alpha}    \big]  & \dots &    e_{(2n+2)1} \big[ \underset{\alpha < n+1, \beta = n+1, n+2}{\mathrm{span}} e_{\beta \alpha}    \big]  \\       \dots & \vdots & \vdots & \vdots  \\   \dots & \vdots & \vdots & \vdots  \\   \dots &     e_{(n+3)(2n+2)} \big[ \underset{\alpha < n+1, \beta = n+1, n+2}{\mathrm{span}} e_{\beta \alpha}    \big]  & \dots & e_{(2n+2)(2n+2)} \big[ \underset{\alpha < n+1, \beta = n+1, n+2}{\mathrm{span}} e_{\beta \alpha}    \big]     \end{array}        
\end{align*} }

 { \tiny  \begin{align*} +  \begin{array}{cccccccccccccccccccccccccccccccccccc}      e_{11}       \big[ \underset{\alpha < n+1}{\mathrm{span}} e_{\beta^{\prime}\alpha} \big]         & \dots &   e_{n1}       \big[ \underset{\alpha < n+1}{\mathrm{span}} e_{\beta^{\prime}\alpha} \big]          & 0 & \dots \\  \vdots & \vdots &  \vdots & 0 & \dots \\     e_{1(2n+2) }       \big[ \underset{\alpha < n+1}{\mathrm{span}} e_{\beta^{\prime}\alpha} \big]         & \dots &   e_{n(2n+2)}       \big[ \underset{\alpha < n+1}{\mathrm{span}} e_{\beta^{\prime}\alpha} \big]          & 0 & \dots               \end{array}        
\end{align*} }

 { \tiny  \begin{align*}   \begin{array}{cccccccccccccccccccccccccccccccccccc}  \dots &   e_{(n+3) 1}       \big[ \underset{\alpha < n+1}{\mathrm{span}} e_{\beta^{\prime}\alpha} \big]       & \dots &  e_{(2n+2) 1}       \big[ \underset{\alpha < n+1}{\mathrm{span}} e_{\beta^{\prime}\alpha} \big]  \\    \dots &  \vdots & \vdots & \vdots     \\ \dots &   \vdots & \vdots & \vdots   \\   \dots & e_{(n+3)(2n+1)} \big[ \underset{\alpha < n+1}{\mathrm{span}} e_{\beta^{\prime} \alpha}  \big]  & \dots  & e_{(2n+1)(2n+1)} \big[ \underset{\alpha < n+1}{\mathrm{span}} e_{\beta^{\prime} \alpha}  \big]  \end{array}        
\end{align*} }

 { \tiny  \begin{align*} -  \begin{array}{cccccccccccccccccccccccccccccccccccc}           e_{11} e_{\alpha^{\prime} \beta^{\prime}} &   e_{21} e_{\alpha^{\prime} \beta^{\prime}} & \dots &  e_{(2n+2)1} e_{\alpha^{\prime} \beta^{\prime}} \\ \vdots & \vdots & \vdots & \vdots \\      e_{1(2n+2) } e_{\alpha^{\prime} \beta^{\prime}} &   e_{2(2n+2)} e_{\alpha^{\prime} \beta^{\prime}} & \dots &  e_{(2n+2)(2n+2)} e_{\alpha^{\prime} \beta^{\prime}} \end{array}  
\end{align*} }

 { \tiny  \begin{align*} -  \begin{array}{cccccccccccccccccccccccccccccccccccc}  e_{\beta^{\prime} \alpha^{\prime}} & e_{\beta^{\prime} \alpha^{\prime}} & \dots & e_{\beta^{\prime} \alpha^{\prime}} & e_{\beta^{\prime} \alpha^{\prime}} \big[  {\mathrm{span}}\text{ }  e_{1 (n+1)}     \big]  & e_{\beta^{\prime} \alpha^{\prime}} \big[  {\mathrm{span}} \text{ }  e_{1 (n+2)}     \big] & \dots \\ \vdots & \vdots & \vdots & \vdots & \vdots & \vdots & \dots \\ e_{\beta^{\prime} \alpha^{\prime}} & e_{\beta^{\prime} \alpha^{\prime}} & \dots & e_{\beta^{\prime} \alpha^{\prime}} & e_{\beta^{\prime} (2n+2) } \big[  {\mathrm{span}}\text{ }  e_{(2n+2)  (n+1)}     \big]  & e_{\beta^{\prime} (2n+2) } \big[  {\mathrm{span}} \text{ }  e_{(2n+2)  (n+2)}     \big] & \dots  \end{array}  
\end{align*} }

 { \tiny  \begin{align*}   \begin{array}{cccccccccccccccccccccccccccccccccccc}  \dots &    e_{\beta^{\prime} \alpha^{\prime}} \big[  {\mathrm{span}}\text{ }  e_{1 (2n+1)}     \big]  & e_{\beta^{\prime} \alpha^{\prime}} \big[  {\mathrm{span}} \text{ }  e_{1 (2n+2)}     \big]      \\  \dots  &  \vdots & \vdots    \\ \dots &      e_{\beta^{\prime} \alpha^{\prime}} \big[  {\mathrm{span}}\text{ }  e_{(2n+2) (2n+1)}     \big]  & e_{\beta^{\prime} \alpha^{\prime}} \big[  {\mathrm{span}} \text{ }  e_{(2n+2)  (2n+2)}     \big]         \end{array}  ,
\end{align*} }

\noindent corresponding to $\mathscr{R}_1$

\item[$\bullet$] \textit{Tensor product of elementary matrices for $\mathscr{R}_2$ and $\mathscr{R}_3$}. Disregarding prefactors, one has that,

\begin{align*}
    \bigg[ \underset{\alpha \neq n+1, n+2}{\sum} e_{\alpha \alpha}  +  \underset{\alpha = n+1 , n+2}{\sum} e_{\alpha \alpha} \bigg]  \otimes \bigg[ \underset{\alpha \neq n+1, n+2}{\sum} e_{\alpha \alpha}   + e_{\alpha^{\prime} \alpha^{\prime}} \bigg]  \\ \\ = \bigg[  \underset{\alpha \neq n+1, n+2}{\sum} e_{\alpha \alpha} \bigg]  \otimes  \bigg[ \underset{\alpha \neq n+1, n+2}{\sum} e_{\alpha \alpha}  \bigg] +   \underset{\alpha = n+1 , n+2}{\sum} \bigg[ e_{\alpha \alpha} \otimes e_{\alpha^{\prime} \alpha^{\prime}} \bigg]  \\ \\ =    \bigg[  \underset{\alpha \neq n+1, n+2}{\sum} e_{\alpha \alpha} \bigg]  \otimes  \bigg[ \underset{\alpha \neq n+1, n+2}{\sum} e_{\alpha \alpha}  \bigg] +   \bigg[ \underset{\alpha = n+1 , n+2}{\sum}  e_{\alpha \alpha} \bigotimes \underset{\alpha = n+1 , n+2}{\sum}  e_{\alpha^{\prime} \alpha^{\prime}} \bigg]    \end{align*}

   \begin{align*} =    \underset{\alpha \neq n+1, n+2}{\sum} \big[ e_{\alpha \alpha} \otimes e_{\alpha \alpha} \big]  +   \underset{\alpha = n+1 , n+2}{\sum} \big[ e_{\alpha \alpha} \big]  \otimes e_{\alpha^{\prime} \alpha^{\prime}}   \\ \\   =   \bigg[ \underset{\alpha \neq n+1, n+2}{\mathrm{span}}     e_{\alpha \alpha} \bigg]  \bigotimes  \bigg[ \underset{\alpha \neq n+1, n+2}{\mathrm{span}}   e_{\alpha \alpha} \bigg]   + \bigg[ \underset{\alpha = n+1 , n+2}{\mathrm{span}} e_{\alpha \alpha}   \bigg]  \otimes e_{\alpha^{\prime} \alpha^{\prime}}  \\ \\ =         \underset{\alpha \neq n+1, n+2}{\mathrm{span}}      \big[ e_{\alpha \alpha} \otimes e_{\alpha \alpha} \big]  + \bigg[ \underset{\alpha = n+1 , n+2}{\mathrm{span}} e_{\alpha \alpha}   \bigg]  \otimes e_{\alpha^{\prime} \alpha^{\prime}}\end{align*}

  { \tiny  \begin{align*} =  \begin{array}{cccccccccccccccccccccccccccccccccccc}  e_{1,1}    \big[ \underset{\alpha \neq n+1, n+2}{\mathrm{span}}       e_{\alpha \alpha} \big]  & \dots & \dots  & e_{n,1} \big[ \underset{\alpha \neq n+1, n+2}{\mathrm{span}}       e_{\alpha \alpha} \big] & 0   \\ e_{1,2} \big[ \underset{\alpha \neq n+1, n+2}{\mathrm{span}}       e_{\alpha \alpha} \big]   & e_{2,2} \big[ \underset{\alpha \neq n+1, n+2}{\mathrm{span}}       e_{\alpha \alpha} \big]  & \dots & e_{n-1,2}  \big[ \underset{\alpha \neq n+1, n+2}{\mathrm{span}}       e_{\alpha \alpha} \big] & \dots  \\ \vdots & \dots &  \ddots & \dots & \dots    \\ e_{1,n-1}
 \big[ \underset{\alpha \neq n+1, n+2}{\mathrm{span}}       e_{\alpha \alpha} \big]    & \dots &  e_{n-1,n-1} \big[ \underset{\alpha \neq n+1, n+2}{\mathrm{span}}       e_{\alpha \alpha} \big] & \ddots & \dots   \\          e_{1,n+3}    \big[ \underset{\alpha \neq n+1, n+2}{\mathrm{span}}       e_{\alpha \alpha} \big]  &   \dots & \dots  & e_{n+3, n+3} \big[ \underset{\alpha \neq n+1, n+2}{\mathrm{span}}       e_{\alpha \alpha} \big]   & \dots        \end{array}        
\end{align*} }

 { \tiny  \begin{align*}   \begin{array}{cccccccccccccccccccccccccccccccccccc} 0 &  e_{n+3,1} \big[ \underset{\alpha \neq n+1, n+2}{\mathrm{span}}       e_{\alpha \alpha} \big] & \dots &    e_{2n+2,1}    \big[ \underset{\alpha \neq n+1, n+2}{\mathrm{span}}       e_{\alpha \alpha} \big]     \\    \dots &   e_{n+3,2} \big[ \underset{\alpha \neq n+1, n+2}{\mathrm{span}}       e_{\alpha \alpha} \big]  & \dots &    e_{n+3,2n+2}    \big[ \underset{\alpha \neq n+1, n+2}{\mathrm{span}}       e_{\alpha \alpha} \big]         \\ \vdots  &    \vdots & \vdots  & \vdots      \\ \dots &   e_{n+3,2n+2}   \big[ \underset{\alpha \neq n+1, n+2}{\mathrm{span}}       e_{\alpha \alpha} \big]  & \dots &      e_{2n+2,2n+2}   \big[ \underset{\alpha \neq n+1, n+2}{\mathrm{span}}       e_{\alpha \alpha} \big]  \end{array}      
\end{align*} }

  { \tiny  \begin{align*} +   \begin{array}{cccccccccccccccccccccccccccccccccccc}  0 & \dots & \dots  & 0 & e_{n+1,1} \big[ \underset{\alpha \neq n+1, n+2}{\mathrm{span}}       e_{\alpha \alpha}  \big] & \dots  \\ 0  & 0 & \dots & 0  & \dots & \dots \\ \vdots & \dots &  \ddots & \dots & \dots  & \dots  \\ 0  & \dots & 0  & \ddots & \dots  & \dots \\          0  &   \dots & \dots  & 0    & \dots     & \dots   \end{array}        
\end{align*} }

 { \tiny  \begin{align*}   \begin{array}{cccccccccccccccccccccccccccccccccccc}  \dots &   e_{n+2,1} \big[ \underset{\alpha \neq n+1, n+2}{\mathrm{span}}       e_{\alpha \alpha} \big] & 0 & \dots &    0     \\   \dots &   0  &  0  & \dots &    0        \\ \dots &  \vdots  &    \vdots & \vdots  & \vdots      \\ \dots & 0   & \dots &   \dots &    0  \end{array}      
\end{align*} }

 { \tiny  \begin{align*} =  \begin{array}{cccccccccccccccccccccccccccccccccccc}  e_{1,1}    \big[ \underset{\alpha \neq n+1, n+2}{\mathrm{span}}       e_{\alpha \alpha} \big]  & \dots & \dots  & e_{n,1} \big[ \underset{\alpha \neq n+1, n+2}{\mathrm{span}}       e_{\alpha \alpha} \big] & \dots   \\ e_{1,2} \big[ \underset{\alpha \neq n+1, n+2}{\mathrm{span}}       e_{\alpha \alpha} \big]   & e_{2,2} \big[ \underset{\alpha \neq n+1, n+2}{\mathrm{span}}       e_{\alpha \alpha} \big]  & \dots & e_{n-1,2}  \big[ \underset{\alpha \neq n+1, n+2}{\mathrm{span}}       e_{\alpha \alpha} \big] & \dots  \\ \vdots & \dots &  \ddots & \dots & \dots    \\ e_{1,n-1}
 \big[ \underset{\alpha \neq n+1, n+2}{\mathrm{span}}       e_{\alpha \alpha} \big]    & \dots &  e_{n-1,n-1} \big[ \underset{\alpha \neq n+1, n+2}{\mathrm{span}}       e_{\alpha \alpha} \big] & \ddots & \dots   \\          e_{1,n+3}    \big[ \underset{\alpha \neq n+1, n+2}{\mathrm{span}}       e_{\alpha \alpha} \big]  &   \dots & \dots  & e_{n+3, n+3} \big[ \underset{\alpha \neq n+1, n+2}{\mathrm{span}}       e_{\alpha \alpha} \big]   & \dots        \end{array}        
\end{align*} }

 { \tiny  \begin{align*}   \begin{array}{cccccccccccccccccccccccccccccccccccc} \dots & e_{n+2,1} \big[ \underset{\alpha \neq n+1, n+2}{\mathrm{span}}       e_{\alpha \alpha} \big] &  e_{n+3,1} \big[ \underset{\alpha \neq n+1, n+2}{\mathrm{span}}       e_{\alpha \alpha} \big] & \dots &    e_{2n+2,1}    \big[ \underset{\alpha \neq n+1, n+2}{\mathrm{span}}       e_{\alpha \alpha} \big]     \\ \dots &     \dots &   e_{n+3,2} \big[ \underset{\alpha \neq n+1, n+2}{\mathrm{span}}       e_{\alpha \alpha} \big]  & \dots &    e_{2n+2,2}    \big[ \underset{\alpha \neq n+1, n+2}{\mathrm{span}}       e_{\alpha \alpha} \big]         \\ \dots &   \vdots  &    \vdots & \vdots  & \vdots      \\ \dots &   \dots &   e_{n+3,2n+2}    \big[ \underset{\alpha \neq n+1, n+2}{\mathrm{span}}       e_{\alpha \alpha} \big]  & \dots &      e_{2n+2,2n+2}   \big[ \underset{\alpha \neq n+1, n+2}{\mathrm{span}}       e_{\alpha \alpha} \big]  \end{array}      
\end{align*} }

\noindent corresponding to $\mathscr{R}_2$, and,

\begin{align*}
    \bigg[  \underset{\alpha \text{ } \text{or} \text{ } \beta, \beta^{\prime} \neq n+1, n+2}{\underset{\alpha \neq \beta, \beta^{\prime}}{\sum}}  e_{\alpha \alpha} +\underset{\alpha = n+1, n+2}{\sum} e_{\alpha \alpha} \bigg]       \otimes \bigg[   e_{\beta\beta} + e_{\alpha \alpha}  \bigg] \\ \\ =           \underset{\alpha \text{ } \text{or} \text{ } \beta, \beta^{\prime} \neq n+1, n+2}{\underset{\alpha \neq \beta, \beta^{\prime}}{\sum}}  e_{\alpha \alpha}  \otimes e_{\beta\beta} +  \underset{\alpha = n+1, n+2}{\sum} e_{\alpha \alpha} \otimes e_{\alpha \alpha} \\ \\ =    \underset{\alpha \text{ } \text{or} \text{ } \beta, \beta^{\prime} \neq n+1, n+2}{\underset{\alpha \neq \beta, \beta^{\prime}}{\mathrm{span}}}  \big[ e_{\alpha \alpha}  \otimes e_{\beta\beta} \big]  +  \underset{\alpha = n+1, n+2}{\mathrm{span}} \big[ e_{\alpha \alpha} \otimes e_{\alpha \alpha} \big] 
\end{align*}

{\tiny \begin{align*}
=    \begin{array}{cccccccccccccccccccccccccccccccccccc}       e_{11} \big[    \underset{\alpha \text{ } \text{or} \text{ } \beta, \beta^{\prime} \neq n+1, n+2}{\underset{\alpha \neq \beta, \beta^{\prime}}{\mathrm{span}}} e_{\beta \beta}       \big] & \ddots & & & & &  \dots  \\  & e_{22} \big[    \underset{\alpha \text{ } \text{or} \text{ } \beta, \beta^{\prime} \neq n+1, n+2}{\underset{\alpha \neq \beta, \beta^{\prime}}{\mathrm{span}}} e_{\beta \beta} \big] & & & & & \dots  \\ & & \ddots & & & &  \dots  \\ \vdots & \vdots & \vdots \\ & & &  e_{nn } \big[ \underset{\alpha \text{ } \text{or} \text{ } \beta, \beta^{\prime} \neq n+1, n+2}{\underset{\alpha \neq \beta, \beta^{\prime}}{\mathrm{span}}} e_{\beta \beta}      \big]  & \ddots &  & \dots  \\ & & & &   0 & \ddots & \dots  \\ & & & & & 0 & \ddots \\ & & & & &  & \dots  \end{array} 
\end{align*}  }

{\tiny \begin{align*}
   \begin{array}{cccccccccccccccccccccccccccccccccccc}      \dots \\ \dots   \\   \dots \\ \dots    \\   \dots & \ddots  \\ \dots &  e_{(2n+2 )(2n+2)}  \big[ \underset{\alpha \text{ } \text{or} \text{ } \beta, \beta^{\prime} \neq n+1, n+2}{\underset{\alpha \neq \beta, \beta^{\prime}}{\mathrm{span}}} e_{\beta \beta}      \big]  \end{array} , 
\end{align*}  }

{\tiny \begin{align*}
+     \begin{array}{cccccccccccccccccccccccccccccccccccc}     0  & \ddots & & & & &  \dots  \\  & 0 & & & & & \dots  \\ & & \ddots & & & &  \dots  \\ \vdots & \vdots & \vdots \\ & & & 0  & \ddots &  & \dots  \\ & & & &   e_{(n+1)(n+1)}  \big[ \underset{\alpha \text{ } \text{or} \text{ } \beta, \beta^{\prime} \neq n+1, n+2}{\underset{\alpha \neq \beta, \beta^{\prime}}{\mathrm{span}}} e_{\beta \beta}      \big] & \ddots & \dots  \\ & & & & & e_{(n+2)(n+2)}  \big[ \underset{\alpha \text{ } \text{or} \text{ } \beta, \beta^{\prime} \neq n+1, n+2}{\underset{\alpha \neq \beta, \beta^{\prime}}{\mathrm{span}}} e_{\beta \beta}      \big]  & \ddots \\ & & & & &  & \dots  \end{array} 
\end{align*}  }

{\tiny \begin{align*}
   \begin{array}{cccccccccccccccccccccccccccccccccccc}      \dots \\ \dots   \\   \dots \\ \dots    \\   \dots & \ddots  \\ \dots & 0   \end{array} , 
\end{align*}  }

\noindent corresponding to $\mathscr{R}_3$.

\bigskip

\item[$\bullet$] \textit{Tensor product of elementary matrices for $\mathscr{R}_4$}. Disregarding constant prefactors, one has that,

   \begin{align*}
     \bigg[ \underset{\alpha, \beta \neq n+1, n+2}{\sum}  e_{\alpha \beta} +  b^+_a \big( u \big)  \underset{\alpha \neq n+1, n+2, \beta = n+1, n+2}{\sum} e_{\alpha \beta}  +  b^-_a \big( u \big)   \underset{\alpha \neq n+1, n+2, \beta = n+1, n+2}{\sum}   e_{\alpha \beta} \bigg] \\ \otimes \bigg[                 \underset{\alpha, \beta \neq n+1, n+2}{\sum}     e_{\alpha^{\prime} \beta^{\prime}}  +   \underset{\alpha \neq n+1, n+2, \beta = n+1,n+2}{\sum}         e_{\alpha^{\prime} \beta^{\prime}}    +  b^-_a \big( u \big) \underset{\alpha \neq n+1, n+2, \beta = n+1, n+2}{\sum} e_{\alpha^{\prime} \beta} \bigg] \\ \\ \approx \bigg[ \underset{\alpha, \beta \neq n+1, n+2}{\sum}  e_{\alpha \beta} +  b^+_a \big( u \big)  \underset{\alpha \neq n+1, n+2, \beta = n+1, n+2}{\sum} e_{\alpha \beta}  +  \underset{\alpha \neq n+1, n+2, \beta = n+1, n+2}{\sum}   e_{\alpha \beta} \bigg] \\  \otimes \bigg[                 \underset{\alpha, \beta \neq n+1, n+2}{\sum}     e_{\alpha^{\prime} \beta^{\prime}}  +   \underset{\alpha \neq n+1, n+2, \beta = n+1,n+2}{\sum}         e_{\alpha^{\prime} \beta^{\prime}}    +   \underset{\alpha \neq n+1, n+2, \beta = n+1, n+2}{\sum} e_{\alpha^{\prime} \beta} \bigg]   \end{align*}

     \begin{align*}  \approx              \bigg[ \bigg[ \underset{\alpha, \beta \neq n+1, n+2}{\sum}  e_{\alpha \beta} \bigg] \bigotimes \bigg[   \underset{\alpha, \beta \neq n+1, n+2}{\sum}     e_{\alpha^{\prime} \beta^{\prime}}   \bigg] \bigg]  + 2 \bigg[ \bigg[    \underset{\alpha \neq n+1, n+2, \beta = n+1, n+2}{\sum} e_{\alpha \beta}     \bigg] \\ \bigotimes  \bigg[   \underset{\alpha, \beta \neq n+1, n+2}{\sum}     e_{\alpha^{\prime} \beta^{\prime}}  \bigg]  \bigg]  \\  +  2  \bigg[ \bigg[ \underset{\alpha \neq n+1, n+2, \beta = n+1, n+2}{\sum}   e_{\alpha \beta} \bigg] \bigotimes  \bigg[  \underset{\alpha, \beta \neq n+1, n+2}{\sum}     e_{\alpha^{\prime} \beta^{\prime}}  \bigg]  \bigg]  + \bigg[ \bigg[   \underset{\alpha, \beta \neq n+1, n+2}{\sum}  e_{\alpha \beta}        \bigg] \\  \bigotimes  \bigg[     \underset{\alpha \neq n+1, n+2, \beta = n+1,n+2}{\sum}         e_{\alpha^{\prime} \beta^{\prime}}        \bigg]  \bigg]  \\ + \bigg[ \bigg[       \underset{\alpha, \beta \neq n+1, n+2}{\sum}  e_{\alpha \beta}    \bigg] \bigotimes  \bigg[      \underset{\alpha \neq n+1, n+2, \beta = n+1, n+2}{\sum} e_{\alpha^{\prime} \beta}      \bigg]  \bigg]   + 2 \bigg[ \bigg[  \underset{\alpha \neq n+1, n+2, \beta = n+1, n+2}{\sum}   e_{\alpha \beta}      \bigg] \\ \bigotimes  \bigg[      \underset{\alpha \neq n+1, n+2, \beta = n+1,n+2}{\sum}         e_{\alpha^{\prime} \beta^{\prime}}     \bigg]  \bigg]            \\  + 2 \bigg[ \bigg[  \underset{\alpha \neq n+1, n+2, \beta = n+1, n+2}{\sum}   e_{\alpha \beta}      \bigg] \bigotimes  \bigg[   \underset{\alpha \neq n+1, n+2, \beta = n+1, n+2}{\sum} e_{\alpha^{\prime} \beta}        \bigg]  \bigg]  \\ \\  =            \bigg[ \bigg[ \underset{\alpha, \beta \neq n+1, n+2}{\mathrm{span}}  e_{\alpha \beta} \bigg] \bigotimes \bigg[   \underset{\alpha, \beta \neq n+1, n+2}{\mathrm{span}}     e_{\alpha^{\prime} \beta^{\prime}}   \bigg] \bigg]   + 2 \bigg[ \bigg[    \underset{\alpha \neq n+1, n+2, \beta = n+1, n+2}{\mathrm{span}} e_{\alpha \beta}     \bigg]  \\ \bigotimes  \bigg[   \underset{\alpha, \beta \neq n+1, n+2}{\mathrm{span}}     e_{\alpha^{\prime} \beta^{\prime}}  \bigg]  \bigg]     \\   +  2  \bigg[ \bigg[ \underset{\alpha \neq n+1, n+2, \beta = n+1, n+2}{\mathrm{span}}   e_{\alpha \beta} \bigg] \bigotimes  \bigg[  \underset{\alpha, \beta \neq n+1, n+2}{\mathrm{span}}     e_{\alpha^{\prime} \beta^{\prime}}  \bigg]  \bigg]    + \bigg[ \bigg[   \underset{\alpha, \beta \neq n+1, n+2}{\mathrm{span}}  e_{\alpha \beta}        \bigg] \\ \bigotimes  \bigg[     \underset{\alpha \neq n+1, n+2, \beta = n+1,n+2}{\mathrm{span}}         e_{\alpha^{\prime} \beta^{\prime}}        \bigg]  \bigg]  \\ + \bigg[ \bigg[       \underset{\alpha, \beta \neq n+1, n+2}{\mathrm{span}}  e_{\alpha \beta}    \bigg] \bigotimes  \bigg[      \underset{\alpha \neq n+1, n+2, \beta = n+1, n+2}{\mathrm{span}} e_{\alpha^{\prime} \beta}      \bigg]  \bigg]    + 2 \bigg[ \bigg[  \underset{\alpha \neq n+1, n+2, \beta = n+1, n+2}{\mathrm{span}}   e_{\alpha \beta}      \bigg] \\ \bigotimes  \bigg[      \underset{\alpha \neq n+1, n+2, \beta = n+1,n+2}{\mathrm{span}}         e_{\alpha^{\prime} \beta^{\prime}}     \bigg]  \bigg]            \\  + 2 \bigg[ \bigg[  \underset{\alpha \neq n+1, n+2, \beta = n+1, n+2}{\mathrm{span}}   e_{\alpha \beta}      \bigg] \bigotimes  \bigg[   \underset{\alpha \neq n+1, n+2, \beta = n+1, n+2}{\mathrm{span}} e_{\alpha^{\prime} \beta}        \bigg]  \bigg]       
 \end{align*}

 \begin{align*} =   \underset{\mathscr{R}_{4,0}}{\underbrace{\bigg[ \bigg[ \underset{\alpha, \beta \neq n+1, n+2}{\mathrm{span}}  e_{\alpha \beta} \bigg] \bigotimes \bigg[   \underset{\alpha, \beta \neq n+1, n+2}{\mathrm{span}}     e_{\alpha^{\prime} \beta^{\prime}}   \bigg] \bigg]}} \\     \underset{\mathscr{R}_{4,1}}{\underbrace{+  2  \bigg[ \bigg[ \underset{\alpha \neq n+1, n+2, \beta = n+1, n+2}{\mathrm{span}}   e_{\alpha \beta} \bigg] \bigotimes  \bigg[  \underset{\alpha, \beta \neq n+1, n+2}{\mathrm{span}}     e_{\alpha^{\prime} \beta^{\prime}}  \bigg]  \bigg]}} \\   \underset{\mathscr{R}_{4,2}}{\underbrace{+ 2 \bigg[ \bigg[    \underset{\alpha \neq n+1, n+2, \beta = n+1, n+2}{\mathrm{span}} e_{\alpha \beta}     \bigg] \bigotimes  \bigg[   \underset{\alpha, \beta \neq n+1, n+2}{\mathrm{span}}     e_{\alpha^{\prime} \beta^{\prime}}  \bigg]  \bigg]}}  \\   \underset{\mathscr{R}_{4,3}}{\underbrace{+ \bigg[ \bigg[   \underset{\alpha, \beta \neq n+1, n+2}{\mathrm{span}}  e_{\alpha \beta}        \bigg] \bigotimes  \bigg[     \underset{\alpha \neq n+1, n+2, \beta = n+1,n+2}{\mathrm{span}}         e_{\alpha^{\prime} \beta^{\prime}}        \bigg]  \bigg]}}      \\ \underset{\mathscr{R}_{4,4}}{\underbrace{+ \bigg[ \bigg[       \underset{\alpha, \beta \neq n+1, n+2}{\mathrm{span}}  e_{\alpha \beta}    \bigg] \bigotimes  \bigg[      \underset{\alpha \neq n+1, n+2, \beta = n+1, n+2}{\mathrm{span}} e_{\alpha^{\prime} \beta}      \bigg]  \bigg]}}  \\   \underset{\mathscr{R}_{4,5}}{\underbrace{+ 2 \bigg[ \bigg[  \underset{\alpha \neq n+1, n+2, \beta = n+1, n+2}{\mathrm{span}}   e_{\alpha \beta}      \bigg] \bigotimes  \bigg[      \underset{\alpha \neq n+1, n+2, \beta = n+1,n+2}{\mathrm{span}}         e_{\alpha^{\prime} \beta^{\prime}}     \bigg]  \bigg]}}           \\    \underset{\mathscr{R}_{4,6}}{\underbrace{+ 2 \bigg[ \bigg[  \underset{\alpha \neq n+1, n+2, \beta = n+1, n+2}{\mathrm{span}}   e_{\alpha \beta}      \bigg] \bigotimes  \bigg[   \underset{\alpha \neq n+1, n+2, \beta = n+1, n+2}{\mathrm{span}} e_{\alpha^{\prime} \beta}        \bigg]  \bigg]}}       
    \end{align*}

 { \tiny  \begin{align*} =  \begin{array}{cccccccccccccccccccccccccccccccccccc} 
      e_{11} \big[ \underset{\alpha , \beta \neq n+1, n+2}{\mathrm{span}} e_{\alpha^{\prime} \beta^{\prime}}  \big] & \dots &  \dots &  e_{n1} \big[ \underset{\alpha , \beta \neq n+1, n+2}{\mathrm{span}} e_{\alpha^{\prime} \beta^{\prime}}  \big] & \dots   \\  e_{12} \big[ \underset{\alpha , \beta \neq n+1, n+2}{\mathrm{span}} e_{\alpha^{\prime} \beta^{\prime}}  \big] & e_{22} \big[ \underset{\alpha , \beta \neq n+1, n+2}{\mathrm{span}} e_{\alpha^{\prime} \beta^{\prime}}  \big]   & \dots & e_{n2} \big[ \underset{\alpha , \beta \neq n+1, n+2}{\mathrm{span}} e_{\alpha^{\prime} \beta^{\prime}}  \big] & \dots  \\ \vdots & \vdots & \vdots & \vdots  \\  e_{1n} \big[ \underset{\alpha , \beta \neq n+1, n+2}{\mathrm{span}} e_{\alpha^{\prime} \beta^{\prime}}  \big] & \dots &  \dots &  e_{nn} \big[ \underset{\alpha , \beta \neq n+1, n+2}{\mathrm{span}} e_{\alpha^{\prime} \beta^{\prime}}  \big]  & \dots \\ 0 & \dots & \dots & 0 & \dots \\     0 & \dots & \dots & 0 & \dots  \\  e_{1(n+3)} \big[ \underset{\alpha , \beta \neq n+1, n+2}{\mathrm{span}} e_{\alpha^{\prime} \beta^{\prime}}  \big] & \dots &  \dots &  e_{n(n+3)} \big[ \underset{\alpha , \beta \neq n+1, n+2}{\mathrm{span}} e_{\alpha^{\prime} \beta^{\prime}}  \big]  & \dots  \\  \vdots & \vdots & \vdots & \vdots \\ e_{1(2n+2)} \big[ \underset{\alpha , \beta \neq n+1, n+2}{\mathrm{span}} e_{\alpha^{\prime} \beta^{\prime}}  \big] & \dots &  \dots &  e_{n(2n+2)} \big[ \underset{\alpha , \beta \neq n+1, n+2}{\mathrm{span}} e_{\alpha^{\prime} \beta^{\prime}}  \big]  & \dots    \end{array} 
\end{align*}  }

  { \tiny  \begin{align*}   \begin{array}{cccccccccccccccccccccccccccccccccccc} 
   \dots &  0 & 0 &   e_{(n+3)1} \big[ \underset{\alpha , \beta \neq n+1, n+2}{\mathrm{span}} e_{\alpha^{\prime} \beta^{\prime}}  \big] & \dots &  e_{(2n+2)1} \big[ \underset{\alpha , \beta \neq n+1, n+2}{\mathrm{span}} e_{\alpha^{\prime} \beta^{\prime}}  \big]  \\ \dots &   0 & 0 &    e_{(n+3)2} \big[ \underset{\alpha , \beta \neq n+1, n+2}{\mathrm{span}} e_{\alpha^{\prime} \beta^{\prime}}  \big] & \dots &  e_{(2n+2)2} \big[ \underset{\alpha , \beta \neq n+1, n+2}{\mathrm{span}} e_{\alpha^{\prime} \beta^{\prime}}  \big]   \\ \dots &    0 & 0 &   e_{(n+3)3} \big[ \underset{\alpha , \beta \neq n+1, n+2}{\mathrm{span}} e_{\alpha^{\prime} \beta^{\prime}}  \big] & \dots &  e_{(2n+2)3} \big[ \underset{\alpha , \beta \neq n+1, n+2}{\mathrm{span}} e_{\alpha^{\prime} \beta^{\prime}}  \big] \\   \dots &     0 & 0 &        \vdots & \vdots   & \vdots    \\ \dots &     0 & 0 &       e_{(n+3)(2n+1)} \big[ \underset{\alpha , \beta \neq n+1, n+2}{\mathrm{span}} e_{\alpha^{\prime} \beta^{\prime}}  \big] & \dots &  e_{(2n+2)(2n+1)} \big[ \underset{\alpha , \beta \neq n+1, n+2}{\mathrm{span}} e_{\alpha^{\prime} \beta^{\prime}}  \big]  \\ \dots &   0 & 0 &    e_{(n+3)(2n+2)} \big[ \underset{\alpha , \beta \neq n+1, n+2}{\mathrm{span}} e_{\alpha^{\prime} \beta^{\prime}}  \big] & \dots &  e_{(2n+2)(2n+2)} \big[ \underset{\alpha , \beta \neq n+1, n+2}{\mathrm{span}} e_{\alpha^{\prime} \beta^{\prime}}  \big]            \end{array}   
\end{align*}  } 

{ \tiny  \begin{align*} + 6 \begin{array}{cccccccccccccccccccccccccccccccccccc} 
      e_{11} \big[ \underset{\alpha , \beta \neq n+1, n+2}{\mathrm{span}} e_{\alpha^{\prime} \beta^{\prime}}  \big] & \dots &  \dots &  e_{n1} \big[ \underset{\alpha , \beta \neq n+1, n+2}{\mathrm{span}} e_{\alpha^{\prime} \beta^{\prime}}  \big] & \dots   \\  e_{12} \big[ \underset{\alpha , \beta \neq n+1, n+2}{\mathrm{span}} e_{\alpha^{\prime} \beta^{\prime}}  \big] & e_{22} \big[ \underset{\alpha , \beta \neq n+1, n+2}{\mathrm{span}} e_{\alpha^{\prime} \beta^{\prime}}  \big]   & \dots & e_{n2} \big[ \underset{\alpha , \beta \neq n+1, n+2}{\mathrm{span}} e_{\alpha^{\prime} \beta^{\prime}}  \big] & \dots  \\ \vdots & \vdots & \vdots & \vdots  \\  e_{1n} \big[ \underset{\alpha , \beta \neq n+1, n+2}{\mathrm{span}} e_{\alpha^{\prime} \beta^{\prime}}  \big] & \dots &  \dots &  e_{nn} \big[ \underset{\alpha , \beta \neq n+1, n+2}{\mathrm{span}} e_{\alpha^{\prime} \beta^{\prime}}  \big]  & \dots \\ 0 & \dots & \dots & 0 & \dots \\     0 & \dots & \dots & 0 & \dots  \\  e_{1(n+3)} \big[ \underset{\alpha , \beta \neq n+1, n+2}{\mathrm{span}} e_{\alpha^{\prime} \beta^{\prime}}  \big] & \dots &  \dots &  e_{n(n+3)} \big[ \underset{\alpha , \beta \neq n+1, n+2}{\mathrm{span}} e_{\alpha^{\prime} \beta^{\prime}}  \big]  & \dots  \\  \vdots & \vdots & \vdots & \vdots \\ e_{1(2n+2)} \big[ \underset{\alpha , \beta \neq n+1, n+2}{\mathrm{span}} e_{\alpha^{\prime} \beta^{\prime}}  \big] & \dots &  \dots &  e_{n(2n+2)} \big[ \underset{\alpha , \beta \neq n+1, n+2}{\mathrm{span}} e_{\alpha^{\prime} \beta^{\prime}}  \big]  & \dots    \end{array} 
\end{align*}  }

  { \tiny  \begin{align*}   \begin{array}{cccccccccccccccccccccccccccccccccccc} 
   \dots & 0 & 0 &   e_{(n+3)1} \big[ \underset{\alpha , \beta \neq n+1, n+2}{\mathrm{span}} e_{\alpha^{\prime} \beta^{\prime}}  \big] & \dots  \\ \dots &   e_{(n+3) (n+1)} \big[ \underset{\alpha, \beta \neq n+1, n+2} e_{\alpha^{\prime} \beta^{\prime}} \big] &  e_{(n+3) (n+2)} \big[ \underset{\alpha, \beta \neq n+1, n+2} e_{\alpha^{\prime} \beta^{\prime}} \big] &    e_{(n+3)2} \big[ \underset{\alpha , \beta \neq n+1, n+2}{\mathrm{span}} e_{\alpha^{\prime} \beta^{\prime}}  \big] & \dots  \\ \dots &    0 & 0 &   e_{(n+3)3} \big[ \underset{\alpha , \beta \neq n+1, n+2}{\mathrm{span}} e_{\alpha^{\prime} \beta^{\prime}}  \big] & \dots \\   \dots &     0 & 0 &        \vdots & \vdots      \\ \dots &     0 & 0 &       e_{(n+3)(2n+1)} \big[ \underset{\alpha , \beta \neq n+1, n+2}{\mathrm{span}} e_{\alpha^{\prime} \beta^{\prime}}  \big] & \dots   \\ \dots &   0 & 0 &    e_{(n+3)(2n+2)} \big[ \underset{\alpha , \beta \neq n+1, n+2}{\mathrm{span}} e_{\alpha^{\prime} \beta^{\prime}}  \big] & \dots          \end{array}   
\end{align*}  }

 {\tiny \begin{align*}
    \begin{array}{cccccccccccccccccccccccccccccccccccc}         \dots &     e_{(2n+2)1} \big[ \underset{\alpha , \beta \neq n+1, n+2}{\mathrm{span}} e_{\alpha^{\prime} \beta^{\prime}}  \big]      \\ \dots &      e_{(2n+2)2} \big[ \underset{\alpha , \beta \neq n+1, n+2}{\mathrm{span}} e_{\alpha^{\prime} \beta^{\prime}}  \big]      \\  \dots &      e_{(2n+2)3} \big[ \underset{\alpha , \beta \neq n+1, n+2}{\mathrm{span}} e_{\alpha^{\prime} \beta^{\prime}}  \big] \\ \dots & 0 \\ \vdots & \vdots \\ \dots & 0    \\ \dots &   e_{(2n+2)(2n+1)} \big[ \underset{\alpha , \beta \neq n+1, n+2}{\mathrm{span}} e_{\alpha^{\prime} \beta^{\prime}}  \big]    \\   \dots &  e_{(2n+2)(2n+2)} \big[ \underset{\alpha , \beta \neq n+1, n+2}{\mathrm{span}} e_{\alpha^{\prime} \beta^{\prime}}  \big]                          \end{array}  
\end{align*}  }

{\tiny \begin{align*}
+    \begin{array}{cccccccccccccccccccccccccccccccccccc}                     \underset{\alpha, \beta \neq n+1, n+2}{\mathrm{span}}  e_{\alpha \beta}       \big[       \underset{\alpha \neq n+1, n+2, \beta = n+1, n+2}{\mathrm{span}} \big[ e_{\alpha^{\prime} \beta^{\prime}} +  e_{\alpha^{\prime} \beta}      \big]     \big]        &    \dots &     \dots & \dots \\     \underset{\alpha, \beta \neq n+1, n+2}{\mathrm{span}}  e_{\alpha \beta}       \big[       \underset{\alpha \neq n+1, n+2, \beta = n+1, n+2}{\mathrm{span}} \big[ e_{\alpha^{\prime} \beta^{\prime}} +  e_{\alpha^{\prime} \beta}      \big]     \big]        &    \dots    & \dots & \dots  \\ \vdots & \vdots & \vdots & \vdots \\ 0     & \underset{\alpha, \beta \neq n+1, n+2}{\mathrm{span}}  e_{\alpha \beta}       \big[       \underset{\alpha \neq n+1, n+2, \beta = n+1, n+2}{\mathrm{span}} \big[ e_{\alpha^{\prime} \beta^{\prime}} +  e_{\alpha^{\prime} \beta}      \big]     \big]   & \dots  & \dots   \\ 0  & \underset{\alpha, \beta \neq n+1, n+2}{\mathrm{span}}  e_{\alpha \beta}       \big[       \underset{\alpha \neq n+1, n+2, \beta = n+1, n+2}{\mathrm{span}} \big[ e_{\alpha^{\prime} \beta^{\prime}} +  e_{\alpha^{\prime} \beta}      \big]     \big]   & \dots  & \dots   \\     \underset{\alpha, \beta \neq n+1, n+2}{\mathrm{span}}  e_{\alpha \beta}       \big[       \underset{\alpha \neq n+1, n+2, \beta = n+1, n+2}{\mathrm{span}} \big[ e_{\alpha^{\prime} \beta^{\prime}} +  e_{\alpha^{\prime} \beta}      \big]     \big] & \dots & \dots & \dots & \vdots   \\ \vdots & \vdots & \vdots & \dots & \vdots   \\ \underset{\alpha, \beta \neq n+1, n+2}{\mathrm{span}}  e_{\alpha \beta}       \big[       \underset{\alpha \neq n+1, n+2, \beta = n+1, n+2}{\mathrm{span}} \big[ e_{\alpha^{\prime} \beta^{\prime}} +  e_{\alpha^{\prime} \beta}      \big]     \big] & \dots & \dots                             \end{array}  
\end{align*}  }

{\tiny \begin{align*}
    \begin{array}{cccccccccccccccccccccccccccccccccccc}     \dots &   \underset{\alpha, \beta \neq n+1, n+2}{\mathrm{span}}  e_{\alpha \beta}       \big[       \underset{\alpha \neq n+1, n+2, \beta = n+1, n+2}{\mathrm{span}} \big[ e_{\alpha^{\prime} \beta^{\prime}} +  e_{\alpha^{\prime} \beta}      \big]     \big]  \dots  & \dots & \dots \\   \dots  & \dots & \dots \\ \dots &    \underset{\alpha, \beta \neq n+1, n+2}{\mathrm{span}}  e_{\alpha \beta}       \big[       \underset{\alpha \neq n+1, n+2, \beta = n+1, n+2}{\mathrm{span}} \big[ e_{\alpha^{\prime} \beta^{\prime}} +  e_{\alpha^{\prime} \beta}      \big]     \big] &  \dots  & \dots & \dots \\      \dots & \dots &    \dots   &       \underset{\alpha, \beta \neq n+1, n+2}{\mathrm{span}}  e_{\alpha \beta}       \big[       \underset{\alpha \neq n+1, n+2, \beta = n+1, n+2}{\mathrm{span}} \big[ e_{\alpha^{\prime} \beta^{\prime}} +  e_{\alpha^{\prime} \beta}      \big]     \big]   & \dots      \\   \dots & \dots &         \dots &       \underset{\alpha, \beta \neq n+1, n+2}{\mathrm{span}}  e_{\alpha \beta}       \big[       \underset{\alpha \neq n+1, n+2, \beta = n+1, n+2}{\mathrm{span}} \big[ e_{\alpha^{\prime} \beta^{\prime}} +  e_{\alpha^{\prime} \beta}      \big]     \big]        & \dots  \\ \dots & \dots &    & \dots  \\     \dots & \dots &         \dots  & \dots & \dots \\ \dots & \dots &    \dots  & \dots & \dots \\ \dots & \dots &     \dots  & \dots & \dots \\                     \end{array}  
\end{align*}  }  

 { \tiny  \begin{align*}   \begin{array}{cccccccccccccccccccccccccccccccccccc} 
    0 & 0 &        \underset{\alpha, \beta \neq n+1, n+2}{\mathrm{span}}  e_{\alpha \beta}       \big[       \underset{\alpha \neq n+1, n+2, \beta = n+1, n+2}{\mathrm{span}} \big[ e_{\alpha^{\prime} \beta^{\prime}} +  e_{\alpha^{\prime} \beta}      \big]     \big]        &    \dots &    \underset{\alpha, \beta \neq n+1, n+2}{\mathrm{span}}  e_{\alpha \beta}       \big[       \underset{\alpha \neq n+1, n+2, \beta = n+1, n+2}{\mathrm{span}} \big[ e_{\alpha^{\prime} \beta^{\prime}} +  e_{\alpha^{\prime} \beta}      \big]     \big]      \\  \vdots & \vdots & \vdots & \vdots  & \vdots  \\     0 & 0 &        \underset{\alpha, \beta \neq n+1, n+2}{\mathrm{span}}  e_{\alpha \beta}       \big[       \underset{\alpha \neq n+1, n+2, \beta = n+1, n+2}{\mathrm{span}} \big[ e_{\alpha^{\prime} \beta^{\prime}} +  e_{\alpha^{\prime} \beta}      \big]     \big]        &    \dots &    \underset{\alpha, \beta \neq n+1, n+2}{\mathrm{span}}  e_{\alpha \beta}       \big[       \underset{\alpha \neq n+1, n+2, \beta = n+1, n+2}{\mathrm{span}} \big[ e_{\alpha^{\prime} \beta^{\prime}} +  e_{\alpha^{\prime} \beta}      \big]     \big]        \end{array}   
\end{align*}  } 

{ \tiny  \begin{align*} + 2 \begin{array}{cccccccccccccccccccccccccccccccccccc} 
      e_{11} \big[ \underset{\alpha , \beta \neq n+1, n+2}{\mathrm{span}} e_{\alpha^{\prime} \beta^{\prime}}  \big] & \dots &  \dots &  e_{n1} \big[ \underset{\alpha , \beta \neq n+1, n+2}{\mathrm{span}} e_{\alpha^{\prime} \beta}  \big] & \dots   \\  e_{12} \big[ \underset{\alpha , \beta \neq n+1, n+2}{\mathrm{span}} e_{\alpha^{\prime} \beta}  \big] & e_{22} \big[ \underset{\alpha , \beta \neq n+1, n+2}{\mathrm{span}} e_{\alpha^{\prime} \beta^{\prime}}  \big]   & \dots & e_{n2} \big[ \underset{\alpha , \beta \neq n+1, n+2}{\mathrm{span}} e_{\alpha^{\prime} \beta}  \big] & \dots  \\ \vdots & \vdots & \vdots & \vdots  \\  e_{1n} \big[ \underset{\alpha , \beta \neq n+1, n+2}{\mathrm{span}} e_{\alpha^{\prime} \beta}  \big] & \dots &  \dots &  e_{nn} \big[ \underset{\alpha , \beta \neq n+1, n+2}{\mathrm{span}} e_{\alpha^{\prime} \beta}  \big]  & \dots \\ 0 & \dots & \dots & 0 & \dots \\     0 & \dots & \dots & 0 & \dots  \\  e_{1(n+3)} \big[ \underset{\alpha , \beta \neq n+1, n+2}{\mathrm{span}} e_{\alpha^{\prime} \beta}  \big] & \dots &  \dots &  e_{n(n+3)} \big[ \underset{\alpha , \beta \neq n+1, n+2}{\mathrm{span}} e_{\alpha^{\prime} \beta}  \big]  & \dots  \\  \vdots & \vdots & \vdots & \vdots \\ e_{1(2n+2)} \big[ \underset{\alpha , \beta \neq n+1, n+2}{\mathrm{span}} e_{\alpha^{\prime} \beta} \big] & \dots &  \dots &  e_{n(2n+2)} \big[ \underset{\alpha , \beta \neq n+1, n+2}{\mathrm{span}} e_{\alpha^{\prime} \beta}  \big]  & \dots    \end{array} 
\end{align*}  }

  { \tiny  \begin{align*}   \begin{array}{cccccccccccccccccccccccccccccccccccc} 
   \dots & 0 & 0 &   e_{(n+3)1} \big[ \underset{\alpha , \beta \neq n+1, n+2}{\mathrm{span}} e_{\alpha^{\prime} \beta}  \big] & \dots  \\ \dots &   e_{(n+3) (n+1)} \big[ \underset{\alpha, \beta \neq n+1, n+2} e_{\alpha^{\prime} \beta^{\prime}} \big] &  e_{(n+3) (n+2)} \big[ \underset{\alpha, \beta \neq n+1, n+2} e_{\alpha^{\prime} \beta} \big] &    e_{(n+3)2} \big[ \underset{\alpha , \beta \neq n+1, n+2}{\mathrm{span}} e_{\alpha^{\prime} \beta}  \big] & \dots  \\ \dots &    0 & 0 &   e_{(n+3)3} \big[ \underset{\alpha , \beta \neq n+1, n+2}{\mathrm{span}} e_{\alpha^{\prime} \beta} \big] & \dots \\   \dots &     0 & 0 &        \vdots & \vdots      \\ \dots &     0 & 0 &       e_{(n+3)(2n+1)} \big[ \underset{\alpha , \beta \neq n+1, n+2}{\mathrm{span}} e_{\alpha^{\prime} \beta}  \big] & \dots   \\ \dots &   0 & 0 &    e_{(n+3)(2n+2)} \big[ \underset{\alpha , \beta \neq n+1, n+2}{\mathrm{span}} e_{\alpha^{\prime} \beta} \big] & \dots          \end{array}   
\end{align*}  }

 {\tiny \begin{align*}
    \begin{array}{cccccccccccccccccccccccccccccccccccc}         \dots &     e_{(2n+2)1} \big[ \underset{\alpha , \beta \neq n+1, n+2}{\mathrm{span}} e_{\alpha^{\prime} \beta}  \big]      \\ \dots &      e_{(2n+2)2} \big[ \underset{\alpha , \beta \neq n+1, n+2}{\mathrm{span}} e_{\alpha^{\prime} \beta}  \big]      \\  \dots &      e_{(2n+2)3} \big[ \underset{\alpha , \beta \neq n+1, n+2}{\mathrm{span}} e_{\alpha^{\prime} \beta}  \big] \\ \dots & 0 \\ \vdots & \vdots \\ \dots & 0    \\ \dots &   e_{(2n+2)(2n+1)} \big[ \underset{\alpha , \beta \neq n+1, n+2}{\mathrm{span}} e_{\alpha^{\prime} \beta}  \big]    \\   \dots &  e_{(2n+2)(2n+2)} \big[ \underset{\alpha , \beta \neq n+1, n+2}{\mathrm{span}} e_{\alpha^{\prime} \beta} \big]                          \end{array}  , 
\end{align*}  }  
 
\noindent corresponding to $\mathscr{R}_4$, from the observation that,

\begin{align*}
 \mathscr{R}_{4,1} + \mathscr{R}_{4,2} + \mathscr{R}_{4,5} = 6   \bigg[ \bigg[  \underset{\alpha \neq n+1, n+2, \beta = n+1, n+2}{\mathrm{span}}   e_{\alpha \beta}      \bigg] \bigotimes  \bigg[      \underset{\alpha \neq n+1, n+2, \beta = n+1,n+2}{\mathrm{span}}         e_{\alpha^{\prime} \beta^{\prime}}     \bigg]  \bigg] , \\  \\     \mathscr{R}_{4,3} + \mathscr{R}_{4,4} =           \bigg[ \bigg[   \underset{\alpha, \beta \neq n+1, n+2}{\mathrm{span}}  e_{\alpha \beta}        \bigg] \bigotimes  \bigg[     \underset{\alpha \neq n+1, n+2, \beta = n+1,n+2}{\mathrm{span}}         e_{\alpha^{\prime} \beta^{\prime}}        \bigg]  \bigg]  +          \bigg[ \bigg[       \underset{\alpha, \beta \neq n+1, n+2}{\mathrm{span}}  e_{\alpha \beta}    \bigg] \\ \bigotimes  \bigg[      \underset{\alpha \neq n+1, n+2, \beta = n+1, n+2}{\mathrm{span}} e_{\alpha^{\prime} \beta}      \bigg]  \bigg] \\ =     \bigg[   \underset{\alpha, \beta \neq n+1, n+2}{\mathrm{span}}  e_{\alpha \beta}        \bigg]  \bigotimes \bigg[       \underset{\alpha \neq n+1, n+2, \beta = n+1, n+2}{\mathrm{span}} \big[ e_{\alpha^{\prime} \beta^{\prime}} +  e_{\alpha^{\prime} \beta}      \big]     \bigg]     . 
\end{align*}

\item[$\bullet$] \textit{Tensor product of elementary matrices for $\mathscr{R}_5$}. Disregarding nonconstant prefactors, one has that,

\begin{align*}
       \bigg[  \underset{\alpha \neq n+1, n+2, \beta = n+1, n+2}{\sum} b^+_{\alpha} \big( u \big) e_{\beta^{\prime} \alpha^{\prime}}  +  \underset{\alpha \neq n+1, n+2, \beta = n+1, n+2}{\sum} b^-_{\alpha} \big( u \big) e_{\beta \alpha^{\prime}} \bigg] \\ \otimes       \big[ e_{\beta \alpha} \big] \\ \\ =     \bigg[  \underset{\alpha \neq n+1, n+2, \beta = n+1, n+2}{\sum} b^+_{\alpha} \big( u \big) e_{\beta^{\prime} \alpha^{\prime}}  \bigg]   \otimes       \big[ e_{\beta \alpha} \big]  + \bigg[  \underset{\alpha \neq n+1, n+2, \beta = n+1, n+2}{\sum} b^-_{\alpha} \big( u \big) e_{\beta \alpha^{\prime}} \bigg] \\ \otimes       \big[ e_{\beta \alpha} \big]   \\ \\ = \bigg[  \bigg[  \underset{\alpha \neq n+1, n+2, \beta = n+1, n+2}{\mathrm{span}} b^+_{\alpha} \big( u \big) e_{\beta^{\prime} \alpha^{\prime}}  \bigg]   \otimes       \big[ e_{\beta \alpha} \big] \bigg]  +  \bigg[ \bigg[  \underset{\alpha \neq n+1, n+2, \beta = n+1, n+2}{\mathrm{span}} b^-_{\alpha} \big( u \big) e_{\beta \alpha^{\prime}} \bigg] \end{align*}

\begin{align*} \otimes       \big[ e_{\beta \alpha} \big] \bigg] 
\\ \\ 
= \bigg[  \bigg[  \underset{\alpha \neq n+1, n+2, \beta = n+1, n+2}{\mathrm{span}} b^+_{\alpha} \big( u \big) e_{\beta^{\prime} \alpha^{\prime}}  \bigg]   \otimes   \underset{\alpha \neq n+1, n+2, \beta = n+1, n+2}{\mathrm{span}}      \big[ e_{\beta \alpha} \big] \bigg]  \\ +  \bigg[ \bigg[  \underset{\alpha \neq n+1, n+2, \beta = n+1, n+2}{\mathrm{span}} b^-_{\alpha} \big( u \big) e_{\beta \alpha^{\prime}} \bigg]  \otimes   \underset{\alpha \neq n+1, n+2, \beta = n+1, n+2}{\mathrm{span}}      \big[ e_{\beta \alpha} \big] \bigg] 
\end{align*}

 { \tiny  \begin{align*} =  
\begin{array}{cccccccccccccccccccccccccccccccccccc} 
   b^+_{1} \big( u \big) e_{11}  \big[ \underset{\alpha \neq n+1, n+2, \beta = n+1, n+2}{\mathrm{span}}      \big[ e_{\beta \alpha} \big]  \big] & \dots &    b^+_{n} \big( u \big) e_{n1}  \big[ \underset{\alpha \neq n+1, n+2, \beta = n+1, n+2}{\mathrm{span}}      \big[ e_{\beta \alpha} \big]  \big] & \dots \\      \vdots & \vdots  & \vdots & \dots   \\ \\   b^+_{2n+2} \big( u \big) e_{1(2n+2)}  \big[ \underset{\alpha \neq n+1, n+2, \beta = n+1, n+2}{\mathrm{span}}      \big[ e_{\beta \alpha} \big]  \big] & \dots &    b^+_{n} \big( u \big) e_{n(n+2)}  \big[ \underset{\alpha \neq n+1, n+2, \beta = n+1, n+2}{\mathrm{span}}      \big[ e_{\beta \alpha} \big]  \big] & \dots   \end{array}  
\end{align*}  }

{\tiny \begin{align*}
    \begin{array}{cccccccccccccccccccccccccccccccccccc}     \dots & 0 & 0 &    b^+_{n+3} \big( u \big) e_{(n+3)1}  \big[ \underset{\alpha \neq n+1, n+2, \beta = n+1, n+2}{\mathrm{span}}      \big[ e_{\beta \alpha} \big]  \big] & \dots &       b^+_{2n+2} \big( u \big) e_{(2n+2)1}  \big[ \underset{\alpha \neq n+1, n+2, \beta = n+1, n+2}{\mathrm{span}}      \big[ e_{\beta \alpha} \big]  \big]     \\ \dots & \vdots & \vdots & \vdots & \vdots  \\ \dots & 0 & 0 &  b^+_{2n+2} \big( u \big) e_{(n+3)(2n+2)}  \big[ \underset{\alpha \neq n+1, n+2, \beta = n+1, n+2}{\mathrm{span}}      \big[ e_{\beta \alpha} \big]  \big] & \dots & b^+_{2n+2} \big( u \big) e_{(2n+2)(2n+2)}  \big[ \underset{\alpha \neq n+1, n+2, \beta = n+1, n+2}{\mathrm{span}}      \big[ e_{\beta \alpha} \big]  \big]   \\          \end{array}  
\end{align*}  }

 { \tiny  \begin{align*} +   \begin{array}{cccccccccccccccccccccccccccccccccccc} 
   b^-_{1} \big( u \big) e_{11}  \big[ \underset{\alpha \neq n+1, n+2, \beta = n+1, n+2}{\mathrm{span}}      \big[ e_{\beta \alpha} \big]  \big] & \dots &    b^-_{n} \big( u \big) e_{n1}  \big[ \underset{\alpha \neq n+1, n+2, \beta = n+1, n+2}{\mathrm{span}}      \big[ e_{\beta \alpha} \big]  \big] & \dots \\      \vdots & \vdots  & \vdots & \dots   \\ \\   b^-_{2n+2} \big( u \big) e_{1(2n+2)}  \big[ \underset{\alpha \neq n+1, n+2, \beta = n+1, n+2}{\mathrm{span}}      \big[ e_{\beta \alpha} \big]  \big] & \dots &    b^-_{n} \big( u \big) e_{n(n+2)}  \big[ \underset{\alpha \neq n+1, n+2, \beta = n+1, n+2}{\mathrm{span}}      \big[ e_{\beta \alpha} \big]  \big] & \dots   \end{array}  
\end{align*}  }

{\tiny \begin{align*}
    \begin{array}{cccccccccccccccccccccccccccccccccccc}     \dots & 0 & 0 &    b^-_{n+3} \big( u \big) e_{(n+3)1}  \big[ \underset{\alpha \neq n+1, n+2, \beta = n+1, n+2}{\mathrm{span}}      \big[ e_{\beta \alpha} \big]  \big] & \dots &       b^-_{2n+2} \big( u \big) e_{(2n+2)1}  \big[ \underset{\alpha \neq n+1, n+2, \beta = n+1, n+2}{\mathrm{span}}      \big[ e_{\beta \alpha} \big]  \big]     \\ \dots & \vdots & \vdots & \vdots & \vdots  \\ \dots & 0 & 0 &  b^-_{2n+2} \big( u \big) e_{(n+3)(2n+2)}  \big[ \underset{\alpha \neq n+1, n+2, \beta = n+1, n+2}{\mathrm{span}}      \big[ e_{\beta \alpha} \big]  \big] & \dots & b^-_{2n+2} \big( u \big) e_{(2n+2)(2n+2)}  \big[ \underset{\alpha \neq n+1, n+2, \beta = n+1, n+2}{\mathrm{span}}      \big[ e_{\beta \alpha} \big]  \big]   \\          \end{array}  , 
\end{align*}  }

\noindent corresponding to $\mathscr{R}_5$. 
    
\bigskip

\item[$\bullet$] \textit{Tensor product of elementary matrices from $\mathscr{R}_1$ to $\mathscr{R}_5$}. From the previous results for the finite-dimensional representations obtained for $\mathscr{R}_1, \mathscr{R}_2, \mathscr{R}_3, \mathscr{R}_4$ and $\mathscr{R}_5$, one has that,

\begin{align*}
 \mathscr{R}_1 +  \mathscr{R}_2 +  \mathscr{R}_3 +  \mathscr{R}_4 +  \mathscr{R}_5 ,
\end{align*}

\noindent approximately equals,

 { \tiny  \begin{align*}  \begin{array}{cccccccccccccccccccccccccccccccccccc}        e_{11} \big[ \underset{\alpha < n+1, \beta = n+1, n+2}{\mathrm{span}} e_{\beta \alpha}    \big]  & \dots &  e_{n1} \big[ \underset{\alpha < n+1, \beta = n+1, n+2}{\mathrm{span}} e_{\beta \alpha}    \big] & 0 & 0 &  e_{(n+3)1} \big[ \underset{\alpha < n+1, \beta = n+1, n+2}{\mathrm{span}} e_{\beta \alpha}    \big]   & \dots    \\     \vdots   & \dots &  \vdots  & 0 & 0 &  \vdots   & \dots  \\    \vdots   & \dots &  \vdots  &  \vdots  &  \vdots  &  \vdots   & \dots           \\     e_{1(2n+2)} \big[ \underset{\alpha < n+1, \beta = n+1, n+2}{\mathrm{span}} e_{\beta \alpha}    \big]  & \dots &  e_{n(2n+2)} \big[ \underset{\alpha < n+1, \beta = n+1, n+2}{\mathrm{span}} e_{\beta \alpha}    \big] & 0 & 0 &  e_{(n+3)(2n+2)} \big[ \underset{\alpha < n+1, \beta = n+1, n+2}{\mathrm{span}} e_{\beta \alpha}    \big]   & \dots      \end{array}      
\end{align*} }

 { \tiny  \begin{align*}   \begin{array}{cccccccccccccccccccccccccccccccccccc}                                   \dots &     e_{(n+4)1} \big[ \underset{\alpha < n+1, \beta = n+1, n+2}{\mathrm{span}} e_{\beta \alpha}    \big]  & \dots &    e_{(2n+2)1} \big[ \underset{\alpha < n+1, \beta = n+1, n+2}{\mathrm{span}} e_{\beta \alpha}    \big]  \\       \dots & \vdots & \vdots & \vdots  \\   \dots & \vdots & \vdots & \vdots  \\   \dots &     e_{(n+3)(2n+2)} \big[ \underset{\alpha < n+1, \beta = n+1, n+2}{\mathrm{span}} e_{\beta \alpha}    \big]  & \dots & e_{(2n+2)(2n+2)} \big[ \underset{\alpha < n+1, \beta = n+1, n+2}{\mathrm{span}} e_{\beta \alpha}    \big]     \end{array}        
\end{align*} }

 { \tiny  \begin{align*} +  \begin{array}{cccccccccccccccccccccccccccccccccccc}      e_{11}       \big[ \underset{\alpha < n+1}{\mathrm{span}} e_{\beta^{\prime}\alpha} \big]         & \dots &   e_{n1}       \big[ \underset{\alpha < n+1}{\mathrm{span}} e_{\beta^{\prime}\alpha} \big]          & 0 & \dots \\  \vdots & \vdots &  \vdots & 0 & \dots \\     e_{1(2n+2) }       \big[ \underset{\alpha < n+1}{\mathrm{span}} e_{\beta^{\prime}\alpha} \big]         & \dots &   e_{n(2n+2)}       \big[ \underset{\alpha < n+1}{\mathrm{span}} e_{\beta^{\prime}\alpha} \big]          & 0 & \dots               \end{array}        
\end{align*} }

 { \tiny  \begin{align*}   \begin{array}{cccccccccccccccccccccccccccccccccccc}  \dots &   e_{(n+3) 1}       \big[ \underset{\alpha < n+1}{\mathrm{span}} e_{\beta^{\prime}\alpha} \big]       & \dots &  e_{(2n+2) 1}       \big[ \underset{\alpha < n+1}{\mathrm{span}} e_{\beta^{\prime}\alpha} \big]  \\    \dots &  \vdots & \vdots & \vdots     \\ \dots &   \vdots & \vdots & \vdots   \\   \dots & e_{(n+3)(2n+1)} \big[ \underset{\alpha < n+1}{\mathrm{span}} e_{\beta^{\prime} \alpha}  \big]  & \dots  & e_{(2n+1)(2n+1)} \big[ \underset{\alpha < n+1}{\mathrm{span}} e_{\beta^{\prime} \alpha}  \big]  \end{array}        
\end{align*} }

 { \tiny  \begin{align*} -  \begin{array}{cccccccccccccccccccccccccccccccccccc}           e_{11} e_{\alpha^{\prime} \beta^{\prime}} &   e_{21} e_{\alpha^{\prime} \beta^{\prime}} & \dots &  e_{(2n+2)1} e_{\alpha^{\prime} \beta^{\prime}} \\ \vdots & \vdots & \vdots & \vdots \\      e_{1(2n+2) } e_{\alpha^{\prime} \beta^{\prime}} &   e_{2(2n+2)} e_{\alpha^{\prime} \beta^{\prime}} & \dots &  e_{(2n+2)(2n+2)} e_{\alpha^{\prime} \beta^{\prime}} \end{array}  
\end{align*} }

 { \tiny  \begin{align*} -  \begin{array}{cccccccccccccccccccccccccccccccccccc}  e_{\beta^{\prime} \alpha^{\prime}} & e_{\beta^{\prime} \alpha^{\prime}} & \dots & e_{\beta^{\prime} \alpha^{\prime}} & e_{\beta^{\prime} \alpha^{\prime}} \big[  {\mathrm{span}}\text{ }  e_{1 (n+1)}     \big]  & e_{\beta^{\prime} \alpha^{\prime}} \big[  {\mathrm{span}} \text{ }  e_{1 (n+2)}     \big] & \dots \\ \vdots & \vdots & \vdots & \vdots & \vdots & \vdots & \dots \\ e_{\beta^{\prime} \alpha^{\prime}} & e_{\beta^{\prime} \alpha^{\prime}} & \dots & e_{\beta^{\prime} \alpha^{\prime}} & e_{\beta^{\prime} (2n+2) } \big[  {\mathrm{span}}\text{ }  e_{(2n+2)  (n+1)}     \big]  & e_{\beta^{\prime} (2n+2) } \big[  {\mathrm{span}} \text{ }  e_{(2n+2)  (n+2)}     \big] & \dots  \end{array}  
\end{align*} }

 { \tiny  \begin{align*}   \begin{array}{cccccccccccccccccccccccccccccccccccc}  \dots &    e_{\beta^{\prime} \alpha^{\prime}} \big[  {\mathrm{span}}\text{ }  e_{1 (2n+1)}     \big]  & e_{\beta^{\prime} \alpha^{\prime}} \big[  {\mathrm{span}} \text{ }  e_{1 (2n+2)}     \big]      \\  \dots  &  \vdots & \vdots    \\ \dots &      e_{\beta^{\prime} \alpha^{\prime}} \big[  {\mathrm{span}}\text{ }  e_{(2n+2) (2n+1)}     \big]  & e_{\beta^{\prime} \alpha^{\prime}} \big[  {\mathrm{span}} \text{ }  e_{(2n+2)  (2n+2)}     \big]         \end{array}  
\end{align*} }

 { \tiny  \begin{align*} +  \begin{array}{cccccccccccccccccccccccccccccccccccc}  e_{1,1}    \big[ \underset{\alpha \neq n+1, n+2}{\mathrm{span}}       e_{\alpha \alpha} \big]  & \dots & \dots  & e_{n,1} \big[ \underset{\alpha \neq n+1, n+2}{\mathrm{span}}       e_{\alpha \alpha} \big] & \dots   \\ e_{1,2} \big[ \underset{\alpha \neq n+1, n+2}{\mathrm{span}}       e_{\alpha \alpha} \big]   & e_{2,2} \big[ \underset{\alpha \neq n+1, n+2}{\mathrm{span}}       e_{\alpha \alpha} \big]  & \dots & e_{n-1,2}  \big[ \underset{\alpha \neq n+1, n+2}{\mathrm{span}}       e_{\alpha \alpha} \big] & \dots  \\ \vdots & \dots &  \ddots & \dots & \dots    \\ e_{1,n-1}
 \big[ \underset{\alpha \neq n+1, n+2}{\mathrm{span}}       e_{\alpha \alpha} \big]    & \dots &  e_{n-1,n-1} \big[ \underset{\alpha \neq n+1, n+2}{\mathrm{span}}       e_{\alpha \alpha} \big] & \ddots & \dots   \\          e_{1,n+3}    \big[ \underset{\alpha \neq n+1, n+2}{\mathrm{span}}       e_{\alpha \alpha} \big]  &   \dots & \dots  & e_{n+3, n+3} \big[ \underset{\alpha \neq n+1, n+2}{\mathrm{span}}       e_{\alpha \alpha} \big]   & \dots        \end{array}        
\end{align*} }

 { \tiny  \begin{align*}   \begin{array}{cccccccccccccccccccccccccccccccccccc} e_{n+2,1} \big[ \underset{\alpha \neq n+1, n+2}{\mathrm{span}}       e_{\alpha \alpha} \big] &  e_{n+3,1} \big[ \underset{\alpha \neq n+1, n+2}{\mathrm{span}}       e_{\alpha \alpha} \big] & \dots &    e_{2n+2,1}    \big[ \underset{\alpha \neq n+1, n+2}{\mathrm{span}}       e_{\alpha \alpha} \big]     \\    \dots &   e_{n+3,2} \big[ \underset{\alpha \neq n+1, n+2}{\mathrm{span}}       e_{\alpha \alpha} \big]  & \dots &    e_{2n+2,2}    \big[ \underset{\alpha \neq n+1, n+2}{\mathrm{span}}       e_{\alpha \alpha} \big]         \\ \vdots  &    \vdots & \vdots  & \vdots      \\ \dots &   e_{n+3,2n+2}   \big[ \underset{\alpha \neq n+1, n+2}{\mathrm{span}}       e_{\alpha \alpha} \big]  & \dots &      e_{2n+2,2n+2}   \big[ \underset{\alpha \neq n+1, n+2}{\mathrm{span}}       e_{\alpha \alpha} \big]  \end{array}      
\end{align*} }

{\tiny \begin{align*}
+   \begin{array}{cccccccccccccccccccccccccccccccccccc}       e_{11} \big[    \underset{\alpha \text{ } \text{or} \text{ } \beta, \beta^{\prime} \neq n+1, n+2}{\underset{\alpha \neq \beta, \beta^{\prime}}{\mathrm{span}}} e_{\beta \beta}       \big] & \ddots & & & & &  \dots  \\  & e_{22} \big[    \underset{\alpha \text{ } \text{or} \text{ } \beta, \beta^{\prime} \neq n+1, n+2}{\underset{\alpha \neq \beta, \beta^{\prime}}{\mathrm{span}}} e_{\beta \beta} \big] & & & & & \dots  \\ & & \ddots & & & &  \dots  \\ \vdots & \vdots & \vdots \\ & & &  e_{nn } \big[ \underset{\alpha \text{ } \text{or} \text{ } \beta, \beta^{\prime} \neq n+1, n+2}{\underset{\alpha \neq \beta, \beta^{\prime}}{\mathrm{span}}} e_{\beta \beta}      \big]  & \ddots &  & \dots  \\ & & & &   0 & \ddots & \dots  \\ & & & & & 0 & \ddots \\ & & & & &  & \dots  \end{array} 
\end{align*}  }

{\tiny \begin{align*}
   \begin{array}{cccccccccccccccccccccccccccccccccccc}      \dots \\ \dots   \\   \dots \\ \dots    \\   \dots & \ddots  \\ \dots &  e_{(2n+2 )(2n+2)}  \big[ \underset{\alpha \text{ } \text{or} \text{ } \beta, \beta^{\prime} \neq n+1, n+2}{\underset{\alpha \neq \beta, \beta^{\prime}}{\mathrm{span}}} e_{\beta \beta}      \big]  \end{array} , 
\end{align*}  }

{\tiny \begin{align*}
+     \begin{array}{cccccccccccccccccccccccccccccccccccc}     0  & \ddots & & & & &  \dots  \\  & 0 & & & & & \dots  \\ & & \ddots & & & &  \dots  \\ \vdots & \vdots & \vdots \\ & & & 0  & \ddots &  & \dots  \\ & & & &   e_{(n+1)(n+1)}  \big[ \underset{\alpha \text{ } \text{or} \text{ } \beta, \beta^{\prime} \neq n+1, n+2}{\underset{\alpha \neq \beta, \beta^{\prime}}{\mathrm{span}}} e_{\beta \beta}      \big] & \ddots & \dots  \\ & & & & & e_{(n+2)(n+2)}  \big[ \underset{\alpha \text{ } \text{or} \text{ } \beta, \beta^{\prime} \neq n+1, n+2}{\underset{\alpha \neq \beta, \beta^{\prime}}{\mathrm{span}}} e_{\beta \beta}      \big]  & \ddots \\ & & & & &  & \dots  \end{array} 
\end{align*}  }

{\tiny \begin{align*}
   \begin{array}{cccccccccccccccccccccccccccccccccccc}      \dots \\ \dots   \\   \dots \\ \dots    \\   \dots & \ddots  \\ \dots & 0   \end{array}  
\end{align*}  }

 { \tiny  \begin{align*} +   \begin{array}{cccccccccccccccccccccccccccccccccccc} 
      e_{11} \big[ \underset{\alpha , \beta \neq n+1, n+2}{\mathrm{span}} e_{\alpha^{\prime} \beta^{\prime}}  \big] & \dots &  \dots &  e_{n1} \big[ \underset{\alpha , \beta \neq n+1, n+2}{\mathrm{span}} e_{\alpha^{\prime} \beta^{\prime}}  \big] & \dots   \\  e_{12} \big[ \underset{\alpha , \beta \neq n+1, n+2}{\mathrm{span}} e_{\alpha^{\prime} \beta^{\prime}}  \big] & e_{22} \big[ \underset{\alpha , \beta \neq n+1, n+2}{\mathrm{span}} e_{\alpha^{\prime} \beta^{\prime}}  \big]   & \dots & e_{n2} \big[ \underset{\alpha , \beta \neq n+1, n+2}{\mathrm{span}} e_{\alpha^{\prime} \beta^{\prime}}  \big] & \dots  \\ \vdots & \vdots & \vdots & \vdots  \\  e_{1n} \big[ \underset{\alpha , \beta \neq n+1, n+2}{\mathrm{span}} e_{\alpha^{\prime} \beta^{\prime}}  \big] & \dots &  \dots &  e_{nn} \big[ \underset{\alpha , \beta \neq n+1, n+2}{\mathrm{span}} e_{\alpha^{\prime} \beta^{\prime}}  \big]  & \dots \\ 0 & \dots & \dots & 0 & \dots \\     0 & \dots & \dots & 0 & \dots  \\  e_{1(n+3)} \big[ \underset{\alpha , \beta \neq n+1, n+2}{\mathrm{span}} e_{\alpha^{\prime} \beta^{\prime}}  \big] & \dots &  \dots &  e_{n(n+3)} \big[ \underset{\alpha , \beta \neq n+1, n+2}{\mathrm{span}} e_{\alpha^{\prime} \beta^{\prime}}  \big]  & \dots  \\  \vdots & \vdots & \vdots & \vdots \\ e_{1(2n+2)} \big[ \underset{\alpha , \beta \neq n+1, n+2}{\mathrm{span}} e_{\alpha^{\prime} \beta^{\prime}}  \big] & \dots &  \dots &  e_{n(2n+2)} \big[ \underset{\alpha , \beta \neq n+1, n+2}{\mathrm{span}} e_{\alpha^{\prime} \beta^{\prime}}  \big]  & \dots    \end{array} 
\end{align*}  }

  { \tiny  \begin{align*}   \begin{array}{cccccccccccccccccccccccccccccccccccc} 
   \dots &  0 & 0 &   e_{(n+3)1} \big[ \underset{\alpha , \beta \neq n+1, n+2}{\mathrm{span}} e_{\alpha^{\prime} \beta^{\prime}}  \big] & \dots &  e_{(2n+2)1} \big[ \underset{\alpha , \beta \neq n+1, n+2}{\mathrm{span}} e_{\alpha^{\prime} \beta^{\prime}}  \big]  \\ \dots &   0 & 0 &    e_{(n+3)2} \big[ \underset{\alpha , \beta \neq n+1, n+2}{\mathrm{span}} e_{\alpha^{\prime} \beta^{\prime}}  \big] & \dots &  e_{(2n+2)2} \big[ \underset{\alpha , \beta \neq n+1, n+2}{\mathrm{span}} e_{\alpha^{\prime} \beta^{\prime}}  \big]   \\ \dots &    0 & 0 &   e_{(n+3)3} \big[ \underset{\alpha , \beta \neq n+1, n+2}{\mathrm{span}} e_{\alpha^{\prime} \beta^{\prime}}  \big] & \dots &  e_{(2n+2)3} \big[ \underset{\alpha , \beta \neq n+1, n+2}{\mathrm{span}} e_{\alpha^{\prime} \beta^{\prime}}  \big] \\   \dots &     0 & 0 &        \vdots & \vdots   & \vdots    \\ \dots &     0 & 0 &       e_{(n+3)(2n+1)} \big[ \underset{\alpha , \beta \neq n+1, n+2}{\mathrm{span}} e_{\alpha^{\prime} \beta^{\prime}}  \big] & \dots &  e_{(2n+2)(2n+1)} \big[ \underset{\alpha , \beta \neq n+1, n+2}{\mathrm{span}} e_{\alpha^{\prime} \beta^{\prime}}  \big]  \\ \dots &   0 & 0 &    e_{(n+3)(2n+2)} \big[ \underset{\alpha , \beta \neq n+1, n+2}{\mathrm{span}} e_{\alpha^{\prime} \beta^{\prime}}  \big] & \dots &  e_{(2n+2)(2n+2)} \big[ \underset{\alpha , \beta \neq n+1, n+2}{\mathrm{span}} e_{\alpha^{\prime} \beta^{\prime}}  \big]            \end{array}   
\end{align*}  } 

{ \tiny  \begin{align*} + 6 \begin{array}{cccccccccccccccccccccccccccccccccccc} 
      e_{11} \big[ \underset{\alpha , \beta \neq n+1, n+2}{\mathrm{span}} e_{\alpha^{\prime} \beta^{\prime}}  \big] & \dots &  \dots &  e_{n1} \big[ \underset{\alpha , \beta \neq n+1, n+2}{\mathrm{span}} e_{\alpha^{\prime} \beta^{\prime}}  \big] & \dots   \\  e_{12} \big[ \underset{\alpha , \beta \neq n+1, n+2}{\mathrm{span}} e_{\alpha^{\prime} \beta^{\prime}}  \big] & e_{22} \big[ \underset{\alpha , \beta \neq n+1, n+2}{\mathrm{span}} e_{\alpha^{\prime} \beta^{\prime}}  \big]   & \dots & e_{n2} \big[ \underset{\alpha , \beta \neq n+1, n+2}{\mathrm{span}} e_{\alpha^{\prime} \beta^{\prime}}  \big] & \dots  \\ \vdots & \vdots & \vdots & \vdots  \\  e_{1n} \big[ \underset{\alpha , \beta \neq n+1, n+2}{\mathrm{span}} e_{\alpha^{\prime} \beta^{\prime}}  \big] & \dots &  \dots &  e_{nn} \big[ \underset{\alpha , \beta \neq n+1, n+2}{\mathrm{span}} e_{\alpha^{\prime} \beta^{\prime}}  \big]  & \dots \\ 0 & \dots & \dots & 0 & \dots \\     0 & \dots & \dots & 0 & \dots  \\  e_{1(n+3)} \big[ \underset{\alpha , \beta \neq n+1, n+2}{\mathrm{span}} e_{\alpha^{\prime} \beta^{\prime}}  \big] & \dots &  \dots &  e_{n(n+3)} \big[ \underset{\alpha , \beta \neq n+1, n+2}{\mathrm{span}} e_{\alpha^{\prime} \beta^{\prime}}  \big]  & \dots  \\  \vdots & \vdots & \vdots & \vdots \\ e_{1(2n+2)} \big[ \underset{\alpha , \beta \neq n+1, n+2}{\mathrm{span}} e_{\alpha^{\prime} \beta^{\prime}}  \big] & \dots &  \dots &  e_{n(2n+2)} \big[ \underset{\alpha , \beta \neq n+1, n+2}{\mathrm{span}} e_{\alpha^{\prime} \beta^{\prime}}  \big]  & \dots    \end{array} 
\end{align*}  }

  { \tiny  \begin{align*}   \begin{array}{cccccccccccccccccccccccccccccccccccc} 
   \dots & 0 & 0 &   e_{(n+3)1} \big[ \underset{\alpha , \beta \neq n+1, n+2}{\mathrm{span}} e_{\alpha^{\prime} \beta^{\prime}}  \big] & \dots  \\ \dots &   e_{(n+3) (n+1)} \big[ \underset{\alpha, \beta \neq n+1, n+2} e_{\alpha^{\prime} \beta^{\prime}} \big] &  e_{(n+3) (n+2)} \big[ \underset{\alpha, \beta \neq n+1, n+2} e_{\alpha^{\prime} \beta^{\prime}} \big] &    e_{(n+3)2} \big[ \underset{\alpha , \beta \neq n+1, n+2}{\mathrm{span}} e_{\alpha^{\prime} \beta^{\prime}}  \big] & \dots  \\ \dots &    0 & 0 &   e_{(n+3)3} \big[ \underset{\alpha , \beta \neq n+1, n+2}{\mathrm{span}} e_{\alpha^{\prime} \beta^{\prime}}  \big] & \dots \\   \dots &     0 & 0 &        \vdots & \vdots      \\ \dots &     0 & 0 &       e_{(n+3)(2n+1)} \big[ \underset{\alpha , \beta \neq n+1, n+2}{\mathrm{span}} e_{\alpha^{\prime} \beta^{\prime}}  \big] & \dots   \\ \dots &   0 & 0 &    e_{(n+3)(2n+2)} \big[ \underset{\alpha , \beta \neq n+1, n+2}{\mathrm{span}} e_{\alpha^{\prime} \beta^{\prime}}  \big] & \dots          \end{array}   
\end{align*}  }

 {\tiny \begin{align*}
    \begin{array}{cccccccccccccccccccccccccccccccccccc}         \dots &     e_{(2n+2)1} \big[ \underset{\alpha , \beta \neq n+1, n+2}{\mathrm{span}} e_{\alpha^{\prime} \beta^{\prime}}  \big]      \\ \dots &      e_{(2n+2)2} \big[ \underset{\alpha , \beta \neq n+1, n+2}{\mathrm{span}} e_{\alpha^{\prime} \beta^{\prime}}  \big]      \\  \dots &      e_{(2n+2)3} \big[ \underset{\alpha , \beta \neq n+1, n+2}{\mathrm{span}} e_{\alpha^{\prime} \beta^{\prime}}  \big] \\ \dots & 0 \\ \vdots & \vdots \\ \dots & 0    \\ \dots &   e_{(2n+2)(2n+1)} \big[ \underset{\alpha , \beta \neq n+1, n+2}{\mathrm{span}} e_{\alpha^{\prime} \beta^{\prime}}  \big]    \\   \dots &  e_{(2n+2)(2n+2)} \big[ \underset{\alpha , \beta \neq n+1, n+2}{\mathrm{span}} e_{\alpha^{\prime} \beta^{\prime}}  \big]                          \end{array}  
\end{align*}  }

{\tiny \begin{align*}
+    \begin{array}{cccccccccccccccccccccccccccccccccccc}                     \underset{\alpha, \beta \neq n+1, n+2}{\mathrm{span}}  e_{\alpha \beta}       \big[       \underset{\alpha \neq n+1, n+2, \beta = n+1, n+2}{\mathrm{span}} \big[ e_{\alpha^{\prime} \beta^{\prime}} +  e_{\alpha^{\prime} \beta}      \big]     \big]        &    \dots &     \dots & \dots \\     \underset{\alpha, \beta \neq n+1, n+2}{\mathrm{span}}  e_{\alpha \beta}       \big[       \underset{\alpha \neq n+1, n+2, \beta = n+1, n+2}{\mathrm{span}} \big[ e_{\alpha^{\prime} \beta^{\prime}} +  e_{\alpha^{\prime} \beta}      \big]     \big]        &    \dots    & \dots & \dots  \\ \vdots & \vdots & \vdots & \vdots \\ 0     & \underset{\alpha, \beta \neq n+1, n+2}{\mathrm{span}}  e_{\alpha \beta}       \big[       \underset{\alpha \neq n+1, n+2, \beta = n+1, n+2}{\mathrm{span}} \big[ e_{\alpha^{\prime} \beta^{\prime}} +  e_{\alpha^{\prime} \beta}      \big]     \big]   & \dots  & \dots   \\ 0  & \underset{\alpha, \beta \neq n+1, n+2}{\mathrm{span}}  e_{\alpha \beta}       \big[       \underset{\alpha \neq n+1, n+2, \beta = n+1, n+2}{\mathrm{span}} \big[ e_{\alpha^{\prime} \beta^{\prime}} +  e_{\alpha^{\prime} \beta}      \big]     \big]   & \dots  & \dots   \\     \underset{\alpha, \beta \neq n+1, n+2}{\mathrm{span}}  e_{\alpha \beta}       \big[       \underset{\alpha \neq n+1, n+2, \beta = n+1, n+2}{\mathrm{span}} \big[ e_{\alpha^{\prime} \beta^{\prime}} +  e_{\alpha^{\prime} \beta}      \big]     \big] & \dots & \dots & \dots & \vdots   \\ \vdots & \vdots & \vdots & \dots & \vdots   \\ \underset{\alpha, \beta \neq n+1, n+2}{\mathrm{span}}  e_{\alpha \beta}       \big[       \underset{\alpha \neq n+1, n+2, \beta = n+1, n+2}{\mathrm{span}} \big[ e_{\alpha^{\prime} \beta^{\prime}} +  e_{\alpha^{\prime} \beta}      \big]     \big] & \dots & \dots                             \end{array}  
\end{align*}  }

{\tiny \begin{align*}
    \begin{array}{cccccccccccccccccccccccccccccccccccc}     \dots &   \underset{\alpha, \beta \neq n+1, n+2}{\mathrm{span}}  e_{\alpha \beta}       \big[       \underset{\alpha \neq n+1, n+2, \beta = n+1, n+2}{\mathrm{span}} \big[ e_{\alpha^{\prime} \beta^{\prime}} +  e_{\alpha^{\prime} \beta}      \big]     \big]  \dots  & \dots & \dots \\   \dots  & \dots & \dots \\ \dots &    \underset{\alpha, \beta \neq n+1, n+2}{\mathrm{span}}  e_{\alpha \beta}       \big[       \underset{\alpha \neq n+1, n+2, \beta = n+1, n+2}{\mathrm{span}} \big[ e_{\alpha^{\prime} \beta^{\prime}} +  e_{\alpha^{\prime} \beta}      \big]     \big] &  \dots  & \dots & \dots \\      \dots & \dots &    \dots   &       \underset{\alpha, \beta \neq n+1, n+2}{\mathrm{span}}  e_{\alpha \beta}       \big[       \underset{\alpha \neq n+1, n+2, \beta = n+1, n+2}{\mathrm{span}} \big[ e_{\alpha^{\prime} \beta^{\prime}} +  e_{\alpha^{\prime} \beta}      \big]     \big]   & \dots      \\   \dots & \dots &         \dots &       \underset{\alpha, \beta \neq n+1, n+2}{\mathrm{span}}  e_{\alpha \beta}       \big[       \underset{\alpha \neq n+1, n+2, \beta = n+1, n+2}{\mathrm{span}} \big[ e_{\alpha^{\prime} \beta^{\prime}} +  e_{\alpha^{\prime} \beta}      \big]     \big]        & \dots  \\ \dots & \dots &    & \dots  \\     \dots & \dots &         \dots  & \dots & \dots \\ \dots & \dots &    \dots  & \dots & \dots \\ \dots & \dots &     \dots  & \dots & \dots \\                     \end{array}  
\end{align*}  }  

 { \tiny  \begin{align*}   \begin{array}{cccccccccccccccccccccccccccccccccccc} 
    0 & 0 &        \underset{\alpha, \beta \neq n+1, n+2}{\mathrm{span}}  e_{\alpha \beta}       \big[       \underset{\alpha \neq n+1, n+2, \beta = n+1, n+2}{\mathrm{span}} \big[ e_{\alpha^{\prime} \beta^{\prime}} +  e_{\alpha^{\prime} \beta}      \big]     \big]        &    \dots &    \underset{\alpha, \beta \neq n+1, n+2}{\mathrm{span}}  e_{\alpha \beta}       \big[       \underset{\alpha \neq n+1, n+2, \beta = n+1, n+2}{\mathrm{span}} \big[ e_{\alpha^{\prime} \beta^{\prime}} +  e_{\alpha^{\prime} \beta}      \big]     \big]      \\  \vdots & \vdots & \vdots & \vdots  & \vdots  \\     0 & 0 &        \underset{\alpha, \beta \neq n+1, n+2}{\mathrm{span}}  e_{\alpha \beta}       \big[       \underset{\alpha \neq n+1, n+2, \beta = n+1, n+2}{\mathrm{span}} \big[ e_{\alpha^{\prime} \beta^{\prime}} +  e_{\alpha^{\prime} \beta}      \big]     \big]        &    \dots &    \underset{\alpha, \beta \neq n+1, n+2}{\mathrm{span}}  e_{\alpha \beta}       \big[       \underset{\alpha \neq n+1, n+2, \beta = n+1, n+2}{\mathrm{span}} \big[ e_{\alpha^{\prime} \beta^{\prime}} +  e_{\alpha^{\prime} \beta}      \big]     \big]        \end{array}   
\end{align*}  } 

{ \tiny  \begin{align*} + 2 \begin{array}{cccccccccccccccccccccccccccccccccccc} 
      e_{11} \big[ \underset{\alpha , \beta \neq n+1, n+2}{\mathrm{span}} e_{\alpha^{\prime} \beta^{\prime}}  \big] & \dots &  \dots &  e_{n1} \big[ \underset{\alpha , \beta \neq n+1, n+2}{\mathrm{span}} e_{\alpha^{\prime} \beta}  \big] & \dots   \\  e_{12} \big[ \underset{\alpha , \beta \neq n+1, n+2}{\mathrm{span}} e_{\alpha^{\prime} \beta}  \big] & e_{22} \big[ \underset{\alpha , \beta \neq n+1, n+2}{\mathrm{span}} e_{\alpha^{\prime} \beta^{\prime}}  \big]   & \dots & e_{n2} \big[ \underset{\alpha , \beta \neq n+1, n+2}{\mathrm{span}} e_{\alpha^{\prime} \beta}  \big] & \dots  \\ \vdots & \vdots & \vdots & \vdots  \\  e_{1n} \big[ \underset{\alpha , \beta \neq n+1, n+2}{\mathrm{span}} e_{\alpha^{\prime} \beta}  \big] & \dots &  \dots &  e_{nn} \big[ \underset{\alpha , \beta \neq n+1, n+2}{\mathrm{span}} e_{\alpha^{\prime} \beta}  \big]  & \dots \\ 0 & \dots & \dots & 0 & \dots \\     0 & \dots & \dots & 0 & \dots  \\  e_{1(n+3)} \big[ \underset{\alpha , \beta \neq n+1, n+2}{\mathrm{span}} e_{\alpha^{\prime} \beta}  \big] & \dots &  \dots &  e_{n(n+3)} \big[ \underset{\alpha , \beta \neq n+1, n+2}{\mathrm{span}} e_{\alpha^{\prime} \beta}  \big]  & \dots  \\  \vdots & \vdots & \vdots & \vdots \\ e_{1(2n+2)} \big[ \underset{\alpha , \beta \neq n+1, n+2}{\mathrm{span}} e_{\alpha^{\prime} \beta} \big] & \dots &  \dots &  e_{n(2n+2)} \big[ \underset{\alpha , \beta \neq n+1, n+2}{\mathrm{span}} e_{\alpha^{\prime} \beta}  \big]  & \dots    \end{array} 
\end{align*}  }

  { \tiny  \begin{align*}   \begin{array}{cccccccccccccccccccccccccccccccccccc} 
   \dots & 0 & 0 &   e_{(n+3)1} \big[ \underset{\alpha , \beta \neq n+1, n+2}{\mathrm{span}} e_{\alpha^{\prime} \beta}  \big] & \dots  \\ \dots &   e_{(n+3) (n+1)} \big[ \underset{\alpha, \beta \neq n+1, n+2} e_{\alpha^{\prime} \beta^{\prime}} \big] &  e_{(n+3) (n+2)} \big[ \underset{\alpha, \beta \neq n+1, n+2} e_{\alpha^{\prime} \beta} \big] &    e_{(n+3)2} \big[ \underset{\alpha , \beta \neq n+1, n+2}{\mathrm{span}} e_{\alpha^{\prime} \beta}  \big] & \dots  \\ \dots &    0 & 0 &   e_{(n+3)3} \big[ \underset{\alpha , \beta \neq n+1, n+2}{\mathrm{span}} e_{\alpha^{\prime} \beta} \big] & \dots \\   \dots &     0 & 0 &        \vdots & \vdots      \\ \dots &     0 & 0 &       e_{(n+3)(2n+1)} \big[ \underset{\alpha , \beta \neq n+1, n+2}{\mathrm{span}} e_{\alpha^{\prime} \beta}  \big] & \dots   \\ \dots &   0 & 0 &    e_{(n+3)(2n+2)} \big[ \underset{\alpha , \beta \neq n+1, n+2}{\mathrm{span}} e_{\alpha^{\prime} \beta} \big] & \dots          \end{array}   
\end{align*}  }

 {\tiny \begin{align*}
    \begin{array}{cccccccccccccccccccccccccccccccccccc}         \dots &     e_{(2n+2)1} \big[ \underset{\alpha , \beta \neq n+1, n+2}{\mathrm{span}} e_{\alpha^{\prime} \beta}  \big]      \\ \dots &      e_{(2n+2)2} \big[ \underset{\alpha , \beta \neq n+1, n+2}{\mathrm{span}} e_{\alpha^{\prime} \beta}  \big]      \\  \dots &      e_{(2n+2)3} \big[ \underset{\alpha , \beta \neq n+1, n+2}{\mathrm{span}} e_{\alpha^{\prime} \beta}  \big] \\ \dots & 0 \\ \vdots & \vdots \\ \dots & 0    \\ \dots &   e_{(2n+2)(2n+1)} \big[ \underset{\alpha , \beta \neq n+1, n+2}{\mathrm{span}} e_{\alpha^{\prime} \beta}  \big]    \\   \dots &  e_{(2n+2)(2n+2)} \big[ \underset{\alpha , \beta \neq n+1, n+2}{\mathrm{span}} e_{\alpha^{\prime} \beta} \big]                          \end{array}   
\end{align*}  }

 { \tiny  \begin{align*} +  \begin{array}{cccccccccccccccccccccccccccccccccccc} 
      e_{11} \big[ \underset{\alpha , \beta \neq n+1, n+2}{\mathrm{span}} e_{\alpha^{\prime} \beta^{\prime}}  \big] & \dots &  \dots &  e_{n1} \big[ \underset{\alpha , \beta \neq n+1, n+2}{\mathrm{span}} e_{\alpha^{\prime} \beta^{\prime}}  \big] & \dots   \\  e_{12} \big[ \underset{\alpha , \beta \neq n+1, n+2}{\mathrm{span}} e_{\alpha^{\prime} \beta^{\prime}}  \big] & e_{22} \big[ \underset{\alpha , \beta \neq n+1, n+2}{\mathrm{span}} e_{\alpha^{\prime} \beta^{\prime}}  \big]   & \dots & e_{n2} \big[ \underset{\alpha , \beta \neq n+1, n+2}{\mathrm{span}} e_{\alpha^{\prime} \beta^{\prime}}  \big] & \dots  \\ \vdots & \vdots & \vdots & \vdots  \\  e_{1n} \big[ \underset{\alpha , \beta \neq n+1, n+2}{\mathrm{span}} e_{\alpha^{\prime} \beta^{\prime}}  \big] & \dots &  \dots &  e_{nn} \big[ \underset{\alpha , \beta \neq n+1, n+2}{\mathrm{span}} e_{\alpha^{\prime} \beta^{\prime}}  \big]  & \dots \\ 0 & \dots & \dots & 0 & \dots \\     0 & \dots & \dots & 0 & \dots  \\  e_{1(n+3)} \big[ \underset{\alpha , \beta \neq n+1, n+2}{\mathrm{span}} e_{\alpha^{\prime} \beta^{\prime}}  \big] & \dots &  \dots &  e_{n(n+3)} \big[ \underset{\alpha , \beta \neq n+1, n+2}{\mathrm{span}} e_{\alpha^{\prime} \beta^{\prime}}  \big]  & \dots  \\  \vdots & \vdots & \vdots & \vdots \\ e_{1(2n+2)} \big[ \underset{\alpha , \beta \neq n+1, n+2}{\mathrm{span}} e_{\alpha^{\prime} \beta^{\prime}}  \big] & \dots &  \dots &  e_{n(2n+2)} \big[ \underset{\alpha , \beta \neq n+1, n+2}{\mathrm{span}} e_{\alpha^{\prime} \beta^{\prime}}  \big]  & \dots    \end{array} 
\end{align*}  }

  { \tiny  \begin{align*}   \begin{array}{cccccccccccccccccccccccccccccccccccc} 
   \dots &  0 & 0 &   e_{(n+3)1} \big[ \underset{\alpha , \beta \neq n+1, n+2}{\mathrm{span}} e_{\alpha^{\prime} \beta^{\prime}}  \big] & \dots &  e_{(2n+2)1} \big[ \underset{\alpha , \beta \neq n+1, n+2}{\mathrm{span}} e_{\alpha^{\prime} \beta^{\prime}}  \big]  \\ \dots &   0 & 0 &    e_{(n+3)2} \big[ \underset{\alpha , \beta \neq n+1, n+2}{\mathrm{span}} e_{\alpha^{\prime} \beta^{\prime}}  \big] & \dots &  e_{(2n+2)2} \big[ \underset{\alpha , \beta \neq n+1, n+2}{\mathrm{span}} e_{\alpha^{\prime} \beta^{\prime}}  \big]   \\ \dots &    0 & 0 &   e_{(n+3)3} \big[ \underset{\alpha , \beta \neq n+1, n+2}{\mathrm{span}} e_{\alpha^{\prime} \beta^{\prime}}  \big] & \dots &  e_{(2n+2)3} \big[ \underset{\alpha , \beta \neq n+1, n+2}{\mathrm{span}} e_{\alpha^{\prime} \beta^{\prime}}  \big] \\   \dots &     0 & 0 &        \vdots & \vdots   & \vdots    \\ \dots &     0 & 0 &       e_{(n+3)(2n+1)} \big[ \underset{\alpha , \beta \neq n+1, n+2}{\mathrm{span}} e_{\alpha^{\prime} \beta^{\prime}}  \big] & \dots &  e_{(2n+2)(2n+1)} \big[ \underset{\alpha , \beta \neq n+1, n+2}{\mathrm{span}} e_{\alpha^{\prime} \beta^{\prime}}  \big]  \\ \dots &   0 & 0 &    e_{(n+3)(2n+2)} \big[ \underset{\alpha , \beta \neq n+1, n+2}{\mathrm{span}} e_{\alpha^{\prime} \beta^{\prime}}  \big] & \dots &  e_{(2n+2)(2n+2)} \big[ \underset{\alpha , \beta \neq n+1, n+2}{\mathrm{span}} e_{\alpha^{\prime} \beta^{\prime}}  \big]            \end{array}   
\end{align*}  } 

{ \tiny  \begin{align*} + 4 \begin{array}{cccccccccccccccccccccccccccccccccccc} 
      e_{11} \big[ \underset{\alpha , \beta \neq n+1, n+2}{\mathrm{span}} e_{\alpha^{\prime} \beta^{\prime}}  \big] & \dots &  \dots &  e_{n1} \big[ \underset{\alpha , \beta \neq n+1, n+2}{\mathrm{span}} e_{\alpha^{\prime} \beta^{\prime}}  \big] & \dots   \\  e_{12} \big[ \underset{\alpha , \beta \neq n+1, n+2}{\mathrm{span}} e_{\alpha^{\prime} \beta^{\prime}}  \big] & e_{22} \big[ \underset{\alpha , \beta \neq n+1, n+2}{\mathrm{span}} e_{\alpha^{\prime} \beta^{\prime}}  \big]   & \dots & e_{n2} \big[ \underset{\alpha , \beta \neq n+1, n+2}{\mathrm{span}} e_{\alpha^{\prime} \beta^{\prime}}  \big] & \dots  \\ \vdots & \vdots & \vdots & \vdots  \\  e_{1n} \big[ \underset{\alpha , \beta \neq n+1, n+2}{\mathrm{span}} e_{\alpha^{\prime} \beta^{\prime}}  \big] & \dots &  \dots &  e_{nn} \big[ \underset{\alpha , \beta \neq n+1, n+2}{\mathrm{span}} e_{\alpha^{\prime} \beta^{\prime}}  \big]  & \dots \\ 0 & \dots & \dots & 0 & \dots \\     0 & \dots & \dots & 0 & \dots  \\  e_{1(n+3)} \big[ \underset{\alpha , \beta \neq n+1, n+2}{\mathrm{span}} e_{\alpha^{\prime} \beta^{\prime}}  \big] & \dots &  \dots &  e_{n(n+3)} \big[ \underset{\alpha , \beta \neq n+1, n+2}{\mathrm{span}} e_{\alpha^{\prime} \beta^{\prime}}  \big]  & \dots  \\  \vdots & \vdots & \vdots & \vdots \\ e_{1(2n+2)} \big[ \underset{\alpha , \beta \neq n+1, n+2}{\mathrm{span}} e_{\alpha^{\prime} \beta^{\prime}}  \big] & \dots &  \dots &  e_{n(2n+2)} \big[ \underset{\alpha , \beta \neq n+1, n+2}{\mathrm{span}} e_{\alpha^{\prime} \beta^{\prime}}  \big]  & \dots    \end{array} 
\end{align*}  }

  { \tiny  \begin{align*}   \begin{array}{cccccccccccccccccccccccccccccccccccc} 
   \dots & 0 & 0 &   e_{(n+3)1} \big[ \underset{\alpha , \beta \neq n+1, n+2}{\mathrm{span}} e_{\alpha^{\prime} \beta^{\prime}}  \big] & \dots  \\ \dots &   e_{(n+3) (n+1)} \big[ \underset{\alpha, \beta \neq n+1, n+2} e_{\alpha^{\prime} \beta^{\prime}} \big] &  e_{(n+3) (n+2)} \big[ \underset{\alpha, \beta \neq n+1, n+2} e_{\alpha^{\prime} \beta^{\prime}} \big] &    e_{(n+3)2} \big[ \underset{\alpha , \beta \neq n+1, n+2}{\mathrm{span}} e_{\alpha^{\prime} \beta^{\prime}}  \big] & \dots  \\ \dots &    0 & 0 &   e_{(n+3)3} \big[ \underset{\alpha , \beta \neq n+1, n+2}{\mathrm{span}} e_{\alpha^{\prime} \beta^{\prime}}  \big] & \dots \\   \dots &     0 & 0 &        \vdots & \vdots      \\ \dots &     0 & 0 &       e_{(n+3)(2n+1)} \big[ \underset{\alpha , \beta \neq n+1, n+2}{\mathrm{span}} e_{\alpha^{\prime} \beta^{\prime}}  \big] & \dots   \\ \dots &   0 & 0 &    e_{(n+3)(2n+2)} \big[ \underset{\alpha , \beta \neq n+1, n+2}{\mathrm{span}} e_{\alpha^{\prime} \beta^{\prime}}  \big] & \dots          \end{array}   
\end{align*}  }

 {\tiny \begin{align*}
    \begin{array}{cccccccccccccccccccccccccccccccccccc}         \dots &     e_{(2n+2)1} \big[ \underset{\alpha , \beta \neq n+1, n+2}{\mathrm{span}} e_{\alpha^{\prime} \beta^{\prime}}  \big]      \\ \dots &      e_{(2n+2)2} \big[ \underset{\alpha , \beta \neq n+1, n+2}{\mathrm{span}} e_{\alpha^{\prime} \beta^{\prime}}  \big]      \\  \dots &      e_{(2n+2)3} \big[ \underset{\alpha , \beta \neq n+1, n+2}{\mathrm{span}} e_{\alpha^{\prime} \beta^{\prime}}  \big] \\ \dots & 0 \\ \vdots & \vdots \\ \dots & 0    \\ \dots &   e_{(2n+2)(2n+1)} \big[ \underset{\alpha , \beta \neq n+1, n+2}{\mathrm{span}} e_{\alpha^{\prime} \beta^{\prime}}  \big]    \\   \dots &  e_{(2n+2)(2n+2)} \big[ \underset{\alpha , \beta \neq n+1, n+2}{\mathrm{span}} e_{\alpha^{\prime} \beta^{\prime}}  \big]                          \end{array}  
\end{align*}  }  

\end{itemize}

 { \tiny  \begin{align*} +   \begin{array}{cccccccccccccccccccccccccccccccccccc} 
   b^+_{1} \big( u \big) e_{11}  \big[ \underset{\alpha \neq n+1, n+2, \beta = n+1, n+2}{\mathrm{span}}      \big[ e_{\beta \alpha} \big]  \big] & \dots &    b^+_{n} \big( u \big) e_{n1}  \big[ \underset{\alpha \neq n+1, n+2, \beta = n+1, n+2}{\mathrm{span}}      \big[ e_{\beta \alpha} \big]  \big] & \dots \\      \vdots & \vdots  & \vdots & \dots   \\ \\   b^+_{2n+2} \big( u \big) e_{1(2n+2)}  \big[ \underset{\alpha \neq n+1, n+2, \beta = n+1, n+2}{\mathrm{span}}      \big[ e_{\beta \alpha} \big]  \big] & \dots &    b^+_{n} \big( u \big) e_{n(n+2)}  \big[ \underset{\alpha \neq n+1, n+2, \beta = n+1, n+2}{\mathrm{span}}      \big[ e_{\beta \alpha} \big]  \big] & \dots   \end{array}  
\end{align*}  }

{\tiny \begin{align*}
    \begin{array}{cccccccccccccccccccccccccccccccccccc}     \dots & 0 & 0 &    b^+_{n+3} \big( u \big) e_{(n+3)1}  \big[ \underset{\alpha \neq n+1, n+2, \beta = n+1, n+2}{\mathrm{span}}      \big[ e_{\beta \alpha} \big]  \big] & \dots &       b^+_{2n+2} \big( u \big) e_{(2n+2)1}  \big[ \underset{\alpha \neq n+1, n+2, \beta = n+1, n+2}{\mathrm{span}}      \big[ e_{\beta \alpha} \big]  \big]     \\ \dots & \vdots & \vdots & \vdots & \vdots  \\ \dots & 0 & 0 &  b^+_{2n+2} \big( u \big) e_{(n+3)(2n+2)}  \big[ \underset{\alpha \neq n+1, n+2, \beta = n+1, n+2}{\mathrm{span}}      \big[ e_{\beta \alpha} \big]  \big] & \dots & b^+_{2n+2} \big( u \big) e_{(2n+2)(2n+2)}  \big[ \underset{\alpha \neq n+1, n+2, \beta = n+1, n+2}{\mathrm{span}}      \big[ e_{\beta \alpha} \big]  \big]   \\          \end{array}  
\end{align*}  }

 { \tiny  \begin{align*} +   \begin{array}{cccccccccccccccccccccccccccccccccccc} 
   b^-_{1} \big( u \big) e_{11}  \big[ \underset{\alpha \neq n+1, n+2, \beta = n+1, n+2}{\mathrm{span}}      \big[ e_{\beta \alpha} \big]  \big] & \dots &    b^-_{n} \big( u \big) e_{n1}  \big[ \underset{\alpha \neq n+1, n+2, \beta = n+1, n+2}{\mathrm{span}}      \big[ e_{\beta \alpha} \big]  \big] & \dots \\      \vdots & \vdots  & \vdots & \dots   \\ \\   b^-_{2n+2} \big( u \big) e_{1(2n+2)}  \big[ \underset{\alpha \neq n+1, n+2, \beta = n+1, n+2}{\mathrm{span}}      \big[ e_{\beta \alpha} \big]  \big] & \dots &    b^-_{n} \big( u \big) e_{n(n+2)}  \big[ \underset{\alpha \neq n+1, n+2, \beta = n+1, n+2}{\mathrm{span}}      \big[ e_{\beta \alpha} \big]  \big] & \dots   \end{array}%
  , 
\end{align*}  }

\subsubsection{Similarities with objects of vertex models}

\noindent Before obtaining an approximation of infinite dimensional representations of the transfer matrix, we draw the attention of the reader to several closely defined objects in vertex models, particularly the 6-vertex model (\citet{keating2022}), from which adaptations in several closely related models can be obtained (\citet{Rigas1}, \citet{Rigas2}, \citet{Rigas3}).

\bigskip

\noindent 

\noindent For non-symmetric weights, the weights of the six-vertex model admit the parametrization,

\begin{align*}
   a_1 \equiv      a \text{ }  \mathrm{exp} \big(  H + V \big)   \text{, } \\ 
   a_2 \equiv   a  \text{ }  \mathrm{exp} \big( - H - V \big)   \text{, } \\  b_1 \equiv  \text{ }  \mathrm{exp} \big( H - V \big)    \text{, } \\ b_2 \equiv \text{ }  \mathrm{exp} \big( - H + V \big)   \text{, } \\ c_1 \equiv  c \lambda  \text{, } \\ c_2 \equiv c \lambda^{-1} \text{, } 
\end{align*}

\noindent for $a_1 \equiv a_2 \equiv a$, $b_1 \equiv b_2 \equiv b$, $c_1 \equiv c_2 \equiv c$, and $\lambda \geq 1$, and external fields $H,V$. From such a parametrization of the weights as given above, one can form the so-called $R$-matrix, for the standard basis of $\textbf{C}^2$, with,

\[
R \equiv R \big( u , H , V \big) \equiv 
  \begin{bmatrix}
      a \text{ }  \mathrm{exp} \big(  H + V \big)    & 0 & 0 & 0  \\
    0 & b \text{ } \mathrm{exp} \big( H - V \big) & c & 0  \\0 & c & b \text{ }  \mathrm{exp} \big( - H + V \big) & 0 \\ 0 & 0 & 0 & a \text{ }  \mathrm{exp} \big( - H - V \big) \\ 
  \end{bmatrix} \text{, }
\]

\noindent in the tensor product basis $e_1 \otimes e_1$, $e_1 \otimes e_2$, $e_2 \otimes e_1$, $e_2 \otimes e_2$, for $e_1 \equiv \big[ 1 \text{  } 0 \big]^{\mathrm{T}}$ and $e_1 \equiv \big[ 0 \text{  } 1 \big]^{\mathrm{T}}$. For vanishing external fields $H \equiv V \equiv 0$, if we denote $R \big( u \big) \equiv R \big( u , 0 , 0 \big)$, the matrix above admits the identity,

\begin{align*}
     R \big( u , H , V \big) =    \bigg[    \mathrm{diag} \big[ \mathrm{exp} \big( \frac{H}{2} \big) , \mathrm{exp} \big( - \frac{H}{2} \big)  \big] \otimes      \mathrm{diag} \big[ \mathrm{exp} \big( \frac{V}{2} \big) , \mathrm{exp} \big( - \frac{V}{2} \big)  \big]    \bigg] R \big( u \big) \\ \times \bigg[     \mathrm{diag} \big[ \mathrm{exp} \big( \frac{H}{2} \big) , \mathrm{exp} \big( - \frac{H}{2} \big)  \big]  \otimes    \mathrm{diag} \big[ \mathrm{exp} \big( \frac{V}{2} \big) , \mathrm{exp} \big( - \frac{V}{2} \big)  \big]          \bigg]   \\ \equiv  \big( D^H \otimes D^V \big) R \big( u \big) \big( D^H \otimes D^V \big)      \text{, } 
\end{align*}

\noindent over $\textbf{C}^2 \otimes \textbf{C}^2 \otimes \textbf{C}^2$. In widely celebrated work, [1], Baxter demonstrated that the $R$ matrix satisfies the Yang-Baxter equation,

\begin{align*}
R_{12} \big( u \big) R_{13} \big( u + v \big) R_{23} \big( v \big) =  R_{23} \big( v \big) R_{13} \big( u + v \big) R_{12} \big( u \big)    \text{. } 
\end{align*}

\noindent For $\Delta < -1$, Baxter's parametrization, for the weights, given $0 < u < \eta$,

\begin{align*}
        a \equiv   \mathrm{sinh} \big( \eta - u \big) \text{, } b \equiv \mathrm{sinh} \big( u \big)\text{, }     c     \equiv \mathrm{sinh} \big( \eta \big)  \text{, } 
\end{align*}

\noindent allows one to classify the eigenvalues of the transfer matrix which is related to the Bethe equations. Besides this fact, equipped with $u$, the horizontal lines of the square lattice, $H$ and $V$, two external fields, the partition function over the $N \times M$ torus, $\textbf{T}_{NM}$, can be expressed with the summation,

\begin{align*}
      Z_{\textbf{T}_{MN}}   \big( u , H \big) \equiv    Z_{\textbf{T}_{MN}}  = \overset{N}{\underset{n=0}{\sum}}  \mathrm{exp} \big( M \big( N - 2 n \big) V \big)    Z_{\textbf{T}_{MN}}^n \big( u , H \big)         \text{, } 
\end{align*}

\noindent for the semigrand canonical partition function,

\begin{align*}
 Z_{\textbf{T}_{MN}}^n \big( u , H \big)  \equiv   \underset{\{ \alpha_i \}}{\underset{ i \geq 1}{\sum}}   \big( \Lambda_{\{ \alpha_i \}} \big( u , H \big) \big)^M            \text{, } 
\end{align*}

\noindent for countably many solutions $\big\{ \alpha_i \big\}$ to the Bethe equation, 

\begin{align*}
     \overset{N}{\underset{k=1}{\prod}}   \frac{\mathrm{sinh} \big(   \frac{\eta}{2} + i \alpha_j - v_k  \big)}{\mathrm{sinh} \big(     \frac{\eta}{2} - i \alpha_j + v_k    \big)}       = \mathrm{exp} \big( 2 H N          \big) \overset{n}{\underset{m=1, m \neq j}{\prod} }            \frac{\mathrm{sinh} \big(   i \big( \alpha_j - \alpha_m \big)+ \eta    \big)}{\mathrm{sinh} \big(   i \big( \alpha_j - \alpha_m \big) - \eta      \big)}      \text{, } 
\end{align*}

\noindent and for the eigenvalue $\Lambda_{\{ \alpha_i \}}$,

\begin{align*}
\Lambda \equiv  \Lambda_{\{ \alpha_i \}} \equiv \mathrm{exp} \big( N H \big) \overset{N}{\underset{k=1}{\prod}}  \mathrm{sinh} \big( \eta - u + v_k \big) \overset{n}{\underset{j=1}{\prod}} \frac{\mathrm{sinh}\big( \frac{\eta}{2} + u - i \alpha_j \big) }{\mathrm{sinh} \big( \frac{\eta}{2} - u + i \alpha_j \big) } + \mathrm{exp} \big( - NH \big) \\ \times \overset{n}{\underset{k=1}{\prod}} \mathrm{sinh} \big( u - v_k \big)   \overset{n}{\underset{j=1}{\prod}}  \frac{\mathrm{sinh}\big( \frac{3 \eta}{2} - u + i \alpha_j  \big) }{\mathrm{sinh} \big( u - \frac{\eta}{2} - i \alpha_j  \big) } \text{, }
\end{align*}

\noindent with corresponding eigenvector,

\begin{align*}
\overset{n}{\underset{i=1}{\prod}}              B \big( \alpha_i \big) \ket{\Downarrow}
  \equiv      \overset{n}{\underset{i=1}{\prod}}              B \big( \alpha_i \big) \bigg[ \overset{n}{\underset{i=1}{\bigotimes}} \ket{\downarrow}
 \bigg]            \equiv   \overset{n}{\underset{i=1}{\prod}}              B \big( \alpha_i \big) \bigg[  \ket{\downarrow}
 {\otimes}   \overset{n-2}{\cdots} \otimes  \ket{\downarrow}
 \bigg]     \equiv  \overset{n}{\underset{i=1}{\prod}}              B \big( \alpha_i \big) \bigg[  {{0}\choose{1}} 
 {\otimes}   \overset{n-2}{\cdots} \\ \otimes  {{0}\choose{1} }
 \bigg] \text{, } 
\end{align*}

\noindent of the transfer matrix,

\begin{align*}
   t \big( u , \big\{ v_k \big\} , H , V ) : \big( \textbf{C}^2 \big)^{\otimes N} \longrightarrow \big( \textbf{C}^2 \big)^{\otimes N}        \text{, } 
\end{align*}

\noindent which, explicitly, is proportional to the trace of the quantum monodromy matrix, which is given by,

\begin{align*}
  t \big( u , \big\{ v_k \big\} , H , V \big) \equiv \overset{N}{\underset{i=1}{\prod}}         D_i^{2V}    \mathrm{Tr}_a \big[       T_a \big( u , \big\{ v_k \big\} , H , 0 \big)       \big]          \text{, } 
\end{align*}

\noindent for the quantum monodromy matrix,

\begin{align*}
 T_a \big( u , \big\{ v_k \big\} , H , 0 \big) : \textbf{C}^2 \otimes \big( \textbf{C}^2 \big)^{\otimes N} \longrightarrow \textbf{C}^2 \otimes \big( \textbf{C}^2 \big)^{\otimes N}      \mapsto    \overset{N}{\underset{i=1}{\prod}}   \mathrm{diag} \big( \mathrm{exp} \big( 2H \big) ,  \mathrm{exp} \big( 2 H \big)  \big)   \\ \times     R_{ia} \big( u - v_i \big)      \text{, } 
\end{align*}

\noindent where each $v_i$ is chosen so that each $u-v_i$ is a spectral parameter given at site $i$. In the next section, to apply the formalism first introduced for Hamiltonian systems in the nonlinear Schrodinger's equation to the quantum monodromy and transfer matrices of the six-vertex model, observe that the product of diagonal matrices with each $R_{ia}$ is equivalent to,

\begin{align*}
     \mathrm{diag} \big( \mathrm{exp} \big( 2H \big) , \mathrm{exp} \big( 2 H \big)  \big)  \bigg[    R_{1a} \big( u - v_1 \big)   \cdots  \times  R_{(N-1)a} \big( u - v_{N-1} \big) \bigg]   \mathrm{diag} \big( \mathrm{exp} \big( 2H \big) , \mathrm{exp} \big( 2 H \big)  \big)  \\ \times   R_{Na} \big( u - v_N \big) \text{. } 
\end{align*}

\noindent To formulate the Hamiltonian flow for the six-vertex model, from the statement of the Bethe equations, and their eigenvalues, introduce the functions $\psi^{\pm}_u \big( \alpha + i u \big)  \equiv  \psi_{\pm} \big( \alpha + i u\big)$, which are given by, [7],

\begin{align*}
     \psi_{+} \big( \alpha + i u\big) = \mathrm{log} \big[ \big|       \frac{\mathrm{exp} \big(  \eta + 2 u   \big)  - \mathrm{exp} \big(    2 i \alpha    \big) }{\mathrm{exp} \big( \eta - 2 i \alpha \big)  - \mathrm{exp} \big(   2u      \big)} \big| \big] = \mathrm{log}\big[  \big| \frac{\mathrm{sinh} \big(    \frac{\eta}{2} + u - i \alpha    \big) }{\mathrm{sinh} \big(    \frac{\eta}{2} - u + i \alpha    \big) }  \big| \big] \\ \equiv \mathrm{log}\big[   \frac{\big| \mathrm{sinh}  \big(    \frac{\eta}{2} + u - i \alpha    \big) \big| }{ \big|\mathrm{sinh} \big(    \frac{\eta}{2} - u + i \alpha    \big) \big| }   \big]      \end{align*}

     \begin{align*} =   \mathrm{log}\big[  \frac{\mathrm{sinh} \big(    \frac{\eta}{2} + u - i \alpha    \big) }{\mathrm{sinh} \big(    \frac{\eta}{2} - u + i \alpha    \big) }  \big]    \text{, } \\  \\    \psi_{-} \big( \alpha + i u\big)   = \mathrm{log} \big[   \big|   \frac{\mathrm{exp} \big(  2 \eta + 2 i \alpha \big)  - \mathrm{exp} \big( 2 u - \eta \big)}{\mathrm{exp} \big( 2 u  \big)  - \mathrm{exp} \big( \eta + 2 i \alpha  \big)} \big|  \big]  \\ \equiv  \mathrm{log} \big[     \frac{\big| \mathrm{exp} \big(  2 \eta + 2 i \alpha \big)  - \mathrm{exp} \big( 2 u - \eta \big) \big| }{ \big| \mathrm{exp} \big( 2 u  \big)  - \mathrm{exp} \big( \eta + 2 i \alpha  \big) \big| }   \big]   \equiv \mathrm{log} \big[  \frac{ \big| \mathrm{sinh} \big(     \frac{3 \eta}{2} - u + i \alpha     \big) \big| }{ \big| \mathrm{sinh} \big(    u - \frac{\eta}{2} - i \alpha     \big) \big| }     \big]  \\   =  \mathrm{log} \big[  \frac{\mathrm{sinh} \big(     \frac{3 \eta}{2} - u + i \alpha     \big) }{\mathrm{sinh} \big(    u - \frac{\eta}{2} - i \alpha     \big) }     \big]    \text{, } 
\end{align*}

\noindent as well as the relation,

\begin{align*}
    \Theta \big( \alpha - \beta \big) \equiv \frac{1+\mathrm{exp} \big(   i p \big( \alpha \big) + i p \big( \beta \big)        \big) - 2 \Delta \mathrm{exp} \big(     i p \big( \alpha \big)             \big) }{1+\mathrm{exp} \big(     i p \big( \alpha \big) + i p \big( \beta \big)                \big) - 2 \Delta \mathrm{exp} \big(             i p \big( \beta \big)    \big)} = - \frac{\mathrm{sinh} \big( i \alpha - i \beta + \eta \big) }{\mathrm{sinh} \big( i \alpha - i \beta - \eta \big) }   \text{, } 
\end{align*}

\noindent for,

\begin{align*}
    p \big( \alpha \big) = \mathrm{log} \big[  \big|       \frac{\mathrm{sinh} \big(   \frac{\eta}{2} + i \alpha      \big) }{\mathrm{sinh} \big(  \frac{\eta}{2} - i \alpha   \big) }    \big|    \big]  \equiv \mathrm{log} \big[         \frac{\mathrm{sinh} \big(   \frac{\eta}{2} + i \alpha      \big) }{\mathrm{sinh} \big(  \frac{\eta}{2} - i \alpha   \big) }       \big]    \text{, } 
\end{align*}

\noindent from which the statement of the Bethe equations is equivalent to, in terms of a parameterization of $\psi_{-} \big( \alpha + v_k \big)$, and $\mathrm{exp} \big( \Theta \big( \alpha_j - \alpha_m \big)\big)$,

\begin{align*}
       \overset{N}{\underset{k=1}{\prod}} {\mathrm{exp} \big(   \psi_{+} \big(  \alpha + v_k      \big) \big) }  = \frac{1}{\mathrm{exp} \big( 2 H N          \big)}         {\underset{m=1, m \neq j}{\prod} }           \frac{\big( -1 \big)^m }{\mathrm{exp} \big( \Theta \big( \alpha_j - \alpha_m \big) \big) }           \\  = \frac{1}{\mathrm{exp} \big( 2 H N          \big)} 
 \overset{n}{\underset{m=1, m \neq j}{\prod} }            \frac{\mathrm{sinh} \big(   i \big( \alpha_j - \alpha_m \big) - \eta      \big)}{\mathrm{sinh} \big(   i \big( \alpha_j - \alpha_m \big)+ \eta    \big) }           \text{, } 
\end{align*}

\noindent while the statement for the eigenvalue of the Bethe equations is equivalent to, given some $\big\{ \alpha_i \big\}$, in terms of a parameterization of $\psi_{\pm} \big( \alpha + v_k \big)$,

\begin{align*}
     \Lambda \big( \psi_{\pm} \big( \alpha + i u \big) \big)  \equiv  \Lambda_{\{ \alpha_i \}} \equiv \mathrm{exp} \big( N H \big) \overset{N}{\underset{k=1}{\prod}}  \mathrm{sinh} \big( \eta - u + v_k \big) \\ \times \overset{n}{\underset{j=1}{\prod}}   \mathrm{exp} \big(  \psi_{+} \big( \alpha_j + i u\big)  \big)           + \mathrm{exp} \big( - NH \big) \overset{n}{\underset{k=1}{\prod}} \mathrm{sinh} \big( u - v_k \big)   \overset{n}{\underset{j=1}{\prod}}  \mathrm{exp} \big(   \psi_{-} \big( \alpha_j + i u\big)      \big)  \text{. } 
\end{align*}

\noindent To introduce the Hamiltonian formulation of the six-vertex model, in which given a height function representation, from the set of all possible asymptotic height functions $\mathcal{H}_{L,q}$, for Lipchitz $h \sim \mathcal{H}_{L,q}$ and $x , x^{\prime} \in \big[ 0 , L \big]$, with,

\begin{align*}
    \big| h\big(x \big) -   h   \big(  x^{\prime}      \big) \big| < \big| x - x^{\prime} \big|  \text{, }
\end{align*}

\noindent and periodic, with,

\begin{align*}
   h \big( L , y \big) = h \big( 0 , y \big) + q   \text{, } 
\end{align*}

\noindent for $h : \big[ 0 , L \big] \longrightarrow \textbf{R}$, and fixed $0 < q < 1$. From such a sampling of the height function, as well as another periodic function $\pi \big( x \big)$ in $x$, the pair $\big(  \pi  \big( x \big) , h\big( x \big) \big)$ can be identified with the cotangent space $T^{*} \mathcal{H}_{L,q}$, while the flow of the Hamiltonian can be identified with the pair $\big( \pi \big( x , y \big) , h \big( x , y \big) \big)$, in which,

\begin{align*}
        H_u \big( \pi \big( x , y \big) , h \big( x , y \big) \big) \equiv H_u \big( \pi , h \big) = {\int}_{[0,L]}   \mathcal{H}_{u - v ( x) } \big[  \partial_x h \big( x \big)  ,    \pi \big( x \big)   \big]  \mathrm{d} x    \text{, } 
\end{align*}

\noindent over $T^{*} \mathcal{H}_{L,q}$, for a solution $h \big( x , y \big)$ to the Euler Lagrange equations.

\subsubsection{Computation of identities between tensor products of elementary matrices in the Jimbo R-matrix}

\noindent We now state each claim below for computing $\mathscr{R}_1, \mathscr{R}_2, \mathscr{R}_3, \mathscr{R}_4, \mathscr{R}_5$:

\begin{itemize}
    \item[$\bullet$] \textbf{Claim 1} (\textit{consolidating tensor product terms dependent upon $\alpha$ and $\beta$}). One has that,

\begin{align*}
\mathscr{R}_1 \equiv  \bigg[   -\frac{1}{2} \big( e^{4 \eta} - 1 \big) \big( e^{2u}- e^{4n \eta} \big) \big( e^u + 1 \big)    - \big( e^u - 1 \big) \bigg] \underset{\alpha < n+1, \beta = n+1, n+2}{\sum}  e_{\alpha \beta} \otimes \big( e_{\beta \alpha} \\ + e_{\beta^{\prime} \alpha} \big) \\        - \bigg[ \frac{1}{2} \big( e^{4 \eta} - 1 \big) \big( e^{2u}- e^{4n \eta} \big) \big( e^u + 1 \big) e^u + \big( e^u - 1 \big) e^u  \bigg]       \underset{ \alpha > n+2 , \beta = n+1 , n+2}{\sum} e_{\beta^{\prime} \alpha^{\prime}} \otimes \big( e_{\alpha^{\prime} \beta^{\prime}}  \\  + e_{\alpha^{\prime}\beta}   \big)         .
\end{align*}

\noindent \textit{Proof of Claim 1}. By direct computation, disregarding prefactors to tensor products of the elementary matrices $e$,

\begin{align*}
   e_{\alpha \beta} \otimes e_{\beta \alpha} + e_{\alpha \beta} \otimes e_{\beta^{\prime}\alpha} =     e_{\alpha \beta} \otimes \big( e_{\beta \alpha}    + e_{\beta^{\prime}\alpha} \big)       , \\ \\ e_{\beta^{\prime} \alpha^{\prime}} \otimes e_{\alpha^{\prime} \beta^{\prime}} + e_{\beta^{\prime}\alpha^{\prime}} \otimes e_{\alpha^{\prime} \beta} = e_{\beta^{\prime} \alpha^{\prime}} \otimes \big( e_{\alpha^{\prime} \beta^{\prime}} + e_{\alpha^{\prime} \beta} \big)  , 
\end{align*}

\noindent readily implies that the first desired claim above holds, from which we conclude the argument. \boxed{}

    \item[$\bullet$] \textbf{Claim 2} (\textit{consolidating tensors product of elementary matrices from $n+1$ and $n+2$ terms in the summation}). One has that,

\begin{align*}
\mathscr{R}_2 \equiv  \big( e^{2u} - e^{4 \eta} \big) \big( e^{2u} - e^{4n \eta}\big) c^{+} \big( u \big) \bigg[ \underset{\alpha \neq n+1, n+2}{\sum} e_{\alpha \alpha}  +  \underset{\alpha = n+1 , n+2}{\sum} e_{\alpha \alpha} \bigg]  \otimes \bigg[ \underset{\alpha \neq n+1, n+2}{\sum} e_{\alpha \alpha} \\  + e_{\alpha^{\prime} \alpha^{\prime}} \bigg]    ,
\end{align*}

\noindent and also that,
    
\begin{align*}
 \mathscr{R}_3 \equiv    e^{2 \eta} \big( e^{2 u} - 1  \big) \big( e^{2u} - e^{4n \eta  } \big)  c^{-} \big( u \big) \bigg[  \underset{\alpha \text{ } \text{or} \text{ } \beta, \beta^{\prime} \neq n+1, n+2}{\underset{\alpha \neq \beta, \beta^{\prime}}{\sum}}  e_{\alpha \alpha} +\underset{\alpha = n+1, n+2}{\sum} e_{\alpha \alpha} \bigg]      \\ \otimes \bigg[   e_{\beta\beta} + e_{\alpha \alpha}  \bigg] .
\end{align*}

\noindent \textit{Proof of Claim 2}. By direct computation, disregarding prefactors, observe,

\begin{align*}
  \bigg[ \underset{\alpha \neq n+1, n+2}{\sum} e_{\alpha \alpha}  +  \underset{\alpha = n+1 , n+2}{\sum} e_{\alpha \alpha} \bigg]  \otimes \bigg[ \underset{\alpha \neq n+1, n+2}{\sum} e_{\alpha \alpha}  + e_{\alpha^{\prime} \alpha^{\prime}} \bigg] =       \bigg[ \underset{\alpha \neq n+1, n+2}{\sum} e_{\alpha \alpha}  \bigg] \\ \otimes \bigg[ \underset{\alpha \neq n+1, n+2}{\sum} e_{\alpha \alpha} \bigg] +  \bigg[  \underset{\alpha = n+1 , n+2}{\sum} e_{\alpha \alpha} \bigg]  \otimes e_{\alpha^{\prime}\alpha^{\prime}}  , 
\end{align*}

\noindent corresponding to the first statement provided in the claim above, and,

\begin{align*}
  \bigg[  \underset{\alpha \text{ } \text{or} \text{ } \beta, \beta^{\prime} \neq n+1, n+2}{\underset{\alpha \neq \beta, \beta^{\prime}}{\sum}}  e_{\alpha \alpha} +\underset{\alpha = n+1, n+2}{\sum} e_{\alpha \alpha} \bigg]       \otimes \bigg[   e_{\beta\beta} + e_{\alpha \alpha}  \bigg]  \\ =  \bigg[  \underset{\alpha \text{ } \text{or} \text{ } \beta, \beta^{\prime} \neq n+1, n+2}{\underset{\alpha \neq \beta, \beta^{\prime}}{\sum}}  e_{\alpha \alpha} \bigg] \otimes e_{\beta \beta}  +  \bigg[ \underset{\alpha = n+1, n+2}{\sum} e_{\alpha \alpha} \bigg] \otimes e_{\alpha\alpha} , 
\end{align*}

\noindent corresponding to the second statement provided in the claim above, from which we conclude the argument. \boxed{}

    \item[$\bullet$] \textbf{Claim 3} (\textit{consolidating tensor product terms dependent upon $\alpha$ and $\beta$}). One has that,

    \begin{align*}
\mathscr{R}_4 \equiv     \bigg[ \underset{\alpha, \beta \neq n+1, n+2}{\sum} a_{\alpha \beta } \big( u \big)e_{\alpha \beta} + \frac{1}{2} \underset{\alpha \neq n+1, n+2, \beta = n+1, n+2}{\sum} b^+_a \big( u \big) e_{\alpha \beta} \\ +  \frac{1}{2} \underset{\alpha \neq n+1, n+2, \beta = n+1, n+2}{\sum} b^-_a \big( u \big)   e_{\alpha \beta} \bigg] \otimes \bigg[                 \underset{\alpha, \beta \neq n+1, n+2}{\sum}     e_{\alpha^{\prime} \beta^{\prime}} \\ +  \frac{1}{2} \underset{\alpha \neq n+1, n+2, \beta = n+1,n+2}{\sum}         e_{\alpha^{\prime} \beta^{\prime}}    + \frac{1}{2} \underset{\alpha \neq n+1, n+2, \beta = n+1, n+2}{\sum}  b^-_{\alpha} \big( u \big) \\ \times e_{\alpha^{\prime} \beta} \bigg]    .
    \end{align*}

\noindent \textit{Proof of Claim 3}. By direct computation, disregarding prefactors, observe,

\begin{align*}
  \bigg[ \underset{\alpha, \beta \neq n+1, n+2}{\sum} a_{\alpha \beta } \big( u \big)e_{\alpha \beta} + \underset{\alpha \neq n+1, n+2, \beta = n+1, n+2}{\sum} e_{\alpha \beta} \bigg] \otimes \bigg[                 \underset{\alpha, \beta \neq n+1, n+2}{\sum}     e_{\alpha^{\prime} \beta^{\prime}} \\ + \underset{\alpha \neq n+1, n+2, \beta = n+1,n+2}{\sum}         e_{\alpha^{\prime} \beta^{\prime}}  \bigg] \end{align*}
  
  \begin{align*} =  \bigg[ \underset{\alpha, \beta \neq n+1, n+2}{\sum} a_{\alpha \beta } \big( u \big)e_{\alpha \beta}  \bigg] \otimes \bigg[                 \underset{\alpha, \beta \neq n+1, n+2}{\sum}     e_{\alpha^{\prime} \beta^{\prime}}  \bigg]  + \bigg[ \underset{\alpha \neq n+1, n+2, \beta = n+1, n+2}{\sum} e_{\alpha \beta} \bigg] \\ \otimes \bigg[ \underset{\alpha \neq n+1, n+2, \beta = n+1,n+2}{\sum}         e_{\alpha^{\prime} \beta^{\prime}}  \bigg] , 
\end{align*}

\noindent and also from the observation that,

\begin{align*}
    \bigg[ \underset{\alpha, \beta \neq n+1, n+2}{\sum} a_{\alpha \beta } \big( u \big)e_{\alpha \beta} + \underset{\alpha \neq n+1, n+2, \beta = n+1, n+2}{\sum} e_{\alpha \beta} +  \underset{\alpha \neq n+1, n+2, \beta = n+1, n+2}{\sum} e_{\alpha \beta} \bigg] \\ \otimes \bigg[                 \underset{\alpha, \beta \neq n+1, n+2}{\sum}     e_{\alpha^{\prime} \beta^{\prime}}  + \underset{\alpha \neq n+1, n+2, \beta = n+1,n+2}{\sum}         e_{\alpha^{\prime} \beta^{\prime}} + \underset{\alpha \neq n+1, n+2, \beta = n+1,n+2}{\sum}         e_{\alpha^{\prime} \beta^{\prime}}  \end{align*}

    \begin{align*} + \underset{\alpha \neq n+1, n+2, \beta = n+1, n+2}{\sum}  b^-_{\alpha} \big( u \big) e_{\alpha^{\prime} \beta} \bigg] \\ \\  =        \bigg[ \underset{\alpha, \beta \neq n+1, n+2}{\sum} a_{\alpha \beta } \big( u \big)e_{\alpha \beta}  \bigg] \otimes  \bigg[                 \underset{\alpha, \beta \neq n+1, n+2}{\sum}     e_{\alpha^{\prime} \beta^{\prime}}  \bigg] +   \bigg[ \underset{\alpha \neq n+1, n+2, \beta = n+1, n+2}{\sum} e_{\alpha \beta}  \bigg] \\ \otimes \bigg[\underset{\alpha \neq n+1, n+2, \beta = n+1,n+2}{\sum}         e_{\alpha^{\prime} \beta^{\prime}}  \bigg] +  \bigg[ \underset{\alpha \neq n+1, n+2, \beta = n+1, n+2}{\sum} e_{\alpha \beta} \bigg] \\ \otimes \bigg[ \underset{\alpha \neq n+1, n+2, \beta = n+1, n+2}{\sum}  b^-_{\alpha} \big( u \big) e_{\alpha^{\prime} \beta} \bigg] ,
\end{align*}

\noindent from which we conclude the argument. \boxed{}

    \item[$\bullet$] \textbf{Claim 4} (\textit{consolidating tensor product terms dependent upon $\beta^{\prime}$, $\alpha^{\prime}$, $\beta$, and $\alpha$}). One has that,

    \begin{align*}
\mathscr{R}_5 \equiv     \bigg[  \frac{1}{2} \underset{\alpha \neq n+1, n+2, \beta = n+1, n+2}{\sum} b^+_{\alpha} \big( u \big) e_{\beta^{\prime} \alpha^{\prime}}  +   \frac{1}{2} \underset{\alpha \neq n+1, n+2, \beta = n+1, n+2}{\sum} b^-_{\alpha} \big( u \big) e_{\beta \alpha^{\prime}} \bigg] \\ \otimes       \big[ e_{\beta \alpha} \big]         .
    \end{align*}

    \noindent \textit{Proof of Claim 4}. By direct computation, disregarding prefactors, one has that,

\begin{align*}
\bigg[   \underset{\alpha \neq n+1, n+2, \beta = n+1, n+2}{\sum} b^+_{\alpha} \big( u \big) e_{\beta^{\prime} \alpha^{\prime}}  +  \underset{\alpha \neq n+1, n+2, \beta = n+1, n+2}{\sum} b^-_{\alpha} \big( u \big) e_{\beta \alpha^{\prime}} \bigg] \\ \otimes \big[ e_{\beta \alpha} \big] \\  = \bigg[   \underset{\alpha \neq n+1, n+2, \beta = n+1, n+2}{\sum} b^+_{\alpha} \big( u \big) e_{\beta^{\prime} \alpha^{\prime}}  \bigg] \otimes e_{\beta \alpha} +    \underset{\alpha \neq n+1, n+2, \beta = n+1, n+2}{\sum} b^-_{\alpha} \big( u \big) e_{\beta \alpha^{\prime}} \bigg] \\ \otimes  e_{\beta \alpha}   ,
\end{align*}

    \noindent from which we conclude the argument. \boxed{}

\begin{itemize}

\item[$\bullet$] \textbf{Claim 5} (\textit{incorporating previously obtained representations for all components of the Jimbo R-matrix}). The Jimbo R-matrix,

\begin{align*}
    R_J ,
\end{align*}

\noindent which coincides with the representation,

\begin{align*}
R_J   - \underset{1 \leq i \leq 5}{\sum} \mathscr{R}_i + \bigg[ \mathscr{R}_1 + \mathscr{R}_2 + \mathscr{R}_3 + \mathscr{R}_4 + \mathscr{R}_5 \bigg]  ,
\end{align*}

\end{itemize}

\noindent given the finite-dimensional representations provided in the previous \textbf{Claims} above, equals,

 { \tiny  \begin{align*}  
  , 
\end{align*}  }

\noindent \textit{Proof of Claim 5}. Combine the previous claims for each component of $\mathscr{R}$ from the series of \textbf{Claims} above, from which we conclude the argument. \boxed{}

\begin{itemize}
    \item[$\bullet$] \textbf{Claim 6} (\textit{computing partial traces of the matrix product between K-matrices and the Jimbo R-matrices}). One has that,

\[   \left\{\!%
\right. 
\]

\noindent which can be used to approximate,

\begin{align*}
\textbf{T}^{\mathrm{Open}}\big( u \big) \equiv  \mathrm{tr}_0 \bigg[ \mathcal{K}_{+,0} \big( u \big)  \prod_{1 \leq j \leq L} R_{0j} \big( u \big)  \mathcal{K}_{-,0} \big( u \big)   \prod_{1 \leq j^{\prime} \leq L} R_{j^{\prime}0} \big( u \big)  \bigg]  , 
\end{align*}

\noindent the higher-rank open boundary condition $D^{(2)}_3$ transfer matrix, with,

\begin{align*}
 \mathscr{R}_1  \mathscr{T} \mathrm{Tr}_0 \bigg[ \underset{1 \leq j^{\prime} \leq 5}{\sum}  \mathscr{R} \mathscr{T}^{\prime}_j  \bigg]  +   \mathscr{R}_2  \mathscr{T} \mathrm{Tr}_0 \bigg[  \underset{1 \leq j^{\prime} \leq 5}{\sum}  \mathscr{R} \mathscr{T}^{\prime}_j \bigg] +  \mathscr{R}_3  \mathscr{T} \mathrm{Tr}_0 \bigg[  \underset{1 \leq j^{\prime} \leq 5}{\sum}  \mathscr{R} \mathscr{T}^{\prime}_j \bigg]  \\ +    \mathscr{R}_4  \mathscr{T}  \mathrm{Tr}_0 \bigg[ \underset{1 \leq j^{\prime} \leq 5}{\sum}  \mathscr{R} \mathscr{T}^{\prime}_j \bigg]   +   \mathscr{R}_5 \mathscr{T} \mathrm{Tr}_0 \bigg[  \underset{1 \leq j^{\prime} \leq 5}{\sum}  \mathscr{R} \mathscr{T}^{\prime}_j \bigg]  \\ \\ \equiv   \bigg[ \mathscr{R}_1 \mathscr{T}_1  +    \mathscr{R}_1 \mathscr{T}_2  +  \mathscr{R}_1 \mathscr{T}_3  +   \mathscr{R}_1 \mathscr{T}_4   +  \mathscr{R}_1 \mathscr{T}_5  \bigg]  \mathrm{Tr}_0 \bigg[  \underset{1 \leq j^{\prime} \leq 5}{\sum}  \mathscr{R} \mathscr{T}^{\prime}_j \bigg]   \\ + \bigg[ \mathscr{R}_2 \mathscr{T}_1  +    \mathscr{R}_2 \mathscr{T}_2  +  \mathscr{R}_2 \mathscr{T}_3  +   \mathscr{R}_2 \mathscr{T}_4   +  \mathscr{R}_2 \mathscr{T}_5  \bigg]  \mathrm{Tr}_0 \bigg[  \underset{1 \leq j^{\prime} \leq 5}{\sum}  \mathscr{R} \mathscr{T}^{\prime}_j \bigg]   \\ + \bigg[ \mathscr{R}_3 \mathscr{T}_1  +    \mathscr{R}_3 \mathscr{T}_2  +  \mathscr{R}_3 \mathscr{T}_3  +   \mathscr{R}_3 \mathscr{T}_4   +  \mathscr{R}_3 \mathscr{T}_5  +  \mathscr{R}_3 \mathscr{T}_6  \bigg]  \mathrm{Tr}_0 \bigg[  \underset{1 \leq j^{\prime} \leq 5}{\sum}  \mathscr{R} \mathscr{T}^{\prime}_j \bigg]   \\ + \bigg[ \mathscr{R}_4 \mathscr{T}_1  +    \mathscr{R}_4 \mathscr{T}_2  +  \mathscr{R}_4 \mathscr{T}_3  +   \mathscr{R}_4 \mathscr{T}_4   +  \mathscr{R}_4 \mathscr{T}_5  +  \mathscr{R}_4 \mathscr{T}_6  \bigg]  \mathrm{Tr}_0 \bigg[  \underset{1 \leq j^{\prime} \leq 5}{\sum}  \mathscr{R} \mathscr{T}^{\prime}_j \bigg]   \\ + \bigg[ \mathscr{R}_5 \mathscr{T}_1  +    \mathscr{R}_5 \mathscr{T}_2  +  \mathscr{R}_5 \mathscr{T}_3  +   \mathscr{R}_5 \mathscr{T}_4   +  \mathscr{R}_5 \mathscr{T}_5  \bigg]  \mathrm{Tr}_0 \bigg[  \underset{1 \leq j^{\prime} \leq 5}{\sum}  \mathscr{R} \mathscr{T}^{\prime}_j \bigg]  \end{align*}

 \begin{align*}
   \equiv \bigg[ \bigg[ \mathscr{R}_1 \mathscr{T}_1  +    \mathscr{R}_1 \mathscr{T}_2  +  \mathscr{R}_1 \mathscr{T}_3  +   \mathscr{R}_1 \mathscr{T}_4   +  \mathscr{R}_1 \mathscr{T}_5  \bigg]     \\ + \bigg[ \mathscr{R}_2 \mathscr{T}_1  +    \mathscr{R}_2 \mathscr{T}_2  +  \mathscr{R}_2 \mathscr{T}_3  +   \mathscr{R}_2 \mathscr{T}_4   +  \mathscr{R}_2 \mathscr{T}_5  \bigg]    \\ + \bigg[ \mathscr{R}_3 \mathscr{T}_1  +    \mathscr{R}_3 \mathscr{T}_2  +  \mathscr{R}_3 \mathscr{T}_3  +   \mathscr{R}_3 \mathscr{T}_4   +  \mathscr{R}_3 \mathscr{T}_5  +  \mathscr{R}_3 \mathscr{T}_6  \bigg]    \\ + \bigg[ \mathscr{R}_4 \mathscr{T}_1  +    \mathscr{R}_4 \mathscr{T}_2  +  \mathscr{R}_4 \mathscr{T}_3  +   \mathscr{R}_4 \mathscr{T}_4   +  \mathscr{R}_4 \mathscr{T}_5  +  \mathscr{R}_4 \mathscr{T}_6  \bigg]   \\ + \bigg[ \mathscr{R}_5 \mathscr{T}_1  +    \mathscr{R}_5 \mathscr{T}_2  +  \mathscr{R}_5 \mathscr{T}_3  +   \mathscr{R}_5 \mathscr{T}_4   +  \mathscr{R}_5 \mathscr{T}_5  \bigg]  \bigg] \mathrm{Tr}_0 \bigg[  \underset{1 \leq j^{\prime} \leq 5}{\sum}  \mathscr{R} \mathscr{T}^{\prime}_j \bigg]    , 
\end{align*}

\noindent where,

\begin{align*}
  \mathscr{R} \mathscr{T}_1 \equiv   \mathscr{R} \mathscr{T}_1  \big( u \big)   , \mathscr{R} \mathscr{T}_1 \equiv   \mathscr{R} \mathscr{T}^{\prime}_1  \big( u^{\prime} \big)   \\ \vdots \\   \mathscr{R} \mathscr{T}_5 \equiv   \mathscr{R} \mathscr{T}_5  \big( u \big)   , \mathscr{R} \mathscr{T}_5 \equiv   \mathscr{R} \mathscr{T}^{\prime}_5  \big( u^{\prime} \big)  , 
\end{align*}

\noindent and, 

\[ \mathscr{R}_1\mathscr{T} \equiv   \left\{\!\begin{array}{ll@{}>{{}}l} 
\mathscr{R}_1 \mathscr{T}_1 \equiv \textit{First term of } \mathscr{R}_1  , \\ \\ \mathscr{R}_1 \mathscr{T}_2 \equiv \textit{Second term of } \mathscr{R}_1  , \\ \\ \mathscr{R}_1 \mathscr{T}_3 \equiv \textit{Third term of } \mathscr{R}_1 , \\ \\ \mathscr{R}_1 \mathscr{T}_4 \equiv \textit{Fourth term of } \mathscr{R}_1 , \\ \\ \mathscr{R}_1 \mathscr{T}_5 \equiv \textit{Fifth term of } \mathscr{R}_1 .
\end{array}\right.
\]

\[ \mathscr{R}_2\mathscr{T} \equiv   \left\{\!\begin{array}{ll@{}>{{}}l} 
\mathscr{R}_2 \mathscr{T}_1 \equiv \textit{First term of } \mathscr{R}_2  , \\ \\ \mathscr{R}_1 \mathscr{T}_2 \equiv \textit{Second term of } \mathscr{R}_2  , \\ \\ \mathscr{R}_1 \mathscr{T}_3 \equiv \textit{Third term of } \mathscr{R}_2 , \\ \\ \mathscr{R}_2 \mathscr{T}_4 \equiv \textit{Fourth term of } \mathscr{R}_2 , \\ \\ \mathscr{R}_2 \mathscr{T}_5 \equiv \textit{Fifth term of } \mathscr{R}_2 .
\end{array}\right.
\]

\[ \mathscr{R}_3\mathscr{T} \equiv   \left\{\!\begin{array}{ll@{}>{{}}l} 
\mathscr{R}_3 \mathscr{T}_1 \equiv \textit{First term of } \mathscr{R}_3  , \\ \\ \mathscr{R}_3 \mathscr{T}_2 \equiv \textit{Second term of } \mathscr{R}_3 , \\ \\ \mathscr{R}_3 \mathscr{T}_3 \equiv \textit{Third term of } \mathscr{R}_3 , \\ \\ \mathscr{R}_3 \mathscr{T}_4 \equiv \textit{Fourth term of } \mathscr{R}_3 , \\ \\ \mathscr{R}_3 \mathscr{T}_5 \equiv \textit{Fifth term of } \mathscr{R}_3 , \\ \\ \mathscr{R}_3 \mathscr{T}_6 \equiv \textit{Sixth term of } \mathscr{R}_3 ,
\end{array}\right.
\]

\[ \mathscr{R}_4\mathscr{T} \equiv   \left\{\!\begin{array}{ll@{}>{{}}l} 
\mathscr{R}_4 \mathscr{T}_1 \equiv \textit{First term of } \mathscr{R}_4  , \\ \\ \mathscr{R}_4 \mathscr{T}_2 \equiv \textit{Second term of } \mathscr{R}_4  , \\ \\ \mathscr{R}_4 \mathscr{T}_3 \equiv \textit{Third term of } \mathscr{R}_4 , \\ \\ \mathscr{R}_4 \mathscr{T}_4 \equiv \textit{Fourth term of } \mathscr{R}_4 , \\ \\ \mathscr{R}_4 \mathscr{T}_5 \equiv \textit{Fifth term of } \mathscr{R}_4 , \\ \\ \mathscr{R}_4 \mathscr{T}_6 \equiv \textit{Sixth term of } \mathscr{R}_4 , 
\end{array}\right.
\]

\[ \mathscr{R}_5\mathscr{T} \equiv   \left\{\!\begin{array}{ll@{}>{{}}l} 
\mathscr{R}_5 \mathscr{T}_1 \equiv \textit{First term of } \mathscr{R}_5  , \\ \\ \mathscr{R}_5 \mathscr{T}_2 \equiv \textit{Second term of } \mathscr{R}_5  , \\ \\ \mathscr{R}_5 \mathscr{T}_3 \equiv \textit{Third term of } \mathscr{R}_5 , \\ \\ \mathscr{R}_5 \mathscr{T}_4 \equiv \textit{Fourth term of } \mathscr{R}_5 , \\ \\ \mathscr{R}_5 \mathscr{T}_5 \equiv \textit{Fifth term of } \mathscr{R}_5 .
\end{array}\right.
\] 

\noindent \textit{Proof of Claim 5}. By direct computation, under the idenfications,

\begin{align*}
    \mathrm{Tr}_0 \bigg[ \mathcal{K}_{+,0} \big( u \big)  \prod_{1 \leq j \leq L} \big( \mathscr{R}_1\big)_{0j} \big( u \big)  \mathcal{K}_{-,0} \big( u \big)   \prod_{1 \leq j^{\prime} \leq L} \big( \mathscr{R}_1\big)_{j^{\prime}0} \big( u \big)       \bigg]  \\ \\  \equiv   \mathrm{Tr}_0 \bigg[ \mathcal{K}_{+,0} \big( u \big)  \prod_{1 \leq u \leq L} \big( \mathscr{R}_1\big)_{0j} \big( u \big)  \mathcal{K}_{-,0} \big( u \big)   \prod_{1 \leq u^{\prime} \leq L} \big( \mathscr{R}_1\big)_{j^{\prime}0} \big( u \big)       \bigg] \end{align*}

    \begin{align*}  \equiv -  \mathrm{Tr}_0 \bigg[  \mathcal{K}_{+,0} \big( u \big) \bigg[ \bigg[  \frac{1}{2} \big( e^{4 \eta} - 1 \big) \big( e^{2u}- e^{4n \eta} \big) \big( e^u + 1 \big)  \bigg]  \underset{\alpha < n+1, \beta = n+1, n+2}{\sum}  e_{\alpha \beta} \otimes \big( e_{\beta \alpha} \\    + e_{\beta^{\prime} \alpha} \big) \bigg]  \mathcal{K}_{-,0} \big( u \big)   \prod_{1 \leq u^{\prime} \leq L} \big( \mathscr{R}_1\big)_{j^{\prime}0} \big( u \big)     \bigg]  \\ \\    -  \mathrm{Tr}_0 \bigg[  \mathcal{K}_{+,0} \big( u \big) \bigg[ \bigg[  \big( e^u - 1 \big)  \bigg]  \underset{\alpha < n+1, \beta = n+1, n+2}{\sum}  e_{\alpha \beta} \otimes \big( e_{\beta \alpha} \\   + e_{\beta^{\prime} \alpha} \big) \bigg] \mathcal{K}_{-,0} \big( u \big)   \prod_{1 \leq u^{\prime} \leq L} \big( \mathscr{R}_1\big)_{j^{\prime}0} \big( u \big)     \bigg]    \\ -   \mathrm{Tr}_0 \bigg[ \mathcal{K}_{+,0} \big( u \big) \bigg[  \frac{1}{2}  \bigg[ \big( e^{4 \eta} - 1 \big) \big( e^{2u}- e^{4n \eta} \big) \big( e^u + 1 \big) e^u \bigg] \underset{ \alpha > n+2 , \beta = n+1 , n+2}{\sum} e_{\beta^{\prime} \alpha^{\prime}} \otimes \big( e_{\alpha^{\prime} \beta^{\prime}}    \\        + e_{\alpha^{\prime}\beta}   \big)   \bigg]  \mathcal{K}_{-,0} \big( u \big)    \prod_{1 \leq u^{\prime} \leq L} \big( \mathscr{R}_1\big)_{j^{\prime}0} \big( u \big)   \bigg]  \\  - \mathrm{Tr}_0 \bigg[  \mathcal{K}_{+,0} \big( u \big) \bigg[ \bigg[  \big( e^u - 1 \big) e^u     \bigg]    \underset{ \alpha > n+2 , \beta = n+1 , n+2}{\sum} e_{\beta^{\prime} \alpha^{\prime}} \otimes \big( e_{\alpha^{\prime} \beta^{\prime}}    + e_{\alpha^{\prime}\beta}   \big) \bigg] \mathcal{K}_{-,0} \big( u \big) \\  \times   \prod_{1 \leq u^{\prime} \leq L} \big( \mathscr{R}_1\big)_{j^{\prime}0} \big( u \big)    \bigg]                   \\ \\ =    -  \mathrm{Tr}_0 \bigg[  \mathcal{K}_{+,0} \big( u \big) \bigg[ \bigg[  \frac{1}{2} \big( e^{4 \eta} - 1 \big) \big( e^{2u}- e^{4n \eta} \big) \big( e^u + 1 \big)  \bigg]  \underset{\alpha < n+1, \beta = n+1, n+2}{\sum}  e_{\alpha \beta} \bigg]      \\    \times \mathcal{K}_{-,0} \big( u \big)   \prod_{1 \leq u^{\prime} \leq L} \big( \mathscr{R}_1\big)_{j^{\prime}0} \big( u \big)   \bigg]  \\ \times \mathrm{Tr}_0 \bigg[    \mathcal{K}_{+,0} \big( u \big)   \bigg[    \underset{\alpha < n+1, \beta = n+1, n+2}{\sum}   \big( e_{\beta \alpha}  + e_{\beta^{\prime} \alpha} \big) \bigg]  \mathcal{K}_{-,0} \big( u \big)  \prod_{1 \leq u^{\prime} \leq L} \big( \mathscr{R}_1\big)_{j^{\prime}0} \big( u \big)   \bigg]                      \\   -  \mathrm{Tr}_0 \bigg[  \mathcal{K}_{+,0} \big( u \big) \bigg[ \bigg[  \big( e^u - 1 \big)  \bigg]  \underset{\alpha < n+1, \beta = n+1, n+2}{\sum}  e_{\alpha \beta}  \mathcal{K}_{-,0} \big( u \big)   \prod_{1 \leq u^{\prime} \leq L} \big( \mathscr{R}_1\big)_{j^{\prime}0} \big( u \big)   \bigg] \\ \times  \mathrm{Tr}_0 \bigg[  \mathcal{K}_{+,0} \big( u \big) \bigg[  \underset{\alpha < n+1, \beta = n+1, n+2}{\sum}  \big( e_{\beta \alpha}  + e_{\beta^{\prime} \alpha} \big) \bigg] \mathcal{K}_{-,0} \big( u \big)  \prod_{1 \leq u^{\prime} \leq L} \big( \mathscr{R}_1\big)_{j^{\prime}0} \big( u \big)    \bigg]                      
\\ \\ =   1_{(\mathrm{I})}   1_{(\mathrm{II })}   + 1_{(\mathrm{III})} 1_{(\mathrm{IV})}      
         , 
\end{align*}

\noindent corresponding to the first partial trace,

\begin{align*}
    \mathrm{Tr}_0 \bigg[\mathcal{K}_{+,0} \big( u \big)  \prod_{1 \leq j \leq L} \big( \mathscr{R}_2\big)_{0j} \big( u \big)  \mathcal{K}_{-,0} \big( u \big)   \prod_{1 \leq j^{\prime} \leq L} \big( \mathscr{R}_2\big)_{j^{\prime}0} \big( u \big)   \bigg] \\ \\ \equiv    \mathrm{Tr}_0 \bigg[\mathcal{K}_{+,0} \big( u \big) \bigg[ \bigg[  \big( e^{2u} - e^{4 \eta} \big) \big( e^{2u} - e^{4n \eta}\big) c^{+} \big( u \big) \bigg[ \underset{\alpha \neq n+1, n+2}{\sum} e_{\alpha \alpha}  +  \underset{\alpha = n+1 , n+2}{\sum} e_{\alpha \alpha} \bigg]   \bigg]         \\ \otimes   \bigg[ \underset{\alpha \neq n+1, n+2}{\sum} e_{\alpha \alpha}   + e_{\alpha^{\prime} \alpha^{\prime}} \bigg]  \bigg]  \mathcal{K}_{-,0} \big( u \big)   \prod_{1 \leq j^{\prime} \leq L} \big( \mathscr{R}_2\big)_{j^{\prime}0} \big( u \big)  \bigg] \\ \\   =       \mathrm{Tr}_0 \bigg[\mathcal{K}_{+,0} \big( u \big) \bigg[ \bigg[  \big( e^{2u} - e^{4 \eta} \big) \big( e^{2u} - e^{4n \eta}\big) c^{+} \big( u \big) \bigg[ \underset{\alpha \neq n+1, n+2}{\sum} e_{\alpha \alpha}  +  \underset{\alpha = n+1 , n+2}{\sum} e_{\alpha \alpha} \bigg]   \bigg]  \\  \times  \mathcal{K}_{-,0} \big( u \big)   \prod_{1 \leq j^{\prime} \leq L} \big( \mathscr{R}_2\big)_{j^{\prime}0} \big( u \big)    \bigg] \\    \times          \mathrm{Tr}_0 \bigg[\mathcal{K}_{+,0} \big( u \big)           \bigg[  \underset{\alpha \neq n+1, n+2}{\sum} e_{\alpha \alpha}   + e_{\alpha^{\prime} \alpha^{\prime}}  \bigg]   \mathcal{K}_{-,0} \big( u \big)   \prod_{1 \leq j^{\prime} \leq L} \big( \mathscr{R}_2\big)_{j^{\prime}0} \big( u \big)    \bigg] \\ \\  =    \bigg\{     \mathrm{Tr}_0 \bigg[\mathcal{K}_{+,0} \big( u \big) \bigg[ \bigg[  \big( e^{2u} - e^{4 \eta} \big) \big( e^{2u} - e^{4n \eta}\big) c^{+} \big( u \big) \bigg[ \underset{\alpha \neq n+1, n+2}{\sum} e_{\alpha \alpha} \bigg] \\   \times  \mathcal{K}_{-,0} \big( u \big)   \prod_{1 \leq j^{\prime} \leq L} \big( \mathscr{R}_2\big)_{j^{\prime}0} \big( u \big)    \bigg]  \\ +  \mathrm{Tr}_0 \bigg[\mathcal{K}_{+,0} \big( u \big) \bigg[  \big( e^{2u} - e^{4 \eta} \big) \big( e^{2u} - e^{4n \eta}\big) c^{+} \big( u \big) \bigg[    \underset{\alpha = n+1 , n+2}{\sum} e_{\alpha \alpha} \bigg]   \bigg] \\ \times  \mathcal{K}_{-,0} \big( u \big)   \prod_{1 \leq j^{\prime} \leq L} \big( \mathscr{R}_2\big)_{j^{\prime}0} \big( u \big)    \bigg] \bigg\} \\  \times          \mathrm{Tr}_0 \bigg[\mathcal{K}_{+,0} \big( u \big)           \bigg[  \underset{\alpha \neq n+1, n+2}{\sum} e_{\alpha \alpha}   + e_{\alpha^{\prime} \alpha^{\prime}}  \bigg]   \mathcal{K}_{-,0} \big( u \big)   \prod_{1 \leq j^{\prime} \leq L} \big( \mathscr{R}_2\big)_{j^{\prime}0} \big( u \big)    \bigg]     \\ \\ =   \mathrm{Tr}_0 \bigg[\mathcal{K}_{+,0} \big( u \big) \bigg[ \bigg[  \big( e^{2u} - e^{4 \eta} \big) \big( e^{2u} - e^{4n \eta}\big) c^{+} \big( u \big) \bigg[ \underset{\alpha \neq n+1, n+2}{\sum} e_{\alpha \alpha} \bigg]  \end{align*}

    \begin{align*}     \times  \mathcal{K}_{-,0} \big( u \big)   \prod_{1 \leq j^{\prime} \leq L} \big( \mathscr{R}_2\big)_{j^{\prime}0} \big( u \big)    \bigg] \\   \times          \mathrm{Tr}_0 \bigg[\mathcal{K}_{+,0} \big( u \big)           \bigg[  \underset{\alpha \neq n+1, n+2}{\sum} e_{\alpha \alpha}   + e_{\alpha^{\prime} \alpha^{\prime}}  \bigg]   \mathcal{K}_{-,0} \big( u \big)   \prod_{1 \leq j^{\prime} \leq L} \big( \mathscr{R}_2\big)_{j^{\prime}0} \big( u \big)    \bigg]   \\ +   \mathrm{Tr}_0 \bigg[\mathcal{K}_{+,0} \big( u \big) \bigg[  \big( e^{2u} - e^{4 \eta} \big) \big( e^{2u} - e^{4n \eta}\big) c^{+} \big( u \big) \bigg[    \underset{\alpha = n+1 , n+2}{\sum} e_{\alpha \alpha} \bigg]   \bigg] \end{align*}

    \begin{align*}    \times  \mathcal{K}_{-,0} \big( u \big)   \prod_{1 \leq j^{\prime} \leq L} \big( \mathscr{R}_2\big)_{j^{\prime}0} \big( u \big)    \bigg]      \\ \times          \mathrm{Tr}_0 \bigg[\mathcal{K}_{+,0} \big( u \big)           \bigg[  \underset{\alpha \neq n+1, n+2}{\sum} e_{\alpha \alpha}   + e_{\alpha^{\prime} \alpha^{\prime}}  \bigg]   \mathcal{K}_{-,0} \big( u \big)   \prod_{1 \leq j^{\prime} \leq L} \big( \mathscr{R}_2\big)_{j^{\prime}0} \big( u \big)    \bigg] \\ \\ =  2_{(\mathrm{I})}   2_{(\mathrm{II })}   + 2_{(\mathrm{III})} 2_{(\mathrm{IV})}                 , 
\end{align*}

\noindent corresponding to the second partial trace,

\begin{align*}
  \mathrm{Tr}_0 \bigg[ \mathcal{K}_{+,0} \big( u \big)  \prod_{1 \leq j \leq L} \big( \mathscr{R}_3\big)_{0j} \big( u \big)  \mathcal{K}_{-,0} \big( u \big)   \prod_{1 \leq j^{\prime} \leq L} \big( \mathscr{R}_3\big)_{j^{\prime}0} \big( u \big)  \bigg]  \\ \\ \equiv \mathrm{Tr}_0 \bigg[             \mathcal{K}_{+,0} \big( u \big)           \bigg[         e^{2 \eta} \big( e^{2 u} - 1  \big) \big( e^{2u} - e^{4n \eta  } \big)  c^{-} \big( u \big) \bigg[  \underset{\alpha \text{ } \text{or} \text{ } \beta, \beta^{\prime} \neq n+1, n+2}{\underset{\alpha \neq \beta, \beta^{\prime}}{\sum}}  e_{\alpha \alpha} \\  +\underset{\alpha = n+1, n+2}{\sum} e_{\alpha \alpha} \bigg]       \otimes \bigg[   e_{\beta\beta} + e_{\alpha \alpha}  \bigg]               \bigg]              \mathcal{K}_{-,0} \big( u \big)   \prod_{1 \leq j^{\prime} \leq L} \big( \mathscr{R}_3\big)_{j^{\prime}0} \big( u \big)                   \bigg] \\ \\    =    \mathrm{Tr}_0 \bigg[             \mathcal{K}_{+,0} \big( u \big)           \bigg[         e^{2 \eta} \big( e^{2 u} - 1  \big) \big( e^{2u} - e^{4n \eta  } \big)  c^{-} \big( u \big) \bigg[  \underset{\alpha \text{ } \text{or} \text{ } \beta, \beta^{\prime} \neq n+1, n+2}{\underset{\alpha \neq \beta, \beta^{\prime}}{\sum}}  e_{\alpha \alpha} \otimes   \bigg[   e_{\beta\beta} + e_{\alpha \alpha}  \bigg]   \bigg]            \\  \times \mathcal{K}_{-,0} \big( u \big)    \prod_{1 \leq j^{\prime} \leq L} \big( \mathscr{R}_3\big)_{j^{\prime}0} \big( u \big)               \bigg] \\   + \mathrm{Tr}_0 \bigg[             \mathcal{K}_{+,0} \big( u \big)           \bigg[        e^{2 \eta} \big( e^{2 u} - 1  \big) \big( e^{2u} - e^{4n \eta  } \big)  c^{-} \big( u \big)  \bigg[     \underset{\alpha = n+1, n+2}{\sum} e_{\alpha \alpha}    \otimes \bigg[   e_{\beta\beta} + e_{\alpha \alpha}  \bigg] \bigg]            \bigg] \\ \times \mathcal{K}_{-,0} \big( u \big)    \prod_{1 \leq j^{\prime} \leq L} \big( \mathscr{R}_3\big)_{j^{\prime}0} \big( u \big)               \bigg]  \end{align*}

    \begin{align*}    =   \mathrm{Tr}_0 \bigg[             \mathcal{K}_{+,0} \big( u \big)           \bigg[         e^{2 \eta} \big( e^{2 u} - 1  \big) \big( e^{2u} - e^{4n \eta  } \big)  c^{-} \big( u \big) \bigg[  \underset{\alpha \text{ } \text{or} \text{ } \beta, \beta^{\prime} \neq n+1, n+2}{\underset{\alpha \neq \beta, \beta^{\prime}}{\sum}}  e_{\alpha \alpha}\bigg] \bigg]  \\   \times \mathcal{K}_{-,0} \big( u \big)    \prod_{1 \leq j^{\prime} \leq L} \big( \mathscr{R}_3\big)_{j^{\prime}0} \big( u \big)               \bigg] \\   \times  \mathrm{Tr}_0 \bigg[             \mathcal{K}_{+,0} \big( u \big)    \bigg[   \underset{\alpha \text{ } \text{or} \text{ } \beta, \beta^{\prime} \neq n+1, n+2}{\underset{\alpha \neq \beta, \beta^{\prime}}{\sum}}   e_{\beta\beta} + e_{\alpha \alpha}  \bigg]   \bigg]             \bigg]  \\    + \mathrm{Tr}_0 \bigg[    \mathcal{K}_{+,0} \big( u \big)     \bigg[     e^{2 \eta} \big( e^{2 u} - 1  \big) \big( e^{2u} - e^{4n \eta  } \big)  c^{-} \big( u \big)  \bigg[     \underset{\alpha = n+1, n+2}{\sum} e_{\alpha \alpha}   \bigg]   \bigg] \\  \times                 \mathcal{K}_{-,0} \big( u \big)    \prod_{1 \leq j^{\prime} \leq L} \big( \mathscr{R}_3\big)_{j^{\prime}0} \big( u \big)               \bigg]   \\    \times      \mathrm{Tr}_0 \bigg[    \mathcal{K}_{+,0} \big( u \big)     \bigg[ \underset{\alpha = n+1, n+2}{\sum}  e_{\beta\beta} + e_{\alpha \alpha } \bigg]                   \mathcal{K}_{-,0} \big( u \big)    \prod_{1 \leq j^{\prime} \leq L} \big( \mathscr{R}_3\big)_{j^{\prime}0} \big( u \big)               \bigg] \\ \\ =   3_{(\mathrm{I})}   3_{(\mathrm{II })}   + 3_{(\mathrm{III})} 3_{(\mathrm{IV})}         ,
\end{align*}

\noindent corresponding to the third partial trace,

\begin{align*}
  \mathrm{Tr}_0 \bigg[ \mathcal{K}_{+,0} \big( u \big)  \prod_{1 \leq j \leq L} \big( \mathscr{R}_4\big)_{0j} \big( u \big)  \mathcal{K}_{-,0} \big( u \big)   \prod_{1 \leq j^{\prime} \leq L} \big( \mathscr{R}_4\big)_{j^{\prime}0} \big( u \big)  \bigg] \\ \\ \equiv \mathrm{Tr}_0 \bigg[             \mathcal{K}_{+,0} \big( u \big)           \bigg[          \bigg[ \underset{\alpha, \beta \neq n+1, n+2}{\sum} a_{\alpha \beta } \big( u \big)e_{\alpha \beta} + \frac{1}{2} \underset{\alpha \neq n+1, n+2, \beta = n+1, n+2}{\sum} b^+_a \big( u \big) e_{\alpha \beta} \\ +  \frac{1}{2} \underset{\alpha \neq n+1, n+2, \beta = n+1, n+2}{\sum} b^-_a \big( u \big)   e_{\alpha \beta} \bigg] \otimes \bigg[                 \underset{\alpha, \beta \neq n+1, n+2}{\sum}     e_{\alpha^{\prime} \beta^{\prime}} \\   +  \frac{1}{2} \underset{\alpha \neq n+1, n+2, \beta = n+1,n+2}{\sum}         e_{\alpha^{\prime} \beta^{\prime}}    + \frac{1}{2} \underset{\alpha \neq n+1, n+2, \beta = n+1, n+2}{\sum}  b^-_{\alpha} \big( u \big) \\ \times e_{\alpha^{\prime} \beta} \bigg]                                 \bigg]              \mathcal{K}_{-,0} \big( u \big)   \prod_{1 \leq j^{\prime} \leq L} \big( \mathscr{R}_4\big)_{j^{\prime}0} \big( u \big)                   \bigg]   \end{align*}

    \begin{align*}  =   \mathrm{Tr}_0 \bigg[             \mathcal{K}_{+,0} \big( u \big)                     \bigg[ \underset{\alpha, \beta \neq n+1, n+2}{\sum} a_{\alpha \beta } \big( u \big)e_{\alpha \beta} + \frac{1}{2} \underset{\alpha \neq n+1, n+2, \beta = n+1, n+2}{\sum} b^+_a \big( u \big) e_{\alpha \beta} \\   +  \frac{1}{2} \underset{\alpha \neq n+1, n+2, \beta = n+1, n+2}{\sum} b^-_a \big( u \big)   e_{\alpha \beta} \bigg]       \mathcal{K}_{-,0} \big( u \big)   \prod_{1 \leq j^{\prime} \leq L} \big( \mathscr{R}_4\big)_{j^{\prime}0} \big( u \big)                   \bigg]    \\  \times   \mathrm{Tr}_0 \bigg[             \mathcal{K}_{+,0} \big( u \big)      \bigg[                 \underset{\alpha, \beta \neq n+1, n+2}{\sum}     e_{\alpha^{\prime} \beta^{\prime}}  +  \frac{1}{2} \underset{\alpha \neq n+1, n+2, \beta = n+1,n+2}{\sum}         e_{\alpha^{\prime} \beta^{\prime}}   \\  + \frac{1}{2} \underset{\alpha \neq n+1, n+2, \beta = n+1, n+2}{\sum}  b^-_{\alpha} \big( u \big) \\ \times e_{\alpha^{\prime} \beta} \bigg]           \mathcal{K}_{-,0} \big( u \big)   \prod_{1 \leq j^{\prime} \leq L} \big( \mathscr{R}_4\big)_{j^{\prime}0} \big( u \big)                   \bigg]  \\ \\  =  \bigg[  \mathrm{Tr}_0 \bigg[             \mathcal{K}_{+,0} \big( u \big)                     \bigg[  \underset{\alpha, \beta \neq n+1, n+2}{\sum} a_{\alpha \beta } \big( u \big)e_{\alpha \beta}  \bigg]       \mathcal{K}_{-,0} \big( u \big)   \prod_{1 \leq j^{\prime} \leq L} \big( \mathscr{R}_4\big)_{j^{\prime}0} \big( u \big)     \bigg]       \\  +  \mathrm{Tr}_0 \bigg[             \mathcal{K}_{+,0} \big( u \big)                     \bigg[   \frac{1}{2} \underset{\alpha \neq n+1, n+2, \beta = n+1, n+2}{\sum} b^+_a \big( u \big) e_{\alpha \beta}      \bigg]      \mathcal{K}_{-,0} \big( u \big)   \prod_{1 \leq j^{\prime} \leq L} \big( \mathscr{R}_4\big)_{j^{\prime}0} \big( u \big)              \bigg]  \\    +  \mathrm{Tr}_0 \bigg[             \mathcal{K}_{+,0} \big( u \big)                     \bigg[    \frac{1}{2} \underset{\alpha \neq n+1, n+2, \beta = n+1, n+2}{\sum} b^-_a \big( u \big)   e_{\alpha \beta}     \bigg]      \mathcal{K}_{-,0} \big( u \big)   \prod_{1 \leq j^{\prime} \leq L} \big( \mathscr{R}_4\big)_{j^{\prime}0} \big( u \big)              \bigg]   \bigg]      \\ \times \bigg[ \mathrm{Tr}_0 \bigg[             \mathcal{K}_{+,0} \big( u \big)                     \bigg[    \underset{\alpha, \beta \neq n+1, n+2}{\sum}     e_{\alpha^{\prime} \beta^{\prime}}   \bigg]  \mathcal{K}_{-,0} \big( u \big)   \prod_{1 \leq j^{\prime} \leq L} \big( \mathscr{R}_4\big)_{j^{\prime}0} \big( u \big)   \bigg]  \\ \\   +    \mathrm{Tr}_0 \bigg[             \mathcal{K}_{+,0} \big( u \big)                     \bigg[    \frac{1}{2} \underset{\alpha \neq n+1, n+2, \beta = n+1,n+2}{\sum}         e_{\alpha^{\prime} \beta^{\prime}}    \bigg]      \mathcal{K}_{-,0} \big( u \big)   \prod_{1 \leq j^{\prime} \leq L} \big( \mathscr{R}_4\big)_{j^{\prime}0} \big( u \big)    \bigg] \\ +  \mathrm{Tr}_0 \bigg[             \mathcal{K}_{+,0} \big( u \big)                     \bigg[  \frac{1}{2} \underset{\alpha \neq n+1, n+2, \beta = n+1, n+2}{\sum}  b^-_{\alpha} \big( u \big)  e_{\alpha^{\prime} \beta}       \bigg]      \mathcal{K}_{-,0} \big( u \big)   \prod_{1 \leq j^{\prime} \leq L} \big( \mathscr{R}_4\big)_{j^{\prime}0} \big( u \big)              \bigg]    \bigg]              \\ \\ =     \big[ \mathscr{T}_1 + \mathscr{T}_2 + \mathscr{T}_3  \big] \big[ \mathscr{T}^{\prime}_1 + \mathscr{T}^{\prime}_2 + \mathscr{T}^{\prime}_3   \big]    \\ \\ =   \mathscr{T}_1 \mathscr{T}^{\prime}_1  + \mathscr{T}_2 \mathscr{T}^{\prime}_1    + \mathscr{T}_3 \mathscr{T}^{\prime}_1  +   \mathscr{T}_1   \mathscr{T}^{\prime}_2 +            \mathscr{T}_1   \mathscr{T}^{\prime}_3  +  \mathscr{T}_2 \mathscr{T}^{\prime}_2 + \mathscr{T}_2    \mathscr{T}^{\prime}_3   + \mathscr{T}_3 \mathscr{T}^{\prime}_2        \\   + \mathscr{T}_3 \mathscr{T}^{\prime}_3  \\ \\  =      \mathrm{Tr}_0 \bigg[             \mathcal{K}_{+,0} \big( u \big)                     \bigg[  \underset{\alpha, \beta \neq n+1, n+2}{\sum} a_{\alpha \beta } \big( u \big)e_{\alpha \beta}  \bigg]       \mathcal{K}_{-,0} \big( u \big)   \prod_{1 \leq j^{\prime} \leq L} \big( \mathscr{R}_4\big)_{j^{\prime}0} \big( u \big)     \bigg] \\   \times    \mathrm{Tr}_0 \bigg[             \mathcal{K}_{+,0} \big( u \big)                     \bigg[    \underset{\alpha, \beta \neq n+1, n+2}{\sum}     e_{\alpha^{\prime} \beta^{\prime}}   \bigg]  \mathcal{K}_{-,0} \big( u \big)   \prod_{1 \leq j^{\prime} \leq L} \big( \mathscr{R}_4\big)_{j^{\prime}0} \big( u \big)   \bigg]  \end{align*}

    \begin{align*}  +   \mathrm{Tr}_0 \bigg[             \mathcal{K}_{+,0} \big( u \big)                     \bigg[   \frac{1}{2} \underset{\alpha \neq n+1, n+2, \beta = n+1, n+2}{\sum} b^+_a \big( u \big) e_{\alpha \beta}      \bigg]      \mathcal{K}_{-,0} \big( u \big)   \prod_{1 \leq j^{\prime} \leq L} \big( \mathscr{R}_4\big)_{j^{\prime}0} \big( u \big)              \bigg]     \\  \times    \mathrm{Tr}_0 \bigg[             \mathcal{K}_{+,0} \big( u \big)                     \bigg[    \underset{\alpha, \beta \neq n+1, n+2}{\sum}     e_{\alpha^{\prime} \beta^{\prime}}   \bigg]  \mathcal{K}_{-,0} \big( u \big)   \prod_{1 \leq j^{\prime} \leq L} \big( \mathscr{R}_4\big)_{j^{\prime}0} \big( u \big)   \bigg]  \\    +  \mathrm{Tr}_0 \bigg[             \mathcal{K}_{+,0} \big( u \big)                     \bigg[    \frac{1}{2} \underset{\alpha \neq n+1, n+2, \beta = n+1, n+2}{\sum} b^-_a \big( u \big)   e_{\alpha \beta}     \bigg]      \mathcal{K}_{-,0} \big( u \big)   \prod_{1 \leq j^{\prime} \leq L} \big( \mathscr{R}_4\big)_{j^{\prime}0} \big( u \big)              \bigg]     \\   \times    \mathrm{Tr}_0 \bigg[             \mathcal{K}_{+,0} \big( u \big)                     \bigg[    \underset{\alpha, \beta \neq n+1, n+2}{\sum}     e_{\alpha^{\prime} \beta^{\prime}}   \bigg]  \mathcal{K}_{-,0} \big( u \big)   \prod_{1 \leq j^{\prime} \leq L} \big( \mathscr{R}_4\big)_{j^{\prime}0} \big( u \big)   \bigg]  \\    +         \mathrm{Tr}_0 \bigg[             \mathcal{K}_{+,0} \big( u \big)                     \bigg[  \underset{\alpha, \beta \neq n+1, n+2}{\sum} a_{\alpha \beta } \big( u \big)e_{\alpha \beta}  \bigg]       \mathcal{K}_{-,0} \big( u \big)   \prod_{1 \leq j^{\prime} \leq L} \big( \mathscr{R}_4\big)_{j^{\prime}0} \big( u \big)     \bigg] \\  \times \mathrm{Tr}_0 \bigg[             \mathcal{K}_{+,0} \big( u \big)                     \bigg[    \frac{1}{2} \underset{\alpha \neq n+1, n+2, \beta = n+1,n+2}{\sum}         e_{\alpha^{\prime} \beta^{\prime}}    \bigg]      \mathcal{K}_{-,0} \big( u \big)   \prod_{1 \leq j^{\prime} \leq L} \big( \mathscr{R}_4\big)_{j^{\prime}0} \big( u \big)    \bigg]  \\  + \mathrm{Tr}_0 \bigg[             \mathcal{K}_{+,0} \big( u \big)                     \bigg[  \underset{\alpha, \beta \neq n+1, n+2}{\sum} a_{\alpha \beta } \big( u \big)e_{\alpha \beta}  \bigg]       \mathcal{K}_{-,0} \big( u \big)   \prod_{1 \leq j^{\prime} \leq L} \big( \mathscr{R}_4\big)_{j^{\prime}0} \big( u \big)     \bigg]  \\ \times \mathrm{Tr}_0 \bigg[             \mathcal{K}_{+,0} \big( u \big)                     \bigg[  \frac{1}{2} \underset{\alpha \neq n+1, n+2, \beta = n+1, n+2}{\sum}  b^-_{\alpha} \big( u \big)  e_{\alpha^{\prime} \beta}       \bigg]      \mathcal{K}_{-,0} \big( u \big)   \prod_{1 \leq j^{\prime} \leq L} \big( \mathscr{R}_4\big)_{j^{\prime}0} \big( u \big)              \bigg]    \bigg]       \\ +   \mathrm{Tr}_0 \bigg[             \mathcal{K}_{+,0} \big( u \big)                     \bigg[   \frac{1}{2} \underset{\alpha \neq n+1, n+2, \beta = n+1, n+2}{\sum} b^+_a \big( u \big) e_{\alpha \beta}      \bigg]      \mathcal{K}_{-,0} \big( u \big)   \prod_{1 \leq j^{\prime} \leq L} \big( \mathscr{R}_4\big)_{j^{\prime}0} \big( u \big)              \bigg] \\ \times     \mathrm{Tr}_0 \bigg[             \mathcal{K}_{+,0} \big( u \big)                     \bigg[    \frac{1}{2} \underset{\alpha \neq n+1, n+2, \beta = n+1,n+2}{\sum}         e_{\alpha^{\prime} \beta^{\prime}}    \bigg]      \mathcal{K}_{-,0} \big( u \big)   \prod_{1 \leq j^{\prime} \leq L} \big( \mathscr{R}_4\big)_{j^{\prime}0} \big( u \big)    \bigg]      \\ +  \mathrm{Tr}_0 \bigg[             \mathcal{K}_{+,0} \big( u \big)                     \bigg[   \frac{1}{2} \underset{\alpha \neq n+1, n+2, \beta = n+1, n+2}{\sum} b^+_a \big( u \big) e_{\alpha \beta}      \bigg]      \mathcal{K}_{-,0} \big( u \big)   \prod_{1 \leq j^{\prime} \leq L} \big( \mathscr{R}_4\big)_{j^{\prime}0} \big( u \big)              \bigg] \\  \times \mathrm{Tr}_0 \bigg[             \mathcal{K}_{+,0} \big( u \big)                     \bigg[  \frac{1}{2} \underset{\alpha \neq n+1, n+2, \beta = n+1, n+2}{\sum}  b^-_{\alpha} \big( u \big)  e_{\alpha^{\prime} \beta}       \bigg]      \mathcal{K}_{-,0} \big( u \big)   \prod_{1 \leq j^{\prime} \leq L} \big( \mathscr{R}_4\big)_{j^{\prime}0} \big( u \big)              \bigg]   \\ +     \mathrm{Tr}_0 \bigg[             \mathcal{K}_{+,0} \big( u \big)                     \bigg[    \frac{1}{2} \underset{\alpha \neq n+1, n+2, \beta = n+1, n+2}{\sum} b^-_a \big( u \big)   e_{\alpha \beta}     \bigg]      \mathcal{K}_{-,0} \big( u \big)   \prod_{1 \leq j^{\prime} \leq L} \big( \mathscr{R}_4\big)_{j^{\prime}0} \big( u \big)              \bigg]           \\ \times  \mathrm{Tr}_0 \bigg[             \mathcal{K}_{+,0} \big( u \big)                     \bigg[    \frac{1}{2} \underset{\alpha \neq n+1, n+2, \beta = n+1,n+2}{\sum}         e_{\alpha^{\prime} \beta^{\prime}}    \bigg]      \mathcal{K}_{-,0} \big( u \big)   \prod_{1 \leq j^{\prime} \leq L} \big( \mathscr{R}_4\big)_{j^{\prime}0} \big( u \big)    \bigg]      \\   +  \mathrm{Tr}_0 \bigg[             \mathcal{K}_{+,0} \big( u \big)                     \bigg[    \frac{1}{2} \underset{\alpha \neq n+1, n+2, \beta = n+1, n+2}{\sum} b^-_a \big( u \big)   e_{\alpha \beta}     \bigg]      \mathcal{K}_{-,0} \big( u \big)   \prod_{1 \leq j^{\prime} \leq L} \big( \mathscr{R}_4\big)_{j^{\prime}0} \big( u \big)              \bigg]    \end{align*}

    \begin{align*}   \times     \mathrm{Tr}_0 \bigg[             \mathcal{K}_{+,0} \big( u \big)                     \bigg[  \frac{1}{2} \underset{\alpha \neq n+1, n+2, \beta = n+1, n+2}{\sum}  b^-_{\alpha} \big( u \big)  e_{\alpha^{\prime} \beta}       \bigg]      \mathcal{K}_{-,0} \big( u \big)   \prod_{1 \leq j^{\prime} \leq L} \big( \mathscr{R}_4\big)_{j^{\prime}0} \big( u \big)              \bigg]   \\ \\ =   4_{(\mathrm{I})}   4_{(\mathrm{II })}   + 4_{(\mathrm{III})} 4_{(\mathrm{IV})}   +      4_{(\mathrm{V})} 4_{(\mathrm{VI})} + 4_{(\mathrm{VII})} 4_{(\mathrm{VIII})}          + 4_{(\mathrm{VIV})} 4_{(\mathrm{X})}           + 4_{(\mathrm{XI})} 4_{(\mathrm{XII})}  \\  + 4_{(\mathrm{XIII})} 4_{(\mathrm{XIV})}        + 4_{(\mathrm{XV})} 4_{(\mathrm{XVI})}               + 4_{(\mathrm{XVII})} 4_{(\mathrm{XVIII})}                                ,
\end{align*}

\noindent corresponding to the fourth partial trace, and,

\begin{align*}
  \mathrm{Tr}_0 \bigg[ \mathcal{K}_{+,0} \big( u \big)  \prod_{1 \leq j \leq L} \big( \mathscr{R}_5\big)_{0j} \big( u \big)  \mathcal{K}_{-,0} \big( u \big)   \prod_{1 \leq j^{\prime} \leq L} \big( \mathscr{R}_5\big)_{j^{\prime}0} \big( u \big)  \bigg] \\ \\ \equiv \mathrm{Tr}_0 \bigg[             \mathcal{K}_{+,0} \big( u \big)           \bigg[     \bigg[     \frac{1}{2} \underset{\alpha \neq n+1, n+2, \beta = n+1, n+2}{\sum} b^+_{\alpha} \big( u \big) e_{\beta^{\prime} \alpha^{\prime}}  +   \frac{1}{2} \underset{\alpha \neq n+1, n+2, \beta = n+1, n+2}{\sum} b^-_{\alpha} \big( u \big) e_{\beta \alpha^{\prime}} \bigg] \\ \otimes       \big[ e_{\beta \alpha} \big]              \bigg]              \mathcal{K}_{-,0} \big( u \big)   \prod_{1 \leq j^{\prime} \leq L} \big( \mathscr{R}_3\big)_{j^{\prime}0} \big( u \big)                   \bigg] \\  \\ =         \mathrm{Tr}_0 \bigg[             \mathcal{K}_{+,0} \big( u \big)           \bigg[       \frac{1}{2} \underset{\alpha \neq n+1, n+2, \beta = n+1, n+2}{\sum} b^+_{\alpha} \big( u \big) e_{\beta^{\prime} \alpha^{\prime}}  +   \frac{1}{2} \underset{\alpha \neq n+1, n+2, \beta = n+1, n+2}{\sum} b^-_{\alpha} \big( u \big) e_{\beta \alpha^{\prime}} \bigg]       \\   \times \mathcal{K}_{-,0} \big( u \big)   \prod_{1 \leq j^{\prime} \leq L} \big( \mathscr{R}_3\big)_{j^{\prime}0} \big( u \big)                   \bigg]  \\  \times  \mathrm{Tr}_0 \bigg[   \mathcal{K}_{+,0} \big( u \big)  \big[     e_{\beta \alpha}   \big]      \mathcal{K}_{-,0} \big( u \big)   \prod_{1 \leq j^{\prime} \leq L} \big( \mathscr{R}_3\big)_{j^{\prime}0} \big( u \big)                   \bigg]  \\ \\ =  \bigg[ \mathrm{Tr}_0 \bigg[             \mathcal{K}_{+,0} \big( u \big)           \bigg[   \frac{1}{2} \underset{\alpha \neq n+1, n+2, \beta = n+1, n+2}{\sum} b^+_{\alpha} \big( u \big) e_{\beta^{\prime} \alpha^{\prime}} \bigg]     \mathcal{K}_{-,0} \big( u \big)   \prod_{1 \leq j^{\prime} \leq L} \big( \mathscr{R}_3\big)_{j^{\prime}0} \big( u \big)                   \bigg] \\  +  \mathrm{Tr}_0 \bigg[             \mathcal{K}_{+,0} \big( u \big)           \bigg[ \frac{1}{2} \underset{\alpha \neq n+1, n+2, \beta = n+1, n+2}{\sum} b^-_{\alpha} \big( u \big) e_{\beta \alpha^{\prime}} \bigg]   \mathcal{K}_{-,0} \big( u \big)   \prod_{1 \leq j^{\prime} \leq L} \big( \mathscr{R}_3\big)_{j^{\prime}0} \big( u \big)                   \bigg] \bigg] \\    \times  \mathrm{Tr}_0 \bigg[   \mathcal{K}_{+,0} \big( u \big)  \big[     e_{\beta \alpha}   \big]      \mathcal{K}_{-,0} \big( u \big)   \prod_{1 \leq j^{\prime} \leq L} \big( \mathscr{R}_3\big)_{j^{\prime}0} \big( u \big)                   \bigg]  \\ \\ =  \mathrm{Tr}_0 \bigg[             \mathcal{K}_{+,0} \big( u \big)           \bigg[   \frac{1}{2} \underset{\alpha \neq n+1, n+2, \beta = n+1, n+2}{\sum} b^+_{\alpha} \big( u \big) e_{\beta^{\prime} \alpha^{\prime}} \bigg]     \mathcal{K}_{-,0} \big( u \big)   \prod_{1 \leq j^{\prime} \leq L} \big( \mathscr{R}_3\big)_{j^{\prime}0} \big( u \big)                   \bigg] \end{align*}

  \begin{align*}      \times  \mathrm{Tr}_0 \bigg[   \mathcal{K}_{+,0} \big( u \big)  \big[     e_{\beta \alpha}   \big]      \mathcal{K}_{-,0} \big( u \big)   \prod_{1 \leq j^{\prime} \leq L} \big( \mathscr{R}_3\big)_{j^{\prime}0} \big( u \big)                   \bigg]   \\   +  \mathrm{Tr}_0 \bigg[             \mathcal{K}_{+,0} \big( u \big)           \bigg[ \frac{1}{2} \underset{\alpha \neq n+1, n+2, \beta = n+1, n+2}{\sum} b^-_{\alpha} \big( u \big) e_{\beta \alpha^{\prime}} \bigg]   \mathcal{K}_{-,0} \big( u \big)   \prod_{1 \leq j^{\prime} \leq L} \big( \mathscr{R}_3\big)_{j^{\prime}0} \big( u \big)                   \bigg] \\  \times  \mathrm{Tr}_0 \bigg[   \mathcal{K}_{+,0} \big( u \big)  \big[     e_{\beta \alpha}   \big]      \mathcal{K}_{-,0} \big( u \big)   \prod_{1 \leq j^{\prime} \leq L} \big( \mathscr{R}_3\big)_{j^{\prime}0} \big( u \big)                   \bigg]  \\ \\  = 5_{(\mathrm{I})}    5_{(\mathrm{II})}       + 5_{(\mathrm{III})}   5_{(\mathrm{IV})}     ,
\end{align*}

\noindent corresponding to the fifth partial trace, one can readily approximate the desired partial trace,

\begin{align*}
\textbf{T}^{\mathrm{Open}}\big( u \big) \equiv  \mathrm{tr}_0 \bigg[ \mathcal{K}_{+,0} \big( u \big)  \prod_{1 \leq j \leq L} R_{0j} \big( u \big)  \mathcal{K}_{-,0} \big( u \big)   \prod_{1 \leq j^{\prime} \leq L} R_{j^{\prime}0} \big( u \big)  \bigg]  , 
\end{align*}

\noindent from the following approximations, respectively,

\begin{align*}
    \mathrm{Tr}_0 \bigg[ \mathcal{K}_{+,0} \big( u \big)  \bigg[ \prod_{1 \leq j \leq L} \big( \mathscr{R}_1\big)_{0j} \big( u \big) \bigg]  \mathcal{K}_{-,0} \big( u \big)   \bigg[ \prod_{1 \leq j^{\prime} \leq L} \big( \mathscr{R}_1\big)_{j^{\prime}0} \big( u \big) \bigg]       \bigg] \\ \approx  1_{(\mathrm{I})}   1_{(\mathrm{II })}   + 1_{(\mathrm{III})} 1_{(\mathrm{IV})}              , 
\end{align*}

\noindent corresponding to the first partial trace,

\begin{align*}
    \mathrm{Tr}_0 \bigg[\mathcal{K}_{+,0} \big( u \big)  \bigg[ \prod_{1 \leq j \leq L} \big( \mathscr{R}_2\big)_{0j} \big( u \big) \bigg]  \mathcal{K}_{-,0} \big( u \big) \bigg[   \prod_{1 \leq j^{\prime} \leq L} \big( \mathscr{R}_2\big)_{j^{\prime}0} \big( u \big)  \bigg]  \bigg] \\ \approx 2_{(\mathrm{I})}   2_{(\mathrm{II })}   + 2_{(\mathrm{III})} 2_{(\mathrm{IV})}   , 
\end{align*}

\noindent corresponding to the second partial trace,

\begin{align*}
  \mathrm{Tr}_0 \bigg[ \mathcal{K}_{+,0} \big( u \big)  \bigg[ \prod_{1 \leq j \leq L} \big( \mathscr{R}_3\big)_{0j} \big( u \big)  \bigg] \mathcal{K}_{-,0} \big( u \big)   \bigg[ \prod_{1 \leq j^{\prime} \leq L} \big( \mathscr{R}_3\big)_{j^{\prime}0} \big( u \big)  \bigg] \bigg]  \\ \approx  3_{(\mathrm{I})}   3_{(\mathrm{II })}   + 3_{(\mathrm{III})} 3_{(\mathrm{IV})} ,
\end{align*}

\noindent corresponding to the third partial trace,

\begin{align*}
  \mathrm{Tr}_0 \bigg[ \mathcal{K}_{+,0} \big( u \big)  \bigg[ \prod_{1 \leq j \leq L} \big( \mathscr{R}_4\big)_{0j} \big( u \big)  \bigg] \mathcal{K}_{-,0} \big( u \big)   \bigg[ \prod_{1 \leq j^{\prime} \leq L} \big( \mathscr{R}_4\big)_{j^{\prime}0} \big( u \big)  \bigg] \bigg] \\ \approx 4_{(\mathrm{I})}   4_{(\mathrm{II })}   + 4_{(\mathrm{III})} 4_{(\mathrm{IV})}   +      4_{(\mathrm{V})} 4_{(\mathrm{VI})} + 4_{(\mathrm{VII})} 4_{(\mathrm{VIII})}          + 4_{(\mathrm{VIV})} 4_{(\mathrm{X})}           + 4_{(\mathrm{XI})} 4_{(\mathrm{XII})}  \\  + 4_{(\mathrm{XIII})} 4_{(\mathrm{XIV})}        + 4_{(\mathrm{XV})} 4_{(\mathrm{XVI})}               + 4_{(\mathrm{XVII})} 4_{(\mathrm{XVIII})}    ,
\end{align*}

\noindent corresponding to the fourth partial trace, and,

\begin{align*}
  \mathrm{Tr}_0 \bigg[ \mathcal{K}_{+,0} \big( u \big)  \bigg[ \prod_{1 \leq j \leq L} \big( \mathscr{R}_5\big)_{0j} \big( u \big)  \bigg] \mathcal{K}_{-,0} \big( u \big)   \bigg[ \prod_{1 \leq j^{\prime} \leq L} \big( \mathscr{R}_5\big)_{j^{\prime}0} \big( u \big) \bigg]  \bigg] \\ \approx 5_{(\mathrm{I})}    5_{(\mathrm{II})}       + 5_{(\mathrm{III})}   5_{(\mathrm{IV})} ,
\end{align*}

\noindent corresponding to the fifth partial trace.

\end{itemize}

\noindent Cumulatively, the previously obtained approximations for the partial trace of each term of the Jimbo R-matrix above, upon denoting, 

\begin{align*}
   \mathscr{R}_1 \equiv \textit{Restriction of the Jimbo R-matrix to the degree of freedom along the x-axis of } \textbf{Z}^2   \end{align*}

   \begin{align*}
    = \prod_{1 \leq j \leq L} R_{0j} \big( u \big)  ,
\end{align*}
   
   \begin{align*} \mathscr{R}_2 \equiv \textit{Restriction of the Jimbo R-matrix to the degree of freedom along the y-axis of } \textbf{Z}^2   
\end{align*}

\begin{align*}
    = \prod_{1 \leq j^{\prime} \leq L} R_{j^{\prime}0} \big( u \big)  \\ \equiv  \prod_{1 \leq j \leq L} R_{j0} \big( u^{\prime} \big)   ,
\end{align*}

\noindent one has that,

\begin{align*}
\mathrm{Tr}_0 \bigg[ \mathcal{K}_{+,0} \big( u \big)  \mathscr{R}_1 \mathcal{K}_{-,0} \big( u \big)   \mathscr{R}_2     \bigg]  \propto   1_{(\mathrm{I})}   1_{(\mathrm{II })}   + 1_{(\mathrm{III})} 1_{(\mathrm{IV})}    + 2_{(\mathrm{I})}   2_{(\mathrm{II })}   + 2_{(\mathrm{III})} 2_{(\mathrm{IV})}    \\ + 3_{(\mathrm{I})}   3_{(\mathrm{II })}   + 3_{(\mathrm{III})} 3_{(\mathrm{IV})}      +  4_{(\mathrm{I})}   4_{(\mathrm{II })}   + 4_{(\mathrm{III})} 4_{(\mathrm{IV})}   \\ +      4_{(\mathrm{V})} 4_{(\mathrm{VI})} + 4_{(\mathrm{VII})} 4_{(\mathrm{VIII})}          + 4_{(\mathrm{VIV})} 4_{(\mathrm{X})}        \\    + 4_{(\mathrm{XI})} 4_{(\mathrm{XII})}    + 4_{(\mathrm{XIII})} 4_{(\mathrm{XIV})}        + 4_{(\mathrm{XV})} 4_{(\mathrm{XVI})}         \\       + 4_{(\mathrm{XVII})} 4_{(\mathrm{XVIII})}   + 5_{(\mathrm{I})}    5_{(\mathrm{II})}       + 5_{(\mathrm{III})}   5_{(\mathrm{IV})}                 ,
\end{align*}

\noindent implies, due to multiplicativity of the partial trace of $\mathcal{K}_{\pm}$ and $\mathscr{R}$, that,

\begin{align*}
  \mathrm{Tr}_0 \bigg[ \mathcal{K}_{+,0} \big( u \big)    \bigg\{    \big( e^{2u} - e^{4 \eta} \big)  \big( e^{2u} - e^{12 \eta} \big) \underset{\alpha \neq 4,6}{\sum} e_{\alpha \alpha} \otimes e_{\alpha \alpha} +   e^{2\eta} \big( e^{2u} - 1 \big) \big( e^{2u} - e^{12 \eta} \big)\\ \times  \underset{\alpha, \text{ } \textit{or} \text{ } \beta \neq 4,6}{\underset{\alpha \neq \beta , \beta^{\prime}}{\sum}}     e_{\alpha \alpha} \otimes e_{\beta \beta} - \big( e^{4\eta} - 1 \big) \big( e^{2u} - e^{12 \eta} \big) \bigg[    \underset{\alpha, \beta \neq 4,6}{\underset{\alpha < \beta, \alpha \neq \beta^{\prime}}{\sum}}  + e^{2u}   \underset{\alpha, \beta \neq 4,6}{\underset{\alpha >  \beta, \alpha \neq \beta^{\prime}}{\sum}}    \bigg] \\  \times e_{\alpha \beta} \otimes e_{\beta \alpha} - \frac{1}{2}   \big( e^{4 \eta} -1 \big) \big( e^{2u} -  e^{12 \eta} \big) \bigg[ \big( e^u + 1 \big) \bigg[   \underset{\alpha < 4 , \beta = 4,6}{\sum} + e^u   \underset{\alpha > 5 , \beta = 4,6}{\sum}  \bigg] \\ \times \big[ e_{\alpha \beta} \otimes e_{\beta \alpha} +  e_{\beta^{\prime}\alpha^{\prime}} + e_{\alpha^{\prime} \beta^{\prime}} \big] + \big( e^u - 1 \big) \bigg[     \underset{\alpha < 4 , \beta = 4,6}{\sum} + e^u   \underset{\alpha > 5 , \beta = 4,6}{\sum}    \bigg]  \big[ e_{\alpha \beta } \otimes e_{\beta^{\prime}\alpha} \\ + e_{\beta^{\prime} \alpha^{\prime}}  \otimes e_{\alpha^{\prime} \beta}  \big]   \bigg]  + \underset{\alpha, \beta \neq 4,6}{\sum} a_{\alpha \beta} \big( u \big) e_{\alpha \beta} \otimes e_{\alpha^{\prime} \beta^{\prime}} + \frac{1}{2} \underset{\alpha \neq 4,6, \beta = 4,6}{\sum} \big[ b^+_a \big( u \big) \big[ e_{\alpha \beta} \otimes e_{\alpha^{\prime} \beta^{\prime}} \\  + e_{\beta^{\prime} \alpha^{\prime}} \otimes e_{\beta \alpha}  \big]  + b^-_{\alpha} \big( u \big) \big[       e_{\alpha \beta} \otimes e_{\alpha^{\prime} \beta} + e_{\beta \alpha^{\prime}} \otimes e_{\beta \alpha}        \big]     \big]   + \underset{\alpha = 4,6}{\sum} \big[ c^+\big( u \big) e_{\alpha \alpha} \\ + e_{\alpha^{\prime} \alpha^{\prime}} + c^- \big( u \big) e_{\alpha \alpha} \otimes e_{\alpha \alpha} + d^+ \big( u \big)    e_{\alpha \alpha } \otimes e_{\alpha^{\prime} \alpha} + d^- \big( u \big) e_{\alpha \alpha^{\prime}} \otimes e_{\alpha \alpha^{\prime}}     \big]   \bigg\}  \mathcal{K}_{-,0} \big( u^{\prime} \big) \\ \times \bigg\{    \big( e^{2u^{\prime}} - e^{4 \eta} \big)  \big( e^{2u^{\prime}} - e^{12 \eta} \big) \underset{\alpha \neq 4,6}{\sum} e_{\alpha \alpha} \otimes e_{\alpha \alpha} +  e^{2\eta} \big( e^{2u^{\prime}} - 1 \big) \big( e^{2u^{\prime}} - e^{12 \eta} \big)\\ \times  \underset{\alpha, \text{ } \textit{or} \text{ } \beta \neq 4,6}{\underset{\alpha \neq \beta , \beta^{\prime}}{\sum}}     e_{\alpha \alpha} \otimes e_{\beta \beta} - \big( e^{4\eta} - 1 \big) \big( e^{2u^{\prime}} - e^{12 \eta} \big) \bigg[    \underset{\alpha, \beta \neq 4,6}{\underset{\alpha < \beta, \alpha \neq \beta^{\prime}}{\sum}}  + e^{2u^{\prime}}   \underset{\alpha, \beta \neq 4,6}{\underset{\alpha >  \beta, \alpha \neq \beta^{\prime}}{\sum}}   \\   \times e_{\alpha \beta} \otimes e_{\beta \alpha} - \frac{1}{2}   \big( e^{4 \eta} -1 \big) \big( e^{2u^{\prime}} -  e^{12 \eta} \big) \bigg[ \big( e^u + 1 \big) \bigg[   \underset{\alpha < 4 , \beta = 4,6}{\sum} + e^{u^{\prime}}   \underset{\alpha > 5 , \beta = 4,6}{\sum}  \bigg] \\ \times \big[ e_{\alpha \beta} \otimes e_{\beta \alpha} +  e_{\beta^{\prime}\alpha^{\prime}} + e_{\alpha^{\prime} \beta^{\prime}} \big] + \big( e^{u^{\prime}} - 1 \big) \bigg[     \underset{\alpha < 4 , \beta = 4,6}{\sum} + e^{u^{\prime}}   \underset{\alpha > 5 , \beta = 4,6}{\sum}    \bigg]  \big[ e_{\alpha \beta } \otimes e_{\beta^{\prime}\alpha} \\  + e_{\beta^{\prime} \alpha^{\prime}}  \otimes e_{\alpha^{\prime} \beta}  \big]   \bigg]  + \underset{\alpha, \beta \neq 4,6}{\sum} a_{\alpha \beta} \big( u^{\prime} \big) e_{\alpha \beta} \otimes e_{\alpha^{\prime} \beta^{\prime}} + \frac{1}{2} \underset{\alpha \neq 4,6, \beta = 4,6}{\sum} \big[ b^+_a \big( u^{\prime} \big) \big[ e_{\alpha \beta} \otimes e_{\alpha^{\prime} \beta^{\prime}} \\ + e_{\beta^{\prime} \alpha^{\prime}} \otimes e_{\beta \alpha}  \big]  + b^-_{\alpha} \big( u^{\prime} \big) \big[       e_{\alpha \beta} \otimes e_{\alpha^{\prime} \beta} + e_{\beta \alpha^{\prime}} \otimes e_{\beta \alpha}        \big]     \big]   + \underset{\alpha = 4,6}{\sum} \big[ c^+\big( u^{\prime} \big) e_{\alpha \alpha} \end{align*}

  \begin{align*} + e_{\alpha^{\prime} \alpha^{\prime}} + c^- \big( u^{\prime} \big) e_{\alpha \alpha} \otimes e_{\alpha \alpha} + d^+ \big( u^{\prime} \big)    e_{\alpha \alpha } \otimes e_{\alpha^{\prime} \alpha} + d^- \big( u^{\prime} \big) e_{\alpha \alpha^{\prime}} \otimes e_{\alpha \alpha^{\prime}}     \big]     \bigg\}     \bigg]            \\ \\ =   \mathrm{Tr}_0 \bigg[ \mathcal{K}_{+,0} \big( u \big)    \bigg\{    \big( e^{2u} - e^{4 \eta} \big)  \big( e^{2u} - e^{12 \eta} \big) \underset{\alpha \neq 4,6}{\sum} e_{\alpha \alpha} \otimes e_{\alpha \alpha} \bigg\} \mathcal{K}_{-,0} \big( u^{\prime} \big) \bigg\{    \big( e^{2u^{\prime}} - e^{4 \eta} \big)  \big( e^{2u^{\prime}} \\  - e^{12 \eta} \big) \underset{\alpha \neq 4,6}{\sum} e_{\alpha \alpha} \otimes e_{\alpha \alpha}    + e^{2\eta} \big( e^{2u^{\prime}} - 1 \big) \big( e^{2u^{\prime}} - e^{12 \eta} \big) \underset{\alpha, \text{ } \textit{or} \text{ } \beta \neq 4,6}{\underset{\alpha \neq \beta , \beta^{\prime}}{\sum}}     e_{\alpha \alpha} \otimes e_{\beta \beta}  \\  - \big( e^{4\eta} - 1 \big) \big( e^{2u^{\prime}} - e^{12 \eta} \big) \bigg[    \underset{\alpha, \beta \neq 4,6}{\underset{\alpha < \beta, \alpha \neq \beta^{\prime}}{\sum}}    + e^{2u^{\prime}}   \underset{\alpha, \beta \neq 4,6}{\underset{\alpha >  \beta, \alpha \neq \beta^{\prime}}{\sum}}   e_{\alpha \beta} \otimes e_{\beta \alpha} - \frac{1}{2}   \big( e^{4 \eta} -1 \big) \big( e^{2u^{\prime}} -  e^{12 \eta} \big) \\ \times \bigg[ \big( e^u + 1 \big) \bigg[   \underset{\alpha < 4 , \beta = 4,6}{\sum} + e^{u^{\prime}}   \underset{\alpha > 5 , \beta = 4,6}{\sum}  \bigg]  \big[ e_{\alpha \beta} \otimes e_{\beta \alpha} +  e_{\beta^{\prime}\alpha^{\prime}} + e_{\alpha^{\prime} \beta^{\prime}} \big] \\ + \big( e^{u^{\prime}} - 1 \big) \bigg[     \underset{\alpha < 4 , \beta = 4,6}{\sum} + e^{u^{\prime}}   \underset{\alpha > 5 , \beta = 4,6}{\sum}    \bigg]  \big[ e_{\alpha \beta } \otimes e_{\beta^{\prime}\alpha}  + e_{\beta^{\prime} \alpha^{\prime}}  \otimes e_{\alpha^{\prime} \beta}  \big]   \bigg] \bigg]  \\  + \underset{\alpha, \beta \neq 4,6}{\sum} a_{\alpha \beta} \big( u^{\prime} \big) e_{\alpha \beta} \otimes e_{\alpha^{\prime} \beta^{\prime}} + \frac{1}{2} \underset{\alpha \neq 4,6, \beta = 4,6}{\sum} \big[ b^+_a \big( u^{\prime} \big) \big[ e_{\alpha \beta} \otimes e_{\alpha^{\prime} \beta^{\prime}} \\ + e_{\beta^{\prime} \alpha^{\prime}} \otimes e_{\beta \alpha}  \big]  + b^-_{\alpha} \big( u^{\prime} \big) \big[       e_{\alpha \beta} \otimes e_{\alpha^{\prime} \beta} + e_{\beta \alpha^{\prime}} \otimes e_{\beta \alpha}        \big]     \big]   + \underset{\alpha = 4,6}{\sum} \big[ c^+\big( u^{\prime} \big) e_{\alpha \alpha} \\ + e_{\alpha^{\prime} \alpha^{\prime}} + c^- \big( u^{\prime} \big) e_{\alpha \alpha} \otimes e_{\alpha \alpha} + d^+ \big( u^{\prime} \big)    e_{\alpha \alpha } \otimes e_{\alpha^{\prime} \alpha} + d^- \big( u^{\prime} \big) e_{\alpha \alpha^{\prime}} \otimes e_{\alpha \alpha^{\prime}}     \big]     \bigg\}     \bigg]   \\ \\ +  \mathrm{Tr}_0 \bigg[ \mathcal{K}_{+,0} \big( u \big)    \bigg\{    e^{2\eta} \big( e^{2u} - 1 \big) \big( e^{2u} - e^{12 \eta} \big)  \underset{\alpha, \text{ } \textit{or} \text{ } \beta \neq 4,6}{\underset{\alpha \neq \beta , \beta^{\prime}}{\sum}}     e_{\alpha \alpha} \otimes e_{\beta \beta} - \big( e^{4\eta} - 1 \big)  \big( e^{2u} - e^{12 \eta} \big) \bigg[    \underset{\alpha, \beta \neq 4,6}{\underset{\alpha < \beta, \alpha \neq \beta^{\prime}}{\sum}} \\  + e^{2u}   \underset{\alpha, \beta \neq 4,6}{\underset{\alpha >  \beta, \alpha \neq \beta^{\prime}}{\sum}}  \bigg]   e_{\alpha \beta}  \otimes e_{\beta \alpha}       \bigg\} \mathcal{K}_{-,0} \big( u^{\prime} \big) \bigg\{    \big( e^{2u^{\prime}} - e^{4 \eta} \big)  \big( e^{2u^{\prime}}   - e^{12 \eta} \big) \underset{\alpha \neq 4,6}{\sum} e_{\alpha \alpha} \otimes e_{\alpha \alpha}    + e^{2\eta} \big( e^{2u^{\prime}} - 1 \big) \big( e^{2u^{\prime}} \\ - e^{12 \eta} \big) \underset{\alpha, \text{ } \textit{or} \text{ } \beta \neq 4,6}{\underset{\alpha \neq \beta , \beta^{\prime}}{\sum}}     e_{\alpha \alpha} \otimes e_{\beta \beta}           - \big( e^{4\eta} - 1 \big) \big( e^{2u^{\prime}}  - e^{12 \eta} \big) \bigg[    \underset{\alpha, \beta \neq 4,6}{\underset{\alpha < \beta, \alpha \neq \beta^{\prime}}{\sum}}   \end{align*}

  \begin{align*}   + e^{2u^{\prime}}   \underset{\alpha, \beta \neq 4,6}{\underset{\alpha >  \beta, \alpha \neq \beta^{\prime}}{\sum}}   e_{\alpha \beta} \otimes e_{\beta \alpha}  - \frac{1}{2}   \big( e^{4 \eta} -1 \big) \big( e^{2u^{\prime}} -  e^{12 \eta} \big) \\ \times  \bigg[ \big( e^u + 1 \big) \bigg[   \underset{\alpha < 4 , \beta = 4,6}{\sum} + e^{u^{\prime}}   \underset{\alpha > 5 , \beta = 4,6}{\sum}  \bigg]  \big[ e_{\alpha \beta} \otimes e_{\beta \alpha} +  e_{\beta^{\prime}\alpha^{\prime}} + e_{\alpha^{\prime} \beta^{\prime}} \big] \\ + \big( e^{u^{\prime}} - 1 \big) \bigg[     \underset{\alpha < 4 , \beta = 4,6}{\sum} + e^{u^{\prime}}   \underset{\alpha > 5 , \beta = 4,6}{\sum}    \bigg]  \big[ e_{\alpha \beta } \otimes e_{\beta^{\prime}\alpha}  + e_{\beta^{\prime} \alpha^{\prime}}  \otimes e_{\alpha^{\prime} \beta}  \big]  \bigg]  \bigg] \\  + \underset{\alpha, \beta \neq 4,6}{\sum} a_{\alpha \beta} \big( u^{\prime} \big) e_{\alpha \beta} \otimes e_{\alpha^{\prime} \beta^{\prime}} + \frac{1}{2} \underset{\alpha \neq 4,6, \beta = 4,6}{\sum} \big[ b^+_a \big( u^{\prime} \big) \big[ e_{\alpha \beta} \otimes e_{\alpha^{\prime} \beta^{\prime}} \\ + e_{\beta^{\prime} \alpha^{\prime}} \otimes e_{\beta \alpha}  \big]  + b^-_{\alpha} \big( u^{\prime} \big) \big[       e_{\alpha \beta} \otimes e_{\alpha^{\prime} \beta} + e_{\beta \alpha^{\prime}} \otimes e_{\beta \alpha}        \big]     \big]   + \underset{\alpha = 4,6}{\sum} \big[ c^+\big( u^{\prime} \big) e_{\alpha \alpha} \\  + e_{\alpha^{\prime} \alpha^{\prime}} + c^- \big( u^{\prime} \big) e_{\alpha \alpha} \otimes e_{\alpha \alpha} + d^+ \big( u^{\prime} \big)    e_{\alpha \alpha } \otimes e_{\alpha^{\prime} \alpha} + d^- \big( u^{\prime} \big) e_{\alpha \alpha^{\prime}} \otimes e_{\alpha \alpha^{\prime}}     \big]     \bigg\}     \bigg]   \\ \\ + \mathrm{Tr}_0 \bigg[ \mathcal{K}_{+,0} \big( u \big)  \bigg\{ - \frac{1}{2}   \big( e^{4 \eta} -1 \big) \big( e^{2u} -  e^{12 \eta} \big) \bigg[ \big( e^u + 1 \big) \bigg[   \underset{\alpha < 4 , \beta = 4,6}{\sum} + e^u   \underset{\alpha > 5 , \beta = 4,6}{\sum}  \bigg]  \big[ e_{\alpha \beta} \otimes e_{\beta \alpha} \\ +  e_{\beta^{\prime}\alpha^{\prime}} + e_{\alpha^{\prime} \beta^{\prime}} \big] + \big( e^u - 1 \big) \bigg[     \underset{\alpha < 4 , \beta = 4,6}{\sum} + e^u   \underset{\alpha > 5 , \beta = 4,6}{\sum}    \bigg]  \big[ e_{\alpha \beta } \otimes e_{\beta^{\prime}\alpha}  + e_{\beta^{\prime} \alpha^{\prime}}  \otimes e_{\alpha^{\prime} \beta}  \big]   \bigg]   \bigg\} \mathcal{K}_{-,0} \big( u^{\prime} \big) \\ \times \bigg\{    \big( e^{2u^{\prime}} - e^{4 \eta} \big)  \big( e^{2u^{\prime}}   - e^{12 \eta} \big) \underset{\alpha \neq 4,6}{\sum} e_{\alpha \alpha} \otimes e_{\alpha \alpha}    + e^{2\eta} \big( e^{2u^{\prime}} - 1 \big) \big( e^{2u^{\prime}} - e^{12 \eta} \big) \underset{\alpha, \text{ } \textit{or} \text{ } \beta \neq 4,6}{\underset{\alpha \neq \beta , \beta^{\prime}}{\sum}}     e_{\alpha \alpha} \otimes e_{\beta \beta}   \\        - \big( e^{4\eta} - 1 \big) \big( e^{2u^{\prime}}  - e^{12 \eta} \big) \bigg[    \underset{\alpha, \beta \neq 4,6}{\underset{\alpha < \beta, \alpha \neq \beta^{\prime}}{\sum}}    + e^{2u^{\prime}}   \underset{\alpha, \beta \neq 4,6}{\underset{\alpha >  \beta, \alpha \neq \beta^{\prime}}{\sum}}   e_{\alpha \beta} \otimes e_{\beta \alpha}  - \frac{1}{2}   \big( e^{4 \eta} -1 \big) \big( e^{2u^{\prime}} -  e^{12 \eta} \big) \\  \times  \bigg[ \big( e^u + 1 \big) \bigg[   \underset{\alpha < 4 , \beta = 4,6}{\sum} + e^{u^{\prime}}   \underset{\alpha > 5 , \beta = 4,6}{\sum}  \bigg]  \big[ e_{\alpha \beta} \otimes e_{\beta \alpha} +  e_{\beta^{\prime}\alpha^{\prime}} + e_{\alpha^{\prime} \beta^{\prime}} \big] \\ + \big( e^{u^{\prime}} - 1 \big) \bigg[     \underset{\alpha < 4 , \beta = 4,6}{\sum} + e^{u^{\prime}}   \underset{\alpha > 5 , \beta = 4,6}{\sum}    \bigg]  \big[ e_{\alpha \beta } \otimes e_{\beta^{\prime}\alpha}  + e_{\beta^{\prime} \alpha^{\prime}}  \otimes e_{\alpha^{\prime} \beta}  \big]   \bigg] \\  + \underset{\alpha, \beta \neq 4,6}{\sum} a_{\alpha \beta} \big( u^{\prime} \big) e_{\alpha \beta} \otimes e_{\alpha^{\prime} \beta^{\prime}} + \frac{1}{2} \underset{\alpha \neq 4,6, \beta = 4,6}{\sum} \big[ b^+_a \big( u^{\prime} \big) \big[ e_{\alpha \beta} \otimes e_{\alpha^{\prime} \beta^{\prime}} \\ + e_{\beta^{\prime} \alpha^{\prime}} \otimes e_{\beta \alpha}  \big]  + b^-_{\alpha} \big( u^{\prime} \big) \big[       e_{\alpha \beta} \otimes e_{\alpha^{\prime} \beta} + e_{\beta \alpha^{\prime}} \otimes e_{\beta \alpha}        \big]     \big]   + \underset{\alpha = 4,6}{\sum} \big[ c^+\big( u^{\prime} \big) e_{\alpha \alpha} \end{align*}

  \begin{align*} + e_{\alpha^{\prime} \alpha^{\prime}} + c^- \big( u^{\prime} \big) e_{\alpha \alpha} \otimes e_{\alpha \alpha} + d^+ \big( u^{\prime} \big)    e_{\alpha \alpha } \otimes e_{\alpha^{\prime} \alpha} + d^- \big( u^{\prime} \big) e_{\alpha \alpha^{\prime}} \otimes e_{\alpha \alpha^{\prime}}     \big]     \bigg\}     \bigg]                
\\ \\ 
+  \mathrm{Tr}_0 \bigg[ \mathcal{K}_{+,0} \big( u \big)   \bigg\{ \underset{\alpha, \beta \neq 4,6}{\sum} a_{\alpha \beta} \big( u \big) e_{\alpha \beta} \otimes e_{\alpha^{\prime} \beta^{\prime}} + \frac{1}{2} \underset{\alpha \neq 4,6, \beta = 4,6}{\sum} \big[ b^+_a \big( u \big) \big[ e_{\alpha \beta} \otimes e_{\alpha^{\prime} \beta^{\prime}}  + e_{\beta^{\prime} \alpha^{\prime}} \otimes e_{\beta \alpha}  \big]  \\ + b^-_{\alpha} \big( u \big) \big[       e_{\alpha \beta} \otimes e_{\alpha^{\prime} \beta} + e_{\beta \alpha^{\prime}} \otimes e_{\beta \alpha}        \big]     \big]   \bigg\} \mathcal{K}_{-,0} \big( u^{\prime} \big) \bigg\{    \big( e^{2u^{\prime}} - e^{4 \eta} \big)  \big( e^{2u^{\prime}}   - e^{12 \eta} \big) \underset{\alpha \neq 4,6}{\sum} e_{\alpha \alpha} \\  \otimes e_{\alpha \alpha}    + e^{2\eta} \big( e^{2u^{\prime}} - 1 \big) \big( e^{2u^{\prime}} - e^{12 \eta} \big) \underset{\alpha, \text{ } \textit{or} \text{ } \beta \neq 4,6}{\underset{\alpha \neq \beta , \beta^{\prime}}{\sum}}     e_{\alpha \alpha} \otimes e_{\beta \beta}           - \big( e^{4\eta}  - 1 \big) \big( e^{2u^{\prime}}  - e^{12 \eta} \big) \bigg[    \underset{\alpha, \beta \neq 4,6}{\underset{\alpha < \beta, \alpha \neq \beta^{\prime}}{\sum}}  \\   + e^{2u^{\prime}}   \underset{\alpha, \beta \neq 4,6}{\underset{\alpha >  \beta, \alpha \neq \beta^{\prime}}{\sum}}   e_{\alpha \beta} \otimes e_{\beta \alpha}  - \frac{1}{2}   \big( e^{4 \eta} -1 \big) \big( e^{2u^{\prime}} -  e^{12 \eta} \big) \\ \times  \bigg[ \big( e^u + 1 \big) \bigg[   \underset{\alpha < 4 , \beta = 4,6}{\sum} + e^{u^{\prime}}   \underset{\alpha > 5 , \beta = 4,6}{\sum}  \bigg]  \big[ e_{\alpha \beta} \otimes e_{\beta \alpha} +  e_{\beta^{\prime}\alpha^{\prime}} + e_{\alpha^{\prime} \beta^{\prime}} \big] \\ + \big( e^{u^{\prime}} - 1 \big) \bigg[     \underset{\alpha < 4 , \beta = 4,6}{\sum} + e^{u^{\prime}}   \underset{\alpha > 5 , \beta = 4,6}{\sum}    \bigg]  \big[ e_{\alpha \beta } \otimes e_{\beta^{\prime}\alpha}  + e_{\beta^{\prime} \alpha^{\prime}}  \otimes e_{\alpha^{\prime} \beta}  \big]   \bigg] \\  + \underset{\alpha, \beta \neq 4,6}{\sum} a_{\alpha \beta} \big( u^{\prime} \big) e_{\alpha \beta} \otimes e_{\alpha^{\prime} \beta^{\prime}} + \frac{1}{2} \underset{\alpha \neq 4,6, \beta = 4,6}{\sum} \big[ b^+_a \big( u^{\prime} \big) \big[ e_{\alpha \beta} \otimes e_{\alpha^{\prime} \beta^{\prime}} \\ + e_{\beta^{\prime} \alpha^{\prime}} \otimes e_{\beta \alpha}  \big]  + b^-_{\alpha} \big( u^{\prime} \big) \big[       e_{\alpha \beta} \otimes e_{\alpha^{\prime} \beta} + e_{\beta \alpha^{\prime}} \otimes e_{\beta \alpha}        \big]     \big]   + \underset{\alpha = 4,6}{\sum} \big[ c^+\big( u^{\prime} \big) e_{\alpha \alpha} \\ + e_{\alpha^{\prime} \alpha^{\prime}} + c^- \big( u^{\prime} \big) e_{\alpha \alpha} \otimes e_{\alpha \alpha} + d^+ \big( u^{\prime} \big)    e_{\alpha \alpha } \otimes e_{\alpha^{\prime} \alpha} + d^- \big( u^{\prime} \big) e_{\alpha \alpha^{\prime}} \otimes e_{\alpha \alpha^{\prime}}     \big]     \bigg\}     \bigg]         
\\ \\ 
+  \mathrm{Tr}_0 \bigg[ \mathcal{K}_{+,0} \big( u \big)  \bigg\{  \underset{\alpha = 4,6}{\sum} \big[ c^+\big( u \big) e_{\alpha \alpha} \otimes  e_{\alpha^{\prime} \alpha^{\prime}} + c^- \big( u \big) e_{\alpha \alpha} \otimes e_{\alpha \alpha} + d^+ \big( u \big)    e_{\alpha \alpha } \otimes e_{\alpha^{\prime} \alpha}  + d^- \big( u \big) e_{\alpha \alpha^{\prime}} \\ \otimes e_{\alpha \alpha^{\prime}}     \big]   \bigg\} \mathcal{K}_{-,0} \big( u^{\prime} \big) \bigg\{    \big( e^{2u^{\prime}} - e^{4 \eta} \big)  \big( e^{2u^{\prime}}   - e^{12 \eta} \big) \underset{\alpha \neq 4,6}{\sum} e_{\alpha \alpha} \otimes e_{\alpha \alpha}    + e^{2\eta} \big( e^{2u^{\prime}} - 1 \big) \big( e^{2u^{\prime}} \\ - e^{12 \eta} \big) \underset{\alpha, \text{ } \textit{or} \text{ } \beta \neq 4,6}{\underset{\alpha \neq \beta , \beta^{\prime}}{\sum}}     e_{\alpha \alpha} \otimes e_{\beta \beta}           - \big( e^{4\eta} - 1 \big) \big( e^{2u^{\prime}}  - e^{12 \eta} \big) \bigg[    \underset{\alpha, \beta \neq 4,6}{\underset{\alpha < \beta, \alpha \neq \beta^{\prime}}{\sum}}    + e^{2u^{\prime}}   \underset{\alpha, \beta \neq 4,6}{\underset{\alpha >  \beta, \alpha \neq \beta^{\prime}}{\sum}}   e_{\alpha \beta} \otimes e_{\beta \alpha}   \end{align*}

\begin{align*}  - \frac{1}{2}   \big( e^{4 \eta} -1 \big) \big( e^{2u^{\prime}} -  e^{12 \eta} \big) \bigg[ \big( e^u + 1 \big) \bigg[   \underset{\alpha < 4 , \beta = 4,6}{\sum} + e^{u^{\prime}}   \underset{\alpha > 5 , \beta = 4,6}{\sum}  \bigg]  \big[ e_{\alpha \beta} \otimes e_{\beta \alpha} +  e_{\beta^{\prime}\alpha^{\prime}} \\ + e_{\alpha^{\prime} \beta^{\prime}} \big]  + \big( e^{u^{\prime}} - 1 \big) \bigg[     \underset{\alpha < 4 , \beta = 4,6}{\sum} + e^{u^{\prime}}   \underset{\alpha > 5 , \beta = 4,6}{\sum}    \bigg]  \big[ e_{\alpha \beta } \otimes e_{\beta^{\prime}\alpha} \\   + e_{\beta^{\prime} \alpha^{\prime}}  \otimes e_{\alpha^{\prime} \beta}  \big]   \bigg]  + \underset{\alpha, \beta \neq 4,6}{\sum} a_{\alpha \beta} \big( u^{\prime} \big) e_{\alpha \beta} \otimes e_{\alpha^{\prime} \beta^{\prime}} + \frac{1}{2} \underset{\alpha \neq 4,6, \beta = 4,6}{\sum} \big[ b^+_a \big( u^{\prime} \big)\\ \times  \big[ e_{\alpha \beta} \otimes e_{\alpha^{\prime} \beta^{\prime}}  + e_{\beta^{\prime} \alpha^{\prime}} \otimes e_{\beta \alpha}  \big]  + b^-_{\alpha} \big( u^{\prime} \big) \big[       e_{\alpha \beta} \otimes e_{\alpha^{\prime} \beta} + e_{\beta \alpha^{\prime}} \otimes e_{\beta \alpha}        \big]     \big]   + \underset{\alpha = 4,6}{\sum} \big[ c^+\big( u^{\prime} \big) e_{\alpha \alpha} \\ + e_{\alpha^{\prime} \alpha^{\prime}} + c^- \big( u^{\prime} \big) e_{\alpha \alpha} \otimes e_{\alpha \alpha} + d^+ \big( u^{\prime} \big)    e_{\alpha \alpha } \otimes e_{\alpha^{\prime} \alpha} + d^- \big( u^{\prime} \big) e_{\alpha \alpha^{\prime}} \otimes e_{\alpha \alpha^{\prime}}     \big]     \bigg\}     \bigg] .             
\end{align*} 

\noindent Applying the linearity property of the partial trace function to the above superposition implies:

  \begin{itemize}
  \item[$\bullet$] \textbf{First partial trace}
  \end{itemize}

  \begin{align*}
    \mathrm{Tr}_0 \bigg[ \mathcal{K}_{+,0} \big( u \big)    \bigg\{    \big( e^{2u} - e^{4 \eta} \big)  \big( e^{2u} - e^{12 \eta} \big) \underset{\alpha \neq 4,6}{\sum} e_{\alpha \alpha} \otimes e_{\alpha \alpha} \bigg\} \mathcal{K}_{-,0} \big( u^{\prime} \big) \bigg\{  \big( e^{2u^{\prime}} - e^{4 \eta} \big)  \big( e^{2u^{\prime}} \\  - e^{12 \eta} \big) \underset{\alpha \neq 4,6}{\sum} e_{\alpha \alpha} \otimes e_{\alpha \alpha}    \bigg\} \bigg]   \\ \\ \mathrm{Tr}_0 \bigg[ \mathcal{K}_{+,0} \big( u \big)    \bigg\{    \big( e^{2u} - e^{4 \eta} \big)  \big( e^{2u} - e^{12 \eta} \big) \underset{\alpha \neq 4,6}{\sum} e_{\alpha \alpha} \otimes e_{\alpha \alpha} \bigg\} \mathcal{K}_{-,0} \big( u^{\prime} \big) \bigg\{  e^{2\eta} \big( e^{2u^{\prime}} - 1 \big) \big( e^{2u^{\prime}} - e^{12 \eta} \big) \\ \times  \underset{\alpha, \text{ } \textit{or} \text{ } \beta \neq 4,6}{\underset{\alpha \neq \beta , \beta^{\prime}}{\sum}}     e_{\alpha \alpha} \otimes e_{\beta \beta} \bigg\} \bigg]  \\ \\ \mathrm{Tr}_0 \bigg[ \mathcal{K}_{+,0} \big( u \big)    \bigg\{    \big( e^{2u} - e^{4 \eta} \big)  \big( e^{2u} - e^{12 \eta} \big) \underset{\alpha \neq 4,6}{\sum} e_{\alpha \alpha} \otimes e_{\alpha \alpha} \bigg\} \mathcal{K}_{-,0} \big( u^{\prime} \big) \bigg\{   - \big( e^{4\eta} - 1 \big) \big( e^{2u^{\prime}} \\ - e^{12 \eta} \big) \bigg[    \underset{\alpha, \beta \neq 4,6}{\underset{\alpha < \beta, \alpha \neq \beta^{\prime}}{\sum}}      + e^{2u^{\prime}}   \underset{\alpha, \beta \neq 4,6}{\underset{\alpha >  \beta, \alpha \neq \beta^{\prime}}{\sum}}   e_{\alpha \beta} \otimes e_{\beta \alpha} - \frac{1}{2}   \big( e^{4 \eta} -1 \big) \big( e^{2u^{\prime}} -  e^{12 \eta} \big) \\  \times \bigg[ \big( e^u + 1 \big) \bigg[   \underset{\alpha < 4 , \beta = 4,6}{\sum} + e^{u^{\prime}}   \underset{\alpha > 5 , \beta = 4,6}{\sum}  \bigg]  \big[ e_{\alpha \beta} \otimes e_{\beta \alpha} +  e_{\beta^{\prime}\alpha^{\prime}} + e_{\alpha^{\prime} \beta^{\prime}} \big]   \end{align*}

  \begin{align*}  + \big( e^{u^{\prime}} - 1 \big) \bigg[     \underset{\alpha < 4 , \beta = 4,6}{\sum} + e^{u^{\prime}}   \underset{\alpha > 5 , \beta = 4,6}{\sum}    \bigg]  \big[ e_{\alpha \beta } \otimes e_{\beta^{\prime}\alpha}  + e_{\beta^{\prime} \alpha^{\prime}}  \otimes e_{\alpha^{\prime} \beta}  \big]   \bigg]  \bigg\} \bigg]  
\\ \\ 
  \mathrm{Tr}_0 \bigg[ \mathcal{K}_{+,0} \big( u \big)    \bigg\{    \big( e^{2u} - e^{4 \eta} \big)  \big( e^{2u} - e^{12 \eta} \big) \underset{\alpha \neq 4,6}{\sum} e_{\alpha \alpha} \otimes e_{\alpha \alpha} \bigg\} \mathcal{K}_{-,0} \big( u^{\prime} \big) \bigg\{    \underset{\alpha, \beta \neq 4,6}{\sum} a_{\alpha \beta} \big( u^{\prime} \big) e_{\alpha \beta} \otimes e_{\alpha^{\prime} \beta^{\prime}} \\ + \frac{1}{2} \underset{\alpha \neq 4,6, \beta = 4,6}{\sum} \big[ b^+_a \big( u^{\prime} \big) \big[ e_{\alpha \beta} \otimes e_{\alpha^{\prime} \beta^{\prime}}   + e_{\beta^{\prime} \alpha^{\prime}} \otimes e_{\beta \alpha}  \big]  + b^-_{\alpha} \big( u^{\prime} \big) \big[       e_{\alpha \beta} \otimes e_{\alpha^{\prime} \beta} + e_{\beta \alpha^{\prime}} \otimes e_{\beta \alpha}        \big]     \big]  \bigg\} \bigg]    \\       \\   \mathrm{Tr}_0 \bigg[ \mathcal{K}_{+,0} \big( u \big)    \bigg\{    \big( e^{2u} - e^{4 \eta} \big)  \big( e^{2u} - e^{12 \eta} \big) \underset{\alpha \neq 4,6}{\sum} e_{\alpha \alpha} \otimes e_{\alpha \alpha} \bigg\} \mathcal{K}_{-,0} \big( u^{\prime} \big) \bigg\{    \underset{\alpha = 4,6}{\sum} \big[ c^+\big( u^{\prime} \big) e_{\alpha \alpha} \\ + e_{\alpha^{\prime} \alpha^{\prime}} + c^- \big( u^{\prime} \big) e_{\alpha \alpha} \otimes e_{\alpha \alpha} + d^+ \big( u^{\prime} \big)    e_{\alpha \alpha } \otimes e_{\alpha^{\prime} \alpha} + d^- \big( u^{\prime} \big) e_{\alpha \alpha^{\prime}} \otimes e_{\alpha \alpha^{\prime}}     \big]       \bigg\} \bigg] 
  \end{align*}

  \begin{itemize}
  \item[$\bullet$] \textbf{Second partial trace}
  \end{itemize}

 \begin{align*}
    \mathrm{Tr}_0 \bigg[ \mathcal{K}_{+,0} \big( u \big)    \bigg\{    e^{2\eta} \big( e^{2u} - 1 \big) \big( e^{2u} - e^{12 \eta} \big)  \underset{\alpha, \text{ } \textit{or} \text{ } \beta \neq 4,6}{\underset{\alpha \neq \beta , \beta^{\prime}}{\sum}}     e_{\alpha \alpha} \otimes e_{\beta \beta} - \big( e^{4\eta} - 1 \big)  \big( e^{2u} - e^{12 \eta} \big) \bigg[    \underset{\alpha, \beta \neq 4,6}{\underset{\alpha < \beta, \alpha \neq \beta^{\prime}}{\sum}} \\  + e^{2u}   \underset{\alpha, \beta \neq 4,6}{\underset{\alpha >  \beta, \alpha \neq \beta^{\prime}}{\sum}}  \bigg]   e_{\alpha \beta}  \otimes e_{\beta \alpha}       \bigg\} \mathcal{K}_{-,0} \big( u^{\prime} \big) \bigg\{    \big( e^{2u^{\prime}} - e^{4 \eta} \big)  \big( e^{2u^{\prime}}   - e^{12 \eta} \big) \underset{\alpha \neq 4,6}{\sum} e_{\alpha \alpha} \otimes e_{\alpha \alpha}   \bigg\} \bigg]    \\ \\ \mathrm{Tr}_0 \bigg[ \mathcal{K}_{+,0} \big( u \big)    \bigg\{    e^{2\eta} \big( e^{2u} - 1 \big) \big( e^{2u} - e^{12 \eta} \big)  \underset{\alpha, \text{ } \textit{or} \text{ } \beta \neq 4,6}{\underset{\alpha \neq \beta , \beta^{\prime}}{\sum}}     e_{\alpha \alpha} \otimes e_{\beta \beta} - \big( e^{4\eta} - 1 \big)  \big( e^{2u} - e^{12 \eta} \big) \bigg[    \underset{\alpha, \beta \neq 4,6}{\underset{\alpha < \beta, \alpha \neq \beta^{\prime}}{\sum}} \\  + e^{2u}   \underset{\alpha, \beta \neq 4,6}{\underset{\alpha >  \beta, \alpha \neq \beta^{\prime}}{\sum}}  \bigg]   e_{\alpha \beta}  \otimes e_{\beta \alpha}       \bigg\} \mathcal{K}_{-,0} \big( u^{\prime} \big) \bigg\{  e^{2\eta} \big( e^{2u^{\prime}} - 1 \big) \big( e^{2u^{\prime}}  - e^{12 \eta} \big) \underset{\alpha, \text{ } \textit{or} \text{ } \beta \neq 4,6}{\underset{\alpha \neq \beta , \beta^{\prime}}{\sum}}     e_{\alpha \alpha} \otimes e_{\beta \beta}   \end{align*}

    \begin{align*} \mathrm{Tr}_0 \bigg[ \mathcal{K}_{+,0} \big( u \big)    \bigg\{    e^{2\eta} \big( e^{2u} - 1 \big) \big( e^{2u} - e^{12 \eta} \big)  \underset{\alpha, \text{ } \textit{or} \text{ } \beta \neq 4,6}{\underset{\alpha \neq \beta , \beta^{\prime}}{\sum}}     e_{\alpha \alpha} \otimes e_{\beta \beta} - \big( e^{4\eta} - 1 \big)  \big( e^{2u} - e^{12 \eta} \big) \bigg[    \underset{\alpha, \beta \neq 4,6}{\underset{\alpha < \beta, \alpha \neq \beta^{\prime}}{\sum}} \\  + e^{2u}   \underset{\alpha, \beta \neq 4,6}{\underset{\alpha >  \beta, \alpha \neq \beta^{\prime}}{\sum}}  \bigg]   e_{\alpha \beta}  \otimes e_{\beta \alpha}       \bigg\} \mathcal{K}_{-,0} \big( u^{\prime} \big) \bigg\{    - \big( e^{4\eta} - 1 \big) \big( e^{2u^{\prime}}  - e^{12 \eta} \big) \bigg[    \underset{\alpha, \beta \neq 4,6}{\underset{\alpha < \beta, \alpha \neq \beta^{\prime}}{\sum}}   \\  + e^{2u^{\prime}}   \underset{\alpha, \beta \neq 4,6}{\underset{\alpha >  \beta, \alpha \neq \beta^{\prime}}{\sum}}   e_{\alpha \beta} \otimes e_{\beta \alpha}  - \frac{1}{2}   \big( e^{4 \eta} -1 \big) \big( e^{2u^{\prime}} -  e^{12 \eta} \big) \\ \times  \bigg[ \big( e^u + 1 \big) \bigg[   \underset{\alpha < 4 , \beta = 4,6}{\sum} + e^{u^{\prime}}   \underset{\alpha > 5 , \beta = 4,6}{\sum}  \bigg]  \big[ e_{\alpha \beta} \otimes e_{\beta \alpha} +  e_{\beta^{\prime}\alpha^{\prime}} + e_{\alpha^{\prime} \beta^{\prime}} \big] \\ + \big( e^{u^{\prime}} - 1 \big) \bigg[     \underset{\alpha < 4 , \beta = 4,6}{\sum} + e^{u^{\prime}}   \underset{\alpha > 5 , \beta = 4,6}{\sum}    \bigg]  \big[ e_{\alpha \beta } \otimes e_{\beta^{\prime}\alpha}  + e_{\beta^{\prime} \alpha^{\prime}}  \otimes e_{\alpha^{\prime} \beta}  \big]  \bigg]  \bigg] \bigg\} \bigg]    \\ \\  \mathrm{Tr}_0 \bigg[ \mathcal{K}_{+,0} \big( u \big)    \bigg\{    e^{2\eta} \big( e^{2u} - 1 \big) \big( e^{2u} - e^{12 \eta} \big)  \underset{\alpha, \text{ } \textit{or} \text{ } \beta \neq 4,6}{\underset{\alpha \neq \beta , \beta^{\prime}}{\sum}}     e_{\alpha \alpha} \otimes e_{\beta \beta} - \big( e^{4\eta} - 1 \big)  \big( e^{2u} - e^{12 \eta} \big) \bigg[    \underset{\alpha, \beta \neq 4,6}{\underset{\alpha < \beta, \alpha \neq \beta^{\prime}}{\sum}} \\  + e^{2u}   \underset{\alpha, \beta \neq 4,6}{\underset{\alpha >  \beta, \alpha \neq \beta^{\prime}}{\sum}}  \bigg]   e_{\alpha \beta}  \otimes e_{\beta \alpha}       \bigg\} \mathcal{K}_{-,0} \big( u^{\prime} \big) \bigg\{   \underset{\alpha, \beta \neq 4,6}{\sum} a_{\alpha \beta} \big( u^{\prime} \big) e_{\alpha \beta} \otimes e_{\alpha^{\prime} \beta^{\prime}} + \frac{1}{2} \underset{\alpha \neq 4,6, \beta = 4,6}{\sum} \big[ b^+_a \big( u^{\prime} \big) \big[ e_{\alpha \beta} \otimes e_{\alpha^{\prime} \beta^{\prime}} \\ + e_{\beta^{\prime} \alpha^{\prime}} \otimes e_{\beta \alpha}  \big]  + b^-_{\alpha} \big( u^{\prime} \big) \big[       e_{\alpha \beta} \otimes e_{\alpha^{\prime} \beta} + e_{\beta \alpha^{\prime}} \otimes e_{\beta \alpha}        \big]     \big]   \bigg\} \bigg]   \\ \\ \mathrm{Tr}_0 \bigg[ \mathcal{K}_{+,0} \big( u \big)    \bigg\{    e^{2\eta} \big( e^{2u} - 1 \big) \big( e^{2u} - e^{12 \eta} \big)  \underset{\alpha, \text{ } \textit{or} \text{ } \beta \neq 4,6}{\underset{\alpha \neq \beta , \beta^{\prime}}{\sum}}     e_{\alpha \alpha} \otimes e_{\beta \beta} - \big( e^{4\eta} - 1 \big)  \big( e^{2u} - e^{12 \eta} \big) \\ \times \bigg[    \underset{\alpha, \beta \neq 4,6}{\underset{\alpha < \beta, \alpha \neq \beta^{\prime}}{\sum}}   + e^{2u}   \underset{\alpha, \beta \neq 4,6}{\underset{\alpha >  \beta, \alpha \neq \beta^{\prime}}{\sum}}  \bigg]   e_{\alpha \beta}  \otimes e_{\beta \alpha}       \bigg\} \mathcal{K}_{-,0} \big( u^{\prime} \big) \bigg\{   \underset{\alpha = 4,6}{\sum} \big[ c^+\big( u^{\prime} \big) e_{\alpha \alpha}  + e_{\alpha^{\prime} \alpha^{\prime}} \\ + c^- \big( u^{\prime} \big) e_{\alpha \alpha} \otimes e_{\alpha \alpha} + d^+ \big( u^{\prime} \big)    e_{\alpha \alpha } \otimes e_{\alpha^{\prime} \alpha} + d^- \big( u^{\prime} \big) e_{\alpha \alpha^{\prime}} \otimes e_{\alpha \alpha^{\prime}}     \big] \bigg\} \bigg]  
  \end{align*}

  \begin{itemize}
  \item[$\bullet$] \textbf{Third partial trace}
  \end{itemize}

\begin{align*}
        \mathrm{Tr}_0 \bigg[ \mathcal{K}_{+,0} \big( u \big)  \bigg\{ - \frac{1}{2}   \big( e^{4 \eta} -1 \big) \big( e^{2u} -  e^{12 \eta} \big) \bigg[ \big( e^u + 1 \big) \bigg[   \underset{\alpha < 4 , \beta = 4,6}{\sum} + e^u   \underset{\alpha > 5 , \beta = 4,6}{\sum}  \bigg]  \big[ e_{\alpha \beta} \otimes e_{\beta \alpha} \\ +  e_{\beta^{\prime}\alpha^{\prime}} + e_{\alpha^{\prime} \beta^{\prime}} \big] + \big( e^u - 1 \big) \bigg[     \underset{\alpha < 4 , \beta = 4,6}{\sum} + e^u   \underset{\alpha > 5 , \beta = 4,6}{\sum}    \bigg]  \big[ e_{\alpha \beta } \otimes e_{\beta^{\prime}\alpha}  + e_{\beta^{\prime} \alpha^{\prime}}  \otimes e_{\alpha^{\prime} \beta}  \big]   \bigg]   \bigg\} \mathcal{K}_{-,0} \big( u^{\prime} \big) \\ \times \bigg\{    \big( e^{2u^{\prime}} - e^{4 \eta} \big)  \big( e^{2u^{\prime}}   - e^{12 \eta} \big) \underset{\alpha \neq 4,6}{\sum} e_{\alpha \alpha} \otimes e_{\alpha \alpha}    \bigg\}     \bigg]   \\ \\ 
  +  \mathrm{Tr}_0 \bigg[ \mathcal{K}_{+,0} \big( u \big)  \bigg\{ - \frac{1}{2}   \big( e^{4 \eta} -1 \big) \big( e^{2u} -  e^{12 \eta} \big) \bigg[ \big( e^u + 1 \big) \bigg[   \underset{\alpha < 4 , \beta = 4,6}{\sum} + e^u   \underset{\alpha > 5 , \beta = 4,6}{\sum}  \bigg]  \big[ e_{\alpha \beta} \otimes e_{\beta \alpha} \\ +  e_{\beta^{\prime}\alpha^{\prime}} + e_{\alpha^{\prime} \beta^{\prime}} \big] + \big( e^u - 1 \big) \bigg[     \underset{\alpha < 4 , \beta = 4,6}{\sum} + e^u   \underset{\alpha > 5 , \beta = 4,6}{\sum}    \bigg]  \big[ e_{\alpha \beta } \otimes e_{\beta^{\prime}\alpha}  + e_{\beta^{\prime} \alpha^{\prime}}  \otimes e_{\alpha^{\prime} \beta}  \big]   \bigg]   \bigg\} \mathcal{K}_{-,0} \big( u^{\prime} \big)   \\ \times \bigg\{  e^{2\eta} \big( e^{2u^{\prime}} - 1 \big) \big( e^{2u^{\prime}} - e^{12 \eta} \big) \underset{\alpha, \text{ } \textit{or} \text{ } \beta \neq 4,6}{\underset{\alpha \neq \beta , \beta^{\prime}}{\sum}}     e_{\alpha \alpha} \otimes e_{\beta \beta} \bigg\} \bigg]   \\              \\   - \mathrm{Tr}_0 \bigg[ \mathcal{K}_{+,0} \big( u \big)  \bigg\{ - \frac{1}{2}   \big( e^{4 \eta} -1 \big) \big( e^{2u} -  e^{12 \eta} \big) \bigg[ \big( e^u + 1 \big) \bigg[   \underset{\alpha < 4 , \beta = 4,6}{\sum} + e^u   \underset{\alpha > 5 , \beta = 4,6}{\sum}  \bigg]  \big[ e_{\alpha \beta} \otimes e_{\beta \alpha}    \end{align*}

\begin{align*} +  e_{\beta^{\prime}\alpha^{\prime}} + e_{\alpha^{\prime} \beta^{\prime}} \big] + \big( e^u - 1 \big) \bigg[     \underset{\alpha < 4 , \beta = 4,6}{\sum} + e^u   \underset{\alpha > 5 , \beta = 4,6}{\sum}    \bigg]  \big[ e_{\alpha \beta } \otimes e_{\beta^{\prime}\alpha}  + e_{\beta^{\prime} \alpha^{\prime}}  \otimes e_{\alpha^{\prime} \beta}  \big]   \bigg]   \bigg\} \mathcal{K}_{-,0} \big( u^{\prime} \big) \\ \times \bigg\{  \big( e^{4\eta} - 1 \big) \big( e^{2u^{\prime}}  - e^{12 \eta} \big) \bigg[    \underset{\alpha, \beta \neq 4,6}{\underset{\alpha < \beta, \alpha \neq \beta^{\prime}}{\sum}}    + e^{2u^{\prime}}   \underset{\alpha, \beta \neq 4,6}{\underset{\alpha >  \beta, \alpha \neq \beta^{\prime}}{\sum}}   e_{\alpha \beta} \otimes e_{\beta \alpha}  - \frac{1}{2}   \big( e^{4 \eta} -1 \big) \big( e^{2u^{\prime}} -  e^{12 \eta} \big) \\ \times  \bigg[ \big( e^u + 1 \big) \bigg[   \underset{\alpha < 4 , \beta = 4,6}{\sum} + e^{u^{\prime}}   \underset{\alpha > 5 , \beta = 4,6}{\sum}  \bigg]  \big[ e_{\alpha \beta} \otimes e_{\beta \alpha} +  e_{\beta^{\prime}\alpha^{\prime}} + e_{\alpha^{\prime} \beta^{\prime}} \big] \\ + \big( e^{u^{\prime}} - 1 \big) \bigg[     \underset{\alpha < 4 , \beta = 4,6}{\sum} + e^{u^{\prime}}   \underset{\alpha > 5 , \beta = 4,6}{\sum}    \bigg]  \big[ e_{\alpha \beta } \otimes e_{\beta^{\prime}\alpha}  + e_{\beta^{\prime} \alpha^{\prime}}  \otimes e_{\alpha^{\prime} \beta}  \big]   \bigg] \bigg] \bigg\} \bigg] 
\\ \\  \mathrm{Tr}_0 \bigg[ \mathcal{K}_{+,0} \big( u \big)  \bigg\{ - \frac{1}{2}   \big( e^{4 \eta} -1 \big) \big( e^{2u} -  e^{12 \eta} \big) \bigg[ \big( e^u + 1 \big) \bigg[   \underset{\alpha < 4 , \beta = 4,6}{\sum} + e^u   \underset{\alpha > 5 , \beta = 4,6}{\sum}  \bigg]  \big[ e_{\alpha \beta} \otimes e_{\beta \alpha} \end{align*}

        \begin{align*} +  e_{\beta^{\prime}\alpha^{\prime}} + e_{\alpha^{\prime} \beta^{\prime}} \big] + \big( e^u - 1 \big) \bigg[     \underset{\alpha < 4 , \beta = 4,6}{\sum} + e^u   \underset{\alpha > 5 , \beta = 4,6}{\sum}    \bigg]  \big[ e_{\alpha \beta } \otimes e_{\beta^{\prime}\alpha}  + e_{\beta^{\prime} \alpha^{\prime}}  \otimes e_{\alpha^{\prime} \beta}  \big]   \bigg]   \bigg\} \mathcal{K}_{-,0} \big( u^{\prime} \big)  \\ \times \bigg\{    \underset{\alpha = 4,6}{\sum} \big[ c^+\big( u^{\prime} \big) e_{\alpha \alpha}  + e_{\alpha^{\prime} \alpha^{\prime}} + c^- \big( u^{\prime} \big) e_{\alpha \alpha} \otimes e_{\alpha \alpha} + d^+ \big( u^{\prime} \big)    e_{\alpha \alpha } \otimes e_{\alpha^{\prime} \alpha} \\ + d^- \big( u^{\prime} \big) e_{\alpha \alpha^{\prime}} \otimes e_{\alpha \alpha^{\prime}}     \big]    \bigg\} \bigg]       \\ \\    + \mathrm{Tr}_0 \bigg[ \mathcal{K}_{+,0} \big( u \big)  \bigg\{ - \frac{1}{2}   \big( e^{4 \eta} -1 \big) \big( e^{2u} -  e^{12 \eta} \big) \bigg[ \big( e^u + 1 \big) \bigg[   \underset{\alpha < 4 , \beta = 4,6}{\sum} + e^u   \underset{\alpha > 5 , \beta = 4,6}{\sum}  \bigg]  \big[ e_{\alpha \beta} \otimes e_{\beta \alpha} \\ +  e_{\beta^{\prime}\alpha^{\prime}} + e_{\alpha^{\prime} \beta^{\prime}} \big] + \big( e^u - 1 \big) \bigg[     \underset{\alpha < 4 , \beta = 4,6}{\sum} + e^u   \underset{\alpha > 5 , \beta = 4,6}{\sum}    \bigg]  \big[ e_{\alpha \beta } \otimes e_{\beta^{\prime}\alpha}  + e_{\beta^{\prime} \alpha^{\prime}}  \otimes e_{\alpha^{\prime} \beta}  \big]   \bigg]   \bigg\} \mathcal{K}_{-,0} \big( u^{\prime} \big)    \\ \times \bigg\{ \frac{1}{2} \underset{\alpha \neq 4,6, \beta = 4,6}{\sum} \big[ b^+_a \big( u^{\prime} \big) \big[ e_{\alpha \beta} \otimes e_{\alpha^{\prime} \beta^{\prime}}  + e_{\beta^{\prime} \alpha^{\prime}} \otimes e_{\beta \alpha}  \big]  + b^-_{\alpha} \big( u^{\prime} \big) \big[       e_{\alpha \beta}  \otimes e_{\alpha^{\prime} \beta} \\ + e_{\beta \alpha^{\prime}} \otimes e_{\beta \alpha}        \big]     \big] \bigg\} \bigg]      \\ \\     +      \mathrm{Tr}_0 \bigg[ \mathcal{K}_{+,0} \big( u \big)  \bigg\{ - \frac{1}{2}   \big( e^{4 \eta} -1 \big) \big( e^{2u} -  e^{12 \eta} \big) \bigg[ \big( e^u + 1 \big) \bigg[   \underset{\alpha < 4 , \beta = 4,6}{\sum} + e^u   \underset{\alpha > 5 , \beta = 4,6}{\sum}  \bigg]  \big[ e_{\alpha \beta} \otimes e_{\beta \alpha} \\ +  e_{\beta^{\prime}\alpha^{\prime}} + e_{\alpha^{\prime} \beta^{\prime}} \big] + \big( e^u - 1 \big) \bigg[     \underset{\alpha < 4 , \beta = 4,6}{\sum} + e^u   \underset{\alpha > 5 , \beta = 4,6}{\sum}    \bigg]  \big[ e_{\alpha \beta } \otimes e_{\beta^{\prime}\alpha}  + e_{\beta^{\prime} \alpha^{\prime}}  \otimes e_{\alpha^{\prime} \beta}  \big]   \bigg]   \bigg\} \mathcal{K}_{-,0} \big( u^{\prime} \big)  \\ \times \bigg\{ \underset{\alpha, \beta \neq 4,6}{\sum} a_{\alpha \beta} \big( u^{\prime} \big) e_{\alpha \beta} \otimes e_{\alpha^{\prime} \beta^{\prime}}  \bigg\} \bigg]   
\end{align*}

  \begin{itemize}
  \item[$\bullet$] \textbf{Fourth partial trace}
  \end{itemize}

\begin{align*}
  \mathrm{Tr}_0 \bigg[ \mathcal{K}_{+,0} \big( u \big)   \bigg\{ \underset{\alpha, \beta \neq 4,6}{\sum} a_{\alpha \beta} \big( u \big) e_{\alpha \beta} \otimes e_{\alpha^{\prime} \beta^{\prime}} + \frac{1}{2} \underset{\alpha \neq 4,6, \beta = 4,6}{\sum} \big[ b^+_a \big( u \big) \big[ e_{\alpha \beta} \otimes e_{\alpha^{\prime} \beta^{\prime}}  + e_{\beta^{\prime} \alpha^{\prime}} \otimes e_{\beta \alpha}  \big]  \\ + b^-_{\alpha} \big( u \big) \big[       e_{\alpha \beta} \otimes e_{\alpha^{\prime} \beta} + e_{\beta \alpha^{\prime}} \otimes e_{\beta \alpha}        \big]     \big]   \bigg\} \mathcal{K}_{-,0} \big( u^{\prime} \big) \bigg\{    \big( e^{2u^{\prime}} - e^{4 \eta} \big)  \big( e^{2u^{\prime}}   - e^{12 \eta} \big) \underset{\alpha \neq 4,6}{\sum} e_{\alpha \alpha} \\ \otimes e_{\alpha \alpha}        \bigg\}     \bigg]      ,  \\ \\  +  \mathrm{Tr}_0  \bigg[ \mathcal{K}_{+,0} \big( u \big)   \bigg\{ \underset{\alpha, \beta \neq 4,6}{\sum} a_{\alpha \beta} \big( u \big) e_{\alpha \beta} \otimes e_{\alpha^{\prime} \beta^{\prime}} + \frac{1}{2} \underset{\alpha \neq 4,6, \beta = 4,6}{\sum} \big[ b^+_a \big( u \big) \big[ e_{\alpha \beta} \otimes e_{\alpha^{\prime} \beta^{\prime}}  + e_{\beta^{\prime} \alpha^{\prime}} \otimes e_{\beta \alpha}  \big]  \end{align*}

        \begin{align*} + b^-_{\alpha} \big( u \big) \big[       e_{\alpha \beta} \otimes e_{\alpha^{\prime} \beta} + e_{\beta \alpha^{\prime}} \otimes e_{\beta \alpha}        \big]     \big]   \bigg\} \mathcal{K}_{-,0} \big( u^{\prime} \big) \bigg\{     e^{2\eta} \big( e^{2u^{\prime}} - 1 \big) \big( e^{2u^{\prime}} - e^{12 \eta} \big) \underset{\alpha, \text{ } \textit{or} \text{ } \beta \neq 4,6}{\underset{\alpha \neq \beta , \beta^{\prime}}{\sum}}     e_{\alpha \alpha} \otimes e_{\beta \beta}     \bigg\}     \bigg]       ,   \\ \\  - \mathrm{Tr}_0 \bigg[ \mathcal{K}_{+,0} \big( u \big)   \bigg\{ \underset{\alpha, \beta \neq 4,6}{\sum} a_{\alpha \beta} \big( u \big) e_{\alpha \beta} \otimes e_{\alpha^{\prime} \beta^{\prime}} + \frac{1}{2} \underset{\alpha \neq 4,6, \beta = 4,6}{\sum} \big[ b^+_a \big( u \big) \big[ e_{\alpha \beta} \otimes e_{\alpha^{\prime} \beta^{\prime}}  + e_{\beta^{\prime} \alpha^{\prime}} \otimes e_{\beta \alpha}  \big]  \\ + b^-_{\alpha} \big( u \big) \big[       e_{\alpha \beta} \otimes e_{\alpha^{\prime} \beta} + e_{\beta \alpha^{\prime}} \otimes e_{\beta \alpha}        \big]     \big]   \bigg\} \mathcal{K}_{-,0} \big( u^{\prime} \big) \bigg\{  \big( e^{4\eta}  - 1 \big) \big( e^{2u^{\prime}}  - e^{12 \eta} \big) \bigg[    \underset{\alpha, \beta \neq 4,6}{\underset{\alpha < \beta, \alpha \neq \beta^{\prime}}{\sum}} \\  + e^{2u^{\prime}}   \underset{\alpha, \beta \neq 4,6}{\underset{\alpha >  \beta, \alpha \neq \beta^{\prime}}{\sum}}   e_{\alpha \beta} \otimes e_{\beta \alpha}   - \frac{1}{2}   \big( e^{4 \eta} -1 \big) \big( e^{2u^{\prime}} -  e^{12 \eta} \big)  \bigg[ \big( e^u + 1 \big) \bigg[   \underset{\alpha < 4 , \beta = 4,6}{\sum} + e^{u^{\prime}}   \underset{\alpha > 5 , \beta = 4,6}{\sum}  \bigg]  \big[ e_{\alpha \beta} \otimes e_{\beta \alpha} \\  +  e_{\beta^{\prime}\alpha^{\prime}}  + e_{\alpha^{\prime} \beta^{\prime}} \big]   + \big( e^{u^{\prime}} - 1 \big) \bigg[     \underset{\alpha < 4 , \beta = 4,6}{\sum} + e^{u^{\prime}}   \underset{\alpha > 5 , \beta = 4,6}{\sum}    \bigg]  \big[ e_{\alpha \beta } \otimes e_{\beta^{\prime}\alpha} \\  + e_{\beta^{\prime} \alpha^{\prime}}  \otimes e_{\alpha^{\prime} \beta}  \big] \bigg]   \bigg] \bigg\} \bigg]   ,  \\ \\  +  \mathrm{Tr}_0   \bigg[ \mathcal{K}_{+,0} \big( u \big)   \bigg\{ \underset{\alpha, \beta \neq 4,6}{\sum} a_{\alpha \beta} \big( u \big) e_{\alpha \beta} \otimes e_{\alpha^{\prime} \beta^{\prime}} + \frac{1}{2} \underset{\alpha \neq 4,6, \beta = 4,6}{\sum} \big[ b^+_a \big( u \big) \big[ e_{\alpha \beta} \otimes e_{\alpha^{\prime} \beta^{\prime}}  + e_{\beta^{\prime} \alpha^{\prime}} \otimes e_{\beta \alpha}  \big] \\   + b^-_{\alpha} \big( u \big) \big[       e_{\alpha \beta} \otimes e_{\alpha^{\prime} \beta} + e_{\beta \alpha^{\prime}} \otimes e_{\beta \alpha}        \big]     \big]   \bigg\} \mathcal{K}_{-,0} \big( u^{\prime} \big)       \bigg\{  \underset{\alpha, \beta \neq 4,6}{\sum} a_{\alpha \beta} \big( u^{\prime} \big) e_{\alpha \beta} \otimes e_{\alpha^{\prime} \beta^{\prime}}  \bigg\} \bigg]    ,   \\ \\ + \mathrm{Tr}_0 \bigg[ \mathcal{K}_{+,0} \big( u \big)   \bigg\{ \underset{\alpha, \beta \neq 4,6}{\sum} a_{\alpha \beta} \big( u \big) e_{\alpha \beta} \otimes e_{\alpha^{\prime} \beta^{\prime}} + \frac{1}{2} \underset{\alpha \neq 4,6, \beta = 4,6}{\sum} \big[ b^+_a \big( u \big) \big[ e_{\alpha \beta} \otimes e_{\alpha^{\prime} \beta^{\prime}}  + e_{\beta^{\prime} \alpha^{\prime}} \otimes e_{\beta \alpha}  \big]  \\ + b^-_{\alpha} \big( u \big) \big[       e_{\alpha \beta} \otimes e_{\alpha^{\prime} \beta} + e_{\beta \alpha^{\prime}} \otimes e_{\beta \alpha}        \big]     \big]   \bigg\} \mathcal{K}_{-,0} \big( u^{\prime} \big)  \bigg\{  \frac{1}{2} \underset{\alpha \neq 4,6, \beta = 4,6}{\sum} \big[ b^+_a \big( u^{\prime} \big) \big[ e_{\alpha \beta} \otimes e_{\alpha^{\prime} \beta^{\prime}}  + e_{\beta^{\prime} \alpha^{\prime}} \otimes e_{\beta \alpha}  \big] \\  + b^-_{\alpha} \big( u^{\prime} \big) \big[       e_{\alpha \beta} \otimes e_{\alpha^{\prime} \beta}  + e_{\beta \alpha^{\prime}} \otimes e_{\beta \alpha}        \big]     \big]  \bigg\} \bigg] \\ \\  + \mathrm{Tr}_0 \bigg[ \mathcal{K}_{+,0} \big( u \big)   \bigg\{ \underset{\alpha, \beta \neq 4,6}{\sum} a_{\alpha \beta} \big( u \big) e_{\alpha \beta} \otimes e_{\alpha^{\prime} \beta^{\prime}} + \frac{1}{2} \underset{\alpha \neq 4,6, \beta = 4,6}{\sum} \big[ b^+_a \big( u \big) \big[ e_{\alpha \beta} \otimes e_{\alpha^{\prime} \beta^{\prime}}  + e_{\beta^{\prime} \alpha^{\prime}} \otimes e_{\beta \alpha}  \big]   \end{align*}

  \begin{align*} + b^-_{\alpha} \big( u \big) \big[       e_{\alpha \beta} \otimes e_{\alpha^{\prime} \beta} + e_{\beta \alpha^{\prime}} \otimes e_{\beta \alpha}        \big]     \big]   \bigg\} \mathcal{K}_{-,0} \big( u^{\prime} \big)  \bigg\{   \underset{\alpha = 4,6}{\sum} \big[ c^+\big( u^{\prime} \big) e_{\alpha \alpha}  + e_{\alpha^{\prime} \alpha^{\prime}} + c^- \big( u^{\prime} \big) e_{\alpha \alpha} \\ \otimes e_{\alpha \alpha} + d^+ \big( u^{\prime} \big)    e_{\alpha \alpha } \otimes e_{\alpha^{\prime} \alpha} + d^- \big( u^{\prime} \big) e_{\alpha \alpha^{\prime}} \otimes e_{\alpha \alpha^{\prime}}     \big] \bigg\} \bigg] \end{align*}

  \begin{itemize}
  \item[$\bullet$] \textbf{Fifth partial trace}
  \end{itemize}
  
  \begin{align*} \mathrm{Tr}_0 \bigg[ \mathcal{K}_{+,0} \big( u \big)  \bigg\{  \underset{\alpha = 4,6}{\sum} \big[ c^+\big( u \big) e_{\alpha \alpha} \otimes  e_{\alpha^{\prime} \alpha^{\prime}} + c^- \big( u \big) e_{\alpha \alpha} \otimes e_{\alpha \alpha} + d^+ \big( u \big)    e_{\alpha \alpha } \otimes e_{\alpha^{\prime} \alpha}  + d^- \big( u \big) e_{\alpha \alpha^{\prime}} \\ \otimes e_{\alpha \alpha^{\prime}}     \big]   \bigg\} \mathcal{K}_{-,0} \big( u^{\prime} \big) \bigg\{    \big( e^{2u^{\prime}} - e^{4 \eta} \big)  \big( e^{2u^{\prime}}   - e^{12 \eta} \big) \underset{\alpha \neq 4,6}{\sum} e_{\alpha \alpha} \otimes e_{\alpha \alpha}     \bigg\}     \bigg] \\ \\ + \mathrm{Tr}_0 \bigg[ \mathcal{K}_{+,0} \big( u \big)  \bigg\{  \underset{\alpha = 4,6}{\sum} \big[ c^+\big( u \big) e_{\alpha \alpha} \otimes  e_{\alpha^{\prime} \alpha^{\prime}} + c^- \big( u \big) e_{\alpha \alpha} \otimes e_{\alpha \alpha} + d^+ \big( u \big)    e_{\alpha \alpha } \otimes e_{\alpha^{\prime} \alpha}  + d^- \big( u \big) e_{\alpha \alpha^{\prime}} \\ \otimes e_{\alpha \alpha^{\prime}}     \big]   \bigg\} \mathcal{K}_{-,0} \big( u^{\prime} \big) \bigg\{   e^{2\eta} \big( e^{2u^{\prime}} - 1 \big) \big( e^{2u^{\prime}}  - e^{12 \eta} \big) \underset{\alpha, \text{ } \textit{or} \text{ } \beta \neq 4,6}{\underset{\alpha \neq \beta , \beta^{\prime}}{\sum}}     e_{\alpha \alpha} \otimes e_{\beta \beta}            \bigg\}     \bigg] \\ 
+ \mathrm{Tr}_0 \bigg[ \mathcal{K}_{+,0} \big( u \big)  \bigg\{  \underset{\alpha = 4,6}{\sum} \big[ c^+\big( u \big) e_{\alpha \alpha} \otimes  e_{\alpha^{\prime} \alpha^{\prime}} + c^- \big( u \big) e_{\alpha \alpha} \otimes e_{\alpha \alpha} + d^+ \big( u \big)    e_{\alpha \alpha } \otimes e_{\alpha^{\prime} \alpha}  + d^- \big( u \big) e_{\alpha \alpha^{\prime}} \\ \otimes e_{\alpha \alpha^{\prime}}     \big]   \bigg\} \mathcal{K}_{-,0} \big( u^{\prime} \big) \bigg\{     - \big( e^{4\eta} - 1 \big) \big( e^{2u^{\prime}}  - e^{12 \eta} \big) \bigg[    \underset{\alpha, \beta \neq 4,6}{\underset{\alpha < \beta, \alpha \neq \beta^{\prime}}{\sum}}    + e^{2u^{\prime}}   \underset{\alpha, \beta \neq 4,6}{\underset{\alpha >  \beta, \alpha \neq \beta^{\prime}}{\sum}}   e_{\alpha \beta} \otimes e_{\beta \alpha}  \\ - \frac{1}{2}   \big( e^{4 \eta} -1 \big) \big( e^{2u^{\prime}} -  e^{12 \eta} \big) \bigg[ \big( e^u + 1 \big) \bigg[   \underset{\alpha < 4 , \beta = 4,6}{\sum} + e^{u^{\prime}}   \underset{\alpha > 5 , \beta = 4,6}{\sum}  \bigg]  \big[ e_{\alpha \beta} \otimes e_{\beta \alpha} +  e_{\beta^{\prime}\alpha^{\prime}} \\ + e_{\alpha^{\prime} \beta^{\prime}} \big]   \bigg\}     \bigg]  \\ \\  + \mathrm{Tr}_0 \bigg[ \mathcal{K}_{+,0} \big( u \big)  \bigg\{  \underset{\alpha = 4,6}{\sum} \big[ c^+\big( u \big) e_{\alpha \alpha} \otimes  e_{\alpha^{\prime} \alpha^{\prime}} + c^- \big( u \big) e_{\alpha \alpha} \otimes e_{\alpha \alpha} + d^+ \big( u \big)    e_{\alpha \alpha } \otimes e_{\alpha^{\prime} \alpha}  + d^- \big( u \big) e_{\alpha \alpha^{\prime}} \\ \otimes e_{\alpha \alpha^{\prime}}     \big]   \bigg\} \mathcal{K}_{-,0} \big( u^{\prime} \big) \bigg\{  \big( e^{u^{\prime}} - 1 \big) \bigg[     \underset{\alpha < 4 , \beta = 4,6}{\sum} + e^{u^{\prime}}   \underset{\alpha > 5 , \beta = 4,6}{\sum}    \bigg]  \big[ e_{\alpha \beta } \otimes e_{\beta^{\prime}\alpha}    + e_{\beta^{\prime} \alpha^{\prime}}  \otimes e_{\alpha^{\prime} \beta}  \big]  \bigg\}  \bigg]   \\ \\ +  \mathrm{Tr}_0 \bigg[ \mathcal{K}_{+,0} \big( u \big)  \bigg\{  \underset{\alpha = 4,6}{\sum} \big[ c^+\big( u \big) e_{\alpha \alpha} \otimes  e_{\alpha^{\prime} \alpha^{\prime}} + c^- \big( u \big) e_{\alpha \alpha} \otimes e_{\alpha \alpha} + d^+ \big( u \big)    e_{\alpha \alpha } \otimes e_{\alpha^{\prime} \alpha}  + d^- \big( u \big) e_{\alpha \alpha^{\prime}} \end{align*}

\begin{align*}  \otimes e_{\alpha \alpha^{\prime}}     \big]   \bigg\} \mathcal{K}_{-,0} \big( u^{\prime} \big) \bigg\{   \underset{\alpha, \beta \neq 4,6}{\sum} a_{\alpha \beta} \big( u^{\prime} \big) e_{\alpha \beta} \otimes e_{\alpha^{\prime} \beta^{\prime}}   + \frac{1}{2} \underset{\alpha \neq 4,6, \beta = 4,6}{\sum} \big[ b^+_a \big( u^{\prime} \big)   \big[ e_{\alpha \beta} \otimes e_{\alpha^{\prime} \beta^{\prime}} \\  + e_{\beta^{\prime} \alpha^{\prime}} \otimes e_{\beta \alpha}  \big]  + b^-_{\alpha} \big( u^{\prime} \big) \big[       e_{\alpha \beta} \otimes e_{\alpha^{\prime} \beta} + e_{\beta \alpha^{\prime}} \otimes e_{\beta \alpha}        \big]     \big] \\ \\   + \underset{\alpha = 4,6}{\sum} \big[ c^+\big( u^{\prime} \big) e_{\alpha \alpha}  + e_{\alpha^{\prime} \alpha^{\prime}} + c^- \big( u^{\prime} \big) e_{\alpha \alpha} \otimes e_{\alpha \alpha} \\ + d^+ \big( u^{\prime} \big)    e_{\alpha \alpha } \otimes e_{\alpha^{\prime} \alpha} + d^- \big( u^{\prime} \big) e_{\alpha \alpha^{\prime}} \otimes e_{\alpha \alpha^{\prime}}     \big] \bigg\} \bigg]  
\end{align*}

\noindent respectively for each term, upon making use of the previously obtained representations for the Jimbo R-matrix in \textbf{Claims 1}-\textbf{5} above, with $n \equiv 3$. Straightforwardly, from the approximations provided for the partial trace of each $\mathscr{R}_i$ which can be read off from the Jimbo R-matrix, we obtain the desired approximation for the higher-rank transfer matrix under open boundary conditions, upon reading the corresponding entries from each $\mathscr{R}\mathscr{T}$, from which we conclude the argument. \boxed{}

\subsubsection{Computation of the open-boundary condition spectrum, $\mathcal{S}^{\mathrm{Open}}$}

\noindent From the transfer matrix in lower rank, it remains of interest to determine, as was the case for the lower rank spin-chain, for the higher rank spin-chain, counterparts to the collection of functions,

\begin{align*}
 A^{\mathrm{Quasi-periodic}} \big[ u , u^{[1]}_{j} , \gamma \big]   \equiv   A^{\mathrm{Quasi-periodic}} \big( u \big)    \text{, }  \\ B^{\mathrm{Quasi-periodic}}_1 \big[ u , u^{[1]}_j , \gamma \big]  \equiv B^{\mathrm{Quasi-periodic}}_1 \big( u \big) \text{, }   \\ B^{\mathrm{Quasi-periodic}}_2 \big[ u , u^{[2]}_j , \gamma \big] \equiv B^{\mathrm{Quasi-periodic}}_2 \big( u \big)  \text{, }   \\
 B^{\mathrm{Quasi-periodic}}_3 \bigg[ B_2 \big( u \big) , u , u^{[2]}_j , \gamma  \bigg]\equiv B^{\mathrm{Quasi-periodic}}_3 \big( u \big)  \text{, }   \\ B^{\mathrm{Quasi-periodic}}_4 \big[ u , u^{[1]}_j , u^{[2]}_j , \gamma \big]  \equiv  B^{\mathrm{Quasi-periodic}}_4 \big( u  \big) \text{, }   \\  C^{\mathrm{Quasi-periodic}}  \bigg[A \big( u \big) , u , u^{[1]}_j , \gamma \bigg]\equiv   C^{\mathrm{Quasi-periodic}}\big( u \big) \text{, }  
\end{align*}

\noindent from eigenvalues of the transfer matrix. That is, to compute,

\begin{align*}
    \mathrm{det} \big[ \textbf{T}^{\mathrm{Open}} - \lambda \textbf{I} \big] = 0,  \end{align*}

\noindent it remains of interest to approximate eigenvalues of $\mathcal{S}_{\mathrm{Open}}$, through computations of the form,

\begin{align*}
  \mathrm{det} \big[ \textbf{T}^{\mathrm{Open}} - \lambda \textbf{I} \big] \equiv        \mathrm{det} \bigg\{  \bigg[ \underset{j,j^{\prime} \longrightarrow + \infty}{\mathrm{lim}}  \mathrm{Tr}_0 \bigg[ \mathcal{K}_{+,0} \big( u \big)  \prod_{1 \leq j \leq L} R_{0j} \big( u \big)  \times   \mathcal{K}_{-,0} \big( u \big)   \prod_{1 \leq j^{\prime} \leq L} R_{j^{\prime}0} \big( u \big)  \bigg] \bigg]  - \lambda \textbf{I}  \bigg\}  \\  =    \underset{j,j^{\prime} \longrightarrow + \infty}{\mathrm{lim}}  \mathrm{det} \bigg\{  \bigg[  \mathrm{Tr}_0 \bigg[ \mathcal{K}_{+,0} \big( u \big)  \prod_{1 \leq j \leq L} R_{0j} \big( u \big)  \times   \mathcal{K}_{-,0} \big( u \big)   \prod_{1 \leq j^{\prime} \leq L} R_{j^{\prime}0} \big( u \big)  \bigg] \bigg]  - \lambda \textbf{I}  \bigg\} .  \end{align*}

\subsubsection{Local Hamiltonian encoding}

\noindent Recall the expression,

\begin{align*}
   \mathcal{H} \big( k , \kappa , \textbf{K}_{-} , \textbf{K}_{+} \big)   \equiv \mathcal{H} \sim    \underset{1 \leq k \leq N-1}{\sum} h_{k,k+1} + \frac{1}{2\kappa} \bigg[\textbf{K}^{-}_1  \big( 0 \big) \bigg]^{\prime}   +   \frac{1}{\mathrm{tr} \big( \textbf{K}_{+} \big( 0 \big) \big)  }  \mathrm{tr}_0  \big( \textbf{K}_{0,+} \big( 0 \big) \big)  h_{N0} \text{, }  
\end{align*}

\noindent corresponding to the local Hamiltonian encoding. Straightforwardly, the local encoding for the higher rank spin-chain with open boundary conditions takes the form,

\begin{align*}
 \mathcal{H}^{\mathrm{Open}} \equiv \mathcal{H}^{\mathrm{Open}}_{D^{(2)}_3} \big( k , \kappa , \textbf{K}^{\mathrm{Open}}_{-} , \textbf{K}^{\mathrm{Open}}_{+} \big)   \equiv \mathcal{H} \sim    \underset{1 \leq k \leq N-1}{\sum} h_{k,k+1} + \frac{1}{2\kappa} \bigg[\textbf{K}^{\mathrm{Open}}_{-,1}  \big( 0 \big) \bigg]^{\prime}  \\  +   \frac{1}{\mathrm{tr} \big( \textbf{K}^{\mathrm{Open}}_{+} \big( 0 \big) \big)  }  \mathrm{tr}_0 \big( \textbf{K}^{\mathrm{Open}}_{0,+} \big( 0 \big) \big) h_{N0} \text{, }  
\end{align*}

\noindent where,

\begin{align*} \mathcal{H}^{\mathrm{Open}}_{D^{(2)}_3} \big( k , \kappa , \textbf{K}^{\mathrm{Open}}_{-} , \textbf{K}^{\mathrm{Open}}_{+} \big)  \equiv  \mathcal{H}^{\mathrm{Open}}_{D^{(2)}_3} \big( k , \kappa , \mathcal{K}^{\pm}_{\mathrm{Open},36} \big( u \big) \big) \propto  \mathcal{H}^{\mathrm{Open}}_{D^{(2)}_3} \big( k , \kappa , \mathcal{K}^{\pm}_{\mathrm{Open},36} \big( 0  \big) \big)  , 
\end{align*}

\noindent One can compute the local Hamiltonian encoding under open boundary conditions from the observation that it suffices to compute the derivative,

\begin{align*} \bigg[\textbf{K}^{\mathrm{Open}}_{-,1}  \big( 0 \big) \bigg]^{\prime} \equiv  \bigg[ \mathcal{K}^{-}_{\mathrm{Open}, 36} \big( 0 \big) \bigg]^{\prime}   ,
\end{align*}

\noindent in addition to,

\begin{align*}
   \mathrm{tr} \big( \textbf{K}^{\mathrm{Open}}_{+} \big( 0 \big) \big) \equiv  \mathrm{tr} \big( \mathcal{K}^{-}_{\mathrm{Open}, 36} \big( 0 \big)   \big)   , \\ \\   \mathrm{tr}_0 \big( \textbf{K}^{\mathrm{Open}}_{0,+} \big( 0 \big) \big) \equiv  \mathrm{tr} \big(  \mathcal{K}^{-}_{\mathrm{Open}, 36} \big( 0 \big)  \big)  . 
\end{align*}

\noindent The two claims below provide expressions for the computation of the three quantities provided above, which can together be used to construct the desired local Hamiltonian encoding.

\begin{itemize}
    \item[$\bullet$] \textbf{Claim 7} (\textit{computation of the derivative of the open boundary, higher rank, K-matrix}). One has that,

\begin{align*} \bigg[\textbf{K}^{\mathrm{Open}}_{-,1}  \big( 0 \big) \bigg]^{\prime} \equiv  \bigg[ \mathcal{K}^{-}_{\mathrm{Open}, 36} \big( 0 \big) \bigg]^{\prime} \end{align*}

\begin{align*}
=  \begin{bmatrix}
 \mathrm{diag} \big(   \big( k^n_{0,-} \big( 0 \big) \big)^{\prime} , \overset{34}{\cdots} ,  \big(  k^n_{0,-} \big( 0 \big) \big)^{\prime} \big)    & & \\ &  \big( k^n_{1,-}  \big( 0 \big) \big)^{\prime}  &  \big( k^n_{2,-}  \big( 0 \big) \big)^{\prime} \\ & \big(  k^n_{3,-}  \big( 0 \big) \big)^{\prime} & \big( k^n_{4,-}  \big( 0 \big) \big)^{\prime} \\ & & & & \mathrm{diag} \big(  \big(  k^n_{5,-}  \big( 0 \big) \big)^{\prime} , \overset{34}{\cdots} ,  \big(  k^n_{5,-}  \big( 0 \big) \big)^{\prime} \big) 
    \end{bmatrix}    .
\end{align*}

    \noindent \textit{Proof of Claim 7}. By direct computation, observe,

    \begin{align*}
      \bigg[\textbf{K}^{\mathrm{Open}}_{-,1}  \big( 0 \big) \bigg]^{\prime}     \end{align*}

      \begin{align*} \equiv   \begin{bmatrix}
 \mathrm{diag} \big(   k^n_{0,-} \big( 0 \big) , \overset{34}{\cdots} ,   k^n_{0,-} \big( 0 \big) \big)    & & \\ & k^n_{1,-}  \big( 0 \big)  & k^n_{2,-}  \big( 0 \big) \\ &  k^n_{3,-}  \big( 0 \big) & k^n_{4,-}  \big( 0 \big) \\ & & & & \mathrm{diag} \big(   k^n_{5,-}  \big( 0 \big) , \overset{34}{\cdots} ,   k^n_{5,-}  \big( 0 \big)  \big) 
    \end{bmatrix}^{\prime} \\ =    \begin{bmatrix}
 \mathrm{diag} \big(   \big( k^n_{0,-} \big( 0 \big) \big)^{\prime} , \overset{34}{\cdots} ,  \big(  k^n_{0,-} \big( 0 \big) \big)^{\prime} \big)    & & \\ &  \big( k^n_{1,-}  \big( 0 \big) \big)^{\prime}  &  \big( k^n_{2,-}  \big( 0 \big) \big)^{\prime} \\ & \big(  k^n_{3,-}  \big( 0 \big) \big)^{\prime} & \big( k^n_{4,-}  \big( 0 \big) \big)^{\prime} \\ & & & & \mathrm{diag} \big(  \big(  k^n_{5,-}  \big( 0 \big) \big)^{\prime} , \overset{34}{\cdots} ,  \big(  k^n_{5,-}  \big( 0 \big) \big)^{\prime} \big) 
    \end{bmatrix}               ,
    \end{align*}

    \noindent from which we conclude the argument. \boxed{}

    \item[$\bullet$] \textbf{Claim 8} (\textit{computation of the trace of the open boundary, higher-rank K-matrix, in addition to the partial trace of the open boundary, higher rank, K-matrix}). One has that,

\begin{align*}
   \mathrm{tr} \big( \textbf{K}^{\mathrm{Open}}_{+} \big( 0 \big) \big) =  \underset{1 \leq i \leq 36}{\sum} i^{\prime} \big( 0 \big)  \equiv 1^{\prime} \big( 0 \big) + \cdots + 36^{\prime} \big( 0 \big) \\ \equiv        34 k^n_{0,+} \big( u \big)  +      k^n_{1,+} \big( u \big) + k^n_{4,+} \big( u \big)      , \\ \\   \mathrm{tr}_0 \big( \textbf{K}^{\mathrm{Open}}_{0,+} \big( 0 \big) \big)  = \mathrm{tr}_0 \big( \textit{composite systems of + K-matrices} \big) \\  \equiv  \underset{\textit{systems}}{\sum} \mathrm{tr}_0 \big( \textit{composite system of + K-matrices} \big) \end{align*}

   \begin{align*}
  = \underset{\textit{systems}}{\sum} \mathrm{tr} \bigg( \textbf{K}^{\mathrm{Open}}_{0,+} \big( u \big) \bigg|_{\textit{systems}}  \bigg) \bigg|_{u = 0 } 
\end{align*}

    \noindent \textit{Proof of Claim 8}. By direct computation, we first consider the partial trace of $+$ $K$-matrices under open boundary conditions. That is, for,

\begin{align*}
     \mathrm{tr}_0 \big( \textbf{K}^{\mathrm{Open}}_{0,+} \big( 0 \big) \big)  ,
\end{align*}

\noindent one has that,

\begin{align*}
 \mathrm{tr} \big( \textbf{K}^{\mathrm{Open}}_{+} \big( 0 \big) \big) =  \underset{1 \leq i \leq 36}{\sum} i^{\prime} \big( 0 \big)  \equiv 1^{\prime} \big( 0 \big) + \cdots + 36^{\prime} \big( 0 \big) 
\\ 
     \equiv 17 k^n_{0,+} \big( u \big)  +      k^n_{1,+} \big( u \big) + k^n_{4,+} \big( u \big)    + 17 k^n_{0,+} \big( u \big) \equiv 34 k^n_{0,+} \big( u \big)  +      k^n_{1,+} \big( u \big) + k^n_{4,+} \big( u \big)  ,  
\end{align*}

\noindent from which we conclude the argument for the first equality of the above claim. With regards to the second equality provided in the above claim,

\begin{align*}  \mathrm{tr}_0 \big( \textbf{K}^{\mathrm{Open}}_{0,+} \big( 0 \big) \big)  , 
\end{align*}

\noindent observe, straightforwardly,

\begin{align*}
 \mathrm{tr}_0 \big( \textbf{K}^{\mathrm{Open}}_{0,+} \big( 0 \big) \big) \equiv \underset{\textit{systems}}{\sum} \mathrm{tr} \bigg( \textbf{K}^{\mathrm{Open}}_{0,+} \big( u \big) \bigg|_{\textit{systems}}  \bigg) \bigg|_{u = 0 }  \propto   \mathrm{tr} \big( \textbf{K}^{\mathrm{Open}}_{+} \big( 0 \big) \big)  ,
\end{align*}

\noindent hence implying that,

\begin{align*}
 \mathrm{tr}_0 \big( \textbf{K}^{\mathrm{Open}}_{0,+} \big( 0 \big) \big) =  \underset{\textit{systems}}{\sum} \mathrm{tr} \bigg( \textbf{K}^{\mathrm{Open}}_{0,+} \big( u \big) \bigg|_{\textit{systems}} \bigg) \end{align*}

\noindent which, as $u \longrightarrow 0^{+}$, approximately equals,

 \begin{align*}
 \underset{\textit{systems}}{\sum} \mathrm{tr} \bigg( \textbf{K}^{\mathrm{Open}}_{0,+} \big( u \big) \bigg|_{\textit{systems}}  \bigg) \bigg|_{u = 0 }      ,
\end{align*}

\noindent from which we conclude the argument. \boxed{}
    
\end{itemize}

\noindent Altogether, the previously obtained estimates can be straightforwardly put together to approximate,

\begin{align*}
    \mathcal{H} \sim \mathcal{H}^{\mathrm{Open}}_{D^{(2)}_3} \big( k , \kappa , \mathcal{K}^{\pm}_{\mathrm{Open},36} \big( u \big) \big)  . 
\end{align*}

\subsubsection{Derivative of the higher-rank transfer matrix}

\noindent \textbf{Claim 9} (\textit{approximation of the derivative of the transfer matrix under open boundaries}). One has that,

\begin{align*}
  \big[ \textbf{T}^{\mathrm{Open}} \big( u \big)  \big]^{\prime}   \approx \bigg[ \bigg[ \big[ \mathscr{R}_1 \mathscr{T}_1 \big]^{\prime} +    \big[ \mathscr{R}_1 \mathscr{T}_2  \big]^{\prime} +  \big[ \mathscr{R}_1 \mathscr{T}_3  \big]^{\prime}  +   \big[ \mathscr{R}_1 \mathscr{T}_4 \big]^{\prime}  +  \big[\mathscr{R}_1 \mathscr{T}_5 \big]^{\prime} \bigg]      + \bigg[ \big[ \mathscr{R}_2 \mathscr{T}_1  \big]^{\prime} +  \big[   \mathscr{R}_2 \mathscr{T}_2  \big]^{\prime} + \big[  \mathscr{R}_2 \mathscr{T}_3  \big]^{\prime} \\  +   \big[ \mathscr{R}_2 \mathscr{T}_4   \big]^{\prime} +  \big[ \mathscr{R}_2 \mathscr{T}_5 \big]^{\prime} \bigg]     + \bigg[ \big[ \mathscr{R}_3 \mathscr{T}_1 \big]^{\prime}  +    \big[ \mathscr{R}_3 \mathscr{T}_2  \big]^{\prime} + \big[  \mathscr{R}_3 \mathscr{T}_3 \big]^{\prime}  +  \big[  \mathscr{R}_3 \mathscr{T}_4 \big]^{\prime}  +  \big[ \mathscr{R}_3 \mathscr{T}_5  \big]^{\prime} +  \big[ \mathscr{R}_3 \mathscr{T}_6  \big]^{\prime} \bigg]  \\    + \bigg[ \big[ \mathscr{R}_4 \mathscr{T}_1  \big]^{\prime} +   \big[  \mathscr{R}_4 \mathscr{T}_2   \big]^{\prime} +  \big[ \mathscr{R}_4 \mathscr{T}_3  \big]^{\prime} +  \big[  \mathscr{R}_4 \mathscr{T}_4   \big]^{\prime} + \big[ \mathscr{R}_4 \mathscr{T}_5  \big]^{\prime} +  \big[\mathscr{R}_4 \mathscr{T}_6  \big]^{\prime} \bigg]    + \bigg[ \big[ \mathscr{R}_5 \mathscr{T}_1  \big]^{\prime} +  \big[   \mathscr{R}_5 \mathscr{T}_2  \big]^{\prime} \\ +  \big[ \mathscr{R}_5 \mathscr{T}_3  \big]^{\prime} +   \big[ \mathscr{R}_5 \mathscr{T}_4   \big]^{\prime} +  \big[ \mathscr{R}_5 \mathscr{T}_5  \big]^{\prime} \bigg] \bigg]  \mathrm{Tr}_0 \bigg[  \underset{1 \leq j^{\prime} \leq 5}{\sum}  \mathscr{R} \mathscr{T}^{\prime}_j \bigg]      \end{align*}

  \begin{align*}     + \bigg[ \mathscr{R}_3 \mathscr{T}_1  +    \mathscr{R}_3 \mathscr{T}_2  +  \mathscr{R}_3 \mathscr{T}_3  +   \mathscr{R}_3 \mathscr{T}_4   +  \mathscr{R}_3 \mathscr{T}_5  +  \mathscr{R}_3 \mathscr{T}_6  \bigg]     + \bigg[ \mathscr{R}_4 \mathscr{T}_1  +    \mathscr{R}_4 \mathscr{T}_2   +  \mathscr{R}_4 \mathscr{T}_3  +   \mathscr{R}_4 \mathscr{T}_4   \\ +  \mathscr{R}_4 \mathscr{T}_5  +  \mathscr{R}_4 \mathscr{T}_6  \bigg]    + \bigg[ \mathscr{R}_5 \mathscr{T}_1  +    \mathscr{R}_5 \mathscr{T}_2  +  \mathscr{R}_5 \mathscr{T}_3  +   \mathscr{R}_5 \mathscr{T}_4   +  \mathscr{R}_5 \mathscr{T}_5  \bigg]  \bigg] \\ \times \bigg\{ \mathrm{Tr}_0 \bigg[  \underset{1 \leq j^{\prime} \leq 5}{\sum}  \mathscr{R} \mathscr{T}^{\prime}_j \bigg]  \bigg\}^{\prime}    . 
\end{align*}

\noindent \textit{Proof of Claim 9}. By direct computation, observe,

\begin{align*}
  \big[ \textbf{T}^{\mathrm{Open}} \big( u \big)  \big]^{\prime} \approx \bigg\{  \mathscr{R}_1  \mathscr{T} \mathrm{Tr}_0 \bigg[ \underset{1 \leq j^{\prime} \leq 5}{\sum}  \mathscr{R} \mathscr{T}^{\prime}_j  \bigg]  +   \mathscr{R}_2  \mathscr{T} \mathrm{Tr}_0 \bigg[  \underset{1 \leq j^{\prime} \leq 5}{\sum}  \mathscr{R} \mathscr{T}^{\prime}_j \bigg] +  \mathscr{R}_3  \mathscr{T} \mathrm{Tr}_0 \bigg[  \underset{1 \leq j^{\prime} \leq 5}{\sum}  \mathscr{R} \mathscr{T}^{\prime}_j \bigg]  \\ +    \mathscr{R}_4  \mathscr{T}  \mathrm{Tr}_0 \bigg[ \underset{1 \leq j^{\prime} \leq 5}{\sum}  \mathscr{R} \mathscr{T}^{\prime}_j \bigg]   +   \mathscr{R}_5 \mathscr{T} \mathrm{Tr}_0 \bigg[  \underset{1 \leq j^{\prime} \leq 5}{\sum}  \mathscr{R} \mathscr{T}^{\prime}_j \bigg]  \bigg\}^{\prime}  \\ 
    = \bigg\{  \bigg[ \mathscr{R}_1 \mathscr{T}_1  +    \mathscr{R}_1 \mathscr{T}_2  +  \mathscr{R}_1 \mathscr{T}_3  +   \mathscr{R}_1 \mathscr{T}_4   +  \mathscr{R}_1 \mathscr{T}_5  \bigg]  \mathrm{Tr}_0 \bigg[  \underset{1 \leq j^{\prime} \leq 5}{\sum}  \mathscr{R} \mathscr{T}^{\prime}_j \bigg]   \\ + \bigg[ \mathscr{R}_2 \mathscr{T}_1  +    \mathscr{R}_2 \mathscr{T}_2  +  \mathscr{R}_2 \mathscr{T}_3  +   \mathscr{R}_2 \mathscr{T}_4   +  \mathscr{R}_2 \mathscr{T}_5  \bigg]  \mathrm{Tr}_0 \bigg[  \underset{1 \leq j^{\prime} \leq 5}{\sum}  \mathscr{R} \mathscr{T}^{\prime}_j \bigg]   \\ + \bigg[ \mathscr{R}_3 \mathscr{T}_1  +    \mathscr{R}_3 \mathscr{T}_2  +  \mathscr{R}_3 \mathscr{T}_3  +   \mathscr{R}_3 \mathscr{T}_4   +  \mathscr{R}_3 \mathscr{T}_5  +  \mathscr{R}_3 \mathscr{T}_6  \bigg]  \mathrm{Tr}_0 \bigg[  \underset{1 \leq j^{\prime} \leq 5}{\sum}  \mathscr{R} \mathscr{T}^{\prime}_j \bigg]   \\ + \bigg[ \mathscr{R}_4 \mathscr{T}_1  +    \mathscr{R}_4 \mathscr{T}_2  +  \mathscr{R}_4 \mathscr{T}_3  +   \mathscr{R}_4 \mathscr{T}_4   +  \mathscr{R}_4 \mathscr{T}_5  +  \mathscr{R}_4 \mathscr{T}_6  \bigg]  \mathrm{Tr}_0 \bigg[  \underset{1 \leq j^{\prime} \leq 5}{\sum}  \mathscr{R} \mathscr{T}^{\prime}_j \bigg]   \\ + \bigg[ \mathscr{R}_5 \mathscr{T}_1  +    \mathscr{R}_5 \mathscr{T}_2  +  \mathscr{R}_5 \mathscr{T}_3  +   \mathscr{R}_5 \mathscr{T}_4   +  \mathscr{R}_5 \mathscr{T}_5  \bigg]  \mathrm{Tr}_0 \bigg[  \underset{1 \leq j^{\prime} \leq 5}{\sum}  \mathscr{R} \mathscr{T}^{\prime}_j \bigg]  \bigg\}^{\prime} \end{align*}

  \begin{align*}  = \bigg\{ \bigg[ \bigg[ \mathscr{R}_1 \mathscr{T}_1  +    \mathscr{R}_1 \mathscr{T}_2  +  \mathscr{R}_1 \mathscr{T}_3  +   \mathscr{R}_1 \mathscr{T}_4   +  \mathscr{R}_1 \mathscr{T}_5  \bigg]      + \bigg[ \mathscr{R}_2 \mathscr{T}_1  +    \mathscr{R}_2 \mathscr{T}_2  +  \mathscr{R}_2 \mathscr{T}_3 \\  +   \mathscr{R}_2 \mathscr{T}_4   +  \mathscr{R}_2 \mathscr{T}_5  \bigg]    \\ + \bigg[ \mathscr{R}_3 \mathscr{T}_1  +    \mathscr{R}_3 \mathscr{T}_2  +  \mathscr{R}_3 \mathscr{T}_3  +   \mathscr{R}_3 \mathscr{T}_4   +  \mathscr{R}_3 \mathscr{T}_5  +  \mathscr{R}_3 \mathscr{T}_6  \bigg]     + \bigg[ \mathscr{R}_4 \mathscr{T}_1  +    \mathscr{R}_4 \mathscr{T}_2 \\  +  \mathscr{R}_4 \mathscr{T}_3  +   \mathscr{R}_4 \mathscr{T}_4   +  \mathscr{R}_4 \mathscr{T}_5  +  \mathscr{R}_4 \mathscr{T}_6  \bigg]  \\  + \bigg[ \mathscr{R}_5 \mathscr{T}_1  +    \mathscr{R}_5 \mathscr{T}_2  +  \mathscr{R}_5 \mathscr{T}_3  +   \mathscr{R}_5 \mathscr{T}_4   +  \mathscr{R}_5 \mathscr{T}_5  \bigg]  \bigg] \mathrm{Tr}_0 \bigg[  \underset{1 \leq j^{\prime} \leq 5}{\sum}  \mathscr{R} \mathscr{T}^{\prime}_j \bigg] \bigg\}^{\prime} \\  \\   = \bigg\{ \bigg[ \bigg[ \mathscr{R}_1 \mathscr{T}_1  +    \mathscr{R}_1 \mathscr{T}_2  +  \mathscr{R}_1 \mathscr{T}_3  +   \mathscr{R}_1 \mathscr{T}_4   +  \mathscr{R}_1 \mathscr{T}_5  \bigg]      + \bigg[ \mathscr{R}_2 \mathscr{T}_1  +    \mathscr{R}_2 \mathscr{T}_2  +  \mathscr{R}_2 \mathscr{T}_3 \\  +   \mathscr{R}_2 \mathscr{T}_4   +  \mathscr{R}_2 \mathscr{T}_5  \bigg]    \\ + \bigg[ \mathscr{R}_3 \mathscr{T}_1  +    \mathscr{R}_3 \mathscr{T}_2  +  \mathscr{R}_3 \mathscr{T}_3  +   \mathscr{R}_3 \mathscr{T}_4   +  \mathscr{R}_3 \mathscr{T}_5  +  \mathscr{R}_3 \mathscr{T}_6  \bigg]     + \bigg[ \mathscr{R}_4 \mathscr{T}_1  +    \mathscr{R}_4 \mathscr{T}_2 \\  +  \mathscr{R}_4 \mathscr{T}_3  +   \mathscr{R}_4 \mathscr{T}_4   +  \mathscr{R}_4 \mathscr{T}_5  +  \mathscr{R}_4 \mathscr{T}_6  \bigg]   \\ + \bigg[ \mathscr{R}_5 \mathscr{T}_1  +    \mathscr{R}_5 \mathscr{T}_2  +  \mathscr{R}_5 \mathscr{T}_3  +   \mathscr{R}_5 \mathscr{T}_4   +  \mathscr{R}_5 \mathscr{T}_5  \bigg]  \bigg]  \bigg\}^{\prime} \\ \times \mathrm{Tr}_0 \bigg[  \underset{1 \leq j^{\prime} \leq 5}{\sum}  \mathscr{R} \mathscr{T}^{\prime}_j \bigg] \\ + \bigg[ \bigg[ \mathscr{R}_1 \mathscr{T}_1  +    \mathscr{R}_1 \mathscr{T}_2  +  \mathscr{R}_1 \mathscr{T}_3  +   \mathscr{R}_1 \mathscr{T}_4   +  \mathscr{R}_1 \mathscr{T}_5  \bigg]      + \bigg[ \mathscr{R}_2 \mathscr{T}_1  +    \mathscr{R}_2 \mathscr{T}_2  +  \mathscr{R}_2 \mathscr{T}_3 \\  +   \mathscr{R}_2 \mathscr{T}_4   +  \mathscr{R}_2 \mathscr{T}_5  \bigg]   \\ + \bigg[ \mathscr{R}_3 \mathscr{T}_1  +    \mathscr{R}_3 \mathscr{T}_2  +  \mathscr{R}_3 \mathscr{T}_3  +   \mathscr{R}_3 \mathscr{T}_4   +  \mathscr{R}_3 \mathscr{T}_5  +  \mathscr{R}_3 \mathscr{T}_6  \bigg]     + \bigg[ \mathscr{R}_4 \mathscr{T}_1  +    \mathscr{R}_4 \mathscr{T}_2 \\  +  \mathscr{R}_4 \mathscr{T}_3  +   \mathscr{R}_4 \mathscr{T}_4   +  \mathscr{R}_4 \mathscr{T}_5  +  \mathscr{R}_4 \mathscr{T}_6  \bigg]   \\ + \bigg[ \mathscr{R}_5 \mathscr{T}_1  +    \mathscr{R}_5 \mathscr{T}_2  +  \mathscr{R}_5 \mathscr{T}_3  +   \mathscr{R}_5 \mathscr{T}_4   +  \mathscr{R}_5 \mathscr{T}_5  \bigg]  \bigg] \\ \times \bigg\{ \mathrm{Tr}_0 \bigg[  \underset{1 \leq j^{\prime} \leq 5}{\sum}  \mathscr{R} \mathscr{T}^{\prime}_j \bigg]  \bigg\}^{\prime} \\  \end{align*}

    \begin{align*}  =   \bigg\{ \bigg[ \bigg[ \mathscr{R}_1 \mathscr{T}_1  +    \mathscr{R}_1 \mathscr{T}_2  +  \mathscr{R}_1 \mathscr{T}_3  +   \mathscr{R}_1 \mathscr{T}_4   +  \mathscr{R}_1 \mathscr{T}_5  \bigg]      + \bigg[ \mathscr{R}_2 \mathscr{T}_1  +    \mathscr{R}_2 \mathscr{T}_2  +  \mathscr{R}_2 \mathscr{T}_3 \\  +   \mathscr{R}_2 \mathscr{T}_4   +  \mathscr{R}_2 \mathscr{T}_5  \bigg]    \\ + \bigg[ \mathscr{R}_3 \mathscr{T}_1  +    \mathscr{R}_3 \mathscr{T}_2  +  \mathscr{R}_3 \mathscr{T}_3  +   \mathscr{R}_3 \mathscr{T}_4   +  \mathscr{R}_3 \mathscr{T}_5  +  \mathscr{R}_3 \mathscr{T}_6  \bigg]     + \bigg[ \mathscr{R}_4 \mathscr{T}_1  +    \mathscr{R}_4 \mathscr{T}_2  \\  +  \mathscr{R}_4 \mathscr{T}_3  +   \mathscr{R}_4 \mathscr{T}_4   +  \mathscr{R}_4 \mathscr{T}_5  +  \mathscr{R}_4 \mathscr{T}_6  \bigg]   \\ + \bigg[ \mathscr{R}_5 \mathscr{T}_1  +    \mathscr{R}_5 \mathscr{T}_2  +  \mathscr{R}_5 \mathscr{T}_3  +   \mathscr{R}_5 \mathscr{T}_4   +  \mathscr{R}_5 \mathscr{T}_5  \bigg]  \bigg] \mathrm{Tr}_0 \bigg[  \underset{1 \leq j^{\prime} \leq 5}{\sum}  \mathscr{R} \mathscr{T}^{\prime}_j \bigg] \bigg\}^{\prime} \\ \\    = \bigg[ \bigg[ \mathscr{R}_1 \mathscr{T}_1  +    \mathscr{R}_1 \mathscr{T}_2  +  \mathscr{R}_1 \mathscr{T}_3  +   \mathscr{R}_1 \mathscr{T}_4   +  \mathscr{R}_1 \mathscr{T}_5  \bigg]^{\prime}      + \bigg[ \mathscr{R}_2 \mathscr{T}_1  +    \mathscr{R}_2 \mathscr{T}_2  +  \mathscr{R}_2 \mathscr{T}_3 \\  +   \mathscr{R}_2 \mathscr{T}_4   +  \mathscr{R}_2 \mathscr{T}_5  \bigg]^{\prime}    \\ + \bigg[ \mathscr{R}_3 \mathscr{T}_1  +    \mathscr{R}_3 \mathscr{T}_2  +  \mathscr{R}_3 \mathscr{T}_3  +   \mathscr{R}_3 \mathscr{T}_4   +  \mathscr{R}_3 \mathscr{T}_5  +  \mathscr{R}_3 \mathscr{T}_6  \bigg]^{\prime}     + \bigg[ \mathscr{R}_4 \mathscr{T}_1  +    \mathscr{R}_4 \mathscr{T}_2 \\  +  \mathscr{R}_4 \mathscr{T}_3  +   \mathscr{R}_4 \mathscr{T}_4   +  \mathscr{R}_4 \mathscr{T}_5  +  \mathscr{R}_4 \mathscr{T}_6  \bigg]^{\prime}   \\ + \bigg[ \mathscr{R}_5 \mathscr{T}_1  +    \mathscr{R}_5 \mathscr{T}_2  +  \mathscr{R}_5 \mathscr{T}_3  +   \mathscr{R}_5 \mathscr{T}_4   +  \mathscr{R}_5 \mathscr{T}_5  \bigg]^{\prime}  \bigg]  \\ \times \mathrm{Tr}_0 \bigg[  \underset{1 \leq j^{\prime} \leq 5}{\sum}  \mathscr{R} \mathscr{T}^{\prime}_j \bigg]       \\ + \bigg[ \mathscr{R}_3 \mathscr{T}_1  +    \mathscr{R}_3 \mathscr{T}_2  +  \mathscr{R}_3 \mathscr{T}_3  +   \mathscr{R}_3 \mathscr{T}_4   +  \mathscr{R}_3 \mathscr{T}_5  +  \mathscr{R}_3 \mathscr{T}_6  \bigg]     + \bigg[ \mathscr{R}_4 \mathscr{T}_1  +    \mathscr{R}_4 \mathscr{T}_2 \\  +  \mathscr{R}_4 \mathscr{T}_3  +   \mathscr{R}_4 \mathscr{T}_4   +  \mathscr{R}_4 \mathscr{T}_5  +  \mathscr{R}_4 \mathscr{T}_6  \bigg]   \\ + \bigg[ \mathscr{R}_5 \mathscr{T}_1  +    \mathscr{R}_5 \mathscr{T}_2  +  \mathscr{R}_5 \mathscr{T}_3  +   \mathscr{R}_5 \mathscr{T}_4   +  \mathscr{R}_5 \mathscr{T}_5  \bigg]  \bigg] \\ \times \bigg\{ \mathrm{Tr}_0 \bigg[  \underset{1 \leq j^{\prime} \leq 5}{\sum}  \mathscr{R} \mathscr{T}^{\prime}_j \bigg]  \bigg\}^{\prime}  
\end{align*}

\noindent from which we conclude the argument. \boxed{}

\subsubsection{Logarithmic derivative of the higher-rank transfer matrix}

\noindent \textbf{Claim 10} (\textit{the logarithmic derivative of the the transfer matrix under open boundaries}). One has that,

\begin{align*}
   \frac{\mathrm{d}}{\mathrm{d} u} \bigg\{ \mathrm{log} \big( \textbf{T}^{\mathrm{Open}} \big( u \big) \big) \bigg\}  \bigg|_{u \equiv 0} \approx       \frac{\big[ \textbf{T}^{\mathrm{Open}} \big( u \big)\big]^{\prime}}{\textbf{T}^{\mathrm{Open}} \big( u \big) }  \bigg|_{u \equiv 0} \approx   \frac{\big[ \textbf{T}^{\mathrm{Open}} \big( u \big)\big]^{\prime}}{\textbf{T}^{\mathrm{Open}} \big( 0 \big) }  \bigg|_{u \equiv 0} .
\end{align*}

\noindent \textit{Proof of Claim 10}. By direct computation, observe,

\begin{align*}
  \frac{\mathrm{d}}{\mathrm{d} u} \bigg\{ \mathrm{log} \big( \textbf{T}^{\mathrm{Open}} \big( u \big) \big) \bigg\}  \bigg|_{u \equiv 0} \approx     \frac{\mathrm{d}}{\mathrm{d} u} \bigg\{ \mathrm{log} \bigg[ \mathscr{R}_1  \mathscr{T} \mathrm{Tr}_0 \bigg[ \underset{1 \leq j^{\prime} \leq 5}{\sum}  \mathscr{R} \mathscr{T}^{\prime}_j  \bigg]  +   \mathscr{R}_2  \mathscr{T} \mathrm{Tr}_0 \bigg[  \underset{1 \leq j^{\prime} \leq 5}{\sum}  \mathscr{R} \mathscr{T}^{\prime}_j \bigg]  +  \mathscr{R}_3  \mathscr{T} \\ \times \mathrm{Tr}_0 \bigg[  \underset{1 \leq j^{\prime} \leq 5}{\sum}  \mathscr{R} \mathscr{T}^{\prime}_j \bigg]   +    \mathscr{R}_4  \mathscr{T}  \mathrm{Tr}_0 \bigg[ \underset{1 \leq j^{\prime} \leq 5}{\sum}  \mathscr{R} \mathscr{T}^{\prime}_j \bigg]   +   \mathscr{R}_5 \mathscr{T} \mathrm{Tr}_0 \bigg[  \underset{1 \leq j^{\prime} \leq 5}{\sum}  \mathscr{R} \mathscr{T}^{\prime}_j \bigg]    \bigg]  \bigg\}  \bigg|_{u \equiv 0}         \\ \\  =         \bigg[ \mathscr{R}_1  \mathscr{T} \mathrm{Tr}_0 \bigg[ \underset{1 \leq j^{\prime} \leq 5}{\sum}  \mathscr{R} \mathscr{T}^{\prime}_j  \bigg]  +   \mathscr{R}_2  \mathscr{T} \mathrm{Tr}_0 \bigg[  \underset{1 \leq j^{\prime} \leq 5}{\sum}  \mathscr{R} \mathscr{T}^{\prime}_j \bigg]  +  \mathscr{R}_3  \mathscr{T} \mathrm{Tr}_0 \bigg[  \underset{1 \leq j^{\prime} \leq 5}{\sum}  \mathscr{R} \mathscr{T}^{\prime}_j \bigg]    +    \mathscr{R}_4  \mathscr{T}  \mathrm{Tr}_0 \bigg[ \underset{1 \leq j^{\prime} \leq 5}{\sum}  \mathscr{R} \mathscr{T}^{\prime}_j \bigg]  \\  +   \mathscr{R}_5 \mathscr{T} \mathrm{Tr}_0 \bigg[  \underset{1 \leq j^{\prime} \leq 5}{\sum}  \mathscr{R} \mathscr{T}^{\prime}_j \bigg]    \bigg]^{-1} \bigg|_{u \equiv 0} \\ \times \bigg[       \mathscr{R}_1  \mathscr{T} \mathrm{Tr}_0 \bigg[ \underset{1 \leq j^{\prime} \leq 5}{\sum}  \mathscr{R} \mathscr{T}^{\prime}_j  \bigg]  +   \mathscr{R}_2  \mathscr{T} \mathrm{Tr}_0 \bigg[  \underset{1 \leq j^{\prime} \leq 5}{\sum}  \mathscr{R} \mathscr{T}^{\prime}_j \bigg]  +  \mathscr{R}_3  \mathscr{T} \mathrm{Tr}_0 \bigg[  \underset{1 \leq j^{\prime} \leq 5}{\sum}  \mathscr{R} \mathscr{T}^{\prime}_j \bigg]  \\  +    \mathscr{R}_4  \mathscr{T}  \mathrm{Tr}_0 \bigg[ \underset{1 \leq j^{\prime} \leq 5}{\sum}  \mathscr{R} \mathscr{T}^{\prime}_j \bigg]   +   \mathscr{R}_5 \mathscr{T} \mathrm{Tr}_0 \bigg[  \underset{1 \leq j^{\prime} \leq 5}{\sum}  \mathscr{R} \mathscr{T}^{\prime}_j \bigg]      \bigg]^{\prime} \bigg|_{u \equiv 0} \\ \\       \equiv    \bigg[ \mathscr{R}_1  \mathscr{T} \big( 0 \big) \mathrm{Tr}_0 \bigg[ \underset{1 \leq j^{\prime} \leq 5}{\sum}  \mathscr{R} \mathscr{T}^{\prime}_j \big( 0 \big)  \bigg]  +   \mathscr{R}_2  \mathscr{T} \big( 0 \big) \mathrm{Tr}_0 \bigg[  \underset{1 \leq j^{\prime} \leq 5}{\sum}  \mathscr{R} \mathscr{T}^{\prime}_j\big( 0 \big)  \bigg]  +  \mathscr{R}_3  \mathscr{T} \big( 0 \big) \mathrm{Tr}_0 \bigg[  \underset{1 \leq j^{\prime} \leq 5}{\sum}  \mathscr{R} \mathscr{T}^{\prime}_j \big( 0 \big)  \bigg]  \\  +    \mathscr{R}_4  \mathscr{T} \big( 0 \big)  \mathrm{Tr}_0 \bigg[ \underset{1 \leq j^{\prime} \leq 5}{\sum}  \mathscr{R} \mathscr{T}^{\prime}_j \big( 0 \big)  \bigg]   +   \mathscr{R}_5 \mathscr{T}  \big( 0 \big) \mathrm{Tr}_0 \bigg[  \underset{1 \leq j^{\prime} \leq 5}{\sum}  \mathscr{R} \mathscr{T}^{\prime}_j \big( 0 \big) \bigg]    \bigg]^{-1}  \\ \times \bigg[       \mathscr{R}_1  \mathscr{T} \mathrm{Tr}_0 \bigg[ \underset{1 \leq j^{\prime} \leq 5}{\sum}  \mathscr{R} \mathscr{T}^{\prime}_j  \bigg]  +   \mathscr{R}_2  \mathscr{T} \mathrm{Tr}_0 \bigg[  \underset{1 \leq j^{\prime} \leq 5}{\sum}  \mathscr{R} \mathscr{T}^{\prime}_j \bigg]  +  \mathscr{R}_3  \mathscr{T} \mathrm{Tr}_0 \bigg[  \underset{1 \leq j^{\prime} \leq 5}{\sum}  \mathscr{R} \mathscr{T}^{\prime}_j \bigg]  \\  +    \mathscr{R}_4  \mathscr{T}  \mathrm{Tr}_0 \bigg[ \underset{1 \leq j^{\prime} \leq 5}{\sum}  \mathscr{R} \mathscr{T}^{\prime}_j \bigg]   +   \mathscr{R}_5 \mathscr{T} \mathrm{Tr}_0 \bigg[  \underset{1 \leq j^{\prime} \leq 5}{\sum}  \mathscr{R} \mathscr{T}^{\prime}_j \bigg]      \bigg]^{\prime} \bigg|_{u \equiv 0}          \\ \\ =  \frac{\big[ \textbf{T}^{\mathrm{Open}} \big( u \big)\big]^{\prime}}{\textbf{T}^{\mathrm{Open}} \big( 0 \big) }  \bigg|_{u \equiv 0}        , 
\end{align*}

\noindent from which we conclude the argument. \boxed{}

\subsubsection{Classification of sectors of conformal theories from the open boundary root density approach and linearization of the Bethe equations}

\noindent Altogether, the density approximation of the roots to the Bethe equation for anisotropy parameters which almost vanishes falls into the following characterization:

\begin{itemize}
    \item[$\bullet$] 
\underline{Ground state}: $h_1 \equiv h_2 \equiv 0$  \text{ , } 
     \item[$\bullet$] 
\underline{Type I excitation to the ground state}: $h_1 > 0 $, $h_2 \equiv 0$ \text{ , } 
      \item[$\bullet$] \underline{Type II excitation to the ground state}: $h_1 \equiv 0$, $h_2 > 0$ \text{ , } 
      \item[$\bullet$] \underline{Type III excitation to the ground state}: $h_1 , h_2 > 0$ \text{ . } 
\end{itemize}

\noindent Under each set of possible choices for $h_1$ and $h_2$ provided above, one can characterize solutions to the Bethe equations from the density provided earlier with , similar to the arrangement of roots provided in \textit{Figure 1}, \textit{Figure 2}, \textit{Figure 3}, \textit{Figure 4}, and \textit{Figure 5} of (\citet{frahm2023}). Given the fact that a rigorous identification of the sectors of conformal field theories is out of the scope of current methods introduced in this work, we leave this question open to either future work of the author, or of that of the broader scientific community.

\section{Conclusion}

\noindent We constructed the transfer matrix, and several related objects,. for the $D^{(2)}_3$ spin-chain under open boundary conditions. As suggested in (\citet{frahm2023}) and several related works discussed in this work following the introduction, we discussed how boundary conditions are encoded through higher-rank K-matrices, particularly from the fact that such matrices are $36$ dimensional. By leveraging suitable constructions of such higher-rank K-matrices, we formulated equivalent expressions for the transfer matrices through a decomposition of the Jimbo R-matrix into $\mathscr{R}_1, \mathscr{R}_2, \mathscr{R}_3, \mathscr{R}_4$ and $\mathscr{R}_5$. Besides obtaining approximate representations for the transfer matrix, one can apply such computations, along with the previously aforementioned components $\mathscr{R}$ of the Jimbo R-matrix, to obtain approximations for the quantum monodromy matrix. Such objects are central to seminal work on the quantum inverse scattering method, which establishes connections between solutions to the nonlinear Schrodinger equation, in the defocusing and focusing cases, alike, and exact solvability (\citet{fadeev}).

Equipped with such objects, one can determine the roots of the Bethe equations. The concentration, or density, of such roots is subject to future investigation. While the arrangement of roots to the two-level nested Bethe ansatz implies that a root system of the form mentioned in \textit{1.7} should exist, rigorously identifying the expected sectors of conformal field theories still appears to be out of reach. The spacing of roots in the higher rank spin-chain under open boundary conditions, as is the same higher rank model under quasi periodic boundary conditions, raises implications for sectors of conformal field theories. Such field theories can be related to quasi-momentum operators. While it is not possible to define the quasi-momentum operator in the presence of inhomogeneities, the fact that quasi-periodic boundary conditions are not considered in this work implies that one could hope to possibly construct such an operator in the presence of open boundary conditions.




\newpage 

\section{Appendix}

\noindent We provide the closed form representations for each entry of the representation for the transfer matrix of the 6-vertex model below.

\subsection{$A_3 \big( \lambda_{\alpha} \big) $}

\begin{align*}
    \prod_{0 \leq i \leq 3} \mathrm{sin} \big( \lambda_{\alpha} - v_{N-i} + \eta \sigma^z_{N-j} \big)   +  \big( \mathrm{sin} \big( 2 \eta \big) \big)^2 \bigg[    \text{ }         \bigg[ \text{ }  \underset{i \equiv 1, +}{\underset{i \equiv 0 , -}{\prod_{0 \leq i \leq 1}}}      \sigma^{-,+}_{N-i} \bigg] \\ \times   \text{ }  \bigg[  \text{ } \underset{2\leq i \leq 3}{\prod}   \mathrm{sin}  \big( \lambda_{\alpha} - v_{N-i} + \eta \sigma^z_{N-j} \big) \bigg] \\ 
      +     \bigg[ \text{ }  \underset{i \equiv 3 , +}{\underset{i \equiv 2 , -}{\underset{2 \leq i \leq 3}{\prod} }}    \sigma^{-,+}_{N-i}  \bigg]   \bigg[ \text{ }   \underset{{0 \leq i \leq 1}}{\prod}  \mathrm{sin} \big( \lambda_{\alpha} - v_{N-i} + \eta \sigma^z_{N-j} \big)  \bigg] \\ +    \big( \mathrm{sin} \big( 2 \eta \big) \big)^{-1}      \bigg[  \text{ } \underset{i \equiv 3, +}{\underset{i \equiv 1, -}{\underset{i \mathrm{\text{ } odd \text{ }}: \text{ } 1 \leq i \leq 3}{\prod}} }\mathrm{sin} \big( 2 \eta \big) \sigma^{-}_{N-i}   \bigg] \bigg[ \text{ } 
  \underset{i \equiv 2, - \eta}{\underset{i \equiv 0 , + \eta}{\underset{i \text{ } \mathrm{even \text{ }} 
 :\text{ }  0 \leq i \leq 2}{\prod}}} \mathrm{sin} \big( \lambda_{\alpha} - v_{N-i} \pm  \eta \sigma^z_{N-j} \big)  \bigg]  \\ +   \bigg[ \text{ } \underset{i \equiv 2, +}{\underset{i \equiv 1, -}{\prod_{1 \leq i \leq 2}}}   \sigma^{-,+}_{N-i}         \bigg]              \bigg[ \text{ } \underset{i \mathrm{ \text{ } odd}: \text{ } 1 \leq i \leq 3}{\prod}     \mathrm{sin} \big( \lambda_{\alpha} - v_{N-i} + \eta \sigma^z_{N-j} \big)  \bigg] \end{align*}

 \begin{align*}
 + \bigg[ \text{ } \underset{ i \equiv 2 , +}{\underset{i \equiv 0 ,-}{\underset{i \text{ } \mathrm{even}\text{ } : \text{ } 0 \leq i \leq 2}{\prod}  }  }           \sigma^{-,+}_{N-i} \bigg] \text{ } \bigg[ \text{ }  \underset{i\text{ }  \mathrm{odd}\text{ } : \text{ } 1 \leq i \leq 3}{\prod}   \mathrm{sin} \big( \lambda_{\alpha} - v_{N-i} - \eta \sigma^z_{N-j} \big)   \bigg] \\ + 
 \mathrm{sin} \big( 2 \eta \big)  \text{ }  \bigg[ \text{ }  \underset{i \equiv 3 , +}{\underset{i \equiv 2 , -}{\underset{i \equiv 1 , +}{\underset{i \equiv 0 , -}{\underset{0 \leq i \leq 3}{\prod}}  }} }\sigma^{-,+}_{N-i}   \bigg] +   \bigg[ \text{ }    \underset{i \equiv 3 , +}{\underset{i \equiv 1 , -}{\underset{i \text{ } \mathrm{odd} \text{ } :\text{ } 1 \leq i \leq 3}{\prod}  }}    \sigma^{-,+}_{N-i}      \bigg] \text{ } \bigg[ \text{ }         \underset{1 \leq i \leq 2}{\prod}        \mathrm{sin} \big( \lambda_{\alpha} - v_{N-i} - \eta \sigma^z_{N-j} \big)  \bigg]    \text{ }        
 \bigg]   
\end{align*}

\subsection{$B_3 \big( \lambda_{\alpha} \big) $}

  \begin{align*} \big( \mathrm{sin} \big( 2 \eta \big) \big) \text{ } \sigma^{-}_{N-3} \text{ } \bigg[          \text{ }    \underset{0 \leq i \leq 2}{\prod}   \mathrm{sin} \big( \lambda_{\alpha} - v_{N-i} + \eta \sigma^z_{N-j} \big)      \bigg]      +   \big( \mathrm{sin} \big( 2 \eta \big) \big)^3 \bigg[ \text{ }    \big( \mathrm{sin} \big( 2 \eta \big) \big)^{-1} \\  \times     \bigg[ \text{ } \bigg[ \text{ }    \underset{i \equiv 1 , +}{\underset{i \equiv 0 , -}{\underset{0 \leq i \leq 1}{\prod}} }\sigma^{-,+}_{N-j}                       \bigg] \text{ } \bigg[      \text{ }     \underset{i \equiv 3 ,-}{ \underset{i \equiv 2 ,+}{\underset{2 \leq i \leq 3}{\prod } }}  \mathrm{sin} \big( \lambda_{\alpha} - v_{N-i} \pm \eta \sigma^z_{N-j} \big)     \bigg]  \\ 
  + \bigg[ \text{ } \underset{i \equiv 2 , +}{\underset{i \equiv 1 , -}{\underset{1 \leq i \leq 2}{\prod}} }\sigma^{-,+}_{N-i}  \bigg] \text{ } \bigg[ \text{ }          \underset{i \equiv 3 ,+}{ \underset{ i \equiv 1 , -}{\underset{\mathrm{odd}\text{ } i \text{ } : \text{ }  1 \leq i \leq 3}{ \prod}} }           \mathrm{sin} \big( \lambda_{\alpha}- v_{N-i} \pm \eta \sigma^z_{N-j} \big)  \bigg] \\   +   \sigma^{-}_{N-2} \bigg[ \text{ }      \bigg[ \text{ }     \underset{0 \leq i \leq 1}{ \prod}    \mathrm{sin} \big( \lambda_{\alpha} - v_{N-i} + \eta \sigma^z_{N-j}      \big)                 \bigg]  +  \bigg[ \text{ }    \underset{ i \equiv 3 , -}{\underset{i \equiv 1 , +}{\underset{i \equiv 0 , \text{ } \mathrm{odd} \text{ } i \text{ } : \text{ } 1 \leq i \leq 3}{ \prod   }}}  \mathrm{sin} \big( \lambda_{\alpha} - v_{N-i} \pm \eta \sigma^z_{N-j} \big)     \bigg]               \text{ }              \bigg]  \\    
  +  \bigg[ \text{ } \underset{ 1 \leq i \leq 2}{\prod}       \mathrm{sin} \big( \lambda_{\alpha} - v_{N-i} - \eta \sigma^z_{N-j}    \big)   \bigg]   +  \sigma^{-}_N \mathrm{sin} \big( \lambda_{\alpha} - v_{N-2} - \eta \sigma^z_{N-2} \big) \text{ } \bigg[ \text{ }      \underset{1 \leq i \leq 3}{\prod}      \mathrm{sin} \big( \lambda_{\alpha} - v_{N-i} - \eta \sigma^z_{N-j} \big)    \bigg]     \\     +  \big( 1 + \sigma^{-}_{N-1} \big)  \bigg[ \text{  }  \bigg[ \text{ } \underset{i \equiv 2 , -}{\underset{i \equiv  0 , +}{\underset{\mathrm{even}\text{ } i \text{ } : \text{ }0 \leq i \leq 2}{\prod} }  }\mathrm{sin} \big( \lambda_{\alpha} - v_{N-i} \pm \eta \sigma^z_{N-j} \big) \bigg]  \\   +  \bigg[ \text{ } \underset{i \equiv 2 , -}{\underset{i \equiv 0 , +}{{\underset{\mathrm{even}\text{ } i \text{ } : \text{ }0 \leq i \leq 2}{\prod} }}}\mathrm{sin} \big( \lambda_{\alpha} - v_{N-i} \pm \eta \sigma^z_{N-j} \big) \bigg] \text{ } \bigg]   \text{ } \\ +    \mathrm{sin} \big( 2 \eta \big)  \bigg[ \text{ }    \bigg[ \text{ }   \underset{i \equiv 2 , -}{\underset{i \equiv 1 , +}{\underset{i \equiv 0 ,-}{ \underset{0 \leq i \leq 2}{\prod}  }  }  }    \sigma^{-,+}_{N-i}   \bigg] +       \sigma^{-}_{N-2} \bigg[\text{ }  \underset{\mathrm{even} \text{ } i \text{ } : \text{ } 0 \leq i \leq 2}{\prod}  \sigma^{-,+}_{N-i} \bigg] \bigg[ \text{ }  \underset{i \equiv 3 , +}{\underset{i \equiv 1, -}{\underset{\mathrm{odd} \text{ } i \text{ } : 1 \leq i \leq 3}{\prod}} }  \mathrm{sin} \big( \lambda_{\alpha} - v_N \pm \eta \sigma^z_{N-j} \big)  \bigg] \\ + \bigg[ \text{ }             \underset{i \equiv 1 , +}{\underset{i \equiv 0 , 2 , -}{\underset{0 \leq i \leq 2}{\prod}   }}      \sigma^{-,+}_{N-i}                     \bigg] 
 +   \bigg[ \text{ }     \underset{i \equiv 3 , - \eta}{\underset{ i \equiv 0 , 1 , + \eta}{\underset{0 \leq i \leq 1 , i \equiv 3}{\prod} } }            \mathrm{sin} \big( \lambda_{\alpha} - v_{N-i}  \pm \eta \sigma^z_{N-j}  \big)  \bigg] \\  + \bigg[ \text{ } \underset{i \equiv 1 , +}{\underset{i \equiv 0,2,-}{\underset{0 \leq i \leq 2}{\prod} }}   \sigma^{-,+}_{N-i}   \bigg]  +            \bigg[ \text{ }   \underset{i \equiv 3 , - \eta}{\underset{i \equiv 0 , 1 + \eta}{\underset{0 \leq i \leq 1 , i \equiv 3}{\prod}}   } \mathrm{sin} \big(        \lambda_{\alpha} - v_{N-i} \pm \eta \sigma^z_{N-j}              \big)      \bigg]      +\sigma^{-}_{N-2} \bigg[ \text{ }            \underset{\mathrm{even} \text{ } i \text{ } i : \text{ } 0 \leq i \leq 2 }{\prod}  \sigma^{-,+}_{N-i}                   \bigg]  \end{align*}

 \begin{align*}   + 
 \bigg[ \text{ } \underset{i \equiv 2, -}{\underset{i \equiv 1, +}{\underset{i \equiv 0 , -}{ \underset{0 \leq i \leq 2}{\prod}  } }}   \sigma^{-,+}_{N-i}          \bigg] \text{ }  \bigg]  \text{ }   + 
 \bigg[ \text{ } \underset{i \equiv 1 , -}{\underset{i \equiv 0 ,+}{\underset{0 \leq i \leq 1}{\prod} }  }   \sigma^{-,+}_{N-i}    \bigg] \mathrm{sin} \big( \lambda_{\alpha} - v_{N-2} + \eta \sigma^z_{N-2} \big) \\ + \bigg[ \text{ } \underset{i \equiv 2 , +}{\underset{i \equiv 1 ,-}{\underset{1 \leq i \leq 2}{\prod} }  }  \sigma^{-,+}_{N-i} \bigg] \mathrm{sin} \big( \lambda_{\alpha} - v_N + \eta \sigma^z_N \big)   + \bigg[ \text{ }  \underset{i \equiv 2 , +}{\underset{i \equiv 0 ,-}{\underset{\mathrm{even} \text{ } i \text{ } : \text{ } 0 \leq i \leq 2}{ \prod}  }   } \sigma^{-,+}_{N-i} \bigg] \mathrm{sin} \big( \lambda_{\alpha} - v_N + \eta \sigma^z_N \big)    \bigg]  \text{. } 
 \end{align*}

\subsection{$C_3 \big( \lambda_{\alpha} \big) $}

 \begin{align*}
    \big( \mathrm{sin} \big( 2 \eta \big) \big)^3   \mathrm{sin} \big( \lambda_{\alpha} - v_{N-n} - \eta \sigma^z_{N-n} \big) \bigg[ \text{ }          \underset{0 \leq i \leq 3}{\prod}    \mathrm{sin} \big( \lambda_{\alpha} - v_{N-i} - \eta \sigma^z_{N-j}  \big)       \bigg]     +   \big( \mathrm{sin} \big( 2 \eta \big) \big)^2 \\ 
   \times   \bigg[     \sigma^{+}_{N-i}   \bigg[ \mathrm{sin} \big(   \lambda_{\alpha} - v_{N-i} + \eta \sigma^z_{N-i}     \big)                     \bigg[ \text{ }  \underset{i \equiv 2 , -}{\underset{i \equiv 0 , +}{\underset{\mathrm{even \text{ } } i \text{ } : \text{ }  0 \leq i \leq 3}{\prod  }}  }   \sigma^{-,+}_{N-i}   \bigg]  \\ 
   +   \mathrm{sin} \big( \lambda_{\alpha} - v_{N-i} - \eta \sigma^z_{N-i} \big) \bigg[ \text{ }   \underset{i \equiv 3 , +}{\underset{i \equiv 1 , -}{\underset{\mathrm{odd \text{ } } i \text{ } : 1 \leq i \leq 3}{\prod}}}             \sigma^{-,+}_{N-i}        \bigg]       \text{ }          \bigg] \\ 
   +  \mathrm{sin} \big( \lambda_{\alpha} - v_N - \eta \sigma^z_N \big) \text{ } \\ \times  \bigg[ \text{ } \underset{i \equiv 3 , +}{\underset{i \equiv 2 , -}{\underset{i \equiv 1 , +}{\underset{1 \leq i \leq 3}{\prod}  }}  }\sigma^{-,+}_{N-i} \bigg]     +   \sigma^{+}_{N-(n-3)}     \bigg[ \text{ }     \underset{0 \leq i \leq 2}{\prod}    \mathrm{sin} \big( \lambda_{\alpha} - v_{N-i} - \eta \sigma^z_{N-j} \big)               \bigg]   \text{ }     \bigg]  \\ + \mathrm{sin} \big( 2 \eta \big) \bigg[        \sigma^{+}_{N} \bigg[ \text{ }  \underset{1 \leq i \leq 
        3    }{\prod}           \mathrm{sin} \big( \lambda_{\alpha} - v_{N-i} + \eta \sigma^z_{N-j} \big)          \bigg]   +          \mathrm{sin} \big( \lambda_{\alpha} - v_{N-n} + \eta \sigma^z_{N-n} \big)\\   \times  \bigg[     \text{ }   \bigg[ \text{ }    \underset{i \equiv 2 , +}{\underset{i \equiv 0,-}{\underset{\mathrm{even \text{ } } i\text{ }  : \text{ } 0 \leq i \leq 3}{\prod} }  }         \mathrm{sin} \big( \lambda_{\alpha} - v_{N-i} \pm \eta \sigma^z_{N-j} \big)    \bigg]      +       \sigma^{+}_{N-1}  \bigg]  \text{ }        \underset{0 \leq i \leq 3}{\prod} \mathrm{sin} \big( \lambda_{\alpha} - v_{N-i} - \eta \sigma^z_{N-j} \big)      \bigg]          \text{ }               \bigg]                        \text{ }            \bigg]            \text{ }  \\ + 
        \big( \mathrm{sin} \big( 2 \eta \big) \big)^3  \mathrm{sin} \big( \lambda_{\alpha} - v_{N-3} - \eta \sigma^z_{N-3} \big) \bigg[ \text{ }          \underset{0 \leq i \leq 1}{\prod}    \mathrm{sin} \big( \lambda_{\alpha} - v_{N-i} - \eta \sigma^z_{N-j}  \big)       \bigg]    \text{. } 
\end{align*}

\subsection{$D_3 \big( \lambda_{\alpha} \big) $}

\begin{align*}
         \underset{0 \leq i \leq 3}{\prod}    \mathrm{sin} \big( \lambda_{\alpha} - v_{N-i} - \eta \sigma^z_{N-j} \big)  +    \big(       \mathrm{sin} \big( 2 \eta \big)           \big)^2  \bigg[ \text{ }   \bigg[ \text{ } \underset{i \equiv 2 , -}{\underset{i \equiv 0 , +}{\underset{\mathrm{even \text{ } } i: \text{ }  0 \leq i \leq 2}{\prod}  } }    \sigma^{-,+}_{N-i}     \bigg]        \\ \times   \bigg[  \text{ } \underset{i \equiv 3, -}{\underset{i \equiv 1 , +}{\underset{\mathrm{ odd} \text{ } i \text{ } : \text{ } 1 \leq i \leq 3}{\prod}}}\mathrm{sin} \big(   \lambda_{\alpha} - v_{N-i} \pm \eta \sigma^z_{N-j}     \big)        \bigg]  + 
  \bigg[ \text{ } \underset{i \equiv 2 , -}{\underset{i \equiv 1 , +}{\underset{1 \leq i \leq 2}{\prod}}   }    \sigma^{-,+}_{N-i}    \bigg]   \bigg[  \text{ }   \underset{i \equiv 0, i \equiv 3}{\prod}     \mathrm{sin} \big(    \lambda_{\alpha} - v_N - \eta \sigma^z_N  \big)     \bigg]  +  \\     \bigg[ \text{ } \underset{i \equiv 1}{\prod}   \sigma^{-}_{N-i} \bigg]  \text{ } \bigg[  \text{ }   \mathrm{sin} \big( \lambda_{\alpha} - v_{N-2} - \eta \sigma^z_{N-2} \big)     \bigg] \text{ }         \bigg]  
  +           \big( \mathrm{sin} \big( 2 \eta \big) \big)^3   \bigg[ \text{ } \underset{i \equiv 1}{\prod}       \big( \sigma^{-}_{N-i}  \big)^2 \bigg]   \sigma^{+}_{N-i-1} \\ +     \big( \mathrm{sin} \big( 2 \eta \big) \big)^2         \bigg[      \text{ }              \sigma^{+}_N \sigma^{+}_{N-3}    \bigg[ \text{ } \mathrm{sin} \big( \lambda_{\alpha} - v_{N-i} + \eta \sigma^z_{N-j} \big) \bigg]         + 
  \sigma^{-}_{N-3}  \bigg[  \text{ } \underset{i \equiv 2 , + \eta}{\underset{i \equiv 0 , - \eta}{\underset{\mathrm{even \text{ } } i : \text{ } 0 \leq i \leq 2}{\prod} }}           \mathrm{sin} \big( \lambda_{\alpha} - v_{N-i} \pm \eta \sigma^z_{N-j} \big) \bigg] \\  + \sigma^{+}_{N-2} \sigma^{-}_{N-3}  \bigg[ \text{ } \underset{0 \leq i \leq 1}{\prod}  \mathrm{sin} \big( \lambda_{\alpha} - v_{N-i} - \eta \sigma^z_{N-j} \big)   \bigg] \text{ }                                   \bigg]    \text{.
}
\end{align*}

\end{document}